\documentclass[prb,aps,english,twocolumn,notitlepage]{revtex4-2}
\usepackage{epsfig}
\usepackage{amsmath}
\usepackage{mathtools}
\usepackage{xcolor}

\DeclarePairedDelimiterX\braket[2]{\langle}{\rangle}{#1 \delimsize\vert #2}
\usepackage{sidecap}
\usepackage[figuresleft]{rotating}
\usepackage{caption}
\usepackage{float}
\usepackage{appendix}
\usepackage{subfig}
\usepackage{graphicx}
\usepackage{hyperref}
\usepackage{multirow}
\usepackage{multirow}
\usepackage{tikz}
\usetikzlibrary{fit,matrix,calc}
\usepackage{booktabs}
\newcommand{\tikzmark}[1]{\tikz[overlay,remember picture] \node (#1) {};}
\newcommand{\DrawBox}[3][]{%
    \tikz[overlay,remember picture]{
    \draw[black,#1]
      ($(#2)+(-0.25em,2.85ex)$) rectangle
      ($(#3)+(2.1em,-0.8ex)$);}}

\begin{document}
\title{Marginal Influence of Anomalous Josephson Current on Odd-Frequency Spin-Triplet Pairing in Ferromagnetic Josephson Diodes}
\author{Subhajit Pal}
\author{Colin Benjamin} \email{colin.nano@gmail.com}
\affiliation{School of Physical Sciences, National Institute of Science Education \& Research, Jatni-752050, India.}
\affiliation{Homi Bhabha National Institute, Training School Complex, AnushaktiNagar, Mumbai, 400094, India.}

\begin{abstract}
We examine how an anomalous Josephson current influences odd-frequency superconducting correlations in two Josephson junction geometries. The first consists of two ferromagnetic layers between conventional $s$-wave superconductors, with magnetizations along the $x$- and $y$-axes, forming a bilayer junction. The second contains three ferromagnetic layers between two $s$-wave superconductors, with magnetizations along the $x$-, $y$-, and $z$-axes, forming a trilayer junction. Both systems are analyzed in the short and long junction limits. In the bilayer case, where no anomalous Josephson current is present, odd-frequency equal-spin triplet correlations develop pronounced peaks at finite magnetizations in the short junction limit for both tunneling and transparent interfaces. The odd-frequency mixed-spin triplet correlations also exhibit peaks at finite magnetizations for tunneling interfaces, whereas for transparent interfaces they display both peaks and zeros. In the trilayer case, where an anomalous Josephson current exists, similar peaks in both equal- and mixed-spin odd-frequency triplet correlations occur at finite magnetizations for tunneling and transparent interfaces. The spatial profiles of these correlations remain largely unaffected by the anomalous current. The Josephson diode efficiency is finite and reaches its maximum at magnetizations corresponding to the peaks of the anomalous current. Overall, our results show that the anomalous Josephson current has only a marginal influence on odd-frequency spin-triplet pairing. This indicates that the emergence of odd-frequency correlations and the Josephson diode effect are largely independent phenomena, contrary to earlier conjectures. Analysis of the long junction limit leads to the same qualitative conclusions for both configurations.

\end{abstract}
\maketitle
\section{Introduction} Odd-frequency (odd-$\omega$) superconductivity has emerged as a key frontier in condensed matter physics, attracting growing attention in recent years~\cite{linder}. Its defining feature is the sign reversal of the Cooper-pair wave function under the exchange of the electrons\rq{} time coordinates~\cite{linder,tana,jca}, in contrast to conventional even-frequency (even-$\omega$) superconductivity, where electron pairing occurs at equal times~\cite{bar}. Even-$\omega$ pairing is classified into spin-singlet (SS) and spin-triplet (ST) states. Typical examples include $s$- and $d$-wave pairings for even-$\omega$ SS, and $p$-wave pairing for even-$\omega$ ST states~\cite{sigr}. Likewise, odd-$\omega$ pairing may also occur in either SS or ST channels. ST states, whether even- or odd-$\omega$, can be further distinguished as mixed spin-triplet (MST), represented by $\frac{|\uparrow\downarrow\rangle + |\downarrow\uparrow\rangle}{\sqrt{2}}$, or equal spin-triplet (EST), represented by $|\uparrow\uparrow\rangle$ or $|\downarrow\downarrow\rangle$. The SS state, in contrast, is uniquely characterized by $\frac{|\uparrow\downarrow\rangle-|\downarrow\uparrow\rangle}{\sqrt{2}}$. Odd-$\omega$ ST pairing was first identified in superfluid $^{3}He$~\cite{beri} and later realized in disordered superconductors~\cite{trk,bel}. Subsequently, Balatsky and Abrahams proposed odd-$\omega$ SS pairing in systems where time-reversal and parity symmetries are broken~\cite{balat}, and later works showed that odd-$\omega$ MST states can also be induced by magnetic impurities~\cite{kuzz,sis}. Experimental indications of odd-$\omega$ pairing have been reported via phenomena such as the paramagnetic Meissner effect~\cite{adi,ali,yta} and the Kerr effect~\cite{ers,kom}.

Initially, odd-$\omega$ superconductivity was believed to be an intrinsic bulk property~\cite{balat,abr}. However, later studies revealed that it can also be generated at interfaces and surfaces of superconducting junctions~\cite{tanak,vol,yoko,ann,linn,lindd2,pal,voll,berg3,kopu,amb1,cre,burr,lof,vol1,fom,buz,hwa,tamu,berg1,berg2,dutta,pal1,tsi}, as well as in systems driven by time-dependent external fields~\cite{tri1,tri2}. These discoveries have positioned odd-$\omega$ correlations as a promising platform for superconducting spintronics~\cite{LIND}.

Recently, it has been shown that an anomalous Josephson current can arise between even-$\omega$ and odd-$\omega$ superconductors due to the emergence of induced odd-$\omega$ (even-$\omega$) components at their interface~\cite{tak}. This raises an intriguing question: can odd-$\omega$ pairing be induced at the interface of a bulk even-$\omega$ superconductor be influenced by the anomalous Josephson effect or by the Josephson diode effect (JDE)~\cite{blu,mda,jxh,ytan,costa,bpal,btu,jhuc,zhy,rse,sfr,jdp,rseo,zli,pav,jcn,hva,qch,hfl}? In Ref.~\cite{jfl}, the authors speculated that the anomalous Josephson current may affect odd-$\omega$ EST correlations, as they observed an apparent proportionality between the two in a ferromagnetic trilayer junction.
In this {article,} we seek to find whether this speculation is borne out via detailed calculations in a Josephson diode with bulk even-$\omega$ $s$-wave superconductors, wherein both anomalous Josephson current and odd-$\omega$ pairing are induced. {The odd-$\omega$ pairing correlations studied in Ref.~\cite{jfl} used only local coordinates in a long trilayer Josephson junction (JJ). However, in this article, we consider both local as well as non-local coordinates to study odd-$\omega$ pairing correlations  and for both long and short trilayer JJ alongwith comparing the trilayer results with those from a bilayer JJ. Further, the ferromagnetic interfaces in Ref.\cite{jfl} are only in the tunneling regime, we in this article on the other hand deal with both tunneling and transparent ferromagnetic interfaces. Our study therefore is more exhaustive and our conclusions thereby universal.}

In JJs, JDE arises when both time-reversal\cite{ksc} and inversion symmetries are broken\cite{fdo,czc}, resulting in an asymmetry where the absolute value of the maximum Josephson current ($I_c^{+}$) differs from the absolute value of the minimum Josephson current ($I_c^{-}$), i.e., $I_c^{+}\neq I_c^{-}$. The breaking of these symmetries implies that the Josephson current satisfies $I(-\varphi)\neq-I(\varphi)$, leading to an anomalous Josephson current ($I_{an}$) at zero phase difference, $I_{an}=I(\varphi=0)$, where $\varphi$ is the phase difference across the superconductors. Furthermore, the breaking of translational symmetry at the junction interface can induce odd-$\omega$ pairing\cite{tanak,ytk}. To explore the impact of anomalous Josephson current on odd-$\omega$ ST pairing, we consider two setups: (a) a {bilayer} JJ consisting of two ferromagnets with magnetization aligned along the {$x$-, and $y$-axes,} sandwiched between two bulk even-$\omega$ $s$-wave superconductors ({S-$F_x$-$F_y$-S} JJ), and (b) a {trilayer} JJ comprising three ferromagnets with magnetization aligned along $x$-, $y$-, and $z$-axes, embedded between two bulk even-$\omega$ $s$-wave superconductors (S-$F_x$-$F_y$-$F_z$-S JJ). Our findings reveal that, in the {bilayer} setup, wherein anomalous Josephson current and JDE is absent, odd-$\omega$ ST pairing exhibits peaks at finite values of magnetization in both tunneling and transparent regimes. In addition, odd-$\omega$ EST pairing shows zeros at finite magnetization values in the transparent regime. However, for the {trilayer} setup, wherein an anomalous Josephson current emerges, odd-$\omega$ ST pairing displays similar behavior. Further, odd-$\omega$ ST pairing shows the same spatial behavior regardless of the presence or absence of anomalous Josephson current. Across both setups, even-$\omega$ SS pairing exhibits zeros at finite magnetization values. These results underscore the mutual exclusivity of odd-$\omega$ ST pairing and JDE.

{The rest of the paper is organized as follows. In Sec.~II, we introduce the two setups and present the theoretical framework. In Sec.~III, we outline the procedure for calculating the anomalous Josephson current and odd-$\omega$ pairing amplitudes. Sections IV and V are devoted to the discussion of the impact of anomalous Josephson current on odd-$\omega$ ST pairing for tunneling and transparent ferromagnetic interfaces in the short and long junction limits, respectively. In Sec.~VI, we provide a comparative analysis of the results obtained for the two setups and discuss the underlying physical mechanisms governing our findings. Finally, Sec.~VII discusses possible experimental realizations and summarizes the main findings. Additional details, including the explicit wavefunctions for both setups, the Green's function formulation, analytical expressions for the pairing amplitudes, the effect of anomalous Josephson current on the spatial dependence of non-local odd-$\omega$ ST pairing, comparisons of anomalous Josephson current and local/non-local pairing correlations between bilayer and trilayer JJs in both short and long junction limits, are provided in Appendices A-E.}

\section{Model} Our two setups, illustrated in Figs.~1(a) and 1(b), consist of (a) a {bilayer} JJ with two ferromagnets sandwiched between two bulk even-$\omega$ $s$-wave superconductors, and (b) a {trilayer} JJ with three ferromagnets embedded between two bulk even-$\omega$ $s$-wave superconductors. The interfaces between the ferromagnets and the $s$-wave superconductors are represented using $\delta$-like potential barriers with strengths $\mathcal{B}_1$ and $\mathcal{B}_2$. In the {bilayer} setup, the magnetization vectors of {$F_x$, and $F_y$} are oriented {along the $x$-, and $y$-axes, respectively,} with their interface modeled as a $\delta$-like potential barrier of strength $\mathcal{B}$, while in the {trilayer} setup the magnetization vectors of $F_x$, $F_y$, and $F_z$ are aligned along the $x$-, $y$-, and $z$-directions, respectively, with the interfaces between them represented by $\delta$-like potential barriers of strengths $\mathcal{B}_3$ and $\mathcal{B}_4$. In these multiple-barrier systems, the presence of interface barriers leads to repeated scattering of electrons and holes, which introduces disorder effects within the junctions\cite{mel,duh,hek}.
\begin{figure}[ht!]
\centering{\includegraphics[width=0.5\textwidth]{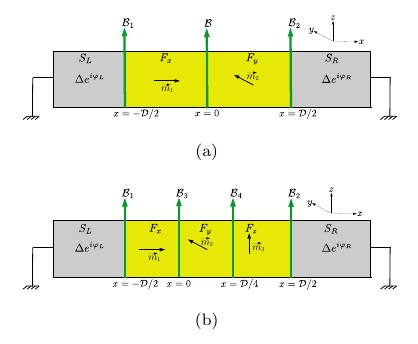}}
\vskip -0.2 in \caption{\small \sl (a) Bilayer JJ with ferromagnet {magnetization vectors oriented along $x$-, and $y$-axes,} embedded between two $s$-wave superconductors, with interfaces at $x=\pm \mathcal{D}/2$, and  at $x=0$ are characterized by $\delta$-like potential barriers with strengths $\mathcal{B}_1$, $\mathcal{B}_2$, and $\mathcal{B}$, respectively, in the short junction limit. (b) Trilayer JJ with ferromagnet magnetization vectors aligned along $x$-, $y$-, and $z$-axes, embedded between two $s$-wave superconductors. Four $\delta$-like potential barriers, with strengths $\mathcal{B}_1$, $\mathcal{B}_2$, $\mathcal{B}_3$ and $\mathcal{B}_4$, are located at the interfaces $x=-\mathcal{D}/2$, $x=\mathcal{D}/2$, $x=0$, and $x=\mathcal{D}/4$, respectively, in the short junction limit.}
\end{figure}

The Bogoliubov-de Gennes (BdG) Hamiltonian for the {bilayer S-$F_x$-$F_y$-S} JJ, as shown in Fig.~1(a), can be written as
follows\cite{enok}:
\begin{equation}
H_{BdG}^{\mbox{{S-$F_x$-$F_y$-S}}}(x)=
\begin{pmatrix}
H_{O}\hat{I} & i\Delta_{s} \hat{\sigma}_{y} \\
-i\Delta_{s}^{*} \hat{\sigma}_{y} & -H_{O}^{*}\hat{I}
\end{pmatrix},
\label{hmff}
\end{equation}
with $H_{O}=-\frac{\hbar^2}{2m^{*}}\frac{\partial^2}{\partial x^2}+\mathcal{B}_1\delta(x+\mathcal{D}/2)+\mathcal{B}_2\delta(x-\mathcal{D}/2)+\mathcal{B}\delta(x)-\vec{m}_{1}.\hat{\sigma}\Theta(x+\mathcal{D}/2)\Theta(-x)-\vec{m}_{2}.\hat{\sigma}\Theta(\mathcal{D}/2-x)\Theta(x)-E_F$. In $H_O$, the first term represents the kinetic energy operator of an electron or hole with effective mass $m^{*}$. The parameters $\mathcal{B}_1$, $\mathcal{B}_2$, and $\mathcal{B}$ denote the strengths of the $\delta$-like potential barriers located at the {S-$F_x$, $F_y$-S, and $F_x$-$F_y$} interfaces, respectively. {The magnetization vectors of the ferromagnets $F_x$, and $F_y$ are expressed as $\vec{m}_{1}.\hat{\sigma}=m_{1}\hat{\sigma}_{x}$, and $\vec{m}_{2}.\hat{\sigma}=m_{2}\hat{\sigma}_{y}$, respectively.} Finally, $E_F$ is the Fermi energy. Here, $\Theta(x)$ represents the Heaviside step function, $\hat{\sigma}$ denotes the Pauli spin matrices, and $\hat{I}$ is the $2\times2$ identity matrix. The superconducting gap $\Delta_s=\Delta[e^{i\varphi_L}\Theta(-x-\mathcal{D}/2)+e^{i\varphi_R}\Theta(x-\mathcal{D}/2)]$, where $\varphi_L$ and $\varphi_R$ are the superconducting phases for left and right superconductors and, $\varphi=\varphi_R-\varphi_L$ represents the phase difference across the superconductors. Further, $\Delta$ denotes the magnitude of the superconducting gap, which varies with temperature via $\Delta=\Delta_0\tanh(1.74\sqrt{T_c/T-1})$, where $T_c$ is the critical temperature\cite{annun}, and $\Delta_0$ is the gap at zero temperature.

The BdG Hamiltonian for the S-$F_x$-$F_y$-$F_z$-S JJ, as depicted in Fig.~1(b), is expressed as\cite{enok,jfl}:
\begin{equation}
H_{BdG}^{\mbox{S-$F_x$-$F_y$-$F_z$-S}}(x)=
\begin{pmatrix}
H_{O}'\hat{I} & i\Delta_{s} \hat{\sigma}_{y} \\
-i\Delta_{s}^{*} \hat{\sigma}_{y} & -H_{O}'^{*}\hat{I}
\end{pmatrix},
\label{hmfff}
\end{equation}
with $H_{O}'=-\frac{\hbar^2}{2m^{*}}\frac{\partial^2}{\partial x^2}+\mathcal{B}_1\delta(x+\mathcal{D}/2)+\mathcal{B}_2\delta(x-\mathcal{D}/2)+\mathcal{B}_3\delta(x)+\mathcal{B}_4\delta(x-\mathcal{D}/4)-\vec{m}_1.\hat{\sigma}\Theta(x+\mathcal{D}/2)\Theta(-x)-\vec{m}_2.\hat{\sigma}\Theta(x)\Theta(\mathcal{D}/4-x)-\vec{m}_3.\hat{\sigma}\Theta(x-\mathcal{D}/4)\Theta(\mathcal{D}/2-x)-E_F$.
In $H_O'$, the parameters $\mathcal{B}_1$, $\mathcal{B}_2$, $\mathcal{B}_3$, and $\mathcal{B}_4$ represent the strengths of
the $\delta$-like potential barriers situated at the S-$F_x$, $F_z$-S, $F_x$-$F_y$, and $F_y$-$F_z$ interfaces, respectively. Further, the magnetization vectors of the ferromagnets $F_x$, $F_y$, and $F_z$ are denoted as $\vec{m}_{1}.\hat{\sigma}=m_{1}\hat{\sigma}_{x}$, $\vec{m}_{2}.\hat{\sigma}=m_{2}\hat{\sigma}_{y}$, and $\vec{m}_{3}.\hat{\sigma}=m_{3}\hat{\sigma}_{z}$, respectively.

In the remainder of this {article,} we use dimensionless parameters: $Z_{1(2)}=\frac{m^{*}\mathcal{B}_{1(2)}}{\hbar^2 k_F}$, $Z_{3(4)}=\frac{m^{*}\mathcal{B}_{3(4)}}{\hbar^2 k_F}$, and $Z=\frac{m^{*}\mathcal{B}}{\hbar^2 k_F}$  to represent interface transparencies\cite{BTK}, where $k_F$ is the Fermi wavevector. By diagonalizing Hamiltonians \eqref{hmff} and \eqref{hmfff}, we derive the wavefunctions corresponding to various scattering processes within the distinct regions of our setups. The detailed expressions for these wavefunctions are provided in {Appendix A.} We choose two JJs (Figs.~1(a,b)), (b) is a Josephson diode wherein anomalous Josephson current flows, while in (a) anomalous Josephson current is absent, so as to explore whether there exists any relation between the anomalous Josephson current and odd-$\omega$ ST superconducting pairing correlations for a short junction.

\section{Method} \subsection{Anomalous Josephson current}  The DC Josephson current is formulated in terms of the Andreev reflection amplitudes, utilizing the Furusaki-Tsukuda formalism\cite{furu,joseph}, and expressed as:
\begin{equation}
\label{eq3}
I(\varphi)=\frac{e\Delta k_BT}{2\hbar}\sum_{\omega_n}\frac{q_{en}^S+q_{hn}^S}{\sqrt{\omega_n^2+\Delta^2}}\Bigg(\frac{s_{1n}+s_{2n}}{q_{en}^S}-\frac{s_{3n}+s_{4n}}{q_{hn}^S}\Bigg),
\end{equation}
where $\omega_n=(2n+1)\pi k_BT$ represent fermionic Matsubara frequencies with $n=0, \pm1, \pm2, \pm3,...$ and  $q_{e(h)}^S$ is the electron-like (hole-like) quasiparticle's wavevector in the superconductors. $q_{en}^{S}$, $q_{hn}^{S}$, and $s_{in}$ ($i=1,2,3,4$) are obtained from $q_{e}^{S}$, $q_{h}^{S}$, and $s_{i}$ by analytically continuing $\omega$ to $i\omega_n$, wherein the amplitude $s_{1(2)}$ corresponds to Andreev reflection process in which an incident spin-up (down) electron originating from the left superconductor is reflected as a spin-down (up) hole, and the amplitude $s_{3(4)}$ represents Andreev reflection process where an incident spin-up (down) hole from the left superconductor is reflected as a spin-down (up) electron. $k_B$ is the Boltzmann constant, and $T$ is temperature. The necessary conditions for the existence of anomalous Josephson current are broken time-reversal and inversion symmetries. Using Eq.~\eqref{eq3},
we compute the anomalous Josephson current as $I_{an}=I(\varphi=0)$. The absolute value of the maximum Josephson current is $I_c^{+}=|\mbox{max}\,\,  I(\varphi)|$ and the absolute value of the minimum Josephson current is $I_c^{-}=|\mbox{min}\,\,  I(\varphi)|$. From $I_c^{+}$ and $I_c^{-}$, we calculate the efficiency\cite{mna,phf} of the Josephson diode shown in Fig.~1(b), as $\eta=\frac{I_c^{+}-I_c^{-}}{I_c^{+}+I_c^{-}}$. Using this procedure we calculate anomalous Josephson current in the Josephson diode setup (Fig.~1(b)), see \cite{mathematica} for details. There is no anomalous Josephson current, but Josephson current only in the setup of Fig.~1(a), as inversion symmetry is not broken, even though time-reversal symmetry is broken in both setups.

\subsection{Odd-frequency pairing} This study aims to determine whether the anomalous Josephson current influences odd-$\omega$ ST pairing. For this purpose, we construct the retarded Green's function, $\Gamma^r(x,\chi,\omega)$, for setups shown in Figs.~1(a), and 1(b) based on the scattering processes at the interface\cite{mcm}. We follow the approach outlined in Refs.~\cite{cayy} and \cite{amb}, with the detailed calculations of $\Gamma^{r}$ provided in {Appendix B.} We concentrate on the anomalous part of $\Gamma^r$, which governs the pairing amplitudes, and in the Matsubara representation is given as\cite{jfl},
\begin{equation}
\sum_{\omega_n>0} \Gamma^r_{eh}(x,\chi,i\omega_n)=i\sum_{\kappa=0}^{3}f_{\kappa}^{r}\sigma_{\kappa}\sigma_2.
\label{green}
\end{equation}
In Eq.~\eqref{green}, the summation is restricted to positive frequencies when calculating odd-$\omega$ pairing amplitudes as these are odd functions of $\omega_n$\cite{jfl}, while for even-$\omega$ we can sum over $\omega_n$ either $>0$ or $<0$. In this {article,} we only do the $\omega_n>0$ sum. Here, $\sigma_0$ is the unit matrix, while $\sigma_\kappa$ ($\kappa=1,2,3$) are the Pauli matrices. The SS ($\uparrow\downarrow-\downarrow\uparrow$) pairing amplitude is given by $f_0^r$, whereas the EST ($\downarrow\downarrow\pm\uparrow\uparrow$) pairing amplitudes are represented by $f_1^r$ and $f_2^r$. Finally, MST ($\uparrow\downarrow+\downarrow\uparrow$) pairing amplitude is given by $f_3^r$. The EST components for spin states $\uparrow\uparrow$ and $\downarrow\downarrow$ are given by $f_{\uparrow\uparrow}=-f_1^r+if_2^r$ and $f_{\downarrow\downarrow}=f_1^r+if_2^r$, respectively. The SS, EST, and MST pairing amplitudes both for even-$\omega$ and odd-$\omega$ are determined via,
\begin{eqnarray}
\label{EVENODD}
f^{E}_{\kappa}(x,\chi,T)&=&\sum_{\omega_n>0}f^{E}_{\kappa}(x,\chi,\omega\rightarrow i\omega_n),\,\,\,\mbox{and}\,\,\,\nonumber\\
f^{O}_{\kappa}(x,\chi,T)&=&\sum_{\omega_n>0}f^{O}_{\kappa}(x,\chi,\omega\rightarrow i\omega_n),
\end{eqnarray}
where,
$f^{E}_{\kappa}(x,\chi,\omega)=\frac{{f^{r}_{\kappa}(x,\chi,\omega)+f^{a}_{\kappa}(x,\chi,-\omega)}}{2}$, and
$f^{O}_{\kappa}(x,\chi,\omega)=\frac{{f^{r}_{\kappa}(x,\chi,\omega)-f^{a}_{\kappa}(x,\chi,-\omega)}}{2}$,
wherein $f_{\kappa}^a$ corresponds to $\Gamma^{a}$ which are the advanced Green's functions and derived from $\Gamma^{a}(x,\chi,\omega)=[\Gamma^{r}(\chi,x,\omega)]^{\dagger}$\cite{cayy}. The even- and odd-$\omega$ EST pairing amplitudes are computed as-
\begin{eqnarray}
\label{eo1}
&&f_{\uparrow\uparrow}^{E}=-f_{1}^{E}+if_{2}^{E}\,\,\,\mbox{and,}\,\,\, f_{\downarrow\downarrow}^{E}=f_{1}^{E}+if_{2}^{E},\nonumber\\&&
f_{\uparrow\uparrow}^{O}=-f_{1}^{O}+if_{2}^{O}\,\,\,\mbox{and,}\,\,\, f_{\downarrow\downarrow}^{O}=f_{1}^{O}+if_{2}^{O}.
\end{eqnarray}
We calculate SS, EST, and MST pairing amplitudes both for even-$\omega$ and odd-$\omega$ using Eqs.~\eqref{green}-\eqref{eo1}; see \cite{mathematica} for further details. Explicit analytical expressions for SS, EST, and MST pairing amplitudes are provided in {Appendix C.}

\section{Anomalous Josephson current and odd-frequency spin-triplet pairing in short junction limit}
In this section, we discuss the effect of anomalous Josephson current on odd-$\omega$ ST pairing for both tunneling and transparent interfaces between ferromagnetic layers.
\vspace{1pt}
{\subsection{Tunneling ferromagnetic interfaces}} We first present the results for the {tunneling ferromagnetic interfaces,} i.e., $Z=Z_3Z_4=2.5$ (where $Z_3=Z_4=1.58$) across the two setups shown in Figs.~1(a) and 1(b).

\subsubsection{{Bilayer S-$F_x$-$F_y$-S Josephson junction}} In this {bilayer} JJ, the anomalous Josephson current vanishes {due to the preservation of inversion symmetry,} and the system does not operate as a Josephson diode. {Time-reversal symmetry is broken in bilayer S-$F_x$-$F_y$-S JJ since the magnetic configuration is not invariant under reversal of the magnetizations of the ferromagnetic layers, i.e., $\vec{m}_{1}\rightarrow-\vec{m}_{1}$ and $\vec{m}_{2}\rightarrow-\vec{m}_{2}$. However, inversion symmetry is preserved in bilayer S-$F_x$-$F_y$-S JJ because the magnetic configuration is invariant under spatial inversion ($x\rightarrow-x$), which reverses the order of the ferromagnetic layers. In Fig.~2, we show the absolute values of local ($x=\chi$) even-$\omega$ SS, odd-$\omega$ EST, and odd-$\omega$ MST correlations in the left superconducting region at $x=-1.6\xi$ ($\xi=\hbar v_{F}/\Delta$ is the superconducting coherence length\cite{smi}), plotted as functions of magnetization $m$ with $m_1=m_2=m$ for the S-$F_x$-$F_y$-S JJ, considering both transparent ($Z_1=Z_2=0$) and disordered ($Z_1\neq Z_2\neq0$) ferromagnet-superconductor (S-$F_x$ and $F_y$-S) interfaces. At $m=0$, the local even-$\omega$ SS pairing exhibits a peak, while the local odd-$\omega$ EST and MST pairings vanish. Notably, at finite $m$ values, the local odd-$\omega$ ST pairing develops peaks, whereas the local even-$\omega$ SS pairing shows zeros.}
\begin{figure}[ht!]
\centering{\includegraphics[width=0.47\textwidth]{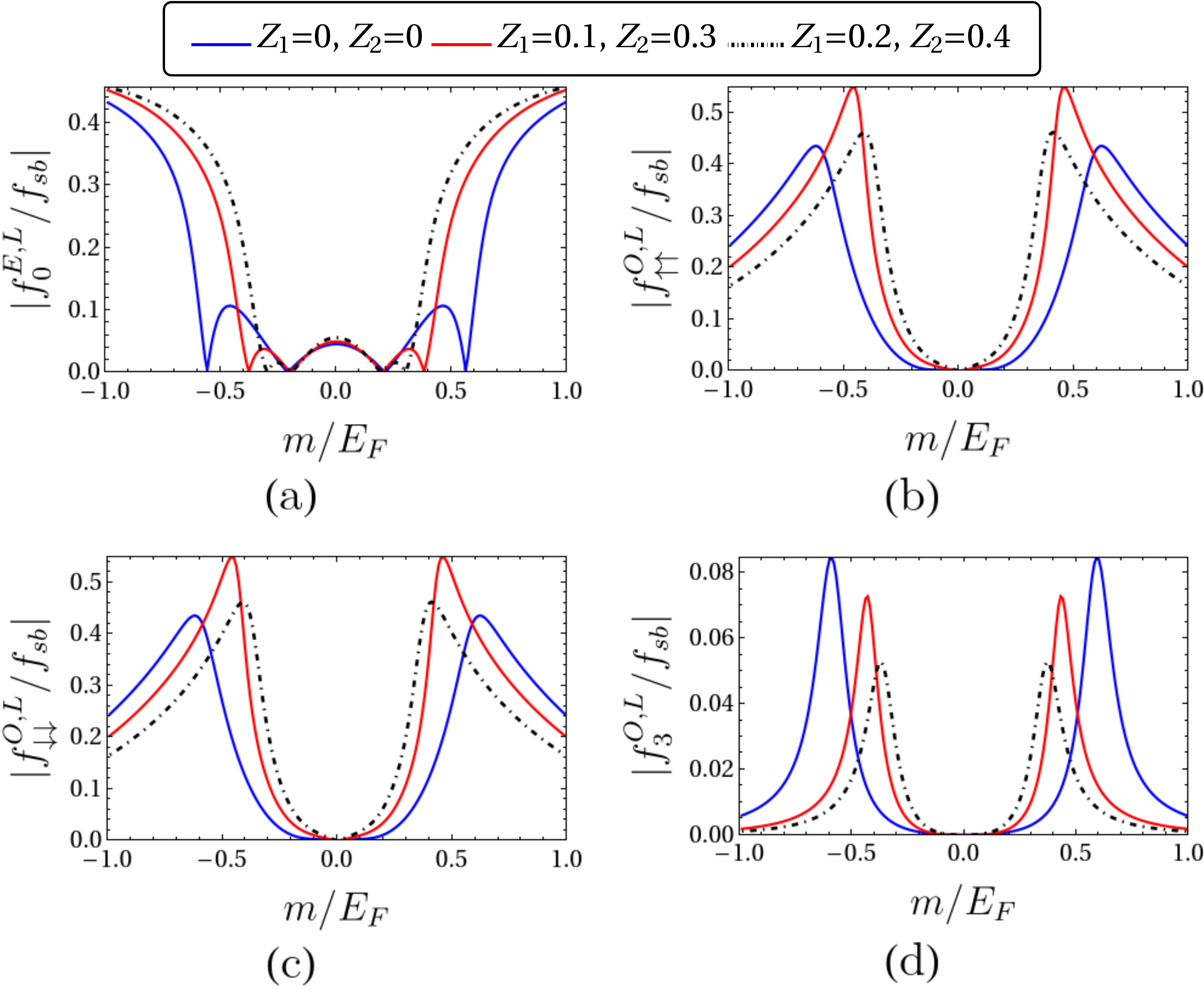}}
\vskip -0.1 in \caption{\small \sl {Absolute values of the local ($x=\chi$) even-$\omega$ SS (a), odd-$\omega$ EST (b,c), and odd-$\omega$ MST (d) pairing amplitudes in the left superconducting region at $x=-1.6\xi$ vs. magnetization $m$ for S-$F_x$-$F_y$-S JJ, considering both transparent and disordered S-$F_x$ and $F_y$-S interfaces in the short junction limit. Parameters: $\varphi=0$, $m_1=m_2=m$, $Z=2.5$, $k_{F}\mathcal{D}=1.5\pi$, $k_{F}\xi=2$, $E_F=100\Delta_0$, $T/T_c=0.002$, $k_{B}T/\Delta_0=0.001$.}}
\end{figure}

In Fig.~3, we present the absolute values of the {non-local ($x\neq\chi$)} even- and odd-$\omega$ SS, EST, and MST pairing amplitudes in the left superconducting region at $x=-1.6\xi$
\begin{figure}[ht!]
\centering{\includegraphics[width=0.47\textwidth]{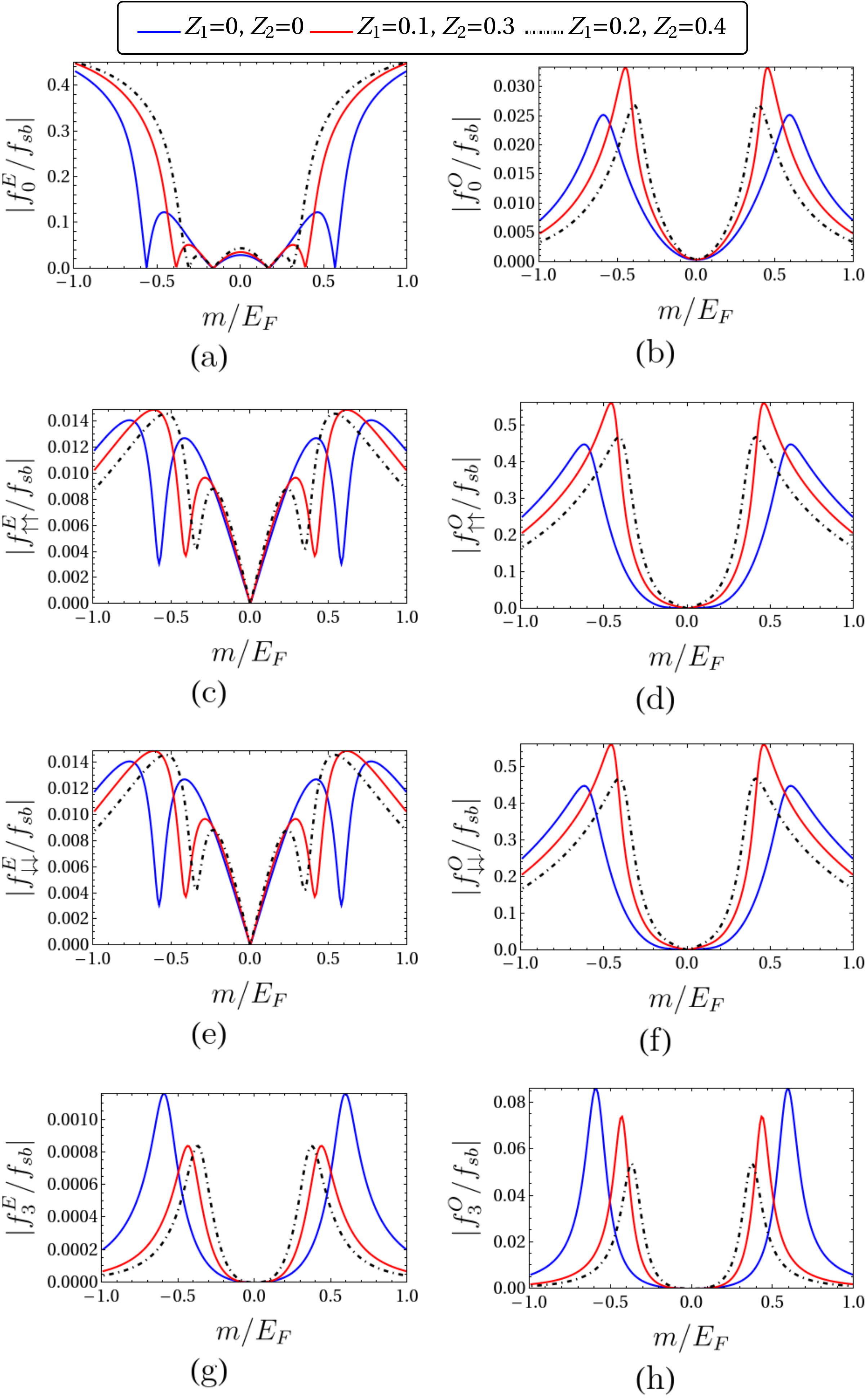}}
\vskip -0.1 in \caption{\small \sl Absolute values of the non-local ($x\neq\chi$) even- and odd-$\omega$ SS (a,b), EST (c,d,e,f), and MST (g,h) pairing amplitudes in the left superconducting region at $x=-1.6\xi$ vs. magnetization $m$ for {S-$F_x$-$F_y$-S} JJ, considering {both transparent and} disordered {S-$F_x$ and $F_y$-S interfaces in the short junction limit.} Parameters: $\varphi=0$, $m_1=m_2=m$, $Z=2.5$, $k_{F}\mathcal{D}=1.5\pi$, $k_{F}\xi=2$, $\chi=0$, $E_F=100\Delta_0$, $T/T_c=0.002$, $k_{B}T/\Delta_0=0.001$.}
\end{figure}
as functions of magnetization $m$ for {S-$F_x$-$F_y$-S} JJ, considering {both transparent} and disordered {S-$F_x$ and $F_y$-S interfaces.} We see that at $m=0$, even-$\omega$ SS pairing shows a peak, while odd-$\omega$ SS pairing exhibits a dip; however, both even- and odd-$\omega$ EST and MST pairings vanish. Importantly, we notice that odd-$\omega$ SS, odd-$\omega$ ST, and even-$\omega$ MST pairings exhibit peaks, even-$\omega$ SS pairing exhibits zeros, and the even-$\omega$ EST pairing exhibits both peaks and {dips} at finite $m$ values.

\subsubsection{Trilayer S-$F_x$-$F_y$-$F_z$-S Josephson junction} In the trilayer JJ-Fig.~1(b), the anomalous Josephson current is finite {due to the simultaneous breaking of time-reversal and inversion symmetries,} and the system acts as a Josephson diode. {In trilayer S-$F_x$-$F_y$-$F_z$-S JJ, time-reversal symmetry is broken as the magnetic configuration is not invariant under magnetization reversal,  i.e., $\vec{m}_{1}\rightarrow-\vec{m}_{1}$, $\vec{m}_{2}\rightarrow-\vec{m}_{2}$ and $\vec{m}_{3}\rightarrow-\vec{m}_{3}$. Inversion symmetry is broken in the trilayer S-$F_x$-$F_y$-$F_z$-S JJ because the magnetic configuration is not invariant under spatial inversion ($x\rightarrow -x$), which reverses the order of the ferromagnetic layers.}
In Figs.~4(a,b), the absolute value of anomalous Josephson current and diode efficiency are plotted as functions of magnetization $m$ of the ferromagnets with $m_1=m_2=m_3=m$ for S-$F_x$-$F_y$-$F_z$-S JJ, considering {both transparent and} disordered {S-$F_x$ and $F_z$-S interfaces.} As shown in Fig.~4(a), for $Z_1=Z_2=0$ ({transparent ferromagnet-superconductor interfaces}), the anomalous Josephson current exhibits peaks around $m\approx\pm0.56E_F$. When disorder is introduced {at ferromagnet-superconductor interfaces,} these peaks shift to $m\approx\pm0.48E_F$ for $Z_1=0.1$, $Z_2=0.3$, and further to $m\approx\pm0.41E_F$ for $Z_1=0.2$, $Z_2=0.4$. Moreover, as shown in Fig.~4(b), when anomalous Josephson current shows peaks, the diode efficiency is finite and becomes maximum. We notice a maximum of around $0.84$\% diode efficiency {for tunneling ferromagnetic interfaces.}
\begin{figure}[ht!]
\centering{\includegraphics[width=0.45\textwidth]{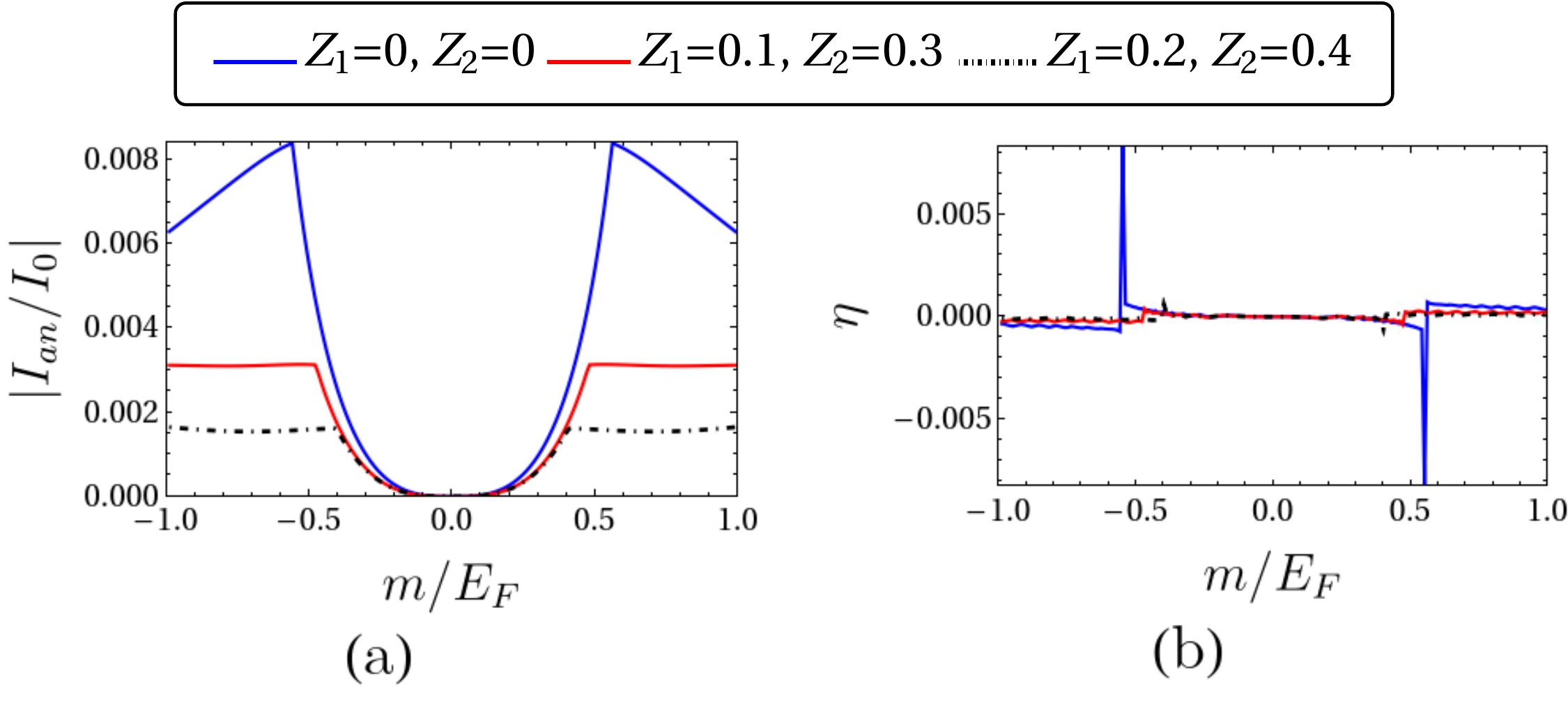}}
\vskip -0.1 in \caption{\small \sl Absolute anomalous Josephson current (a) and diode efficiency (b) as a function of magnetization $m$ for S-$F_x$-$F_y$-$F_z$-S JJ, considering {both transparent and} disordered {S-$F_x$ and $F_z$-S interfaces in the short junction limit.} Parameters: $\varphi=0$, $k_{F}\mathcal{D}=1.5\pi$, $Z_3=Z_4=1.58$, $m_1=m_2=m_3=m$, $E_F=100\Delta_0$, $T/T_c=0.002$, $I_0=e\Delta_0/\hbar$, $k_{B}T/\Delta_0=0.001$.}
\end{figure}

To examine the impact of anomalous Josephson current on ST pairing, {in Fig.~5, we plot the absolute values of local ($x=\chi$) even-$\omega$ SS, odd-$\omega$ EST, and odd-$\omega$ MST correlations in the left superconducting region at $x=-1.6\xi$ as functions of magnetization $m$ for S-$F_x$-$F_y$-$F_z$-S JJ, considering {both transparent and} disordered {S-$F_x$ and $F_z$-S interfaces.} At $m=0$, where the anomalous Josephson current vanishes, the local odd-$\omega$ ST pairing also vanishes, while the local even-$\omega$ SS pairing remains finite and exhibits a peak, consistent with the results in Fig.~2(a). At finite values of $m$, the local even-$\omega$ SS pairing develops zeros, as also seen in Fig.~2(a). Moreover, the local odd-$\omega$ ST pairing shows peaks at specific $m$ values, in agreement with Figs.~2(b,c,d). This behavior is seen in junctions both transparent and disordered ferromagnet-superconductor interfaces. Hence, the local odd-$\omega$ ST pairing exhibits similar characteristics regardless of the presence or absence of the anomalous Josephson current, indicating that the two phenomena are mutually exclusive.}
\begin{figure}[ht!]
\centering{\includegraphics[width=0.47\textwidth]{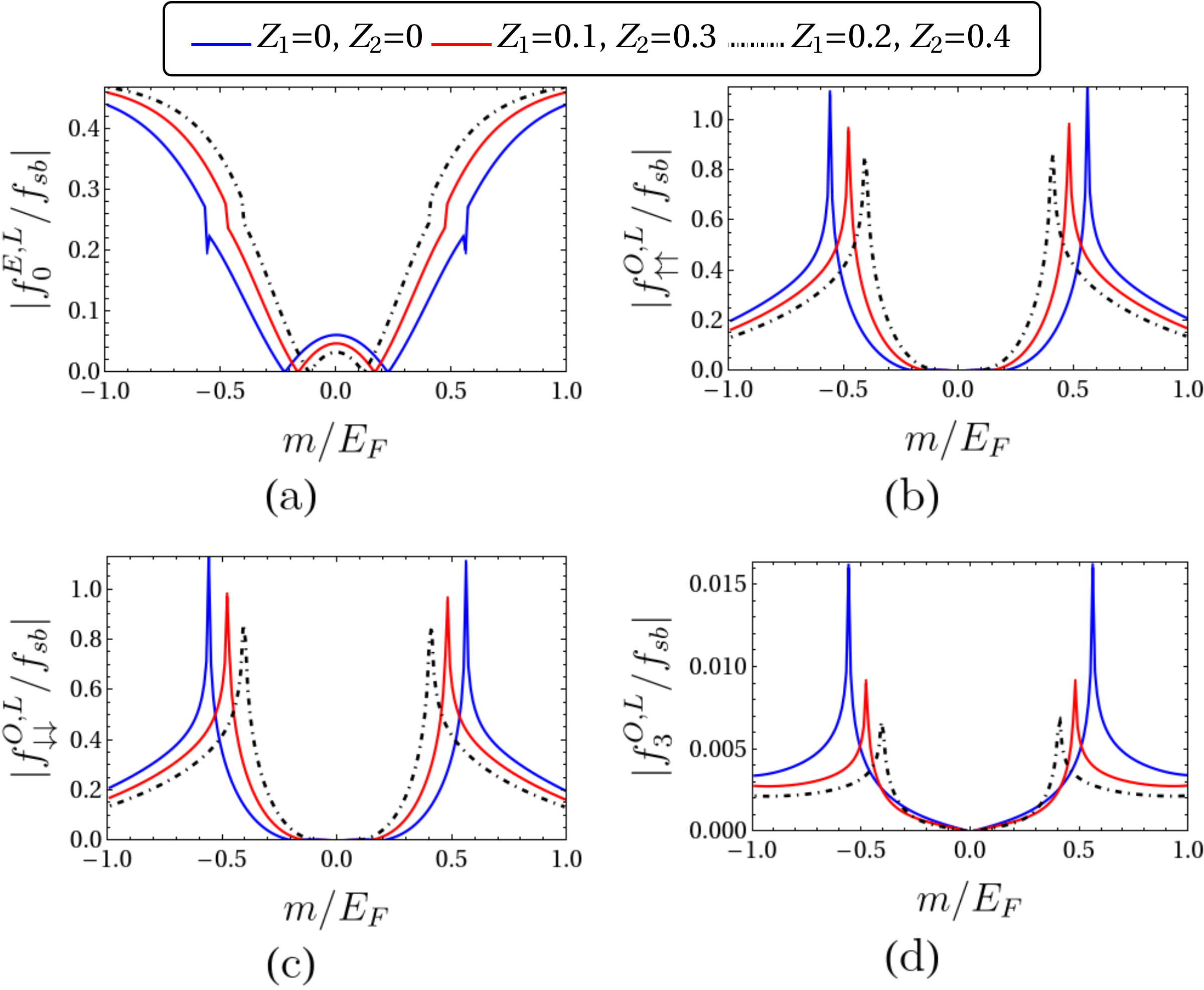}}
\vskip -0.1 in \caption{\small \sl {Absolute values of the local ($x=\chi$) even-$\omega$ SS (a), odd-$\omega$ EST (b,c), and odd-$\omega$ MST (d) pairing amplitudes in the left superconducting region at $x=-1.6\xi$ vs. magnetization $m$ for S-$F_x$-$F_y$-$F_z$-S JJ, considering both transparent and disordered S-$F_x$ and $F_z$-S interfaces in the short junction limit. Parameters: $\varphi=0$, $k_{F}\mathcal{D}=1.5\pi$, $k_{F}\xi=2$, $Z_3=Z_4=1.58$, $m_1=m_2=m_3=m$, $E_F=100\Delta_0$, $T/T_c=0.002$, $k_{B}T/\Delta_0=0.001$.}}
\end{figure}

In Fig.~6, the absolute values of the {non-local ($x\neq\chi$)} even- and odd-$\omega$ SS, EST, and MST pairing amplitudes in the left superconducting region at $x=-1.6\xi$ are plotted as functions of magnetization $m$ for S-$F_x$-$F_y$-$F_z$-S JJ, considering {both transparent and} disordered {S-$F_x$ and $F_z$-S interfaces.} At $m=0$, when the anomalous Josephson current vanishes, even- and odd-$\omega$ ST pairings vanish; however, even-$\omega$ SS pairing exhibits a peak, while odd-$\omega$ SS pairing shows a dip. Similar results are also seen in Fig.~3. Even-$\omega$ SS pairing shows zeros and even-$\omega$ ST pairing exhibits peaks at finite values of $m$, which is again seen in Figs.~3(a,c,e,g). Further, odd-$\omega$ SS and ST pairings exhibit peaks at specific values of $m$, which are also seen in Figs.~3(b,d,f,h). This holds true for junctions with both {transparent} and {disordered ferromagnet-superconductor interfaces.} Thus, {non-local} odd-$\omega$ ST pairing shows similar behavior regardless of the presence or absence of anomalous Josephson current, indicating anomalous Josephson current and {non-local} odd-$\omega$ ST pairing are mutually exclusive effects.
\begin{figure}[ht!]
\centering{\includegraphics[width=0.47\textwidth]{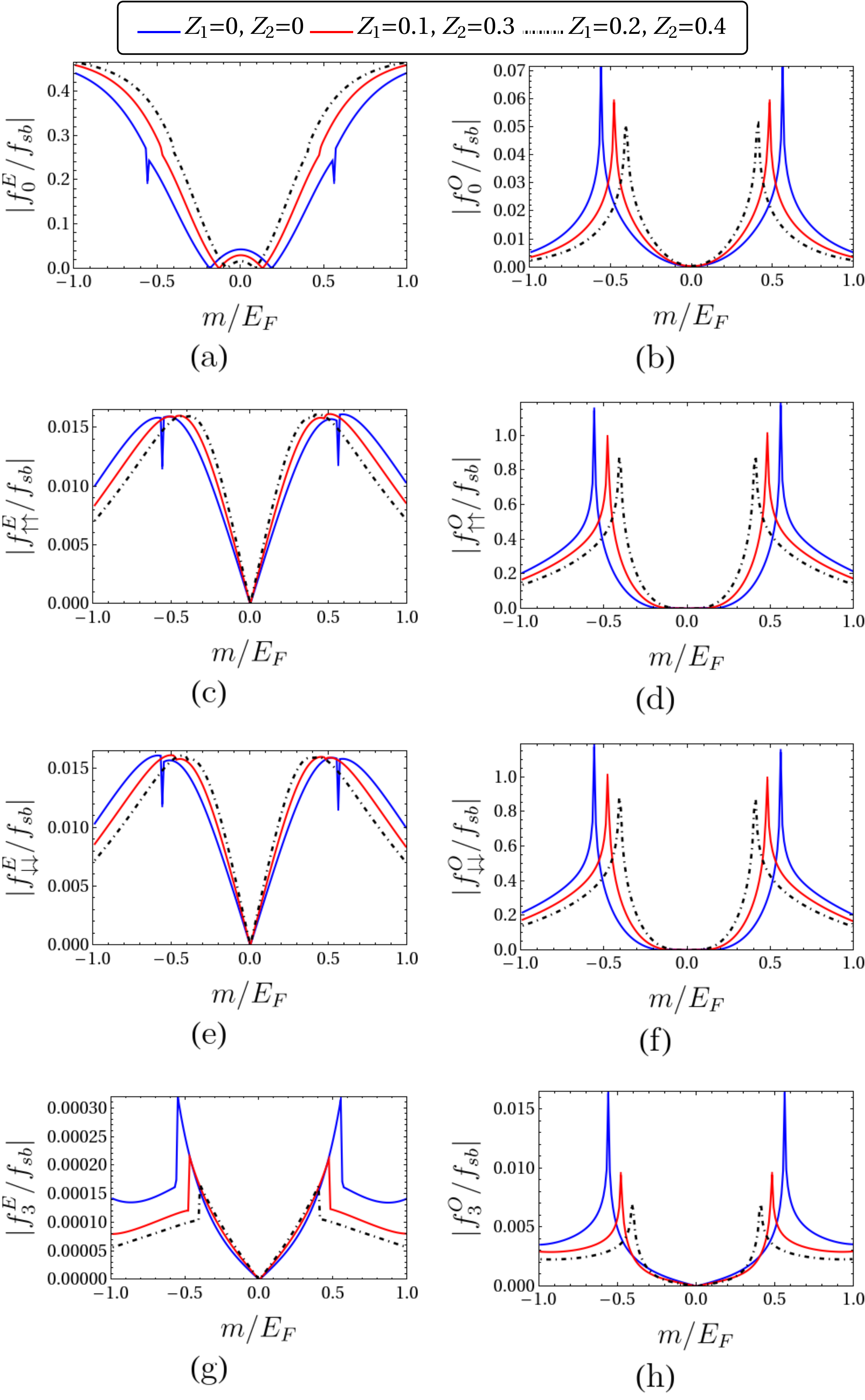}}
\vskip -0.1 in \caption{\small \sl Absolute values of the non-local ($x\neq\chi$) even- and odd-$\omega$ SS (a,b), EST (c,d,e,f), and MST (g,h) pairing amplitudes in the left superconducting region at $x=-1.6\xi$ vs. magnetization $m$ for S-$F_x$-$F_y$-$F_z$-S JJ, considering {both transparent and} disordered {S-$F_x$ and $F_z$-S interfaces in the short junction limit.} Parameters: $\varphi=0$, $k_{F}\mathcal{D}=1.5\pi$, $k_{F}\xi=2$, $\chi=0$, $Z_3=Z_4=1.58$, $m_1=m_2=m_3=m$, $E_F=100\Delta_0$, $T/T_c=0.002$, $k_{B}T/\Delta_0=0.001$.}
\end{figure}
We examine the impact of anomalous Josephson current on the spatial dependence of {non-local} odd-$\omega$ ST pairing in {Appendix D} and find that {non-local} odd-$\omega$ ST pairing behaves similarly irrespective of the presence or absence of anomalous Josephson current. Thus, anomalous Josephson current has no impact on the {non-local} odd-$\omega$ ST pairing {for tunneling ferromagnetic interfaces.}

{\subsection{Transparent ferromagnetic interfaces}} Herein, we present the results for the {transparent ferromagnetic interfaces} with $Z=0$, i.e., $Z_3=Z_4=0$ for all setups depicted in Figs.~1(a) and 1(b).
\vspace{35pt}
{\subsubsection{Bilayer S-$F_x$-$F_y$-S Josephson junction}
In Fig.~7, we present the absolute values of local ($x=\chi$) even-$\omega$ SS, odd-$\omega$ EST, and odd-$\omega$ MST correlations in the left superconducting region at $x=-1.6\xi$, plotted as functions of magnetization $m$ for the S-$F_x$-$F_y$-S junction, considering both transparent and disordered S-$F_x$ and $F_y$-S interfaces. At $m=0$, local even-$\omega$ SS pairing exhibits a peak, while local odd-$\omega$ ST pairing vanishes. At finite magnetization, local odd-$\omega$ EST pairing displays only peaks, whereas local odd-$\omega$ MST pairing shows both peaks and zeros.}
\begin{figure}[ht!]
\centering{\includegraphics[width=0.47\textwidth]{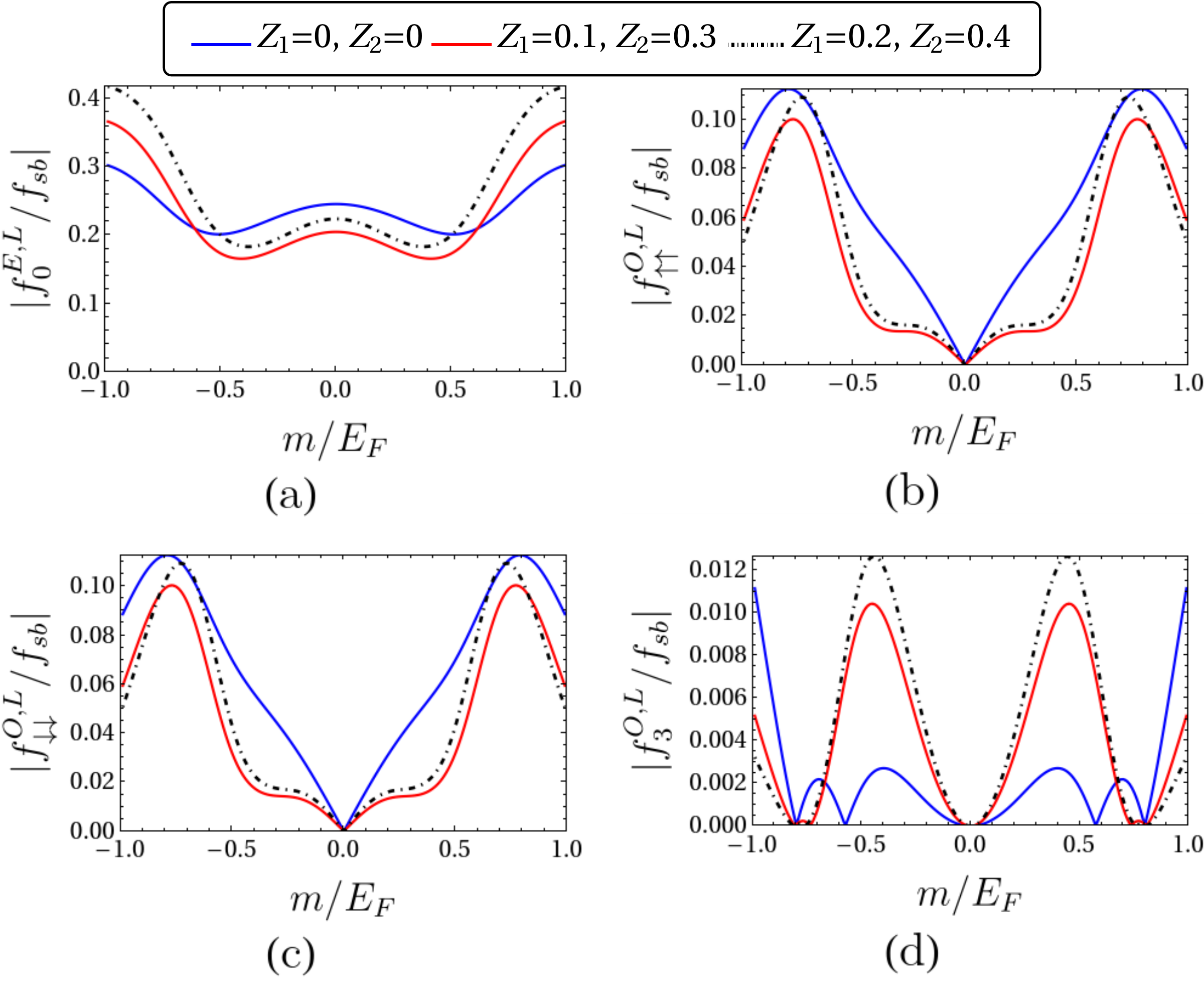}}
\vskip -0.1 in \caption{\small \sl {Absolute values of the local ($x=\chi$) even-$\omega$ SS (a), odd-$\omega$ EST (b,c), and odd-$\omega$ MST (d) pairing amplitudes in the left superconducting region at $x=-1.6\xi$ vs. magnetization $m$ for S-$F_x$-$F_y$-S JJ, considering both transparent and disordered S-$F_x$ and $F_y$-S interfaces in the short junction limit. Parameters: $\varphi=0$, $m_1=m_2=m$, $Z=0$, $k_{F}\mathcal{D}=1.5\pi$, $k_{F}\xi=2$, $E_F=100\Delta_0$, $T/T_c=0.002$, $k_{B}T/\Delta_0=0.001$.}}
\end{figure}

In Fig.~8, we plot the absolute values of the {non-local ($x\neq\chi$)} even- and odd-$\omega$ SS, EST, and MST pairing amplitudes in the left superconducting region at $x=-1.6\xi$ as function of magnetization $m$ with $m_1=m_2=m$ for bilayer {S-$F_x$-$F_y$-S} JJ, considering {both transparent and} disordered {S-$F_x$ and $F_y$-S interfaces.} We see that at $m=0$, even-$\omega$ SS pairing exhibits a peak, while odd-$\omega$ SS pairing shows a dip, but the even- and odd-$\omega$ EST and MST pairings vanish. Furthermore, we notice that even- and odd-$\omega$ EST pairings exhibit {only peaks,} whereas MST pairing displays {both peaks and zeros} at finite values of $m$. Odd-$\omega$ SS pairing exhibits peaks at finite $m$ values.
\begin{figure}[ht!]
\centering{\includegraphics[width=0.47\textwidth]{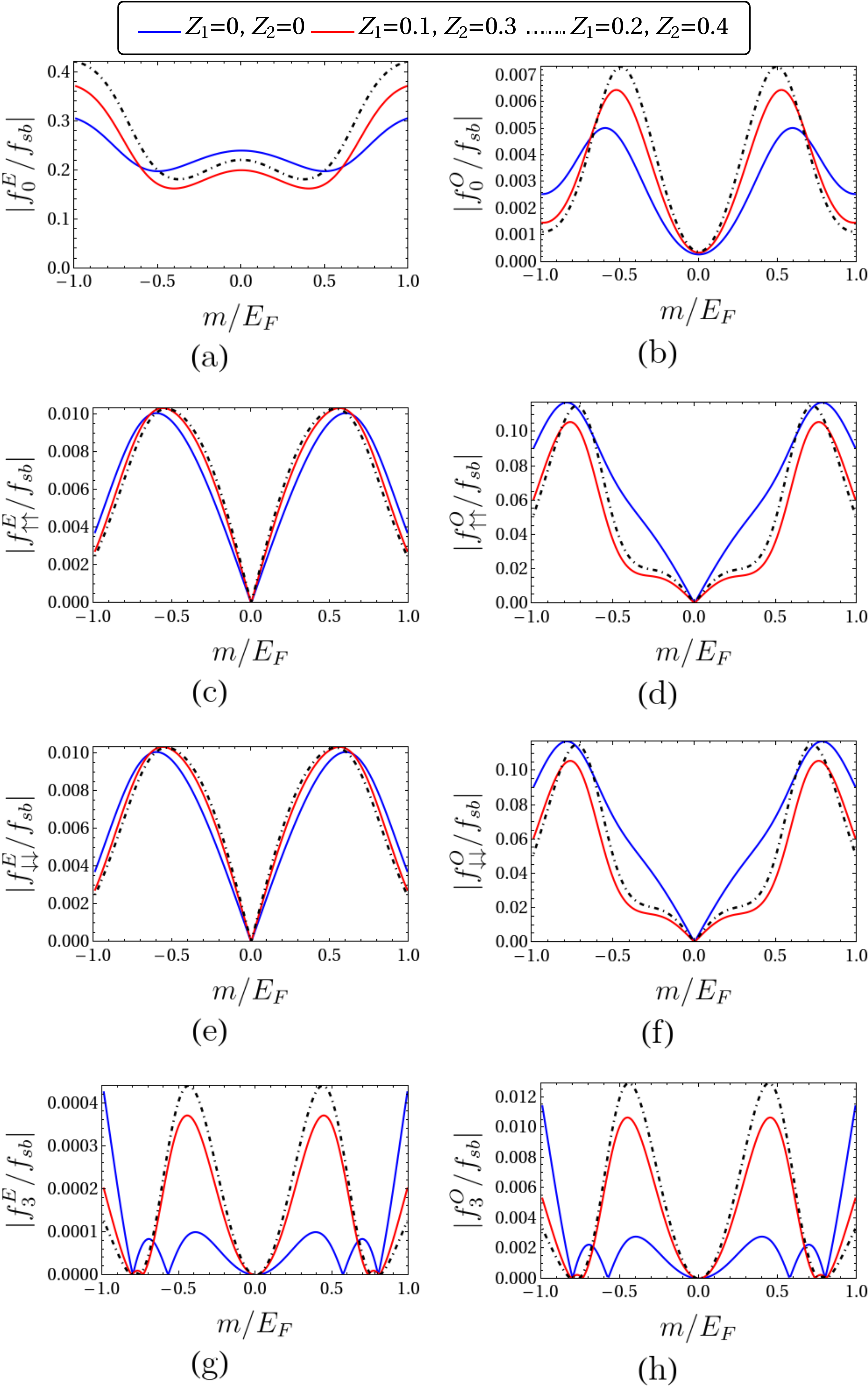}}
\vskip -0.1 in \caption{\small \sl Absolute values of the non-local ($x\neq\chi$) even- and odd-$\omega$ SS (a,b), EST (c,d,e,f), and MST (g,h) pairing amplitudes in the left superconducting region at $x=-1.6\xi$ vs. magnetization $m$ for {S-$F_x$-$F_y$-S} JJ, considering {both transparent and} disordered {S-$F_x$ and $F_y$-S interfaces in the short junction limit.} Parameters: $\varphi=0$, $m_1=m_2=m$, $Z=0$, $k_{F}\mathcal{D}=1.5\pi$, $k_{F}\xi=2$, $\chi=0$, $E_F=100\Delta_0$, $T/T_c=0.002$, $k_{B}T/\Delta_0=0.001$.}
\end{figure}

\subsubsection{Trilayer S-$F_x$-$F_y$-F$_z$-S Josephson junction} In Figs.~9(a,b), we plot the absolute value of anomalous Josephson current and diode efficiency as function of magnetization $m$ of the ferromagnets with $m_1=m_2=m_3=m$ for S-$F_x$-$F_y$-$F_z$-S JJ, considering {both transparent and} disordered {S-$F_x$ and $F_z$-S interfaces.} As seen from Fig.~9(a), anomalous Josephson current exhibits peaks around $m\approx\pm0.66E_F$ and zeros around $m\approx0,\pm0.8E_F$. Furthermore, as shown in Fig.~9(b), the diode efficiency is finite and attains a maximum value of around $4\%$ in the Josephson diode.
\begin{figure}[ht!]
\centering{\includegraphics[width=0.45\textwidth]{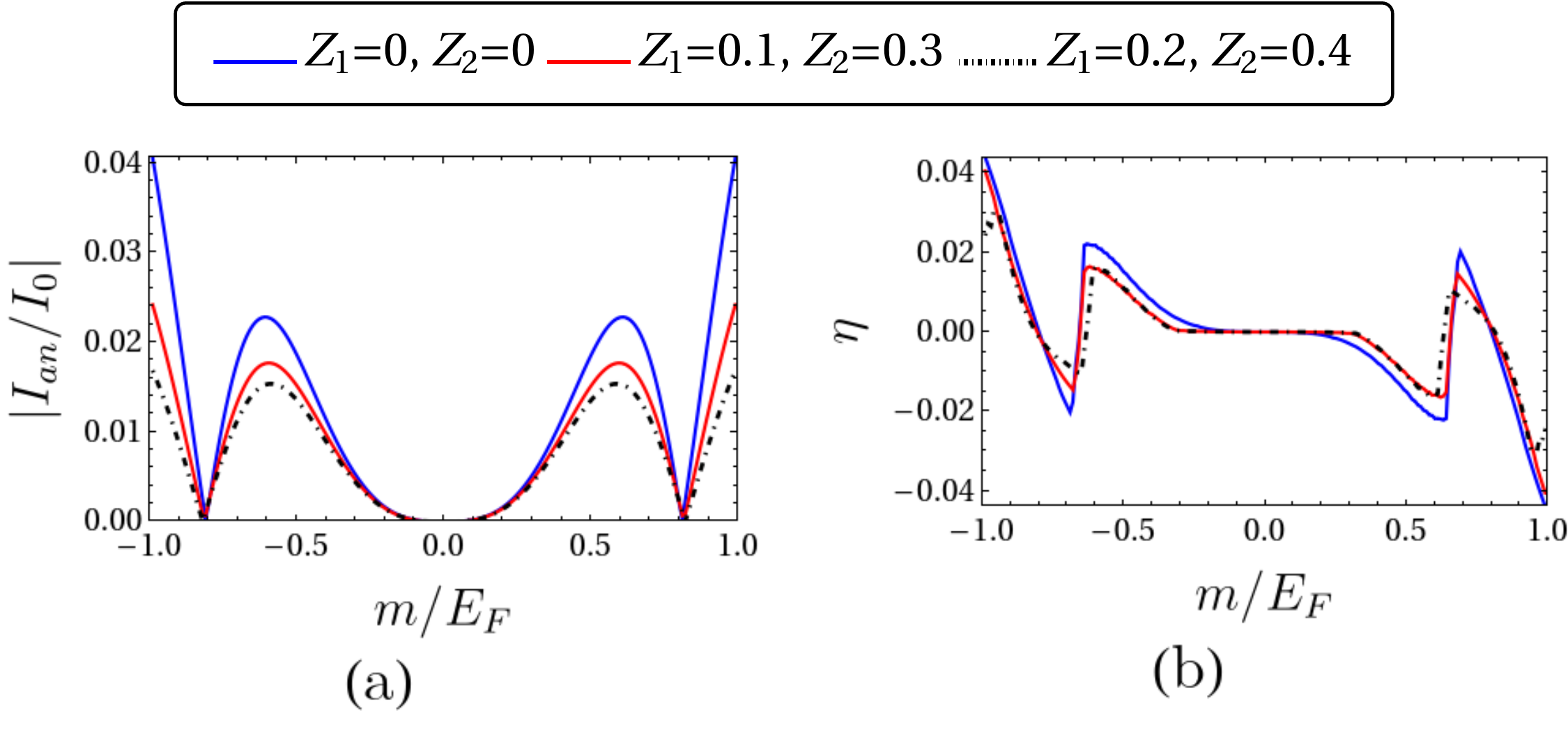}}
\vskip -0.1 in \caption{\small \sl Absolute anomalous Josephson current (a) and diode efficiency (b) as a function of magnetization $m$ for S-$F_x$-$F_y$-$F_z$-S JJ, considering {both transparent and} disordered {S-$F_x$ and $F_z$-S interfaces in the short junction limit.} Parameters: $\varphi=0$, $k_{F}\mathcal{D}=1.5\pi$, $Z_3=Z_4=0$, $m_1=m_2=m_3=m$, $E_F=100\Delta_0$, $T/T_c=0.002$, $I_0=e\Delta_0/\hbar$, $k_{B}T/\Delta_0=0.001$.}
\end{figure}

To explore the effect of anomalous Josephson current on odd-$\omega$ ST pairing, {in Fig.~10, we show the absolute values of local ($x=\chi$) even-$\omega$ SS, odd-$\omega$ EST, and odd-$\omega$ MST correlations in the left superconducting region
at $x=-1.6\xi$, plotted as functions of magnetization $m$ for the S-$F_x$-$F_y$-$F_z$-S junction, considering both transparent and disordered S-$F_x$ and $F_z$-S interfaces. At $m=0$, where the anomalous Josephson current is absent, local odd-$\omega$ ST pairing vanishes, while local even-$\omega$ SS pairing remains finite and exhibits a peak, consistent with Fig.~7. At finite $m$ values, local odd-$\omega$ EST and MST pairings develop peaks, as also seen in Fig.~7 for the S-$F_x$-$F_y$-S junction. We see this behavior for junctions both transparent and disordered ferromagnet-superconductor interfaces. Thus, the local odd-$\omega$ ST pairing remains essentially unchanged by the presence or absence of the anomalous Josephson current, suggesting that the two phenomena are not directly correlated.}
\begin{figure}[ht!]
\centering{\includegraphics[width=0.47\textwidth]{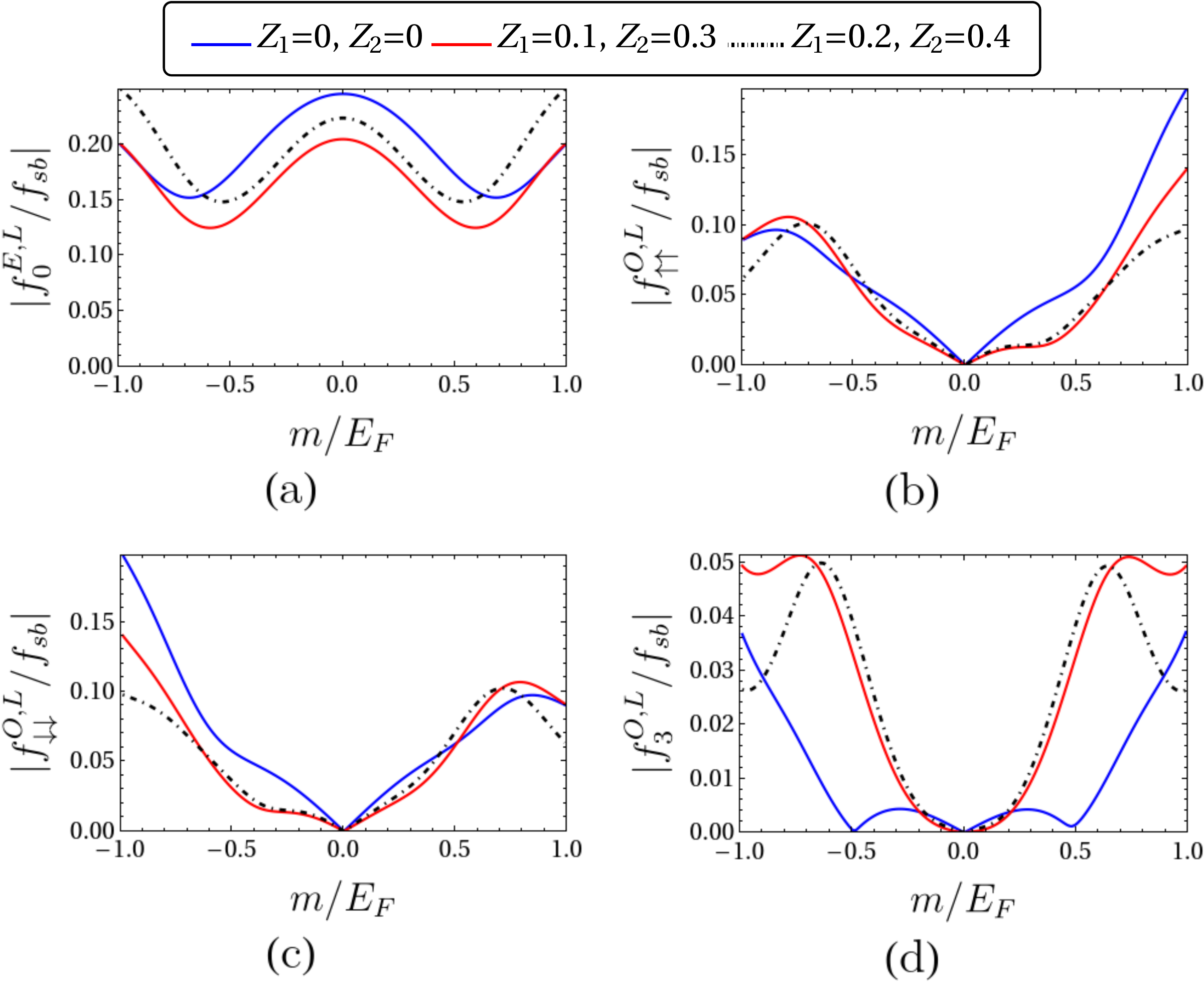}}
\vskip -0.1 in \caption{\small \sl {Absolute values of the local ($x=\chi$) even-$\omega$ SS (a), odd-$\omega$ EST (b,c), and odd-$\omega$ MST (d) pairing amplitudes in the left superconducting region at $x=-1.6\xi$ vs. magnetization $m$ for S-$F_x$-$F_y$-$F_z$-S JJ, considering both transparent and disordered S-$F_x$ and $F_z$-S interfaces in the short junction limit. Parameters: $\varphi=0$, $k_{F}\mathcal{D}=1.5\pi$, $k_{F}\xi=2$, $Z_3=Z_4=0$, $m_1=m_2=m_3=m$, $E_F=100\Delta_0$, $T/T_c=0.002$, $k_{B}T/\Delta_0=0.001$.}}
\end{figure}

In Fig.~11, we plot the absolute values of the {non-local ($x\neq\chi$)} even- and odd-$\omega$ SS, EST, and MST pairing amplitudes in the left superconducting region at $x=-1.6\xi$ as functions of magnetization $m$ for S-$F_x$-$F_y$-$F_z$-S JJ, considering {both transparent and} disordered {S-$F_x$ and $F_z$-S interfaces.} At $m=0$, when anomalous Josephson current is absent, even- and odd-$\omega$ ST pairings vanish; however, even-$\omega$ SS pairing shows a peak, and odd-$\omega$ SS pairing exhibits a dip similar to S-$F_x$-$F_y$-S JJ. Odd-$\omega$ SS, EST, and MST pairings, as well as even-$\omega$ EST and MST pairings, exhibit peaks at finite $m$ values, which are also seen in Fig.~8 for S-$F_x$-$F_y$-S JJ. This behavior persists in junctions both {transparent} and {disordered ferromagnet-superconductor interfaces.} Thus, {non-local} odd-$\omega$ ST pairing exhibits almost same characteristics regardless of the presence of anomalous Josephson current, implying that the anomalous Josephson current has marginal impact on {non-local} odd-$\omega$ ST pairing. We check the effect of anomalous Josephson current on
the spatial dependence of {non-local} odd-$\omega$ ST pairing in {Appendix D} and notice that {non-local} odd-$\omega$ ST pairing shows similar behavior irrespective of the presence or absence of anomalous Josephson current. Thus, anomalous Josephson current exerts a marginal influence on {non-local} odd-$\omega$ ST pairing, even in the {transparent ferromagnetic interfaces.
\begin{figure}[ht!]
\centering{\includegraphics[width=0.49\textwidth]{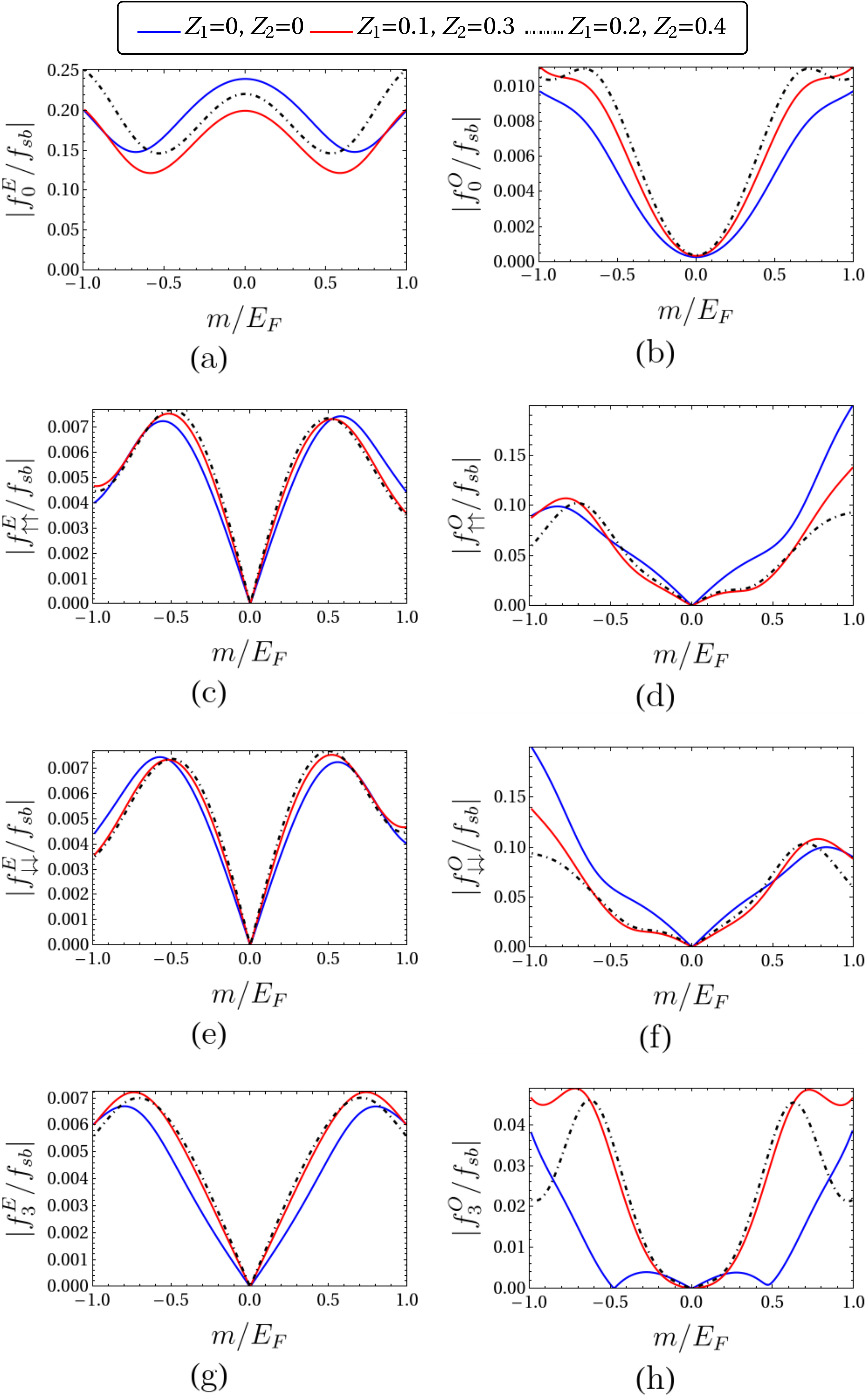}}
\vskip -0.1 in \caption{\small \sl Absolute values of the non-local ($x\neq\chi$) even- and odd-$\omega$ SS (a,b), EST (c,d,e,f), and MST (g,h) pairing amplitudes in the left superconducting region at $x=-1.6\xi$ vs. magnetization $m$ for S-F$_x$-F$_y$-F$_z$-S JJ, considering {both transparent and} disordered {S-$F_x$ and $F_z$-S interfaces in the short junction limit.} Parameters: $\varphi=0$, $k_{F}\mathcal{D}=1.5\pi$, $k_{F}\xi=2$, $\chi=0$, $Z_3=Z_4=0$, $m_1=m_2=m_3=m$, $E_F=100\Delta_0$, $T/T_c=0.002$, $k_{B}T/\Delta_0=0.001$.
}
\end{figure}
We compare anomalous Josephson current, local and non-local pairing amplitudes
between bilayer S-$F_x$-$F_y$-S, and trilayer S-$F_x$-$F_y$-$F_z$-S JJs in tabular form in Appendix E. Here, we focus on the short junction limit, while the corresponding long junction analysis is presented in the next section.} In all our figures, the pairing amplitudes are normalized to the value of the SS pairing amplitude in the bulk superconductors\cite{ltr}, given by $f_{sb}=2\sum_{\omega_n}\frac{\Delta}{\sqrt{\omega_n^2+\Delta^2}}$.
{\section{Effect of anomalous Josephson current on odd-frequency spin-triplet pairing in long junction limit}
In this section, we analysis our results in the long junction limit. For this purpose, we consider bilayer S-$F_x$-$F_y$-S and trilayer S-$F_x$-$F_y$-$F_z$-S JJs as shown in Fig.~12(a) and Fig.~12(b), respectively. In Fig.~12(a), two ferromagnets with lengths $\mathcal{D}_{1}$ and $\mathcal{D}_{2}$ and magnetization vectors aligned along the $x$- and $y$-axes are embedded between two $s$-wave superconductors. In Fig.~12(b), three ferromagnets with lengths $\mathcal{D}_{1}$, $\mathcal{D}_{2}$, and $\mathcal{D}_{3}$, whose magnetization vectors are oriented along the $x$-, $y$-, and $z$-axes, respectively, are sandwiched between two $s$-wave superconductors.
\begin{figure}[ht!]
\centering{\includegraphics[width=0.5\textwidth]{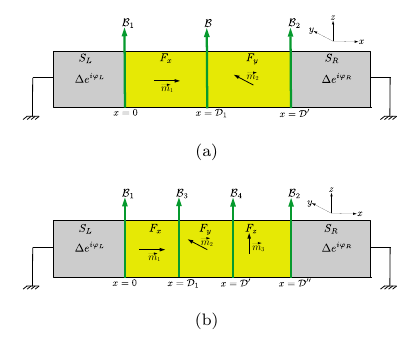}}
\vskip -0.2 in \caption{\small \sl{(a) JJ consisting of two ferromagnets (bilayer) with lengths $\mathcal{D}_{1}$ and $\mathcal{D}_{2}$ and magnetization vectors aligned along $x$- and $y$-axes, sandwiched between two $s$-wave superconductors, in the long junction limit. The interface between two ferromagnets is characterized by $\delta$-like potential barrier with strength $\mathcal{B}$, while the ferromagnet-superconductor interfaces are represented by $\delta$-like potential barriers with strengths $\mathcal{B}_{1}$ and $\mathcal{B}_{2}$. (b) JJ consisting of three ferromagnets (trilayer) with lengths $\mathcal{D}_{1}$, $\mathcal{D}_{2}$, and $\mathcal{D}_{3}$ and magnetization vectors aligned along $x$-, $y$-, and $z$-axes, sandwiched between two $s$-wave superconductors, in the long junction limit. Four $\delta$-like potential barriers, with strengths $\mathcal{B}_1$, $\mathcal{B}_2$, $\mathcal{B}_3$, and $\mathcal{B}_4$, are located at the S-$F_x$, $F_z$-S, $F_x$-$F_y$, and $F_y$-$F_z$ interfaces, respectively. In (a) and (b) $\mathcal{D}^{\prime}=\mathcal{D}_{1}+\mathcal{D}_{2}$ and $\mathcal{D}^{\prime\prime}=\mathcal{D}_{1}+\mathcal{D}_{2}+\mathcal{D}_{3}$.}}
\end{figure}
The ferromagnet-superconductor interfaces are modeled by $\delta$-like potential barriers with strengths $\mathcal{B}_{1}$ and $\mathcal{B}_{2}$. In the bilayer setup, the interface between the two ferromagnetic layers is represented by a $\delta$-like potential barrier with strength $\mathcal{B}$, whereas in the trilayer setup, the interfaces between adjacent ferromagnetic layers are characterized by $\delta$-like potential barriers with strengths $\mathcal{B}_{3}$ and $\mathcal{B}_{4}$.}

{The BdG Hamiltonian for a bilayer S-$F_x$-$F_y$-S JJ in long junction limit, as shown in Fig.~12(a), is expressed as\cite{enok},
\begin{equation}
\mathcal{H}_{BdG}^{\mbox{S-$F_x$-$F_y$-S}}(x)=
\begin{pmatrix}
\mathcal{H}_{P}\hat{I} & i\Delta_{S} \hat{\sigma}_{y} \\
-i\Delta_{S}^{*} \hat{\sigma}_{y} & -\mathcal{H}_{P}^{*}\hat{I}
\end{pmatrix},
\label{hmfxfy}
\end{equation}
with $\mathcal{H}_{P}=-\frac{\hbar^2}{2m^{*}}\frac{\partial^2}{\partial x^2}+\mathcal{B}_{1}\delta(x)+\mathcal{B}_{2}\delta(x-\mathcal{D}^{\prime})+\mathcal{B}\delta(x-\mathcal{D}_{1})-\vec{m}_{1}.\hat{\sigma}\Theta(x)\Theta(\mathcal{D}_{1}-x)-\vec{m}_{2}.\hat{\sigma}\Theta(x-\mathcal{D}_{1})\Theta(\mathcal{D}^{\prime}-x)-E_F$. The different parameters in $\mathcal{H}_{P}$ are defined below Eq.~\eqref{hmff}. Further, $\mathcal{D}^{\prime} =\mathcal{D}_{1}+\mathcal{D}_{2}$, where $\mathcal{D}_{1}$, $\mathcal{D}_{2}$ are units of $\frac{1}{k_F}$ and the superconducting gap is given as $\Delta_S=\Delta[e^{i\varphi_L}\Theta(-x)+e^{i\varphi_R}\Theta(x-\mathcal{D}^{\prime})]$. We introduce dimensionless parameters $Z=\frac{m^{*}\mathcal{B}}{\hbar^2 k_F}$, and $Z_{1,2}=\frac{m^{*}\mathcal{B}_{1,2}}{\hbar^2 k_F}$ to characterize interface transparencies between ferromagnets ($F_x$-$F_y$) and ferromagnet-superconductor (S-$F_x$ and $F_y$-S) interfaces, respectively, where $k_F$ is the Fermi wavevector. $Z_1$ and $Z_2$ at ferromagnet-superconductor interfaces are considered either transparent or
disordered, while $Z$ at ferromagnet-ferromagnet interface is considered to be either tunneling
or transparent.}

{The BdG Hamiltonian for a trilayer S-$F_x$-$F_y$-$F_z$-S JJ in long junction limit, as shown in Fig.~12(b), is expressed as follows\cite{enok,jfl},
\begin{equation}
\mathcal{H}_{BdG}^{\mbox{S-$F_x$-$F_y$-$F_z$-S}}(x)=
\begin{pmatrix}
\mathcal{H}_{P}'\hat{I} & i\Delta_{S}' \hat{\sigma}_{y} \\
-i\Delta_{S}'^{*} \hat{\sigma}_{y} & -\mathcal{H}_{P}'^{*}\hat{I}
\end{pmatrix},
\label{hmfxfyfz}
\end{equation}
with $\mathcal{H}_{P}'=-\frac{\hbar^2}{2m^{*}}\frac{\partial^2}{\partial x^2}+\mathcal{B}_{1}\delta(x)+\mathcal{B}_{2}\delta(x-\mathcal{D}^{\prime\prime})+\mathcal{B}_3\delta(x-\mathcal{D}_{1})+\mathcal{B}_4\delta(x-\mathcal{D}^{\prime})-\vec{m}_1.\hat{\sigma}\Theta(x)\Theta(\mathcal{D}_{1}-x)-\vec{m}_2.\hat{\sigma}\Theta(x-\mathcal{D}_{1})\Theta(\mathcal{D}^{\prime}-x)-\vec{m}_3.\hat{\sigma}\Theta(x-\mathcal{D}^{\prime})\Theta(\mathcal{D}^{\prime\prime}-x)-E_F$. The parameters appearing in $\mathcal{H}_{P}$ are defined below Eq.~\eqref{hmfff}. Further, $\mathcal{D}^{\prime\prime}=\mathcal{D}_{1}+\mathcal{D}_{2}+\mathcal{D}_{3}$, where $\mathcal{D}_{1}$, $\mathcal{D}_{2}$, $\mathcal{D}_{3}$ are units of $\frac{1}{k_F}$ and the superconducting gap is given as $\Delta_{S}'=\Delta[e^{i\varphi_L}\Theta(-x)+e^{i\varphi_R}\Theta(x-\mathcal{D}^{\prime\prime})]$. We introduce dimensionless parameters $Z_{3}=\frac{m^{*}\mathcal{B}_{3}}{\hbar^2 k_F}$, and $Z_{4}=\frac{m^{*}\mathcal{B}_{4}}{\hbar^2 k_F}$  to represent interface transparencies between ferromagnetic ($F_x$-$F_y$ and $F_y$-$F_z$) interfaces. $Z_3$ and $Z_4$ at ferromagnet-ferromagnet interfaces are considered to be either tunneling
or transparent. By diagonalizing Hamiltonians \eqref{hmfxfy} and \eqref{hmfxfyfz}, we derive the wavefunctions in distinct regions of the bilayer S-$F_x$-$F_y$-S and trilayer S-$F_x$-$F_y$-$F_z$-S JJs corresponding to different scattering processes. The explicit forms of these wavefunctions are provided in Appendix A. Using the procedure as discussed in Sec.~III, we calculate anomalous Josephson current, retarded Green's function, and even-/odd-$\omega$ SS and ST pairing amplitudes in the different regions of the S-$F_x$-$F_y$-S and S-$F_x$-$F_y$-$F_z$-S JJs in long junction limit.
\subsection{Tunneling ferromagnetic interfaces}
We first discuss the results for tunneling ferromagnetic interfaces corresponding to the two setups shown in Figs.~12(a) and 12(b).
\subsubsection{Bilayer S-$F_x$-$F_y$-S Josephson junction}
In S-$F_x$-$F_y$-S JJ, the anomalous Josephson current vanishes since inversion symmetry is preserved. In Fig.~13, we plot the absolute values of the local ($x=\chi$) even-$\omega$ SS, odd-$\omega$ EST, and odd-$\omega$ MST pairing amplitudes at the center of the $F_y$ layer as function of the length $\mathcal{D}_{1}$ in a bilayer S-$F_x$-$F_y$-S JJ for tunneling ferromagnetic interface and in long junction limit, considering both transparent and disordered S-$F_x$ and $F_y$-S interfaces. We see that local even-$\omega$ SS pairing is finite and shows a dip at $\mathcal{D}_{1}\approx10\pi$. Local odd-$\omega$ EST pairing exhibits zeros at $\mathcal{D}_{1}\approx10\pi, 20\pi$, while it shows peaks at $\mathcal{D}_{1}\approx5\pi, 15\pi$. Local odd-$\omega$ MST pairing is finite and does not exhibit any peaks/zeros. Further, in the presence of disordered ferromagnet-superconductor interfaces, local even- and odd-$\omega$ pairings show oscillatory behavior.
\begin{figure}[ht!]
\centering{\includegraphics[width=0.47\textwidth]{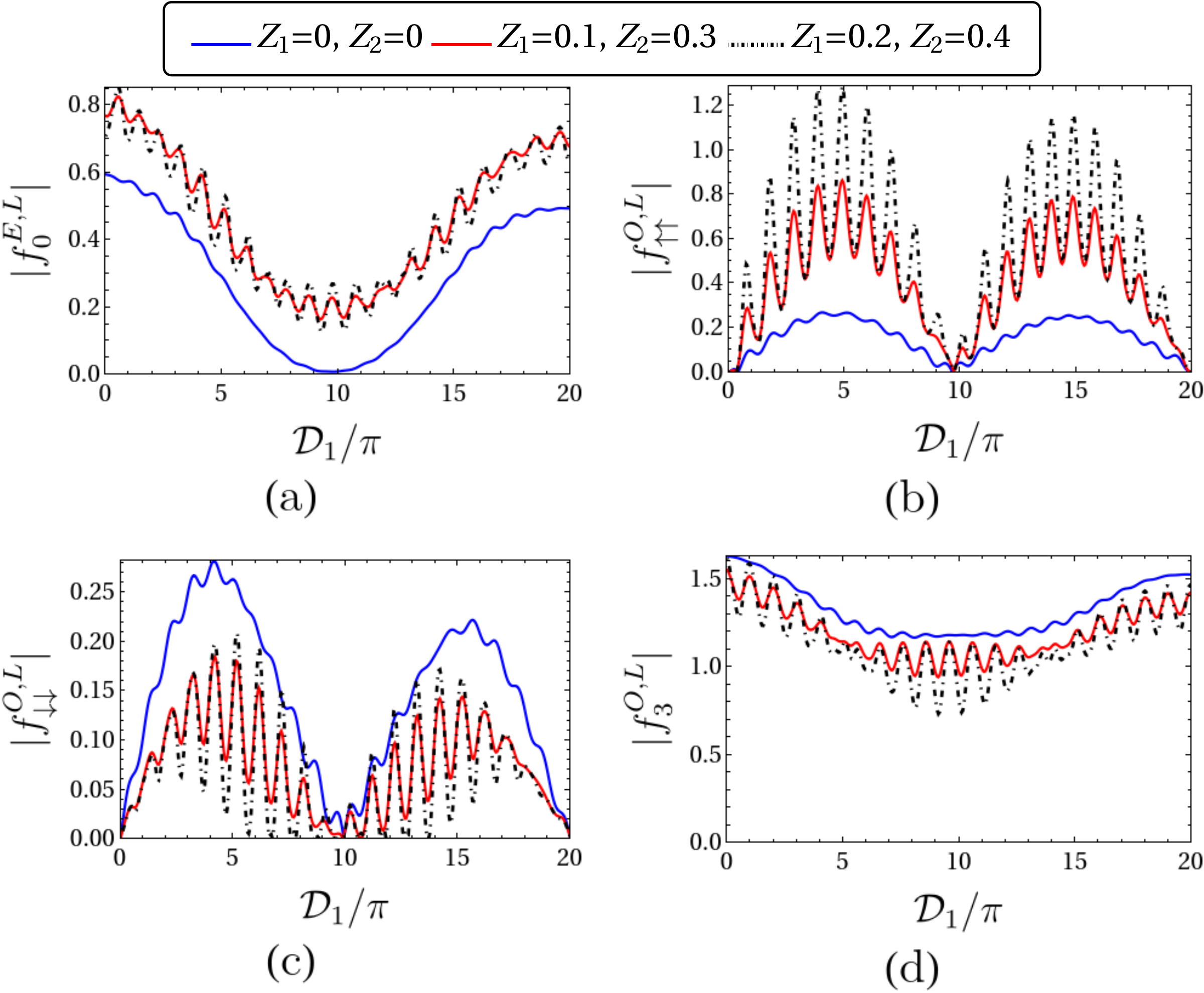}}
\vskip -0.1 in \caption{\small \sl {Absolute values of the local ($x=\chi$) even-$\omega$ SS (a), odd-$\omega$ EST (b,c), and odd-$\omega$ MST (d) pairing amplitudes in the $F_y$ layer at $x=\mathcal{D}_{1}+\frac{\mathcal{D}_{2}}{2}$ vs. length $\mathcal{D}_{1}$ for a bilayer S-$F_x$-$F_y$-S JJ for tunneling ferromagnetic interface in the long junction limit, considering both transparent and disordered S-$F_x$ and $F_y$-S interfaces. Parameters: $\varphi=0$, $m_1=m_2=0.1E_F$, $E_F=1000\Delta_0$, ${Z}=1$, $\mathcal{D}_2=5\pi$, $T/T_c=0.5$.}}
\end{figure}}

{In Fig.~14, we present the absolute values of the non-local ($x\neq\chi$) even- and odd-$\omega$ SS, EST, and MST pairing amplitudes at the center of the $F_y$ layer as function of the length $\mathcal{D}_{1}$ in a bilayer S-$F_x$-$F_y$-S JJ for tunneling ferromagnetic interface and in long junction limit,
\begin{figure}[ht!]
\centering{\includegraphics[width=0.47\textwidth]{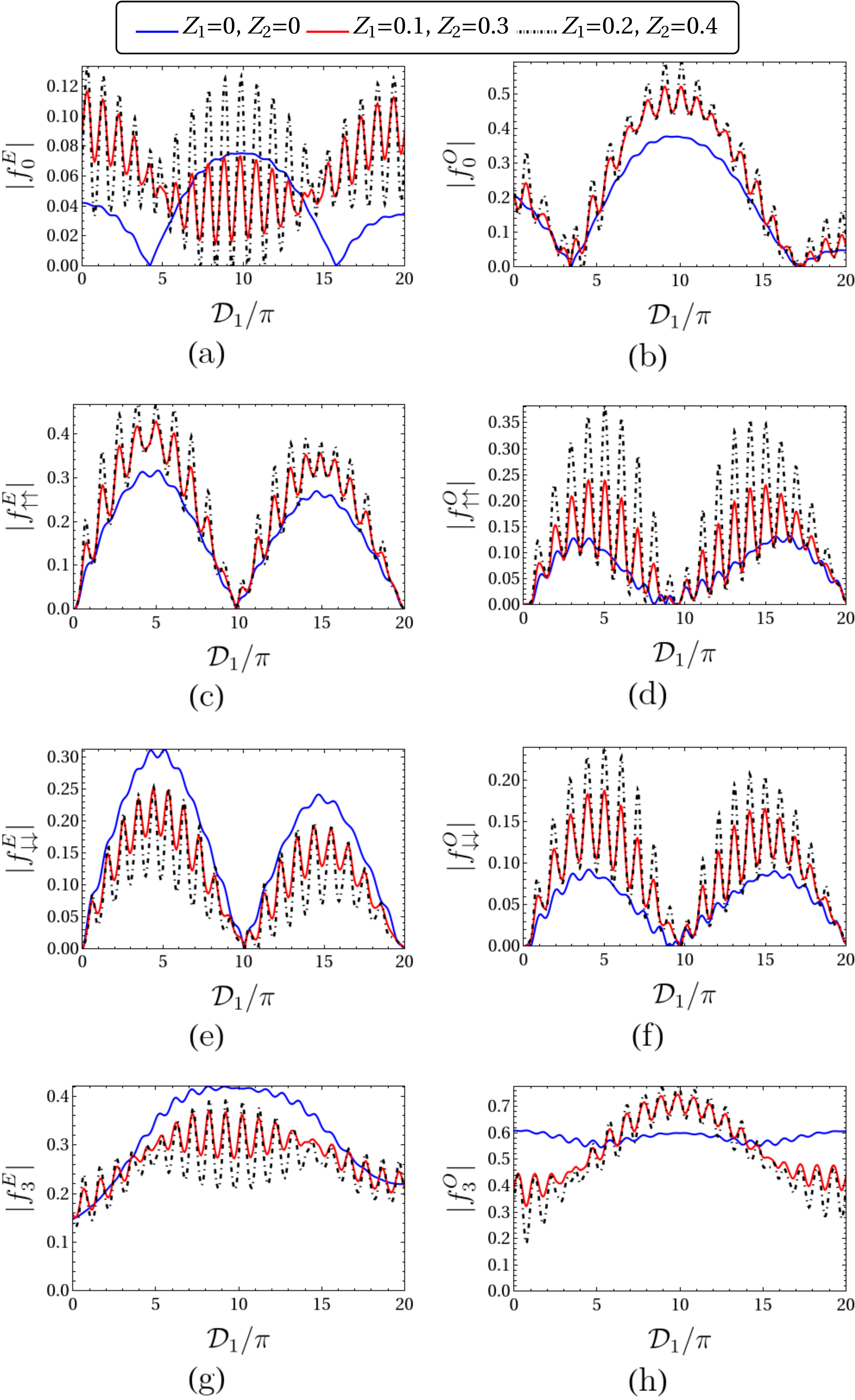}}
\vskip -0.1 in \caption{\small \sl {Absolute values of the non-local ($x\neq\chi$) even- and odd-$\omega$ SS (a,b), EST (c,d,e,f), and MST (g,h) pairing amplitudes in the $F_y$ layer at $x=\mathcal{D}_{1}+\frac{\mathcal{D}_{2}}{2}$ vs. length $\mathcal{D}_{1}$ for a bilayer S-$F_x$-$F_y$-S JJ for tunneling ferromagnetic interface in the long junction limit, considering both transparent and disordered S-$F_x$ and $F_y$-S interfaces. Parameters: $\varphi=0$, $m_1=m_2=0.1E_F$, $E_F=1000\Delta_0$, ${Z}=1$, $\mathcal{D}_2=5\pi$, $T/T_c=0.5$, $\chi=\mathcal{D}_{1}$.}}
\end{figure}
considering both transparent and disordered S-$F_x$ and $F_y$-S interfaces. For transparent ferromagnet-superconductor interfaces, we see that even-$\omega$ SS pairing shows zeros at $\mathcal{D}_{1}\approx4\pi,16\pi$, while it exhibits peak at $\mathcal{D}_{1}\approx10\pi$. Odd-$\omega$ SS pairing exhibits zeros at $\mathcal{D}_{1}\approx3\pi,17\pi$, while it shows peak at $\mathcal{D}_{1}\approx10\pi$. Further, we notice that even- and odd-$\omega$ EST pairings display peaks at $\mathcal{D}_{1}\approx5\pi,15\pi$, while they exhibit zero at $\mathcal{D}_{1}\approx10\pi$. Even- and odd-$\omega$ MST pairings are finite and do not show any peaks/zeros. In the presence of disordered ferromagnet-superconductor interfaces, we notice qualitatively similar behavior; however, disorder induces oscillations in the non-local even- and odd-$\omega$ SS, EST, and MST pairing amplitudes.
\subsubsection{Trilayer S-$F_x$-$F_y$-$F_z$-S Josephson junction}
In Fig.~15, the absolute value of anomalous Josephson current is plotted as function of the length $\mathcal{D}_{1}$ in a trilayer S-$F_x$-$F_y$-$F_z$-S JJ for tunneling ferromagnetic interfaces, again in long junction limit, considering both transparent and disordered S-$F_x$ and $F_z$-S interfaces. We notice that anomalous Josephson current exhibits peaks at $\mathcal{D}_{1}\approx5\pi,15\pi$, while it vanishes at $\mathcal{D}_{1}\approx10\pi,20\pi$. Further, in the presence of disordered ferromagnet-superconductor interfaces, anomalous Josephson current exhibits oscillations.
\begin{figure}[ht!]
\centering{\includegraphics[width=0.35\textwidth]{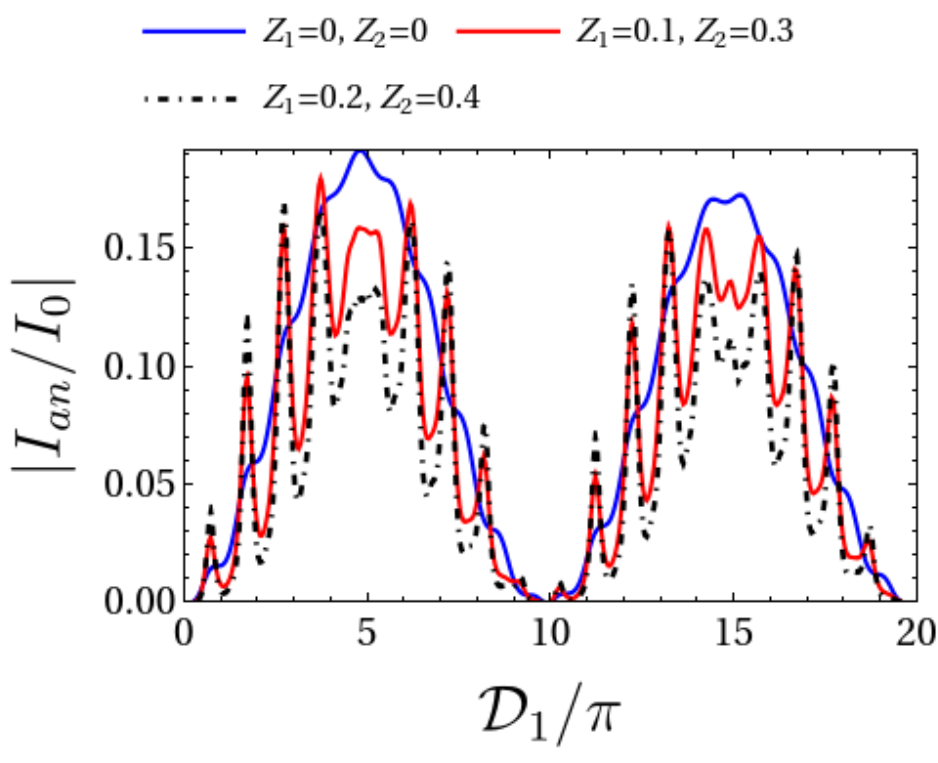}}
\vskip -0.1 in \caption{\small \sl {Absolute value of anomalous Josephson current as function of length $\mathcal{D}_{1}$ in a trilayer S-$F_x$-$F_y$-$F_z$-S JJ for tunneling ferromagnetic interfaces in the long junction limit, with $\mathcal{D}_{3}=\mathcal{D}_{1}$, considering both transparent and disordered S-$F_x$ and $F_z$-S interfaces. Parameters: $\varphi=0$, $m_1=m_2=m_3=0.1E_F$, $E_F=1000\Delta_0$, ${Z}_{3}={Z}_{4}=1$, $\mathcal{D}_2=5\pi$, $T/T_c=0.5$, $I_{0}=e\Delta_{0}/\hbar$.}}
\end{figure}}

{To examine the effect of anomalous Josephson current on odd-$\omega$ ST pairing, in Fig.~16, we plot the absolute values of the local ($x=\chi$) even-$\omega$ SS, odd-$\omega$ EST, and odd-$\omega$ MST pairing amplitudes at the center of the $F_y$ region same as was done in Ref.~\cite{jfl} as function of the length $\mathcal{D}_{1}$ in a trilayer S-$F_x$-$F_y$-$F_z$-S JJ for tunneling ferromagnetic interfaces, again for a long junction, with both transparent and disordered S-$F_x$ and $F_z$-S interfaces. For transparent ferromagnet-superconductor interfaces, we see that local even-$\omega$ SS pairing exhibits zeros at $\mathcal{D}_{1}\approx5\pi,15\pi$, while it shows a peak at $\mathcal{D}_{1}\approx10\pi$. This behavior is not seen in absence of anomalous Josephson current in Fig.~13 for a bilayer setup. Local odd-$\omega$ EST pairing shows zeros at $\mathcal{D}_{1}\approx10\pi,20\pi$, while it exhibits peaks at $\mathcal{D}_{1}\approx5\pi,15\pi$. Similar results are also seen in Figs.~13(b,c). Local odd-$\omega$ MST pairing exhibits zeros at $\mathcal{D}_{1}\approx5\pi,15\pi$, while it displays a peak at $\mathcal{D}_{1}\approx10\pi$ in long junction limit. This behavior is not observed in Fig.~13(d) for a bilayer S-$F_x$-$F_y$-S JJ in long junction limit.
\begin{figure}[ht!]
\centering{\includegraphics[width=0.47\textwidth]{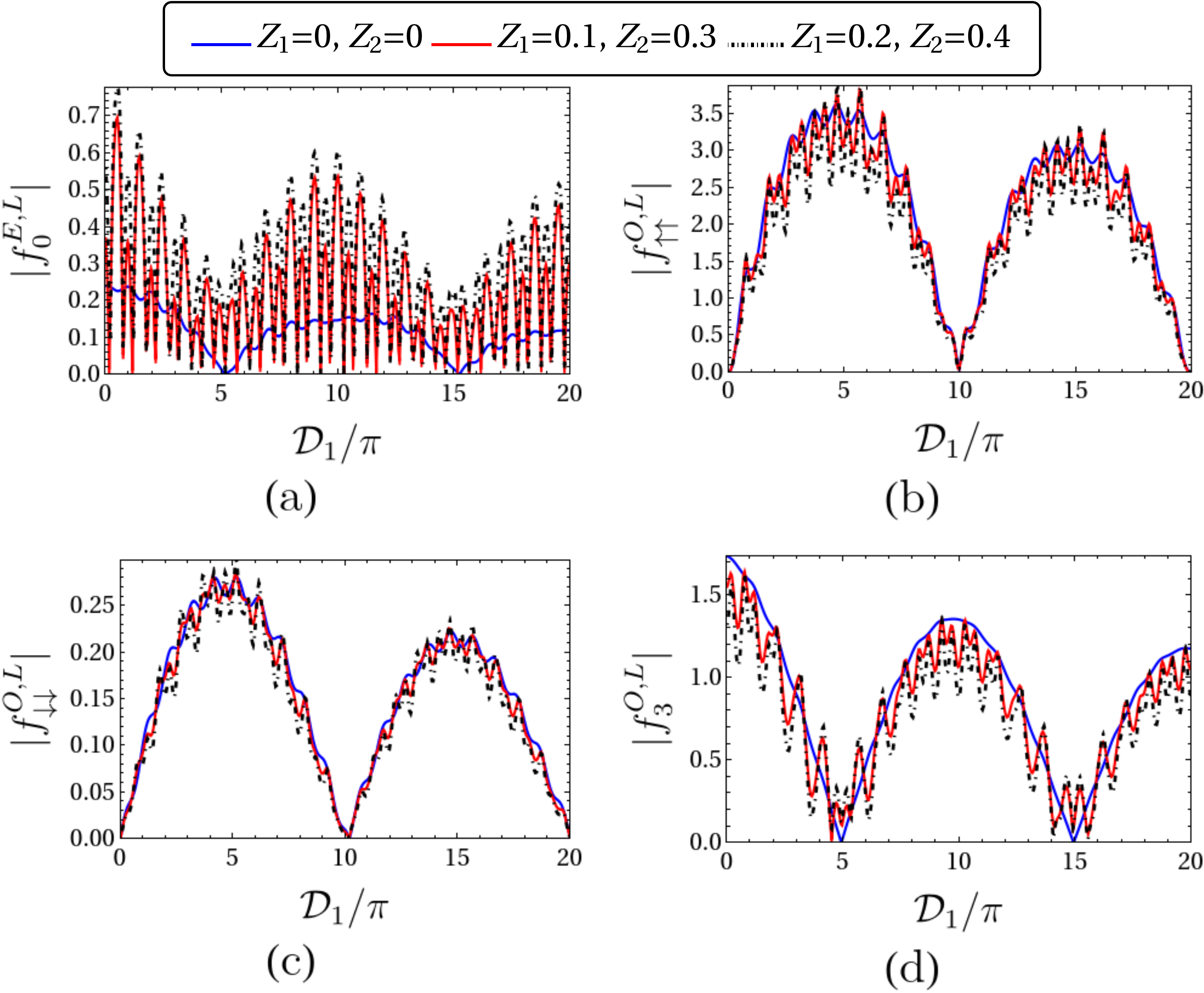}}
\vskip -0.1 in \caption{\small \sl {Absolute values of the local ($x=\chi$) even-$\omega$ SS (a), odd-$\omega$ EST (b,c), and odd-$\omega$ MST (d) pairing amplitudes in the $F_y$ layer at $x=\mathcal{D}_{1}+\frac{\mathcal{D}_{2}}{2}$ vs. length $\mathcal{D}_{1}$ for a trilayer S-$F_x$-$F_y$-$F_z$-S JJ for tunneling ferromagnetic interfaces in the long junction limit, with $\mathcal{D}_{3}=\mathcal{D}_{1}$, considering both transparent and disordered S-$F_x$ and $F_z$-S interfaces. Parameters: $\varphi=0$, $m_1=m_2=m_3=0.1E_F$, $E_F=1000\Delta_0$, ${Z}_{3}={Z}_{4}=1$, $\mathcal{D}_2=5\pi$, $T/T_c=0.5$.}}
\end{figure}
\begin{figure*}[ht!]
\centering{\includegraphics[width=0.83\textwidth]{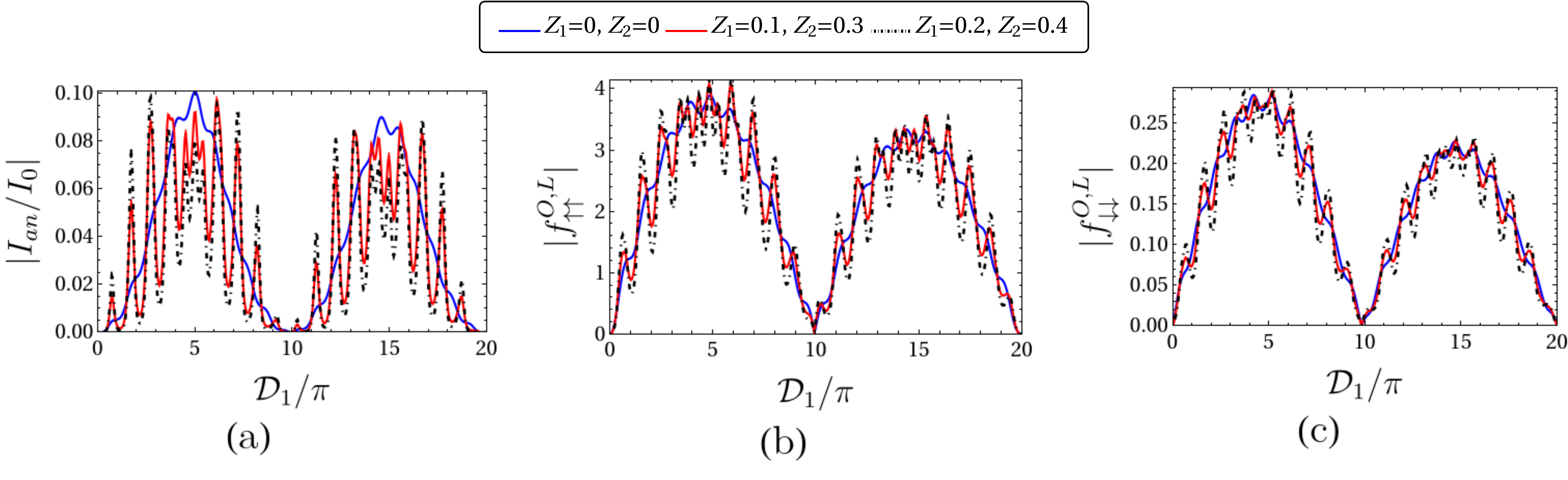}}
\vskip -0.1 in \caption{\small \sl {Absolute value of anomalous Josephson current (a), and absolute values of the local ($x=\chi$) odd-$\omega$ EST pairing amplitudes (b,c) in the $F_y$ layer at $x=\mathcal{D}_{1}+\frac{\mathcal{D}_{2}}{2}$ vs. length $\mathcal{D}_{1}$ for a trilayer S-$F_x$-$F_y$-$F_z$-S JJ for tunneling ferromagnetic interfaces in the long junction limit, with $\mathcal{D}_{3}=\mathcal{D}_{1}$, considering both transparent and disordered S-$F_x$ and $F_z$-S interfaces. Parameters: $\varphi=0$, $m_1=m_2=m_3=0.1E_F$, $E_F=1000\Delta_0$, ${Z}_{3}=1$, ${Z}_{4}=2$, $\mathcal{D}_2=5\pi$, $T/T_c=0.5$, $I_{0}=e\Delta_{0}/\hbar$.}}
\end{figure*}
Local even- and odd-$\omega$ pairings exhibit oscillatory behavior due to disordered ferromagnet-superconductor interfaces. Thus, anomalous Josephson current is clearly proportional to local odd-$\omega$ EST pairing for tunneling ferromagnetic interfaces in a trilayer S-$F_x$-$F_y$-$F_z$-S JJ, but in bilayer S-$F_x$-$F_y$-S JJ there is no relation between anomalous Josephson current and local odd-$\omega$ EST pairing whatsoever, as anomalous Josephson current vanishes itself in a bilayer S-$F_x$-$F_y$-S JJ.}

{For the S-$F_x$-$F_y$-$F_z$-S JJ, we consider two identical interface barriers between the ferromagnetic layers, following Ref.~\cite{jfl}. Such identical barriers give rise to resonant states in the intermediate layer, which may be absent in more general configurations. It has been speculated that these resonant states are responsible for the observed proportionality between the anomalous Josephson current and local odd-$\omega$ EST pairing. To examine the validity of this speculation, in Fig.~17, we plot the absolute values of the anomalous Josephson current
and the local odd-$\omega$ EST pairing as functions of the length $\mathcal{D}_{1}$ in a S-$F_x$-$F_y$-$F_z$-S JJ with non-identical interface barrier strengths, considering both transparent and disordered S-$F_x$ and $F_z$-S interfaces.
As shown in Fig.~17, for transparent ferromagnet-superconductor interfaces, both $|I_{an}|$ and $|f_{\sigma\sigma}^{O,L}|$ with $\sigma\in\{\uparrow,\downarrow\}$ develop  peaks at $\mathcal{D}_{1}\approx5\pi,15\pi$, while they simultaneously vanish at $\mathcal{D}_{1}\approx10\pi$.
In presence of disordered ferromagnet-superconductor interfaces, both $|I_{an}|$ and $|f_{\sigma\sigma}^{O,L}|$ exhibit oscillatory behavior.
These observations demonstrate an apparent proportionality between the anomalous Josephson current and local odd-$\omega$ EST pairing even in the presence of non-identical interface barriers. Therefore, the conjecture that non-identical barriers could eliminate this proportionality is not supported by plots in Fig.~17. The proportionality between anomalous Josephson current and local EST correlations is agnostic to identical or nonidentical interface barriers for tunneling interfaces between the ferromagnetic layers and in long junction limit, while this proportionality disappears for transparent ferromagnetic interfaces and for short junction limit with transparent or tunneling ferromagnetic interfaces in a trilayer JJ.}

{Finally, in Fig.~18, we plot the absolute values of the non-local ($x\neq\chi$) even- and odd-$\omega$ SS, EST, and MST pairing amplitudes at the center of the $F_y$ region as function of the length $\mathcal{D}_1$ in a trilayer S-$F_x$-$F_y$-$F_z$-S JJ for tunneling ferromagnetic interfaces
and in long junction limit,
\begin{figure}[ht!]
\centering{\includegraphics[width=0.439\textwidth]{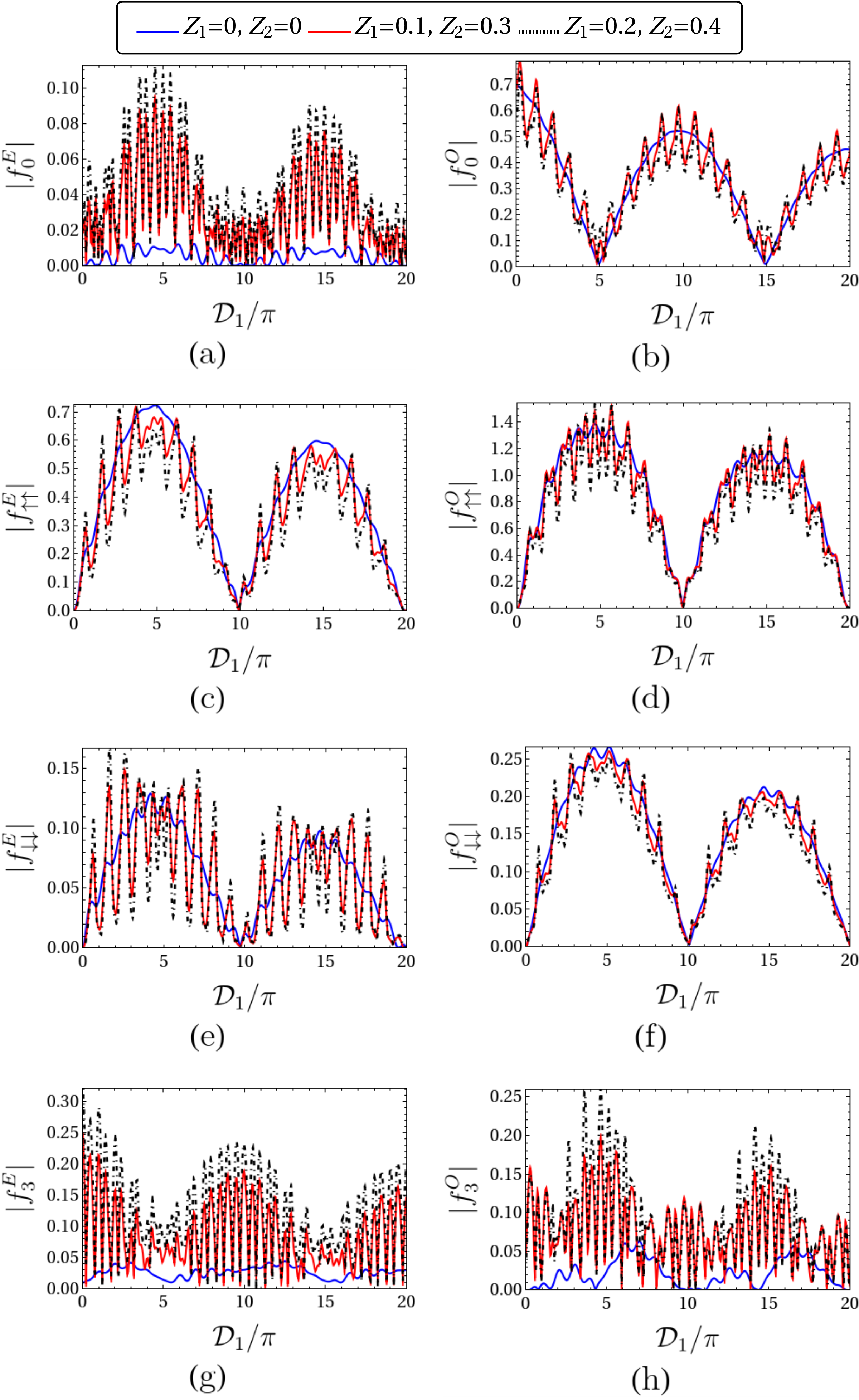}}
\vskip -0.1 in \caption{\small \sl {Absolute values of the non-local ($x\neq\chi$) even- and odd-$\omega$ SS (a,b), EST (c,d,e,f), and MST (g,h) pairing amplitudes in the $F_y$ layer at $x=\mathcal{D}_{1}+\frac{\mathcal{D}_{2}}{2}$ vs. length $\mathcal{D}_{1}$ for a trilayer S-$F_x$-$F_y$-$F_z$-S JJ for tunneling ferromagnetic interfaces in the long junction limit, with $\mathcal{D}_{3}=\mathcal{D}_{1}$, considering both transparent and disordered S-$F_x$ and $F_z$-S interfaces. Parameters: $\varphi=0$, $m_1=m_2=m_3=0.1E_F$, $E_F=1000\Delta_0$, ${Z}_{3}={Z}_{4}=1$, $\mathcal{D}_2=5\pi$, $T/T_c=0.5$, $\chi=\mathcal{D}_{1}$.}}
\end{figure}
considering both transparent and disordered S-$F_x$ and $F_z$-S interfaces. We see that even-$\omega$ SS pairing shows rapid oscillations, which is not observed in absence of anomalous Josephson current in Fig.~14 for a bilayer setup. However, odd-$\omega$ SS pairing exhibits zeros at $\mathcal{D}_{1}\approx5\pi,15\pi$, while it shows a peak at $\mathcal{D}_{1}\approx10\pi$. Similar result is also seen in Fig.~14(b).
Even- and odd-$\omega$ EST pairings exhibit peaks at $\mathcal{D}_{1}\approx5\pi,15\pi$, while they show zero at $\mathcal{D}_{1}\approx10\pi$. This behavior is also seen in Figs.~14(c,d,e,f) in the absence of anomalous Josephson current. Even- and odd-$\omega$ MST pairings exhibit rapid oscillations which is not seen in Figs.~14(g,h) for a bilayer S-$F_x$-$F_y$-S JJ in long junction limit. Further, disordered ferromagnet-superconductor interfaces induce oscillations in odd-$\omega$ SS and even- and odd-$\omega$ EST pairings. Thus, in a trilayer S-$F_x$-$F_y$-$F_z$-S JJ with tunneling ferromagnetic interfaces, the anomalous Josephson current is found to be nearly proportional to the non-local EST pairing. In contrast, a bilayer S-$F_x$-$F_y$-S junction exhibits no correlation between the anomalous Josephson current and non-local EST pairing, as anomalous Josephson current vanishes itself in a bilayer S-$F_x$-$F_y$-S JJ.
\subsection{Transparent ferromagnetic interfaces}
Here, we present the results for transparent ferromagnetic interfaces for the setups shown in Figs.~12(a) and 12(b).
\subsubsection{Bilayer S-$F_x$-$F_y$-S Josephson junction}
In Fig.~19, the absolute values of the local ($x=\chi$) even-$\omega$ SS, odd-$\omega$ EST, and odd-$\omega$ MST pairing amplitudes are plotted at the center of the $F_y$ layer (same as in Ref.~\cite{jfl}) as function of length $\mathcal{D}_{1}$ for a bilayer S-$F_x$-$F_y$-S JJ for transparent ferromagnetic interface, in the long junction limit, considering both transparent and disordered S-$F_x$ and $F_y$-S interfaces. We notice that local even-$\omega$ SS pairing exhibits a dip at length $\mathcal{D}_{1}\approx10\pi$. Local odd-$\omega$ EST pairing shows zeros at $\mathcal{D}_{1}\approx10\pi,20\pi$, while it exhibits peaks at $\mathcal{D}_{1}\approx5\pi, 15\pi$. Local odd-$\omega$ MST pairing does not show any peaks/zeros. Further, local even- and odd-$\omega$ pairings exhibit oscillations due to disordered ferromagnet-superconductor interfaces. In Fig.~19 for a bilayer S-$F_x$-$F_y$-S JJ in the long junction limit too, there is no resemblance between anomalous Josephson current which vanishes and local EST correlations which are finite.
\begin{figure}[ht!]
\centering{\includegraphics[width=0.47\textwidth]{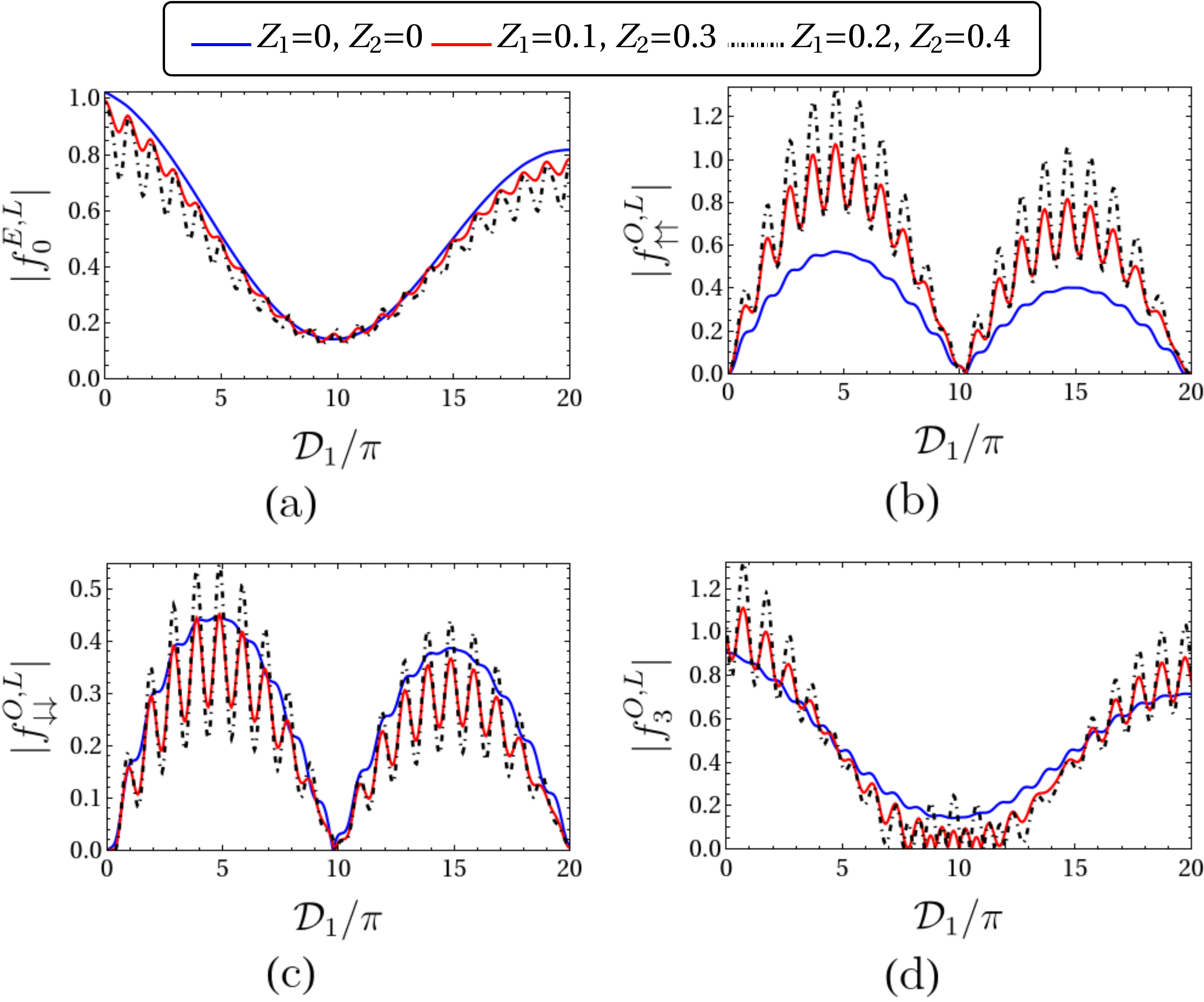}}
\vskip -0.1 in \caption{\small \sl {Absolute values of the local ($x=\chi$) even-$\omega$ SS (a), odd-$\omega$ EST (b,c), and odd-$\omega$ MST (d) pairing amplitudes in the $F_y$ layer at $x=\mathcal{D}_{1}+\frac{\mathcal{D}_{2}}{2}$ vs. length $\mathcal{D}_{1}$ for a bilayer S-$F_x$-$F_y$-S JJ for transparent ferromagnetic interface in the long junction limit, considering both transparent and disordered S-$F_x$ and $F_y$-S interfaces. Parameters: $\varphi=0$, $m_1=m_2=0.1E_F$, $E_F=1000\Delta_0$, ${Z}=0$, $\mathcal{D}_2=5\pi$, $T/T_c=0.5$.}}
\end{figure}}

{In Fig.~20, we present the absolute values of the non-local ($x\neq\chi$) even- and odd-$\omega$ SS, EST, and MST pairing amplitudes at the center of the $F_y$ layer as function of the length $\mathcal{D}_{1}$ in a bilayer S-$F_x$-$F_y$-S JJ
\begin{figure}[ht!]
\centering{\includegraphics[width=0.47\textwidth]{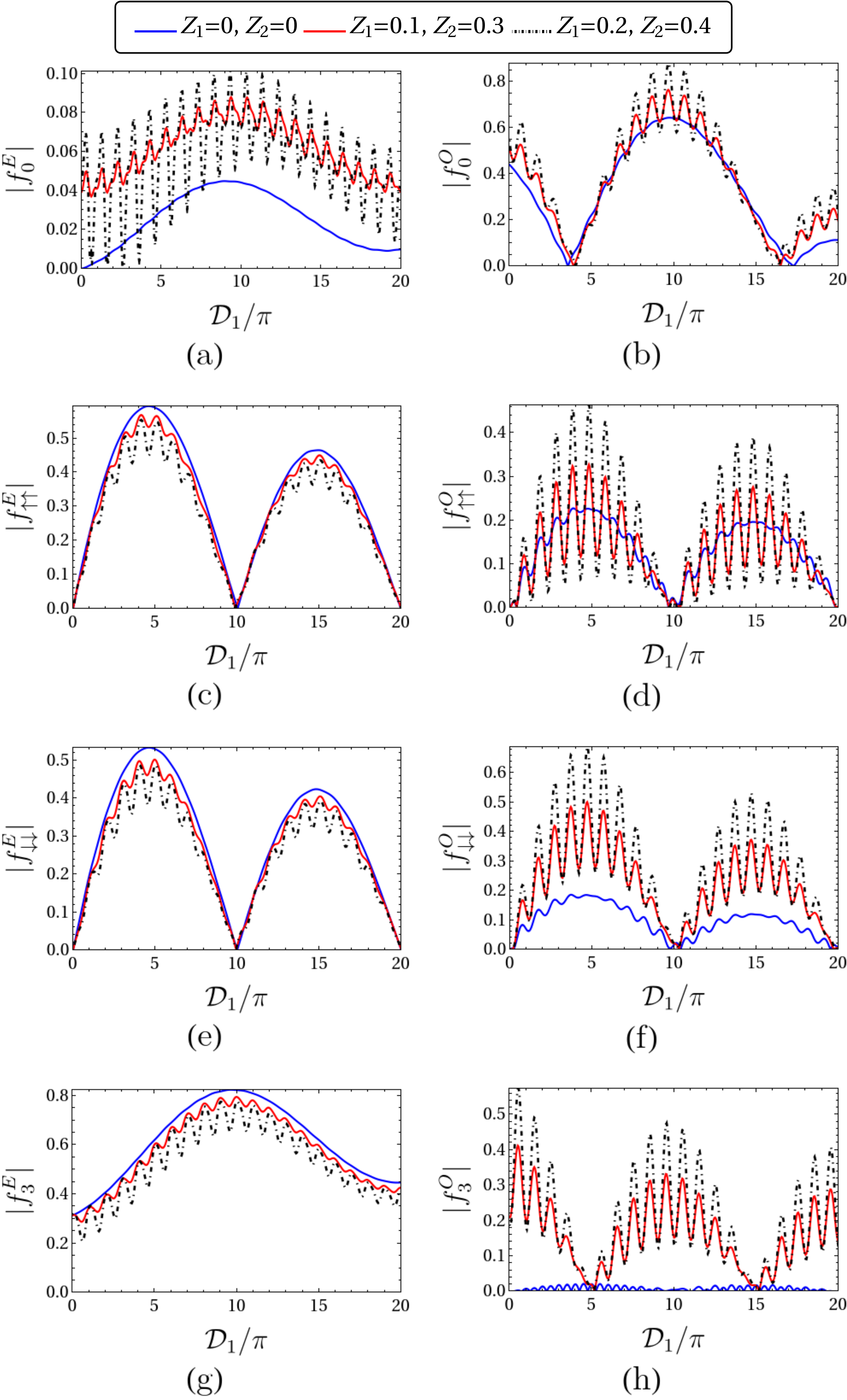}}
\vskip -0.1 in \caption{\small \sl {Absolute values of the non-local ($x\neq\chi$) even- and odd-$\omega$ SS (a,b), EST (c,d,e,f), and MST (g,h) pairing amplitudes in the $F_y$ layer at $x=\mathcal{D}_{1}+\frac{\mathcal{D}_{2}}{2}$ vs. length $\mathcal{D}_{1}$ for a bilayer S-$F_x$-$F_y$-S JJ for transparent ferromagnetic interface in the long junction limit, considering both transparent and disordered S-$F_x$ and $F_y$-S interfaces. Parameters: $\varphi=0$, $m_1=m_2=0.1E_F$, $E_F=1000\Delta_0$, ${Z}=0$, $\mathcal{D}_2=5\pi$, $T/T_c=0.5$, $\chi=\mathcal{D}_{1}$.}}
\end{figure}
for transparent ferromagnetic interface and in long junction limit, considering both transparent and disordered S-$F_x$ and $F_y$-S interfaces. For transparent ferromagnet-superconductor interfaces, we see that even-$\omega$ SS pairing exhibits a peak at $\mathcal{D}_{1}\approx10\pi$. Odd-$\omega$ SS pairing shows zeros at $\mathcal{D}_{1}\approx3\pi,17\pi$, while it exhibits a peak at $\mathcal{D}_{1}\approx10\pi$ in long junction limit. Further, we notice that both even- and odd-$\omega$ EST pairings exhibit peaks at $\mathcal{D}_{1}\approx5\pi,15\pi$, while only even-$\omega$ EST pairing shows zero at $\mathcal{D}_{1}\approx10\pi$. Even-$\omega$ MST pairing exhibits a peak at $\mathcal{D}_{1}\approx10\pi$, while odd-$\omega$ MST pairing shows rapid oscillations. In the presence of disordered ferromagnet-superconductor interfaces, the overall behavior remains qualitatively similar; however, disordered ferromagnet-superconductor interfaces induce oscillations in the non-local even- and odd-$\omega$ SS, EST, and MST pairings.
\subsubsection{Trilayer S-$F_x$-$F_y$-$F_z$-S Josephson junction}
In Fig.~21, we plot the absolute value of anomalous Josephson current as function of length $\mathcal{D}_{1}$ for a trilayer S-$F_x$-$F_y$-$F_z$-S JJ for transparent ferromagnetic interfaces, again for a long junction, considering both transparent and disordered S-$F_x$ and $F_z$-S interfaces. We see that anomalous Josephson current exhibits rapid oscillations as function of length $\mathcal{D}_{1}$. Further, the magnitude of the anomalous Josephson
current is much smaller for transparent as compared to tunneling ferromagnetic interfaces.
\begin{figure}[ht!]
\centering{\includegraphics[width=0.35\textwidth]{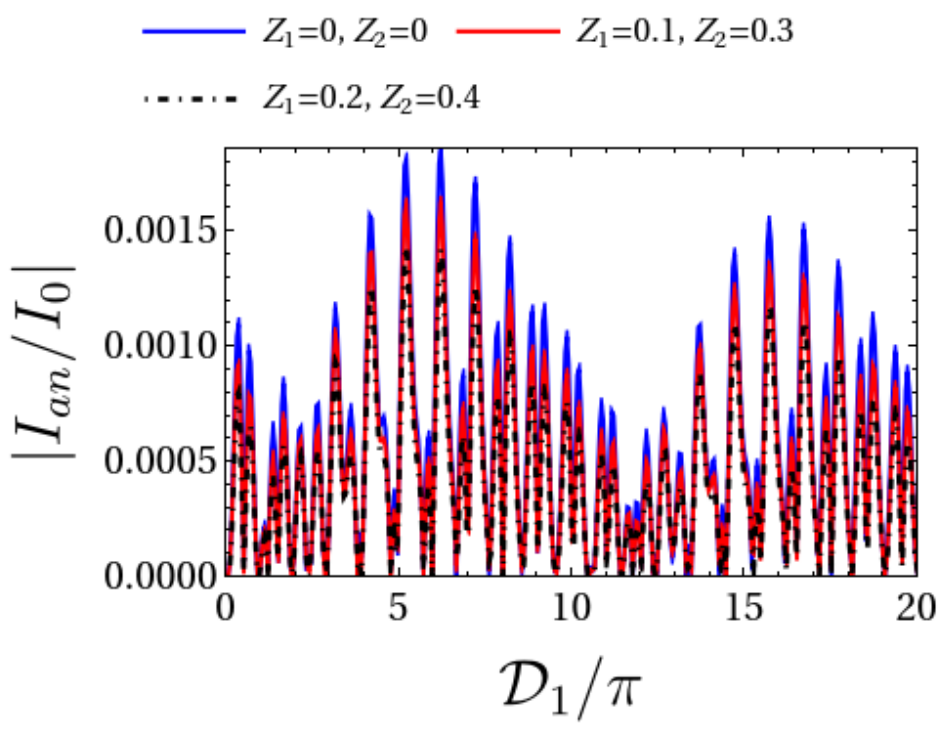}}
\vskip -0.1 in \caption{\small \sl {Absolute value of anomalous Josephson current as function of length $\mathcal{D}_{1}$ in a trilayer S-$F_x$-$F_y$-$F_z$-S JJ for transparent ferromagnetic interfaces in the long junction limit, with $\mathcal{D}_{3}=\mathcal{D}_{1}$, considering both transparent and disordered S-$F_x$ and $F_z$-S interfaces. Parameters: $\varphi=0$, $m_1=m_2=m_3=0.1E_F$, $E_F=1000\Delta_0$, ${Z}_{3}={Z}_{4}=0$, $\mathcal{D}_2=5\pi$, $T/T_c=0.5$, $I_{0}=e\Delta_{0}/\hbar$.}}
\end{figure}}

{To check the impact of anomalous Josephson current on local ($x=\chi$) odd-$\omega$ ST pairing, Fig.~22 shows the absolute values of the local even-$\omega$ SS, odd-$\omega$ EST, and odd-$\omega$ MST pairing amplitudes, plotted at the center of the $F_y$ layer as function of the length $\mathcal{D}_{1}$ in a bilayer S-$F_x$-$F_y$-$F_z$-S JJ for transparent ferromagnetic interfaces and in the long junction limit, considering both transparent and disordered S-$F_x$ and $F_z$-S interfaces.
\begin{figure}[ht!]
\centering{\includegraphics[width=0.47\textwidth]{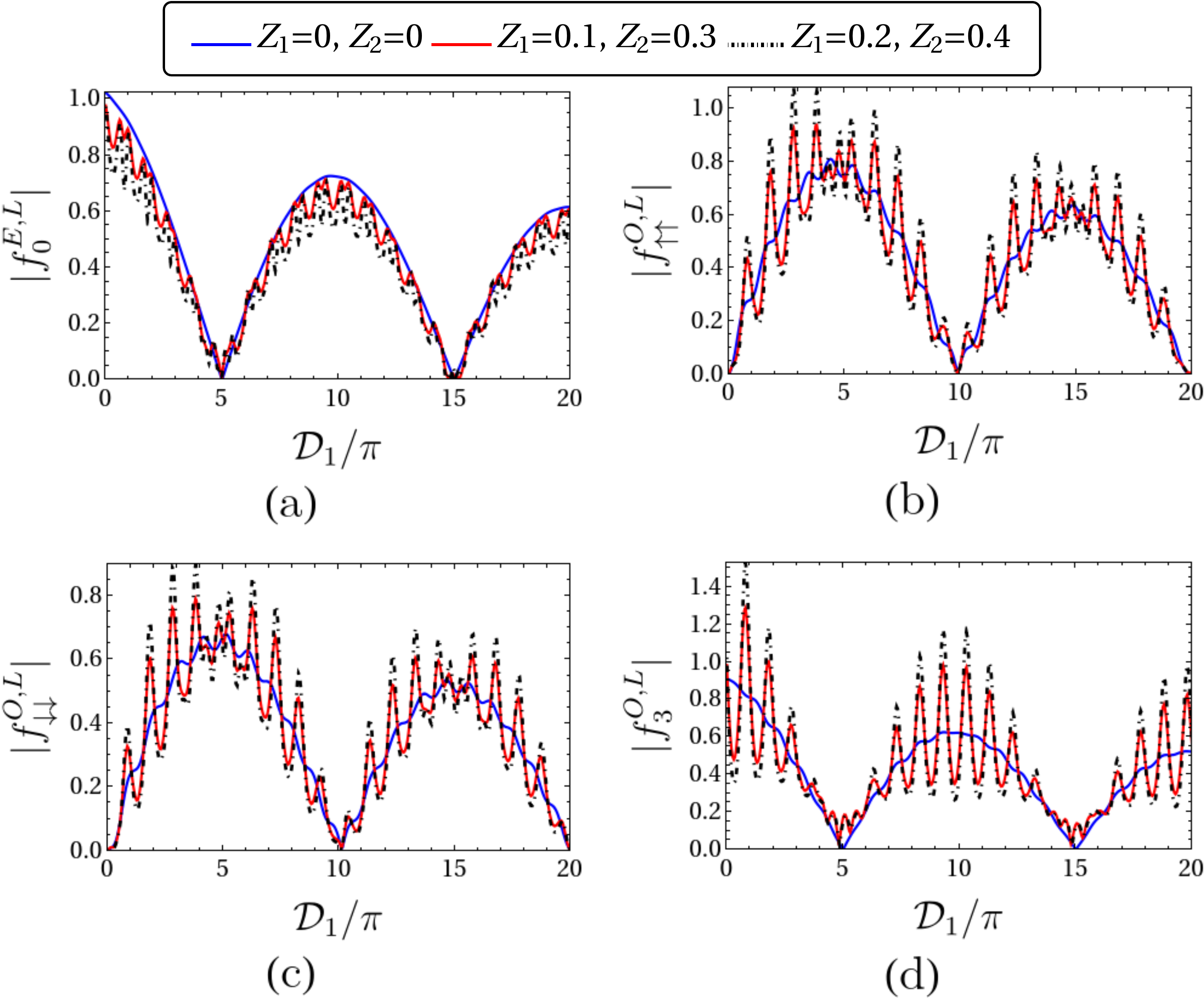}}
\vskip -0.1 in \caption{\small \sl {Absolute values of the local ($x=\chi$) even-$\omega$ SS (a), odd-$\omega$ EST (b,c), and odd-$\omega$ MST (d) pairing amplitudes in the $F_y$ layer at $x=\mathcal{D}_{1}+\frac{\mathcal{D}_{2}}{2}$ vs. length $\mathcal{D}_{1}$ for a trilayer S-$F_x$-$F_y$-$F_z$-S JJ for transparent ferromagnetic interfaces in the long junction limit, and $\mathcal{D}_{3}=\mathcal{D}_{1}$, considering both transparent and disordered S-$F_x$ and $F_z$-S interfaces. Parameters: $\varphi=0$, $m_1=m_2=m_3=0.1E_F$, $E_F=1000\Delta_0$, ${Z}_{3}={Z}_{4}=0$, $\mathcal{D}_2=5\pi$, $T/T_c=0.5$.}}
\end{figure}
We see that local even-$\omega$ SS pairing shows zeros at $\mathcal{D}_{1}\approx5\pi, 15\pi$, while it exhibits a peak at $\mathcal{D}_{1}\approx10\pi$. This behavior is not observed in the absence of anomalous Josephson current. Local odd-$\omega$ EST pairing exhibits zeros at $\mathcal{D}_{1}\approx10\pi, 20\pi$, while it shows peaks at $\mathcal{D}_{1}\approx5\pi, 15\pi$. Similar results are also seen in Figs.~19(b,c). Local odd-$\omega$ MST pairing displays zeros at $\mathcal{D}_{1}\approx5\pi, 15\pi$, while it displays a peak at $\mathcal{D}_{1}\approx10\pi$ in the long junction limit.
Further, disordered ferromagnet-superconductor interfaces induce oscillations in local even- and odd-$\omega$ pairings. Thus in a trilayer S-$F_x$-$F_y$-$F_z$-S junction for transparent ferromagnetic interfaces, anomalous Josephson current and local ST pairing bear no resemblance whatsoever.}

{Finally, Fig.~23 shows the absolute values of the non-local ($x\neq\chi$) even- and odd-$\omega$
\begin{figure}[ht!]
\centering{\includegraphics[width=0.47\textwidth]{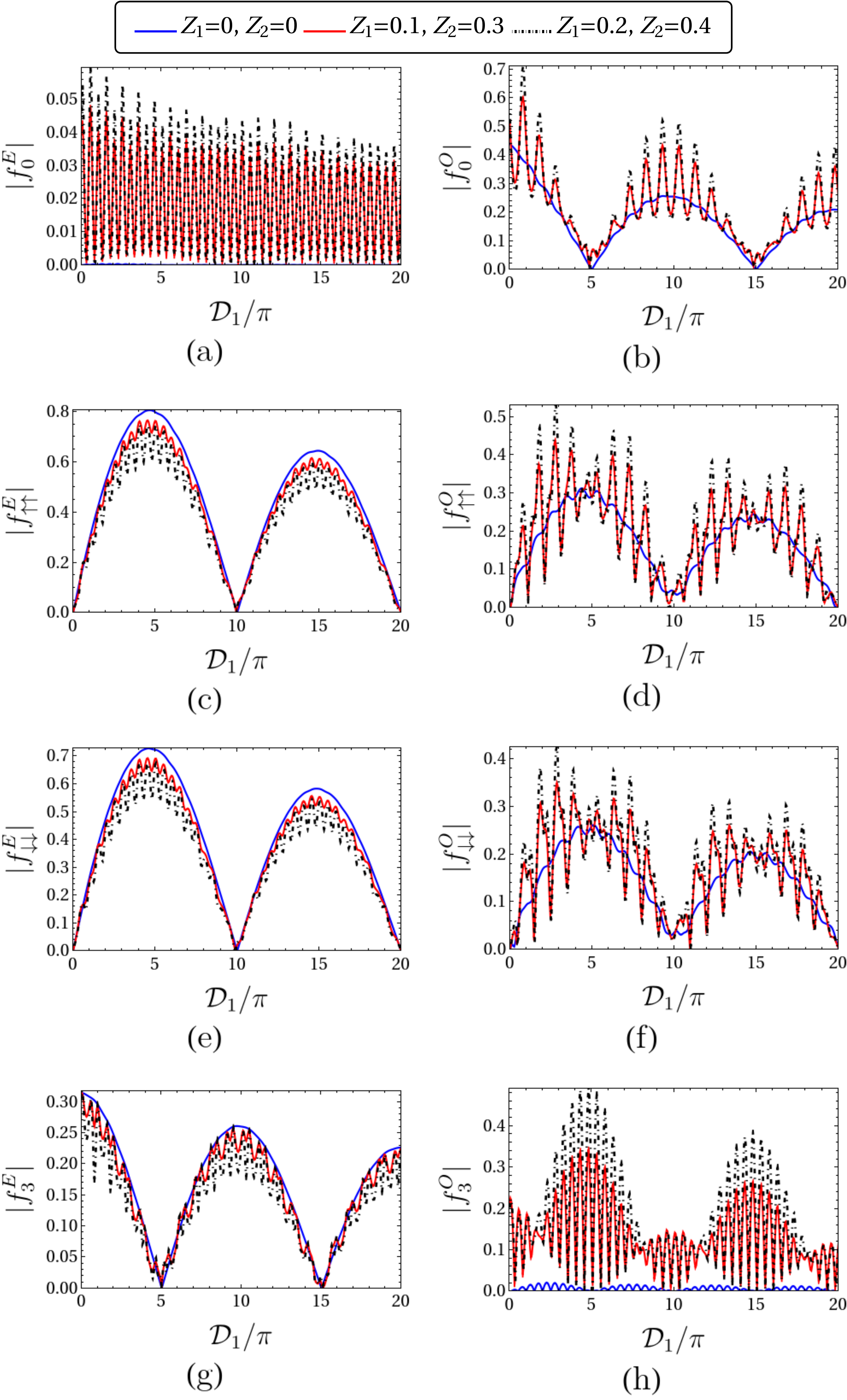}}
\vskip -0.1 in \caption{\small \sl {Absolute values of the non-local ($x\neq\chi$) even- and odd-$\omega$ SS (a,b), EST (c,d,e,f), and MST (g,h) pairing amplitudes in the $F_y$ layer at $x=\mathcal{D}_{1}+\frac{\mathcal{D}_{2}}{2}$ vs. length $\mathcal{D}_{1}$ for a trilayer S-$F_x$-$F_y$-$F_z$-S JJ for transparent ferromagnetic interfaces in the long junction limit, and $\mathcal{D}_{3}=\mathcal{D}_{1}$, considering both transparent and disordered S-$F_x$ and $F_z$-S interfaces. Parameters: $\varphi=0$, $m_1=m_2=m_3=0.1E_F$, $E_F=1000\Delta_0$, ${Z}_{1}={Z}_{2}=0$, $\mathcal{D}_2=5\pi$, $T/T_c=0.5$, $\chi=\mathcal{D}_{1}$.}}
\end{figure}
SS, EST, and MST pairing amplitudes at the center of the $F_y$ layer as function of the length $\mathcal{D}_{1}$ in a trilayer S-$F_x$-$F_y$-$F_z$-S JJ for transparent ferromagnetic interfaces and in long junction limit, considering both transparent and disordered S-$F_x$ and $F_z$-S interfaces. We see that even-$\omega$ SS pairing exhibits rapid oscillations. This behavior is not seen in the absence of anomalous Josephson current. Odd-$\omega$ SS pairing shows zeros at $\mathcal{D}_{1}\approx5\pi,15\pi$, while it exhibits a peak at $\mathcal{D}_{1}\approx10\pi$. Similar result is also noticed in Fig.~20(b). Further, we notice that both even- and odd-$\omega$ EST pairings show peaks at $\mathcal{D}_{1}\approx5\pi,15\pi$, while only even-$\omega$ EST pairing exhibits zero at $\mathcal{D}_{1}\approx10\pi$. This behavior is also seen in Figs.~20(c,d,e,f) in the absence of anomalous Josephson current. Even-$\omega$ MST pairing exhibits a peak at $\mathcal{D}_{1}\approx10\pi$, while it shows zeros at $\mathcal{D}_{1}\approx5\pi,15\pi$. Odd-$\omega$ MST pairing exhibits rapid oscillations, which are also seen in Fig.~20(h). Further, disordered ferromagnet-superconductor interfaces induce oscillations in odd-$\omega$ SS, even- and odd-$\omega$ EST, and even-$\omega$ MST pairings. Thus, in a trilayer S-$F_x$-$F_y$-$F_z$-S JJ for transparent ferromagnetic interfaces, the anomalous Josephson current shows no correlation with non-local EST pairing. We provide a comparison of anomalous Josephson current, local and non-local pairing amplitudes between bilayer S-$F_x$-$F_y$-S, and trilayer S-$F_x$-$F_y$-$F_z$-S JJs in tabular form in Appendix E.}

\section{Analysis}
We compare the anomalous Josephson current, as well as local and non-local correlations in both short and long junction limits across the two setups in Table I. We consider both tunneling and transparent interfaces between the ferromagnetic layers. While in the bilayer JJ, the anomalous Josephson current vanishes, for the trilayer JJ, the anomalous Josephson current is finite and in the short junction limit,  exhibits peaks for tunneling ferromagnetic interfaces, whereas for transparent ferromagnetic interfaces it shows both peaks and zeros. In the long junction limit, the anomalous Josephson current displays both peaks and zeros for tunneling ferromagnetic interfaces, while for transparent ferromagnetic interfaces it undergoes rapid oscillations. Similar to Ref.~\cite{jfl}, we find a correlation between anomalous Josephson current and odd-$\omega$ EST pairing in trilayer JJ for tunneling ferromagnetic interfaces in the long junction limit. We notice that peaks and zeros in anomalous Josephson current coincide with peaks and zeros in odd-$\omega$ EST pairing, and they are nearly proportional. However this is not universally valid and it does not hold for transparent ferromagnetic interfaces in the same trilayer JJ. Further, it does not hold for bilayer JJ in both short and long junction limits and regardless of tunneling or transparent ferromagnetic interfaces. In Ref.~\cite{jfl}, when odd-$\omega$ EST pairing appeared large in presence of a finite anomalous Josephson current, it gave the impression of a causal connection. However, Ref.~\cite{jfl} did not examine whether odd-$\omega$ EST correlations would remain similarly large if the anomalous Josephson current were to be absent or when tunneling ferromagnetic interfaces were replaced with transparent ferromagnetic interfaces or when long junction was replaced with a short junction. In contrast, our article is the first to conduct these controlled comparisons: using identical materials, magnetization strengths, disorder conditions, and interface barriers, we consider
two geometries, one that supports anomalous Josephson current and another that, by symmetry,
cannot. We find that the odd-$\omega$ EST pairing remains largely unchanged between the two cases, despite the presence of anomalous Josephson current in one setup and its absence in the other. We get similar results for both tunneling and transparent ferromagnetic interfaces, as well as in both the short and long junction limits, as summarized in Table I. There are a total of $80$ cases involving both bilayer/trilayer ferromagnetic JJ, short \& long junction limits, tunneling/transparent ferromagnetic
interfaces and local/non-local pairing; out of which only $4$ cases are considered by Ref.~\cite{jfl}. These $4$ cases are marked by bold dashed lines in Table I. However,
in this article, we have done an exhaustive study of all $80$ different cases. In Ref.~\cite{jfl}, only tunneling interfaces are considered between the ferromagnetic layers, while in our article, we consider both tunneling and transparent interfaces between the ferromagnetic layers, and ferromagnet-superconductor interfaces in our article are transparent as well as can be disordered, while in Ref.~\cite{jfl}, they consider ferromagnet-superconductor interfaces to be transparent. In Ref.~\cite{jfl}, they only consider trilayer S-$F_x$-$F_y$-$F_z$-S JJ, to probe the relation between anomalous Josephson current and odd-$\omega$ EST pairing. However, in our article, we choose both bilayer S-$F_x$-$F_y$-S and trilayer S-$F_x$-$F_y$-$F_z$-S JJs to examine the relation between anomalous Josephson current and odd-$\omega$ EST pairing. In Ref.~\cite{jfl}, only the long junction limit is considered, while in our article, we consider both short and long junction limits. Further, in our article we consider both local as well as non-local SS/ST pairing, while Ref.~\cite{jfl} considered only local SS/ST pairing. In our article, for tunneling ferromagnetic interfaces, anomalous Josephson current is nearly proportional to the odd-$\omega$ EST pairing, while for transparent ferromagnetic interfaces, there is no correlation between anomalous Josephson current and odd-$\omega$ EST pairing for a trilayer S-$F_x$-$F_y$-$F_z$-S JJ. In Ref.~\cite{jfl}, anomalous Josephson current is nearly proportional to the odd-$\omega$ EST pairing for tunneling ferromagnetic interfaces and in the long junction limit. In our article, we show odd-$\omega$ EST pairing does not depend on spin precession.
\begin{table*}[ht!]
\scriptsize
\caption{\footnotesize\protect{Summarizing our work: Comparing anomalous Josephson current, local and non-local correlations between bilayer, and trilayer JJs. There are a total of $80$ different cases, out of which only $4$ cases are studied in Ref.~\cite{jfl}. Those $4$ cases are indicated by bold dashed lines. While Ref.~\cite{jfl} reports a near proportionality between anomalous Josephson current and odd-$\omega$ EST pairing for long trilayer junctions with tunneling ferromagnetic interfaces, our systematic analysis across both bilayer and trilayer JJs with both tunneling and transparent ferromagnetic interfaces and in both short and long junction limits reveals no such universal behavior. Text in italic indicates differences between anomalous Josephson current and SS/ST pairing, while text in regular font indicates similarity.}}
{
\begin{tabular}{
|p{0.9cm}
|p{2cm}
|p{2cm}
|p{2cm}
|p{2cm}
|p{1.765cm}
|p{1.765cm}
|p{2.15cm}
|p{2.2cm}|
}
\hline
& \multicolumn{4}{|c|}{Bilayer S-$F_x$-$F_y$-S JJ} 
& \multicolumn{4}{|c|}{Trilayer S-$F_x$-$F_y$-$F_z$-S JJ}\\
\hline
& \multicolumn{2}{|c|}{Short junction limit} 
& \multicolumn{2}{|c|}{Long junction limit} 
& \multicolumn{2}{|c|}{Short junction limit} 
& \multicolumn{2}{|c|}{Long junction limit}\\
\hline
& Tunnel 
& Transparent 
& Tunnel 
& Transparent 
& Tunnel 
& Transparent 
& Tunnel 
& Transparent\\
\hline
\multicolumn{9}{|c|}{ANOMALOUS JOSEPHSON CURRENT}\\\hline
& Absent 
& Absent 
& Absent 
& Absent 
& Finite with peaks at ${m\neq0}$. (Fig.~4(a))
& Finite with peaks and zeros at ${m\neq0}$. (Fig.~9(a))
& \tikzmark{top_left_1}Exhibits peaks at ${\mathcal{D}_{1}\approx5\pi,15\pi}$, zeros at ${\mathcal{D}_{1}\approx 10\pi, 20\pi}$. (Fig.~(15))
& Exhibits rapid oscillations. (Fig.~(21))\\
\hline
\multicolumn{9}{|c|}{LOCAL CORRELATIONS}\\
\hline
Even-$\omega$ SS 
& \textit{Exhibits peak at $\it{m=0}$, zeros at $\it{m\neq0}$. (Fig.~2(a)) }
& \textit{Exhibits peak at $\it{m=0}$. (Fig.~7(a)) }
& \textit{Exhibits dip at $\it{\mathcal{D}_{1}\approx 10\pi}$. (Fig.~13(a)) }
& \textit{Exhibits dip at $\it{\mathcal{D}_{1}\approx 10\pi}$. (Fig.~19(a)) }
& \textit{Exhibits peak at $\it{\footnotesize{m=0}}$, zeros at $\it{\footnotesize{m\neq0}}$. (Fig.~5(a))}
& \textit{Exhibits peak at $\it{\footnotesize{m=0}}$. (Fig.~10(a))}
& \textit{Exhibits peak at $\it{\mathcal{D}_{1}\approx 10\pi}$, zeros at $\it{\mathcal{D}_{1}\approx 5\pi, 15\pi}$, for zero  disorder, and oscillations for finite disorder. (Fig.~16(a))}
& \textit{Exhibits peak at $\it{\footnotesize{\mathcal{D}_{1}\approx10\pi}}$, zeros at $\it{\footnotesize{\mathcal{D}_{1}}\approx 5\pi, 15\pi}$. (Fig.~22(a))}\\
\cline{1-9}
Odd-$\omega$ EST 
& \textit{Exhibits peaks at $\it{\small{m\neq0}}$. (Figs.~2(b,c)) }
& \textit{Exhibits peaks at $\it{\footnotesize{m\neq0}}$. (Figs.~7(b,c))}
& \textit{Exhibits peaks at $\it{\footnotesize{\mathcal{D}_{1}\approx 5\pi, 15\pi}}$, zeros at $\it{\footnotesize{\mathcal{D}_{1}\approx 10\pi, 20\pi}}$. (Figs.~13(b,c))}
& \textit{Exhibits peaks at $\it{\footnotesize{\mathcal{D}_{1}\approx 5\pi, 15\pi}}$, zeros at $\it{\footnotesize{\mathcal{D}_{1}\approx 10\pi, 20\pi}}$. (Figs.~19(b,c))}
& Exhibits peaks at $\footnotesize{m\neq0}$. (Figs.~5(b,c))
&  \textit{Exhibits peaks at $\it{\footnotesize{m\neq0}}$. (Figs.~10(b,c))}
& Exhibits peaks at $\footnotesize{\mathcal{D}_{1}\approx 5\pi, 15\pi}$, zeros at $\footnotesize{\mathcal{D}_{1}\approx 10\pi, 20\pi}$. (Figs.~16(b,c))
& \textit{Exhibits peaks at $\it{\footnotesize{\mathcal{D}_{1}\approx 5\pi, 15\pi}}$, zeros at $\it{\footnotesize{\mathcal{D}_{1}\approx 10\pi, 20\pi}}$. (Figs.~22(b,c))}\\
\cline{1-9}
Odd-$\omega$ MST 
& \textit{Exhibits peaks at $\it{\footnotesize{m\neq0}}$. (Fig.~2(d))}
& \textit{Exhibits both peaks and zeros at $\it{\footnotesize{m\neq0}}$. (Fig.~7(d))}
& \textit{Finite and does not exhibit any peaks. (Fig.~13(d))}
& \textit{Finite and does not exhibit any peaks. (Fig.~19(d))}
& Exhibits peaks at $\footnotesize{m\neq0}$. (Fig.~5(d))
& \textit{Exhibits peaks at $\it{\footnotesize{m\neq0}}$. (Fig.~10(d))}
& \textit{Exhibits peak at $\it{\mathcal{D}_{1}\approx 10\pi}$, zeros at $\it{\mathcal{D}_{1}\approx 5\pi, 15\pi}$. (Fig.~16(d))} \tikzmark{bottom_right_1}
& \textit{Exhibits peak at $\it{\mathcal{D}_{1}\approx 10\pi}$, zeros at $\it{\mathcal{D}_{1}\approx 5\pi, 15\pi}$. (Fig.~22(d))}\\\hline
\multicolumn{9}{|c|}{NON-LOCAL CORRELATIONS}\\\hline
Even-$\omega$ SS 
& \textit{Exhibits peak at $\it{m=0}$ and zeros at $\it{m\neq0}$. (Fig.~3(a))}
& \textit{Exhibits peak at $\it{\footnotesize{m=0}}$. (Fig.~8(a))}
& \textit{Exhibits peak at $\it{\mathcal{D}_{1}\approx 10\pi}$, zeros at $\it{\mathcal{D}_{1}\approx 4\pi, 16\pi}$ for zero disorder and oscillations for finite disorder. (Fig.~14(a))}
& \textit{Exhibits peak at $\it{\mathcal{D}_{1}\approx 10\pi}$. (Fig.~20(a))}
& \textit{Exhibits peak at $\it{\footnotesize{m=0}}$ and zeros at $\it{\footnotesize{m\neq0}}$. (Fig.~6(a))}
& \textit{Exhibits peak at $\it{\footnotesize{m=0}}$. (Fig.~11(a))}
& \textit{Exhibits rapid oscillations. (Fig.~18(a))}
& Exhibits rapid oscillations. (Fig.~23(a))\\
\cline{1-9}
Odd-$\omega$ SS 
& \textit{Exhibits dip at $\it{m=0}$ and peaks at $\it{m\neq0}$. (Fig.~3(b))}
& \textit{Exhibits dip at $\it{m=0}$ and peaks at $\it{m\neq0}$. (Fig.~8(b))}
& \textit{Exhibits peak at $\it{\mathcal{D}_{1}\approx10\pi}$, zeros at $\it{\mathcal{D}_{1}\approx 3\pi, 17\pi}$. (Fig.~14(b))}
& \textit{Exhibits peak at $\it{\mathcal{D}_{1}\approx10\pi}$, zeros at $\it{\mathcal{D}_{1}\approx 3\pi, 17\pi}$. (Fig.~20(b))}
& \textit{Exhibits dip at $\it{m=0}$ and peaks at $\it{m\neq0}$. (Fig.~6(b))}
& \textit{Exhibits dip at $\it{m=0}$ and peaks at $\it{m\neq0}$. (Fig.~11(b))}
& \textit{Exhibits peak at $\it{\mathcal{D}_{1}\approx 10\pi}$, zeros at $\it{\mathcal{D}_{1}\approx 5\pi, 15\pi}$. (Fig.~18(b))}
& \textit{Exhibits peak at $\it{\mathcal{D}_{1}\approx 10\pi}$, zeros at $\it{\mathcal{D}_{1}\approx 5\pi, 15\pi}$. (Fig.~23(b))}\\
\cline{1-9}
Even-$\omega$ EST 
& \textit{Exhibits both peaks and dips at $\it{m\neq0}$. (Figs.~3(c,e))}
& \textit{Exhibits peaks at $\it{m\neq0}$. (Figs.~8(c,e))}
& \textit{Exhibits peaks at $\it{\mathcal{D}_{1}\approx 5\pi, 15\pi}$, zeros at $\it{\mathcal{D}_{1}\approx 10\pi, 20\pi}$. (Figs.~14(c,e))}
& \textit{Exhibits peaks at $\it{\mathcal{D}_{1}\approx 5\pi, 15\pi}$, zeros at $\it{\mathcal{D}_{1}\approx 10\pi, 20\pi}$. (Figs.~20(c,e))}
& Exhibits peaks at ${m\neq0}$. (Figs.~6(c,e))
& \textit{Exhibits peaks at $\it{m\neq0}$. (Figs.~11(c,e))}
& Exhibits peaks at ${\mathcal{D}_{1}\approx5\pi,15\pi}$, zeros at $\mathcal{D}_{1}\approx10\pi,20\pi$. (Figs.~18(c,e))
& \textit{Exhibits peaks at $\it{\mathcal{D}_{1}\approx5\pi,15\pi}$, zeros at $\it{\mathcal{D}_{1}\approx10\pi,20\pi}$. (Figs.~23(c,e))}\\
\cline{1-9}
Odd-$\omega$ EST 
& \textit{Exhibits peaks at $\it{m\neq0}$. (Figs.~3(d,f))}
& \textit{Exhibits peaks at $\it{m\neq0}$. (Figs.~8(d,f))}
& \textit{Exhibits peaks at $\it{\mathcal{D}_{1}\approx5\pi,15\pi}$, zero at $\it{\mathcal{D}_{1}\approx 10\pi}$. (Figs.~14(d,f))}
& \textit{Exhibits peaks at $\it{\mathcal{D}_{1}\approx5\pi,15\pi}$. (Figs.~20(d,f))}
& Exhibits peaks at ${m\neq0}$. (Figs.~6(d,f))
& \textit{Exhibits peaks at $\it{m\neq0}$. (Figs.~11(d,f))}
& Exhibits peaks at ${\mathcal{D}_{1}\approx5\pi,15\pi}$, zero at ${\mathcal{D}_{1}\approx10\pi}$. (Figs.~18(d,f))
& \textit{Exhibits peaks at $\it{\mathcal{D}_{1}\approx5\pi,15\pi}$. (Figs.~23(d,f))}\\
\cline{1-9}
Even-$\omega$ MST 
& \textit{Exhibits peaks at $\it{m\neq0}$. (Fig.~3(g))}
& \textit{Exhibits both peaks and zeros at $\it{m\neq0}$. (Fig.~8(g))}
& \textit{Finite and does not exhibit any peaks. (Fig.~14(g))}
& \textit{Exhibits a peak at $\it{\mathcal{D}_{1}\approx10\pi}$. (Fig.~20(g))}
& Exhibits peaks at ${m\neq0}$. (Fig.~6(g))
& \textit{Exhibits peaks at $\it{m\neq0}$. (Fig.~11(g))}
& \textit{Exhibits rapid oscillations. (Fig.~18(g))}
& \textit{Exhibits peak at $\it{\mathcal{D}_{1}\approx10\pi}$, zeros at $\it{\mathcal{D}_{1}\approx 5\pi, 15\pi}$. (Fig.~23(g))}\\
\cline{1-9}
Odd-$\omega$ MST 
& \textit{Exhibits peaks at $\it{m\neq0}$. (Fig.~3(h))}
& \textit{Exhibits both peaks and zeros at $\it{m\neq0}$. (Fig.~8(h))}
& \textit{Finite and does not exhibit any peaks. (Fig.~14(h))}
& \textit{Exhibits rapid oscillations. (Fig.~20(h))}
& Exhibits peaks at ${m\neq0}$. (Fig.~6(h))
& \textit{Exhibits peaks at $\it{m\neq0}$. (Fig.~11(h))}
& \textit{Exhibits rapid oscillations. (Fig.~18(h))}
&  Exhibits rapid oscillations. (Fig.~23(h)).\\
\hline
\end{tabular}}
\DrawBox[black,dashed,line width=1.9pt]{top_left_1}{bottom_right_1}
\end{table*}
We see odd-$\omega$ EST pairing is finite for a bilayer S-$F_x$-$F_y$-S JJ, wherein anomalous Josephson current vanishes in both short and long junction limits and for both tunneling and transparent ferromagnetic interfaces. Ref.~\cite{jfl} therefore, erroneously concludes that anomalous Josephson current and odd-$\omega$ EST pairing arise due to the spin precession in the first and third ferromagnetic layers in long junction limit for tunneling ferromagnetic interfaces, in our article, we see this explanation does not hold.

{In ferromagnetic JJs, odd-$\omega$ ST pairing emerges due to the breaking of spatial parity at the interfaces in the presence of spin-flip scattering and spin mixing. Spin-flip scattering generates EST pairing\cite{pal}, while spin mixing gives rise to MST pairing\cite{linn}. In contrast, anomalous Josephson current originates from the simultaneous breaking of both time-reversal and inversion symmetries. In the bilayer setup, consisting of two ferromagnetic layers with magnetization vectors aligned along the $x$-, and $y$-axes, embedded between two $s$-wave superconductors, time-reversal symmetry is broken, but inversion symmetry is preserved (see Sec.~IV.A.1), and consequently, the anomalous Josephson current vanishes. In the trilayer setup, wherein three ferromagnetic layers with magnetization vectors oriented along the $x$-,
$y$-, and $z$-axes, sandwiched between two $s$-wave superconductors, both time-reversal and inversion symmetries are broken (see Sec.~IV.A.2), leading to the emergence of an anomalous Josephson current. However, this breaking of inversion symmetry has no significant influence on even-/odd-$\omega$ ST pairing. As a result, the odd-$\omega$ ST correlations remain largely unchanged across the two setups in both short
and long junction limits as well as for tunneling and transparent ferromagnetic interfaces. The breaking of time-reversal and inversion symmetries is not a necessary condition for the emergence of odd-$\omega$ EST pairing. Anomalous Josephson current and odd-$\omega$ ST pairing
are thus fundamentally independent phenomena, as they originate from different physical mechanisms.}

\section{Experimental realization and summary} The experimental realization of the setups depicted in Figs.~1(a), and 1(b) is feasible in a laboratory setting.
{S-$F_x$-$F_y$-S} JJs have been successfully fabricated in experiments for a considerable period\cite{mco}. In Ref.~\cite{mco}, the authors investigated the transport properties of an SFFS junction, where two ferromagnetic wires bridge the superconductors and are separated by a distance much smaller than the superconducting coherence length. They observed that, upon lowering the temperature below the critical temperature of the superconductor, the resistance for the antiparallel alignment of the ferromagnetic wire magnetizations becomes higher than that for the parallel configuration. Inserting an extra ferromagnet between the existing ferromagnets in an {S-$F_x$-$F_y$-S} junction is practically achievable. These setups are particularly viable when using $s$-wave superconductors such as aluminum or lead, ensuring their experimental realizability. Experimental evidence of the anomalous Josephson effect has been reported in nonequilibrium Andreev interferometers\cite{dan}, where both time-reversal and inversion symmetries are broken. The JDE has also been observed in nanowire-based Andreev molecules\cite{zsh}, three-terminal Josephson devices fabricated from an InAs two-dimensional electron gas proximitized by an epitaxial aluminum layer\cite{mgu}, in JJs containing a single magnetic atom\cite{trah}, as well as in van der Waals heterostructures\cite{hwu}. Moreover, experimental signatures of odd-$\omega$ pairing have been detected in systems containing a single magnetic impurity embedded within an $s$-wave superconductor\cite{perr}.

In summary, we have investigated the influence of the anomalous Josephson current on odd-$\omega$ ST superconducting correlations in a Josephson diode. Our results demonstrate that odd-$\omega$ ST pairing persists and develops pronounced peaks at finite magnetization strengths {for both the tunneling and transparent ferromagnetic interfaces}, irrespective of the presence or absence of an anomalous Josephson current. The spatial profile of these correlations remains essentially unchanged under both conditions. Moreover, the anomalous Josephson current and the Josephson diode effect exert no discernible influence on the magnitude or symmetry of odd-$\omega$ correlations. {We get similar results in both short and long junction limits. Ref.~\cite{jfl} considers only the long junction  limit and that too only for the trilayer JJ and also for a specific transport regime, namely tunneling ferromagnetic interfaces with transparent ferromagnet-superconductor interfaces, and reports that the anomalous Josephson current is nearly proportional to the local odd-$\omega$ EST pairing. The non-local pairings aren\rq{}t discussed either. In contrast, this article examines two distinct setups in both short as well as long junction limits and considers both local as well as non-local pairing correlations for tunneling as well as transparent ferromagnetic interfaces, and finds no correlation between the anomalous Josephson current and the odd-$\omega$ EST pairing in any other regime beyond what was reported in Ref.~\cite{jfl}.} We therefore conclude that anomalous Josephson transport and odd-$\omega$ ST superconductivity are largely independent phenomena, and that previously suggested correlations between them are not universally valid.

\section*{Data availability}
The data that support the findings of this article are openly available \cite{mathematica}.

\appendix
\label{appendix}

\begin{widetext}
{\section*{APPENDIX A: Wavefunctions in bilayer S-$F_x$-$F_y$-S and trilayer S-$F_x$-$F_y$-$F_z$-S Josephson junctions}
In this Appendix, we present the explicit wavefunctions for the
bilayer S-$F_x$-$F_y$-S and trilayer S-$F_x$-$F_y$-$F_z$-S JJs in both the short and long junction limits.
\subsection{Short junction limit}}
We diagonalize the BdG Hamiltonians (Eqs.~\eqref{hmff} and \eqref{hmfff}) to derive the wavefunctions corresponding to various scattering processes in the distinct regions of the {bilayer S-$F_x$-$F_y$-S and trilayer S-$F_x$-$F_y$-$F_z$-S JJs, as shown in Figs.~1(a) and 1(b), respectively. The wavefunctions for bilayer S-$F_x$-$F_y$-S JJ are as follows\cite{enok}:}
\begin{equation}
\begin{split}
\phi_{1}(x)&=\begin{cases}
\varphi_{1}^{S_L}e^{iq_{e}^{S}(x+\mathcal{D}/2)}+r_{eh}^{\uparrow\uparrow}\varphi_{3}^{S_L}e^{iq_{h}^{S}(x+\mathcal{D}/2)}+r_{eh}^{\uparrow\downarrow}\varphi_{4}^{S_L}e^{iq_{h}^{S}(x+\mathcal{D}/2)}+r_{ee}^{\uparrow\uparrow}\varphi_{1}^{S_L}e^{-iq_{e}^{S}(x+\mathcal{D}/2)}\\+r_{ee}^{\uparrow\downarrow}\varphi_{2}^{S_L}e^{-iq_{e}^{S}(x+\mathcal{D}/2)}\,, & x<-\mathcal{D}/2\\
a_{ee}^{\uparrow\uparrow}\varphi_{1}^{F_x}e^{iq_{e1}^{\uparrow}(x+\mathcal{D}/2)}+a_{ee}^{\uparrow\downarrow}\varphi_{2}^{F_x}e^{iq_{e1}^{\downarrow}(x+\mathcal{D}/2)}+b_{ee}^{\uparrow\uparrow}\varphi_{1}^{F_x}e^{-iq_{e1}^{\uparrow}x}+b_{ee}^{\uparrow\downarrow}\varphi_{2}^{F_x}e^{-iq_{e1}^{\downarrow}x}\\+c_{eh}^{\uparrow\uparrow}\varphi_{3}^{F_x}e^{iq_{h1}^{\uparrow}x}+c_{eh}^{\uparrow\downarrow}\varphi_{4}^{F_x}e^{iq_{h1}^{\downarrow}x}+d_{eh}^{\uparrow\uparrow}\varphi_{3}^{F_x}e^{-iq_{h1}^{\uparrow}(x+\mathcal{D}/2)}+d_{eh}^{\uparrow\downarrow}\varphi_{4}^{F_x}e^{-iq_{h1}^{\downarrow}(x+\mathcal{D}/2)}\,, & -\mathcal{D}/2<x<0\\
e_{ee}^{\uparrow\uparrow}\varphi_{1}^{F_y}e^{-iq_{e2}^{\uparrow}(x-\mathcal{D}/2)}+e_{ee}^{\uparrow\downarrow}\varphi_{2}^{F_y}e^{-iq_{e2}^{\downarrow}(x-\mathcal{D}/2)}+f_{ee}^{\uparrow\uparrow}\varphi_{1}^{F_y}e^{iq_{e2}^{\uparrow}x}+f_{ee}^{\uparrow\downarrow}\varphi_{2}^{F_y}e^{iq_{e2}^{\downarrow}x}\\+g_{eh}^{\uparrow\uparrow}\varphi_{3}^{F_y}e^{-iq_{h2}^{\uparrow}x}+g_{eh}^{\uparrow\downarrow}\varphi_{4}^{F_y}e^{-iq_{h2}^{\downarrow}x}+h_{eh}^{\uparrow\uparrow}\varphi_{3}^{F_y}e^{iq_{h2}^{\uparrow}(x-\mathcal{D}/2)}+h_{eh}^{\uparrow\downarrow}\varphi_{4}^{F_y}e^{iq_{h2}^{\downarrow}(x-\mathcal{D}/2)}\,, & 0<x<\mathcal{D}/2\\
t_{ee}^{\uparrow\uparrow}\varphi_{1}^{S_{R}}e^{iq_{e}^{S}(x-\mathcal{D}/2)}+t_{ee}^{\uparrow\downarrow}\varphi_{2}^{S_{R}}e^{iq_{e}^{S}(x-\mathcal{D}/2)}+t_{eh}^{\uparrow\uparrow}\varphi_{3}^{S_{R}}e^{-iq_{h}^{S}(x-\mathcal{D}/2)}+t_{eh}^{\uparrow\downarrow}\varphi_{4}^{S_{R}}e^{-iq_{h}^{S}(x-\mathcal{D}/2)}\,, & x>\mathcal{D}/2
\end{cases}\\
\phi_{2}(x)&=\begin{cases}
\varphi_{2}^{S_L}e^{iq_{e}^{S}(x+\mathcal{D}/2)}+r_{eh}^{\downarrow\uparrow}\varphi_{3}^{S_L}e^{iq_{h}^{S}(x+\mathcal{D}/2)}+r_{eh}^{\downarrow\downarrow}\varphi_{4}^{S_L}e^{iq_{h}^{S}(x+\mathcal{D}/2)}+r_{ee}^{\downarrow\uparrow}\varphi_{1}^{S_L}e^{-iq_{e}^{S}(x+\mathcal{D}/2)}\\+r_{ee}^{\downarrow\downarrow}\varphi_{2}^{S_L}e^{-iq_{e}^{S}(x+\mathcal{D}/2)}\,, & x<-\mathcal{D}/2\\
a_{ee}^{\downarrow\uparrow}\varphi_{1}^{F_x}e^{iq_{e1}^{\uparrow}(x+\mathcal{D}/2)}+a_{ee}^{\downarrow\downarrow}\varphi_{2}^{F_x}e^{iq_{e1}^{\downarrow}(x+\mathcal{D}/2)}+b_{ee}^{\downarrow\uparrow}\varphi_{1}^{F_x}e^{-iq_{e1}^{\uparrow}x}+b_{ee}^{\downarrow\downarrow}\varphi_{2}^{F_x}e^{-iq_{e1}^{\downarrow}x}\\+c_{eh}^{\downarrow\uparrow}\varphi_{3}^{F_x}e^{iq_{h1}^{\uparrow}x}+c_{eh}^{\downarrow\downarrow}\varphi_{4}^{F_x}e^{iq_{h1}^{\downarrow}x} +d_{eh}^{\downarrow\uparrow}\varphi_{3}^{F_x}e^{-iq_{h1}^{\uparrow}(x+\mathcal{D}/2)}+d_{eh}^{\downarrow\downarrow}\varphi_{4}^{F_x}e^{-iq_{h1}^{\downarrow}(x+\mathcal{D}/2)}\,, &  -\mathcal{D}/2<x<0\\
e_{ee}^{\downarrow\uparrow}\varphi_{1}^{F_y}e^{-iq_{e2}^{\uparrow}(x-\mathcal{D}/2)}+e_{ee}^{\downarrow\downarrow}\varphi_{2}^{F_y}e^{-iq_{e2}^{\downarrow}(x-\mathcal{D}/2)}+f_{ee}^{\downarrow\uparrow}\varphi_{1}^{F_y}e^{iq_{e2}^{\uparrow}x}+f_{ee}^{\downarrow\downarrow}\varphi_{2}^{F_y}e^{iq_{e2}^{\downarrow}x}\\+g_{eh}^{\downarrow\uparrow}\varphi_{3}^{F_y}e^{-iq_{h2}^{\uparrow}x}+g_{eh}^{\downarrow\downarrow}\varphi_{4}^{F_y}e^{-iq_{h2}^{\downarrow}x}+h_{eh}^{\downarrow\uparrow}\varphi_{3}^{F_y}e^{iq_{h2}^{\uparrow}(x-\mathcal{D}/2)}+h_{eh}^{\downarrow\downarrow}\varphi_{4}^{F_y}e^{iq_{h2}^{\downarrow}(x-\mathcal{D}/2)}\,, & 0<x<\mathcal{D}/2\\
t_{ee}^{\downarrow\uparrow}\varphi_{1}^{S_{R}}e^{iq_{e}^{S}(x-\mathcal{D}/2)}+t_{ee}^{\downarrow\downarrow}\varphi_{2}^{S_{R}}e^{iq_{e}^{S}(x-\mathcal{D}/2)}+t_{eh}^{\downarrow\uparrow}\varphi_{3}^{S_{R}}e^{-iq_{h}^{S}(x-\mathcal{D}/2)}+t_{eh}^{\downarrow\downarrow}\varphi_{4}^{S_{R}}e^{-iq_{h}^{S}(x-\mathcal{D}/2)}\,, & x>\mathcal{D}/2
\end{cases}\\
\phi_{3}(x)&=\begin{cases}
\varphi_{3}^{S_L}e^{-iq_{h}^{S}(x+\mathcal{D}/2)}+r_{hh}^{\uparrow\uparrow}\varphi_{3}^{S_L}e^{iq_{h}^{S}(x+\mathcal{D}/2)}+r_{hh}^{\uparrow\downarrow}\varphi_{4}^{S_L}e^{iq_{h}^{S}(x+\mathcal{D}/2)}+r_{he}^{\uparrow\uparrow}\varphi_{1}^{S_L}e^{-iq_{e}^{S}(x+\mathcal{D}/2)}\\+r_{he}^{\uparrow\downarrow}\varphi_{2}^{S_L}e^{-iq_{e}^{S}(x+\mathcal{D}/2)}\,, & x<-\mathcal{D}/2\\
a_{he}^{\uparrow\uparrow}\varphi_{1}^{F_x}e^{iq_{e1}^{\uparrow}(x+\mathcal{D}/2)}+a_{he}^{\uparrow\downarrow}\varphi_{2}^{F_x}e^{iq_{e1}^{\downarrow}(x+\mathcal{D}/2)}+b_{he}^{\uparrow\uparrow}\varphi_{1}^{F_x}e^{-iq_{e1}^{\uparrow}x}+b_{he}^{\uparrow\downarrow}\varphi_{2}^{F_x}e^{-iq_{e1}^{\downarrow}x}\\+c_{hh}^{\uparrow\uparrow}\varphi_{3}^{F_x}e^{iq_{h1}^{\uparrow}x}+c_{hh}^{\uparrow\downarrow}\varphi_{4}^{F_x}e^{iq_{h1}^{\downarrow}x}
+d_{hh}^{\uparrow\uparrow}\varphi_{3}^{F_x}e^{-iq_{h1}^{\uparrow}(x+\mathcal{D}/2)}+d_{hh}^{\uparrow\downarrow}\varphi_{4}^{F_x}e^{-iq_{h1}^{\downarrow}(x+\mathcal{D}/2)}\,, & -\mathcal{D}/2<x<0\\
e_{he}^{\uparrow\uparrow}\varphi_{1}^{F_y}e^{-iq_{e2}^{\uparrow}(x-\mathcal{D}/2)}+e_{he}^{\uparrow\downarrow}\varphi_{2}^{F_y}e^{-iq_{e2}^{\downarrow}(x-\mathcal{D}/2)}+f_{he}^{\uparrow\uparrow}\varphi_{1}^{F_y}e^{iq_{e2}^{\uparrow}x}+f_{he}^{\uparrow\downarrow}\varphi_{2}^{F_y}e^{iq_{e2}^{\downarrow}x}\\+g_{hh}^{\uparrow\uparrow}\varphi_{3}^{F_y}e^{-iq_{h2}^{\uparrow}x}+g_{hh}^{\uparrow\downarrow}\varphi_{4}^{F_y}e^{-iq_{h2}^{\downarrow}x}+h_{hh}^{\uparrow\uparrow}\varphi_{3}^{F_y}e^{iq_{h2}^{\uparrow}(x-\mathcal{D}/2)}+h_{hh}^{\uparrow\downarrow}\varphi_{4}^{F_y}e^{iq_{h2}^{\downarrow}(x-\mathcal{D}/2)}\,, & 0<x<\mathcal{D}/2\\
t_{he}^{\uparrow\uparrow}\varphi_{1}^{S_{R}}e^{iq_{e}^{S}(x-\mathcal{D}/2)}+t_{he}^{\uparrow\downarrow}\varphi_{2}^{S_{R}}e^{iq_{e}^{S}(x-\mathcal{D}/2)}+t_{hh}^{\uparrow\uparrow}\varphi_{3}^{S_{R}}e^{-iq_{h}^{S}(x-\mathcal{D}/2)}+t_{hh}^{\uparrow\downarrow}\varphi_{4}^{S_{R}}e^{-iq_{h}^{S}(x-\mathcal{D}/2)}\,, & x>\mathcal{D}/2
\end{cases}\\
\phi_{4}(x)&=\begin{cases}
\varphi_{4}^{S_L}e^{-iq_{h}^{S}(x+\mathcal{D}/2)}+r_{hh}^{\downarrow\uparrow}\varphi_{3}^{S_L}e^{iq_{h}^{S}(x+\mathcal{D}/2)}+r_{hh}^{\downarrow\downarrow}\varphi_{4}^{S_L}e^{iq_{h}^{S}(x+\mathcal{D}/2)}+r_{he}^{\downarrow\uparrow}\varphi_{1}^{S_L}e^{-iq_{e}^{S}(x+\mathcal{D}/2)}\\+r_{he}^{\downarrow\downarrow}\varphi_{2}^{S_L}e^{-iq_{e}^{S}(x+\mathcal{D}/2)}\,, & x<-\mathcal{D}/2\\
a_{he}^{\downarrow\uparrow}\varphi_{1}^{F_x}e^{iq_{e1}^{\uparrow}(x+\mathcal{D}/2)}+a_{he}^{\downarrow\downarrow}\varphi_{2}^{F_x}e^{iq_{e1}^{\downarrow}(x+\mathcal{D}/2)}+b_{he}^{\downarrow\uparrow}\varphi_{1}^{F_x}e^{-iq_{e1}^{\uparrow}x}+b_{he}^{\downarrow\downarrow}\varphi_{2}^{F_x}e^{-iq_{e1}^{\downarrow}x}\\+c_{hh}^{\downarrow\uparrow}\varphi_{3}^{F_x}e^{iq_{h1}^{\uparrow}x}+c_{hh}^{\downarrow\downarrow}\varphi_{4}^{F_x}e^{iq_{h1}^{\downarrow}x} +d_{hh}^{\downarrow\uparrow}\varphi_{3}^{F_x}e^{-iq_{h1}^{\uparrow}(x+\mathcal{D}/2)}+d_{hh}^{\downarrow\downarrow}\varphi_{4}^{F_x}e^{-iq_{h1}^{\downarrow}(x+\mathcal{D}/2)}\,, & -\mathcal{D}/2<x<0\\
e_{he}^{\downarrow\uparrow}\varphi_{1}^{F_y}e^{-iq_{e2}^{\uparrow}(x-\mathcal{D}/2)}+e_{he}^{\downarrow\downarrow}\varphi_{2}^{F_y}e^{-iq_{e2}^{\downarrow}(x-\mathcal{D}/2)}+f_{he}^{\downarrow\uparrow}\varphi_{1}^{F_y}e^{iq_{e2}^{\uparrow}x}+f_{he}^{\downarrow\downarrow}\varphi_{2}^{F_y}e^{iq_{e2}^{\downarrow}x}\\+g_{hh}^{\downarrow\uparrow}\varphi_{3}^{F_y}e^{-iq_{h2}^{\uparrow}x}+g_{hh}^{\downarrow\downarrow}\varphi_{4}^{F_y}e^{-iq_{h2}^{\downarrow}x}+h_{hh}^{\downarrow\uparrow}\varphi_{3}^{F_y}e^{iq_{h2}^{\uparrow}(x-\mathcal{D}/2)}+h_{hh}^{\downarrow\downarrow}\varphi_{4}^{F_y}e^{iq_{h2}^{\downarrow}(x-\mathcal{D}/2)}\,, & 0<x<\mathcal{D}/2\\
t_{he}^{\downarrow\uparrow}\varphi_{1}^{S_{R}}e^{iq_{e}^{S}(x-\mathcal{D}/2)}+t_{he}^{\downarrow\downarrow}\varphi_{2}^{S_{R}}e^{iq_{e}^{S}(x-\mathcal{D}/2)}+t_{hh}^{\downarrow\uparrow}\varphi_{3}^{S_{R}}e^{-iq_{h}^{S}(x-\mathcal{D}/2)}+t_{hh}^{\downarrow\downarrow}\varphi_{4}^{S_{R}}e^{-iq_{h}^{S}(x-\mathcal{D}/2)}\,, & x>\mathcal{D}/2
\end{cases}\\
\phi_{5}(x)&=\begin{cases}
\bar{t}_{ee}^{\uparrow\uparrow}\varphi_{1}^{S_L}e^{-iq_{e}^{S}(x+\mathcal{D}/2)}+\bar{t}_{ee}^{\uparrow\downarrow}\varphi_{2}^{S_L}e^{-iq_{e}^{S}(x+\mathcal{D}/2)}+\bar{t}_{eh}^{\uparrow\uparrow}\varphi_{3}^{S_L}e^{iq_{h}^{S}(x+\mathcal{D}/2)}+\bar{t}_{eh}^{\uparrow\downarrow}\varphi_{4}^{S_L}e^{iq_{h}^{S}(x+\mathcal{D}/2)}\,, & x<-\mathcal{D}/2\\
\bar{a}_{ee}^{\uparrow\uparrow}\varphi_{1}^{F_x}e^{iq_{e1}^{\uparrow}(x+\mathcal{D}/2)}+\bar{a}_{ee}^{\uparrow\downarrow}\varphi_{2}^{F_x}e^{iq_{e1}^{\downarrow}(x+\mathcal{D}/2)}+\bar{b}_{ee}^{\uparrow\uparrow}\varphi_{1}^{F_x}e^{-iq_{e1}^{\uparrow}x}+\bar{b}_{ee}^{\uparrow\downarrow}\varphi_{2}^{F_x}e^{-iq_{e1}^{\downarrow}x}\\+\bar{c}_{eh}^{\uparrow\uparrow}\varphi_{3}^{F_x}e^{iq_{h1}^{\uparrow}x}+\bar{c}_{eh}^{\uparrow\downarrow}\varphi_{4}^{F_x}e^{iq_{h1}^{\downarrow}x}+\bar{d}_{eh}^{\uparrow\uparrow}\varphi_{3}^{F_x}e^{-iq_{h1}^{\uparrow}(x+\mathcal{D}/2)}+\bar{d}_{eh}^{\uparrow\downarrow}\varphi_{4}^{F_x}e^{-iq_{h1}^{\downarrow}(x+\mathcal{D}/2)}\,, & -\mathcal{D}/2<x<0\\
\bar{e}_{ee}^{\uparrow\uparrow}\varphi_{1}^{F_y}e^{-iq_{e2}^{\uparrow}(x-\mathcal{D}/2)}+\bar{e}_{ee}^{\uparrow\downarrow}\varphi_{2}^{F_y}e^{-iq_{e2}^{\downarrow}(x-\mathcal{D}/2)}+\bar{f}_{ee}^{\uparrow\uparrow}\varphi_{1}^{F_y}e^{iq_{e2}^{\uparrow}x}+\bar{f}_{ee}^{\uparrow\downarrow}\varphi_{2}^{F_y}e^{iq_{e2}^{\downarrow}x}\\+\bar{g}_{eh}^{\uparrow\uparrow}\varphi_{3}^{F_y}e^{-iq_{h2}^{\uparrow}x}+\bar{g}_{eh}^{\uparrow\downarrow}\varphi_{4}^{F_y}e^{-iq_{h2}^{\downarrow}x}+\bar{h}_{eh}^{\uparrow\uparrow}\varphi_{3}^{F_y}e^{iq_{h2}^{\uparrow}(x-\mathcal{D}/2)}+\bar{h}_{eh}^{\uparrow\downarrow}\varphi_{4}^{F_y}e^{iq_{h2}^{\downarrow}(x-\mathcal{D}/2)}\,, & 0<x<\mathcal{D}/2\\
\varphi_{1}^{S_{R}}e^{-iq_{e}^{S}(x-\mathcal{D}/2)}+\bar{r}_{eh}^{\uparrow\uparrow}\varphi_{3}^{S_{R}}e^{-iq_{h}^{S}(x-\mathcal{D}/2)}+\bar{r}_{eh}^{\uparrow\downarrow}\varphi_{4}^{S_{R}}e^{-iq_{h}^{S}(x-\mathcal{D}/2)}+\bar{r}_{ee}^{\uparrow\uparrow}\varphi_{1}^{S_{R}}e^{iq_{e}^{S}(x-\mathcal{D}/2)}\\+\bar{r}_{ee}^{\uparrow\downarrow}\varphi_{2}^{S_{R}}e^{iq_{e}^{S}(x-\mathcal{D}/2)}\,, & x>\mathcal{D}/2
\end{cases}\\
\end{split}\nonumber
\end{equation}
\newpage
\begin{equation}
\begin{split}
\phi_{6}(x)&=\begin{cases}
\bar{t}_{ee}^{\downarrow\uparrow}\varphi_{1}^{S_L}e^{-iq_{e}^{S}(x+\mathcal{D}/2)}+\bar{t}_{ee}^{\downarrow\downarrow}\varphi_{2}^{S_L}e^{-iq_{e}^{S}(x+\mathcal{D}/2)}+\bar{t}_{eh}^{\downarrow\uparrow}\varphi_{3}^{S_L}e^{iq_{h}^{S}(x+\mathcal{D}/2)}+\bar{t}_{eh}^{\downarrow\downarrow}\varphi_{4}^{S_L}e^{iq_{h}^{S}(x+\mathcal{D}/2)}\,, & x<-\mathcal{D}/2\\
\bar{a}_{ee}^{\downarrow\uparrow}\varphi_{1}^{F_x}e^{iq_{e1}^{\uparrow}(x+\mathcal{D}/2)}+\bar{a}_{ee}^{\downarrow\downarrow}\varphi_{2}^{F_x}e^{iq_{e1}^{\downarrow}(x+\mathcal{D}/2)}+\bar{b}_{ee}^{\downarrow\uparrow}\varphi_{1}^{F_x}e^{-iq_{e1}^{\uparrow}x}+\bar{b}_{ee}^{\downarrow\downarrow}\varphi_{2}^{F_x}e^{-iq_{e1}^{\downarrow}x}\\+\bar{c}_{eh}^{\downarrow\uparrow}\varphi_{3}^{F_x}e^{iq_{h1}^{\uparrow}x}+\bar{c}_{eh}^{\downarrow\downarrow}\varphi_{4}^{F_x}e^{iq_{h1}^{\downarrow}x}+\bar{d}_{eh}^{\downarrow\uparrow}\varphi_{3}^{F_x}e^{-iq_{h1}^{\uparrow}(x+\mathcal{D}/2)}+\bar{d}_{eh}^{\downarrow\downarrow}\varphi_{4}^{F_x}e^{-iq_{h1}^{\downarrow}(x+\mathcal{D}/2)}\,, & -\mathcal{D}/2<x<0\\
\bar{e}_{ee}^{\downarrow\uparrow}\varphi_{1}^{F_y}e^{-iq_{e2}^{\uparrow}(x-\mathcal{D}/2)}+\bar{e}_{ee}^{\downarrow\downarrow}\varphi_{2}^{F_y}e^{-iq_{e2}^{\downarrow}(x-\mathcal{D}/2)}+\bar{f}_{ee}^{\downarrow\uparrow}\varphi_{1}^{F_y}e^{iq_{e2}^{\uparrow}x}+\bar{f}_{ee}^{\downarrow\downarrow}\varphi_{2}^{F_y}e^{iq_{e2}^{\downarrow}x}\\+\bar{g}_{eh}^{\downarrow\uparrow}\varphi_{3}^{F_y}e^{-iq_{h2}^{\uparrow}x}+\bar{g}_{eh}^{\downarrow\downarrow}\varphi_{4}^{F_y}e^{-iq_{h2}^{\downarrow}x}+\bar{h}_{eh}^{\downarrow\uparrow}\varphi_{3}^{F_y}e^{iq_{h2}^{\uparrow}(x-\mathcal{D}/2)}+\bar{h}_{eh}^{\downarrow\downarrow}\varphi_{4}^{F_y}e^{iq_{h2}^{\downarrow}(x-\mathcal{D}/2)}\,, & 0<x<\mathcal{D}/2\\
\varphi_{2}^{S_{R}}e^{-iq_{e}^{S}(x-\mathcal{D}/2)}+\bar{r}_{eh}^{\downarrow\uparrow}\varphi_{3}^{S_{R}}e^{-iq_{h}^{S}(x-\mathcal{D}/2)}+\bar{r}_{eh}^{\downarrow\downarrow}\varphi_{4}^{S_{R}}e^{-iq_{h}^{S}(x-\mathcal{D}/2)}+\bar{r}_{ee}^{\downarrow\uparrow}\varphi_{1}^{S_{R}}e^{iq_{e}^{S}(x-\mathcal{D}/2)}\\+\bar{r}_{ee}^{\downarrow\downarrow}\varphi_{2}^{S_{R}}e^{iq_{e}^{S}(x-\mathcal{D}/2)}\,, & x>\mathcal{D}/2
\end{cases}\\
\phi_{7}(x)&=\begin{cases}
\bar{t}_{he}^{\uparrow\uparrow}\varphi_{1}^{S_L}e^{-iq_{e}^{S}(x+\mathcal{D}/2)}+\bar{t}_{he}^{\uparrow\downarrow}\varphi_{2}^{S_L}e^{-iq_{e}^{S}(x+\mathcal{D}/2)}+\bar{t}_{hh}^{\uparrow\uparrow}\varphi_{3}^{S_L}e^{iq_{h}^{S}(x+\mathcal{D}/2)}+\bar{t}_{hh}^{\uparrow\downarrow}\varphi_{4}^{S_L}e^{iq_{h}^{S}(x+\mathcal{D}/2)}\,, & x<-\mathcal{D}/2\\ \bar{a}_{he}^{\uparrow\uparrow}\varphi_{1}^{F_x}e^{iq_{e1}^{\uparrow}(x+\mathcal{D}/2)}+\bar{a}_{he}^{\uparrow\downarrow}\varphi_{2}^{F_x}e^{iq_{e1}^{\downarrow}(x+\mathcal{D}/2)}+\bar{b}_{he}^{\uparrow\uparrow}\varphi_{1}^{F_x}e^{-iq_{e1}^{\uparrow}x}+\bar{b}_{he}^{\uparrow\downarrow}\varphi_{2}^{F_x}e^{-iq_{e1}^{\downarrow}x}\\+\bar{c}_{hh}^{\uparrow\uparrow}\varphi_{3}^{F_x}e^{iq_{h1}^{\uparrow}x}+\bar{c}_{hh}^{\uparrow\downarrow}\varphi_{4}^{F_x}e^{iq_{h1}^{\downarrow}x}
+\bar{d}_{hh}^{\uparrow\uparrow}\varphi_{3}^{F_x}e^{-iq_{h1}^{\uparrow}(x+\mathcal{D}/2)}+\bar{d}_{hh}^{\uparrow\downarrow}\varphi_{4}^{F_x}e^{-iq_{h1}^{\downarrow}(x+\mathcal{D}/2)}\,, & -\mathcal{D}/2<x<0\\
\bar{e}_{he}^{\uparrow\uparrow}\varphi_{1}^{F_y}e^{-iq_{e2}^{\uparrow}(x-\mathcal{D}/2)}+\bar{e}_{he}^{\uparrow\downarrow}\varphi_{2}^{F_y}e^{-iq_{e2}^{\downarrow}(x-\mathcal{D}/2)}+\bar{f}_{he}^{\uparrow\uparrow}\varphi_{1}^{F_y}e^{iq_{e2}^{\uparrow}x}+\bar{f}_{he}^{\uparrow\downarrow}\varphi_{2}^{F_y}e^{iq_{e2}^{\downarrow}x}\\+\bar{g}_{hh}^{\uparrow\uparrow}\varphi_{3}^{F_y}e^{-iq_{h2}^{\uparrow}x}+\bar{g}_{hh}^{\uparrow\downarrow}\varphi_{4}^{F_y}e^{-iq_{h2}^{\downarrow}x}+\bar{h}_{hh}^{\uparrow\uparrow}\varphi_{3}^{F_y}e^{iq_{h2}^{\uparrow}(x-\mathcal{D}/2)}+\bar{h}_{hh}^{\uparrow\downarrow}\varphi_{4}^{F_y}e^{iq_{h2}^{\downarrow}(x-\mathcal{D}/2)}\,, & 0<x<\mathcal{D}/2\\
\varphi_{3}^{S_{R}}e^{iq_{h}^{S}(x-\mathcal{D}/2)}+\bar{r}_{hh}^{\uparrow\uparrow}\varphi_{3}^{S_{R}}e^{-iq_{h}^{S}(x-\mathcal{D}/2)}+\bar{r}_{hh}^{\uparrow\downarrow}\varphi_{4}^{S_{R}}e^{-iq_{h}^{S}(x-\mathcal{D}/2)}+\bar{r}_{he}^{\uparrow\uparrow}\varphi_{1}^{S_{R}}e^{iq_{e}^{S}(x-\mathcal{D}/2)}\\+\bar{r}_{he}^{\uparrow\downarrow}\varphi_{2}^{S_{R}}e^{iq_{e}^{S}(x-\mathcal{D}/2)}\,, & x>\mathcal{D}/2
\end{cases}\\
\phi_{8}(x)&=\begin{cases}
\bar{t}_{he}^{\downarrow\uparrow}\varphi_{1}^{S_L}e^{-iq_{e}^{S}(x+\mathcal{D}/2)}+\bar{t}_{he}^{\downarrow\downarrow}\varphi_{2}^{S_L}e^{-iq_{e}^{S}(x+\mathcal{D}/2)}+\bar{t}_{hh}^{\downarrow\uparrow}\varphi_{3}^{S_L}e^{iq_{h}^{S}(x+\mathcal{D}/2)}+\bar{t}_{hh}^{\downarrow\downarrow}\varphi_{4}^{S_L}e^{iq_{h}^{S}(x+\mathcal{D}/2)}\,, & x<-\mathcal{D}/2\\
\bar{a}_{he}^{\downarrow\uparrow}\varphi_{1}^{F_x}e^{iq_{e1}^{\uparrow}(x+\mathcal{D}/2)}+\bar{a}_{he}^{\downarrow\downarrow}\varphi_{2}^{F_x}e^{iq_{e1}^{\downarrow}(x+\mathcal{D}/2)}+\bar{b}_{he}^{\downarrow\uparrow}\varphi_{1}^{F_x}e^{-iq_{e1}^{\uparrow}x}+\bar{b}_{he}^{\downarrow\downarrow}\varphi_{2}^{F_x}e^{-iq_{e1}^{\downarrow}x}\\+\bar{c}_{hh}^{\downarrow\uparrow}\varphi_{3}^{F_x}e^{iq_{h1}^{\uparrow}x}+\bar{c}_{hh}^{\downarrow\downarrow}\varphi_{4}^{F_x}e^{iq_{h1}^{\downarrow}x}+\bar{d}_{hh}^{\downarrow\uparrow}\varphi_{3}^{F_x}e^{-iq_{h1}^{\uparrow}(x+\mathcal{D}/2)}+\bar{d}_{hh}^{\downarrow\downarrow}\varphi_{4}^{F_x}e^{-iq_{h1}^{\downarrow}(x+\mathcal{D}/2)}\,, & -\mathcal{D}/2<x<0\\
\bar{e}_{he}^{\downarrow\uparrow}\varphi_{1}^{F_y}e^{-iq_{e2}^{\uparrow}(x-\mathcal{D}/2)}+\bar{e}_{he}^{\downarrow\downarrow}\varphi_{2}^{F_y}e^{-iq_{e2}^{\downarrow}(x-\mathcal{D}/2)}+\bar{f}_{he}^{\downarrow\uparrow}\varphi_{1}^{F_y}e^{iq_{e2}^{\uparrow}x}+\bar{f}_{he}^{\downarrow\downarrow}\varphi_{2}^{F_y}e^{iq_{e2}^{\downarrow}x}\\+\bar{g}_{hh}^{\downarrow\uparrow}\varphi_{3}^{F_y}e^{-iq_{h2}^{\uparrow}x}+\bar{g}_{hh}^{\downarrow\downarrow}\varphi_{4}^{F_y}e^{-iq_{h2}^{\downarrow}x}+\bar{h}_{hh}^{\downarrow\uparrow}\varphi_{3}^{F_y}e^{iq_{h2}^{\uparrow}(x-\mathcal{D}/2)}+\bar{h}_{hh}^{\downarrow\downarrow}\varphi_{4}^{F_y}e^{iq_{h2}^{\downarrow}(x-\mathcal{D}/2)}\,, & 0<x<\mathcal{D}/2\\
\varphi_{4}^{S_{R}}e^{iq_{h}^{S}(x-\mathcal{D}/2)}+\bar{r}_{hh}^{\downarrow\uparrow}\varphi_{3}^{S_{R}}e^{-iq_{h}^{S}(x-\mathcal{D}/2)}+\bar{r}_{hh}^{\downarrow\downarrow}\varphi_{4}^{S_{R}}e^{-iq_{h}^{S}(x-\mathcal{D}/2)}+\bar{r}_{he}^{\downarrow\uparrow}\varphi_{1}^{S_{R}}e^{iq_{e}^{S}(x-\mathcal{D}/2)}\\+\bar{r}_{he}^{\downarrow\downarrow}\varphi_{2}^{S_{R}}e^{iq_{e}^{S}(x-\mathcal{D}/2)}\,, & x>\mathcal{D}/2
\end{cases}
\end{split}
\label{wav}
\end{equation}
where $\varphi_{1}^{S_L}=\begin{pmatrix}
u_0\\
0\\
0\\
v_0                         \end{pmatrix}
$, $\varphi_{2}^{S_L}=\begin{pmatrix}
0\\
-u_0\\
v_0\\
0                         \end{pmatrix}
$, $\varphi_{3}^{S_L}=\begin{pmatrix}
0\\
-v_0\\
u_0\\
0                         \end{pmatrix}
$, $\varphi_{4}^{S_L}=\begin{pmatrix}
v_0\\
0\\
0\\
u_0                         \end{pmatrix}
$, {$\varphi_{1}^{F_x}=\frac{1}{\sqrt{2}}\begin{pmatrix}
   1\\
   1\\
   0\\
   0                      \end{pmatrix}$, $\varphi_{2}^{F_x}=\frac{1}{\sqrt{2}}\begin{pmatrix}
  -1\\
   1\\
   0\\
   0                      \end{pmatrix}
$, $\varphi_{3}^{F_x}=\frac{1}{\sqrt{2}}\begin{pmatrix}
   0\\
   0\\
   1\\
   1                    \end{pmatrix}
$, $\varphi_{4}^{F_x}=\frac{1}{\sqrt{2}}\begin{pmatrix}
   0\\
   0\\
  -1\\
   1                   \end{pmatrix}
$, $\varphi_{1}^{F_y}=\frac{1}{\sqrt{2}}\begin{pmatrix}
1\\
i\\
0\\
0\\                        \end{pmatrix}
$, $\varphi_{2}^{F_y}=\frac{1}{\sqrt{2}}\begin{pmatrix}
i\\
1\\
0\\
0\\                        \end{pmatrix}
$, $\varphi_{3}^{F_y}=\frac{1}{\sqrt{2}}\begin{pmatrix}
0\\
0\\
1\\
-i\\                        \end{pmatrix}
$, $\varphi_{4}^{F_y}=\frac{1}{\sqrt{2}}\begin{pmatrix}
0\\
0\\
-i\\
1\\                        \end{pmatrix}
$,} $\varphi_{1}^{S_R}=\begin{pmatrix}
u_0e^{i\varphi}\\
0\\
0\\
v_0                         \end{pmatrix}
$, $\varphi_{2}^{S_R}=\begin{pmatrix}
0\\
-u_0e^{i\varphi}\\
v_0\\
0                         \end{pmatrix}
$, $\varphi_{3}^{S_R}=\begin{pmatrix}
0\\
-v_0e^{i\varphi}\\
u_0\\
0                         \end{pmatrix}
$, $\varphi_{4}^{S_R}=\begin{pmatrix}
v_0e^{i\varphi}\\
0\\
0\\
u_0                         \end{pmatrix}$.

In Eq.~\eqref{wav}, $\phi_1$, $\phi_2$, $\phi_3$, and $\phi_4$ represent the wavefunctions corresponding to the injection of up-spin electron-like quasiparticles, down-spin electron-like quasiparticles, up-spin hole-like quasiparticles, and down-spin hole-like quasiparticles from the left superconductor, respectively. Similarly, $\phi_5$, $\phi_6$, $\phi_7$, and $\phi_8$ correspond to the wavefunctions for the injection of up-spin electron-like quasiparticles, down-spin electron-like quasiparticles, up-spin hole-like quasiparticles, and down-spin hole-like quasiparticles from the right superconductor, respectively. The terms $r_{pq}^{\sigma\sigma'}$ and $\bar{r}_{pq}^{\sigma\sigma'}$ represent the reflection amplitudes in left and right superconductors, respectively, and the terms $\bar{t}_{pq}^{\sigma\sigma'}$ and $t_{pq}^{\sigma\sigma'}$ denote the corresponding transmission amplitudes, with $\sigma,\sigma'\in\{\uparrow,\downarrow\}$ and $p,q\in\{e,h\}$. $u_0=\sqrt{\frac{\omega+\sqrt{\omega^2-\Delta^2}}{2\omega}}$ and $v_0=\sqrt{\frac{\omega-\sqrt{\omega^2-\Delta^2}}{2\omega}}$ represent the BCS coherence factors. $q_{e1,h1}^{\sigma}=\sqrt{\frac{2m^{*}}{\hbar^2}\big(E_F\pm\omega+\rho_\sigma m_1\big)}$ are the wavevectors in {$F_x$} layer and $q_{e2,h2}^{\sigma}=\sqrt{\frac{2m^{*}}{\hbar^2}\big(E_F\pm\omega+\rho_\sigma m_2\big)}$ represent the wavevectors in {$F_y$} layer with $\rho_{\uparrow(\downarrow)}=+1(-1)$. $q_{e,h}^{S}=\sqrt{\frac{2m^{*}}{\hbar^2}(E_F\pm\sqrt{\omega^2-\Delta^2})}$ denote the wavevectors in superconductor. After diagonalizing the Hamiltonian $\Big(H_{BdG}^{\mbox{{S-$F_x$-$F_y$-S}}}\Big)^{*}(-k)$ instead of $H_{BdG}^{\mbox{{S-$F_x$-$F_y$-S}}}(k)$ (see Eq.~\eqref{hmff}), we will obtain conjugate processes $\tilde{\phi}_{l}$ $(l=1,2,...,8)$ necessary to form $\Gamma^{r}(x,\chi,\omega)$ in next section. We note that $\tilde{\varphi_{i}}^{S_L}=\varphi_{i}^{S_L}$, {$\tilde{\varphi}_{i}^{F_x}=\varphi_{i}^{F_x}$, $\tilde{\varphi_{1}}^{F_y}=\frac{1}{\sqrt{2}}\begin{pmatrix}
   1\\
   -i\\
   0\\
   0                      \end{pmatrix}
$, $\tilde{\varphi_{2}}^{F_y}=\frac{1}{\sqrt{2}}\begin{pmatrix}
 -i\\
 1\\
 0\\
 0\\                                                                                                                                                                                                                                                                                                                                                                                                                                                                                                                                                                                                                                                                                                                                            \end{pmatrix}
$, $\tilde{\varphi_{3}}^{F_y}=\frac{1}{\sqrt{2}}\begin{pmatrix}
 0\\
 0\\
 1\\
 i\\                                                                                                                                                                                                                                                                                                                                                                                                                                                                                                                                                                                                                                                                                                                                            \end{pmatrix}
$, $\tilde{\varphi_{4}}^{F_y}=\frac{1}{\sqrt{2}}\begin{pmatrix}
 0\\
 0\\
 i\\
 1\\                                                                                                                                                                                                                                                                                                                                                                                                                                                                                                                                                                                                                                                                                                                                            \end{pmatrix}
$,} $\tilde{\varphi}_{1}^{S_R}=\begin{pmatrix}
u_0e^{-i\varphi}\\
0\\
0\\
v_0                         \end{pmatrix}
$, $\tilde{\varphi}_{2}^{S_R}=\begin{pmatrix}
0\\
-u_0e^{-i\varphi}\\
v_0\\
0                         \end{pmatrix}
$, $\tilde{\varphi}_{3}^{S_R}=\begin{pmatrix}
0\\
-v_0e^{-i\varphi}\\
u_0\\
0                         \end{pmatrix}
$, $\tilde{\varphi}_{4}^{S_R}=\begin{pmatrix}
v_0e^{-i\varphi}\\
0\\
0\\
u_0                         \end{pmatrix}
$ with $i\in\{1,2,3,4\}$.

The boundary conditions at the {S-$F_x$} interface ($x=-\mathcal{D}/2$) are:
\begin{equation}
\label{eqq8}
\phi_{l}|_{x<-\mathcal{D}/2}=\phi_{l}|_{-\mathcal{D}/2<x<0}\,\,\,\,\,\mbox{and}\,\,\,\,\,\frac{d\phi_{l}|_{-\mathcal{D}/2<x<0}}{dx}-\frac{d\phi_{l}|_{x<-\mathcal{D}/2}}{dx}=\frac{2m^{*}\mathcal{B}_1}{\hbar^2}\phi_{l}(x=-\mathcal{D}/2).
\end{equation}
Similarly, the boundary conditions at the {$F_x$-$F_y$} interface ($x=0$) are:
\begin{equation}
\label{eqq9}
\phi_{l}|_{-\mathcal{D}/2<x<0}=\phi_{l}|_{0<x<\mathcal{D}/2}\,\,\,\,\,\mbox{and}\,\,\,\,\
\frac{d\phi_{l}|_{0<x<\mathcal{D}/2}}{dx}-\frac{d\phi_{l}|_{-\mathcal{D}/2<x<0}}{dx}=\frac{2m^{*}\mathcal{B}}{\hbar^2}\phi_{l}(x=0).
\end{equation}
Finally, the boundary conditions at the {$F_y$-S} interface ($x=\mathcal{D}/2$) are:
\begin{equation}
\label{eqq10}
\phi_{l}|_{0<x<\mathcal{D}/2}=\phi_{l}|_{x>\mathcal{D}/2}\,\,\,\,\,\mbox{and}\,\,\,\,\,\frac{d\phi_{l}|_{x>\mathcal{D}/2}}{dx}-\frac{d\phi_{l}|_{0<x<\mathcal{D}/2}}{dx}=\frac{2m^{*}\mathcal{B}_2}{\hbar^2}\phi_{l}(x=\mathcal{D}/2).
\end{equation}
By applying the above boundary conditions at $x=-\mathcal{D}/2$, $x=0$, and $x=\mathcal{D}/2$, a total of $24$
equations are derived for each type of incident process described in Eq.~\eqref{wav}. The scattering amplitudes, such as $r_{pq}^{\sigma\sigma'}$, $\bar{r}_{pq}^{\sigma\sigma'}$, $\bar{t}_{pq}^{\sigma\sigma'}$, $t_{pq}^{\sigma\sigma'}$, are derived by solving these $24$ equations. $s_{1}=r_{eh}^{\uparrow\downarrow}$, $s_{2}=r_{eh}^{\downarrow\uparrow}$, $s_{3}=r_{he}^{\uparrow\downarrow}$, and $s_{4}=r_{he}^{\downarrow\uparrow}$, obtained by substituting the wavefunctions $\phi_{1}$, $\phi_{2}$, $\phi_{3}$, and $\phi_{4}$, respectively into Eqs.~\eqref{eqq8}-\eqref{eqq10} are used to calculate the Josephson current via Eq.~\eqref{eq3}.

{The wavefunctions for trilayer S-$F_x$-$F_y$-$F_z$-S JJ are as follows:}
\begin{equation}
\begin{split}
\phi_{1}(x)&=\begin{cases}
\varphi_{1}^{S_L}e^{iq_{e}^{S}(x+\mathcal{D}/2)}+r_{eh}^{\prime\uparrow\uparrow}\varphi_{3}^{S_L}e^{iq_{h}^{S}(x+\mathcal{D}/2)}+r_{eh}^{\prime\uparrow\downarrow}\varphi_{4}^{S_L}e^{iq_{h}^{S}(x+\mathcal{D}/2)}+r_{ee}^{\prime\uparrow\uparrow}\varphi_{1}^{S_L}e^{-iq_{e}^{S}(x+\mathcal{D}/2)}\\+r_{ee}^{\prime\uparrow\downarrow}\varphi_{2}^{S_L}e^{-iq_{e}^{S}(x+\mathcal{D}/2)}\,, & x<-\mathcal{D}/2\\
a_{ee}^{\prime\uparrow\uparrow}\varphi_{1}^{F_x}e^{iq_{e1}^{\uparrow}(x+\mathcal{D}/2)}+a_{ee}^{\prime\uparrow\downarrow}\varphi_{2}^{F_x}e^{iq_{e1}^{\downarrow}(x+\mathcal{D}/2)}+b_{ee}^{\prime\uparrow\uparrow}\varphi_{1}^{F_x}e^{-iq_{e1}^{\uparrow}x}+b_{ee}^{\prime\uparrow\downarrow}\varphi_{2}^{F_x}e^{-iq_{e1}^{\downarrow}x}\\+c_{eh}^{\prime\uparrow\uparrow}\varphi_{3}^{F_x}e^{iq_{h1}^{\uparrow}x}+c_{eh}^{\prime\uparrow\downarrow}\varphi_{4}^{F_x}e^{iq_{h1}^{\downarrow}x}
+d_{eh}^{\prime\uparrow\uparrow}\varphi_{3}^{F_x}e^{-iq_{h1}^{\uparrow}(x+\mathcal{D}/2)}+d_{eh}^{\prime\uparrow\downarrow}\varphi_{4}^{F_x}e^{-iq_{h1}^{\downarrow}(x+\mathcal{D}/2)}\,, & -\mathcal{D}/2<x<0\\
e_{ee}^{\prime\uparrow\uparrow}\varphi_{1}^{F_y}e^{-iq_{e2}^{\uparrow}(x-\mathcal{D}/4)}+e_{ee}^{\prime\uparrow\downarrow}\varphi_{2}^{F_y}e^{-iq_{e2}^{\downarrow}(x-\mathcal{D}/4)}+f_{ee}^{\prime\uparrow\uparrow}\varphi_{1}^{F_y}e^{iq_{e2}^{\uparrow}x}+f_{ee}^{\prime\uparrow\downarrow}\varphi_{2}^{F_y}e^{iq_{e2}^{\downarrow}x}\\+g_{eh}^{\prime\uparrow\uparrow}\varphi_{3}^{F_y}e^{-iq_{h2}^{\uparrow}x}+g_{eh}^{\prime\uparrow\downarrow}\varphi_{4}^{F_y}e^{-iq_{h2}^{\downarrow}x} +h_{eh}^{\prime\uparrow\uparrow}\varphi_{3}^{F_y}e^{iq_{h2}^{\uparrow}(x-\mathcal{D}/4)}+h_{eh}^{\prime\uparrow\downarrow}\varphi_{4}^{F_y}e^{iq_{h2}^{\downarrow}(x-\mathcal{D}/4)}\,, & 0<x<\mathcal{D}/4\\
m_{ee}^{\prime\uparrow\uparrow}\varphi_{1}^{F_z}e^{-iq_{e3}^{\uparrow}(x-\mathcal{D}/2)}+m_{ee}^{\prime\uparrow\downarrow}\varphi_{2}^{F_z}e^{-iq_{e3}^{\downarrow}(x-\mathcal{D}/2)}+n_{ee}^{\prime\uparrow\uparrow}\varphi_{1}^{F_z}e^{iq_{e3}^{\uparrow}(x-\mathcal{D}/4)}+n_{ee}^{\prime\uparrow\downarrow}\varphi_{2}^{F_z}e^{iq_{e3}^{\downarrow}(x-\mathcal{D}/4)}\\+o_{eh}^{\prime\uparrow\uparrow}\varphi_{3}^{F_z}e^{-iq_{h3}^{\uparrow}(x-\mathcal{D}/4)}+o_{eh}^{\prime\uparrow\downarrow}\varphi_{4}^{F_z}e^{-iq_{h3}^{\downarrow}(x-\mathcal{D}/4)}
+p_{eh}^{\prime\uparrow\uparrow}\varphi_{3}^{F_z}e^{iq_{h3}^{\uparrow}(x-\mathcal{D}/2)}+p_{eh}^{\prime\uparrow\downarrow}\varphi_{4}^{F_z}e^{iq_{h3}^{\downarrow}(x-\mathcal{D}/2)}\,, & \mathcal{D}/4<x<\mathcal{D}/2\\
t_{ee}^{\prime\uparrow\uparrow}\varphi_{1}^{S_{R}}e^{iq_{e}^{S}(x-\mathcal{D}/2)}+t_{ee}^{\prime\uparrow\downarrow}\varphi_{2}^{S_{R}}e^{iq_{e}^{S}(x-\mathcal{D}/2)}+t_{eh}^{\prime\uparrow\uparrow}\varphi_{3}^{S_{R}}e^{-iq_{h}^{S}(x-\mathcal{D}/2)}+t_{eh}^{\prime\uparrow\downarrow}\varphi_{4}^{S_{R}}e^{-iq_{h}^{S}(x-\mathcal{D}/2)}\,, & x>\mathcal{D}/2
\end{cases}\\
\phi_{2}(x)&=\begin{cases}
\varphi_{2}^{S_L}e^{iq_{e}^{S}(x+\mathcal{D}/2)}+r_{eh}^{\prime\downarrow\uparrow}\varphi_{3}^{S_L}e^{iq_{h}^{S}(x+\mathcal{D}/2)}+r_{eh}^{\prime\downarrow\downarrow}\varphi_{4}^{S_L}e^{iq_{h}^{S}(x+\mathcal{D}/2)}+r_{ee}^{\prime\downarrow\uparrow}\varphi_{1}^{S_L}e^{-iq_{e}^{S}(x+\mathcal{D}/2)}\\+r_{ee}^{\prime\downarrow\downarrow}\varphi_{2}^{S_L}e^{-iq_{e}^{S}(x+\mathcal{D}/2)}\,, & x<-\mathcal{D}/2\\
a_{ee}^{\prime\downarrow\uparrow}\varphi_{1}^{F_x}e^{iq_{e1}^{\uparrow}(x+\mathcal{D}/2)}+a_{ee}^{\prime\downarrow\downarrow}\varphi_{2}^{F_x}e^{iq_{e1}^{\downarrow}(x+\mathcal{D}/2)}+b_{ee}^{\prime\downarrow\uparrow}\varphi_{1}^{F_x}e^{-iq_{e1}^{\uparrow}x}+b_{ee}^{\prime\downarrow\downarrow}\varphi_{2}^{F_x}e^{-iq_{e1}^{\downarrow}x}\\+c_{eh}^{\prime\downarrow\uparrow}\varphi_{3}^{F_x}e^{iq_{h1}^{\uparrow}x}+c_{eh}^{\prime\downarrow\downarrow}\varphi_{4}^{F_x}e^{iq_{h1}^{\downarrow}x} +d_{eh}^{\prime\downarrow\uparrow}\varphi_{3}^{F_x}e^{-iq_{h1}^{\uparrow}(x+\mathcal{D}/2)}+d_{eh}^{\prime\downarrow\downarrow}\varphi_{4}^{F_x}e^{-iq_{h1}^{\downarrow}(x+\mathcal{D}/2)}\,, &  -\mathcal{D}/2<x<0\\
e_{ee}^{\prime\downarrow\uparrow}\varphi_{1}^{F_y}e^{-iq_{e2}^{\uparrow}(x-\mathcal{D}/4)}+e_{ee}^{\prime\downarrow\downarrow}\varphi_{2}^{F_y}e^{-iq_{e2}^{\downarrow}(x-\mathcal{D}/4)}+f_{ee}^{\prime\downarrow\uparrow}\varphi_{1}^{F_y}e^{iq_{e2}^{\uparrow}x}+f_{ee}^{\prime\downarrow\downarrow}\varphi_{2}^{F_y}e^{iq_{e2}^{\downarrow}x}\\+g_{eh}^{\prime\downarrow\uparrow}\varphi_{3}^{F_y}e^{-iq_{h2}^{\uparrow}x}+g_{eh}^{\prime\downarrow\downarrow}\varphi_{4}^{F_y}e^{-iq_{h2}^{\downarrow}x}+h_{eh}^{\prime\downarrow\uparrow}\varphi_{3}^{F_y}e^{iq_{h2}^{\uparrow}(x-\mathcal{D}/4)}+h_{eh}^{\prime\downarrow\downarrow}\varphi_{4}^{F_y}e^{iq_{h2}^{\downarrow}(x-\mathcal{D}/4)}\,, & 0<x<\mathcal{D}/4\\
m_{ee}^{\prime\downarrow\uparrow}\varphi_{1}^{F_z}e^{-iq_{e3}^{\uparrow}(x-\mathcal{D}/2)}+m_{ee}^{\prime\downarrow\downarrow}\varphi_{2}^{F_z}e^{-iq_{e3}^{\downarrow}(x-\mathcal{D}/2)}+n_{ee}^{\prime\downarrow\uparrow}\varphi_{1}^{F_z}e^{iq_{e3}^{\uparrow}(x-\mathcal{D}/4)}+n_{ee}^{\prime\downarrow\downarrow}\varphi_{2}^{F_z}e^{iq_{e3}^{\downarrow}(x-\mathcal{D}/4)}\\+o_{eh}^{\prime\downarrow\uparrow}\varphi_{3}^{F_z}e^{-iq_{h3}^{\uparrow}(x-\mathcal{D}/4)}+o_{eh}^{\prime\downarrow\downarrow}\varphi_{4}^{F_z}e^{-iq_{h3}^{\downarrow}(x-\mathcal{D}/4)}+p_{eh}^{\downarrow\uparrow}\varphi_{3}^{F_z}e^{iq_{h3}^{\uparrow}(x-\mathcal{D}/2)}+p_{eh}^{\prime\downarrow\downarrow}\varphi_{4}^{F_z}e^{iq_{h3}^{\downarrow}(x-\mathcal{D}/2)}\,, & \mathcal{D}/4<x<\mathcal{D}/2\\
t_{ee}^{\prime\downarrow\uparrow}\varphi_{1}^{S_{R}}e^{iq_{e}^{S}(x-\mathcal{D}/2)}+t_{ee}^{\prime\downarrow\downarrow}\varphi_{2}^{S_{R}}e^{iq_{e}^{S}(x-\mathcal{D}/2)}+t_{eh}^{\prime\downarrow\uparrow}\varphi_{3}^{S_{R}}e^{-iq_{h}^{S}(x-\mathcal{D}/2)}+t_{eh}^{\prime\downarrow\downarrow}\varphi_{4}^{S_{R}}e^{-iq_{h}^{S}(x-\mathcal{D}/2)}\,, & x>\mathcal{D}/2
\end{cases}\\
\end{split}\nonumber
\end{equation}
\newpage
\begin{equation}
\begin{split}
\phi_{3}(x)&=\begin{cases}
\varphi_{3}^{S_L}e^{-iq_{h}^{S}(x+\mathcal{D}/2)}+r_{hh}^{\prime\uparrow\uparrow}\varphi_{3}^{S_L}e^{iq_{h}^{S}(x+\mathcal{D}/2)}+r_{hh}^{\prime\uparrow\downarrow}\varphi_{4}^{S_L}e^{iq_{h}^{S}(x+\mathcal{D}/2)}+r_{he}^{\prime\uparrow\uparrow}\varphi_{1}^{S_L}e^{-iq_{e}^{S}(x+\mathcal{D}/2)}\\+r_{he}^{\prime\uparrow\downarrow}\varphi_{2}^{S_L}e^{-iq_{e}^{S}(x+\mathcal{D}/2)}\,, & x<-\mathcal{D}/2\\
a_{he}^{\prime\uparrow\uparrow}\varphi_{1}^{F_x}e^{iq_{e1}^{\uparrow}(x+\mathcal{D}/2)}+a_{he}^{\prime\uparrow\downarrow}\varphi_{2}^{F_x}e^{iq_{e1}^{\downarrow}(x+\mathcal{D}/2)}+b_{he}^{\prime\uparrow\uparrow}\varphi_{1}^{F_x}e^{-iq_{e1}^{\uparrow}x}+b_{he}^{\prime\uparrow\downarrow}\varphi_{2}^{F_x}e^{-iq_{e1}^{\downarrow}x}\\+c_{hh}^{\prime\uparrow\uparrow}\varphi_{3}^{F_x}e^{iq_{h1}^{\uparrow}x}+c_{hh}^{\prime\uparrow\downarrow}\varphi_{4}^{F_x}e^{iq_{h1}^{\downarrow}x} +d_{hh}^{\prime\uparrow\uparrow}\varphi_{3}^{F_x}e^{-iq_{h1}^{\uparrow}(x+\mathcal{D}/2)}+d_{hh}^{\prime\uparrow\downarrow}\varphi_{4}^{F_x}e^{-iq_{h1}^{\downarrow}(x+\mathcal{D}/2)}\,, & -\mathcal{D}/2<x<0\\
e_{he}^{\prime\uparrow\uparrow}\varphi_{1}^{F_y}e^{-iq_{e2}^{\uparrow}(x-\mathcal{D}/4)}+e_{he}^{\prime\uparrow\downarrow}\varphi_{2}^{F_y}e^{-iq_{e2}^{\downarrow}(x-\mathcal{D}/4)}+f_{he}^{\prime\uparrow\uparrow}\varphi_{1}^{F_y}e^{iq_{e2}^{\uparrow}x}+f_{he}^{\prime\uparrow\downarrow}\varphi_{2}^{F_y}e^{iq_{e2}^{\downarrow}x}\\+g_{hh}^{\prime\uparrow\uparrow}\varphi_{3}^{F_y}e^{-iq_{h2}^{\uparrow}x}+g_{hh}^{\prime\uparrow\downarrow}\varphi_{4}^{F_y}e^{-iq_{h2}^{\downarrow}x} +h_{hh}^{\prime\uparrow\uparrow}\varphi_{3}^{F_y}e^{iq_{h2}^{\uparrow}(x-\mathcal{D}/4)}+h_{hh}^{\prime\uparrow\downarrow}\varphi_{4}^{F_y}e^{iq_{h2}^{\downarrow}(x-\mathcal{D}/4)}\,, & 0<x<\mathcal{D}/4\\
m_{he}^{\prime\uparrow\uparrow}\varphi_{1}^{F_z}e^{-iq_{e3}^{\uparrow}(x-\mathcal{D}/2)}+m_{he}^{\prime\uparrow\downarrow}\varphi_{2}^{F_z}e^{-iq_{e3}^{\downarrow}(x-\mathcal{D}/2)}+n_{he}^{\prime\uparrow\uparrow}\varphi_{1}^{F_z}e^{iq_{e3}^{\uparrow}(x-\mathcal{D}/4)}+n_{he}^{\prime\uparrow\downarrow}\varphi_{2}^{F_z}e^{iq_{e3}^{\downarrow}(x-\mathcal{D}/4)}\\+o_{hh}^{\prime\uparrow\uparrow}\varphi_{3}^{F_z}e^{-iq_{h3}^{\uparrow}(x-\mathcal{D}/4)}+o_{hh}^{\prime\uparrow\downarrow}\varphi_{4}^{F_z}e^{-iq_{h3}^{\downarrow}(x-\mathcal{D}/4)}+p_{hh}^{\prime\uparrow\uparrow}\varphi_{3}^{F_z}e^{iq_{h3}^{\uparrow}(x-\mathcal{D}/2)}+p_{hh}^{\prime\uparrow\downarrow}\varphi_{4}^{F_z}e^{iq_{h3}^{\downarrow}(x-\mathcal{D}/2)}\,, & \mathcal{D}/4<x<\mathcal{D}/2\\
t_{he}^{\prime\uparrow\uparrow}\varphi_{1}^{S_{R}}e^{iq_{e}^{S}(x-\mathcal{D}/2)}+t_{he}^{\prime\uparrow\downarrow}\varphi_{2}^{S_{R}}e^{iq_{e}^{S}(x-\mathcal{D}/2)}+t_{hh}^{\prime\uparrow\uparrow}\varphi_{3}^{S_{R}}e^{-iq_{h}^{S}(x-\mathcal{D}/2)}+t_{hh}^{\prime\uparrow\downarrow}\varphi_{4}^{S_{R}}e^{-iq_{h}^{S}(x-\mathcal{D}/2)}\,, & x>\mathcal{D}/2
\end{cases}\\
\phi_{4}(x)&=\begin{cases}
\varphi_{4}^{S_L}e^{-iq_{h}^{S}(x+\mathcal{D}/2)}+r_{hh}^{\prime\downarrow\uparrow}\varphi_{3}^{S_L}e^{iq_{h}^{S}(x+\mathcal{D}/2)}+r_{hh}^{\prime\downarrow\downarrow}\varphi_{4}^{S_L}e^{iq_{h}^{S}(x+\mathcal{D}/2)}+r_{he}^{\prime\downarrow\uparrow}\varphi_{1}^{S_L}e^{-iq_{e}^{S}(x+\mathcal{D}/2)}\\+r_{he}^{\prime\downarrow\downarrow}\varphi_{2}^{S_L}e^{-iq_{e}^{S}(x+\mathcal{D}/2)}\,, & x<-\mathcal{D}/2\\
a_{he}^{\prime\downarrow\uparrow}\varphi_{1}^{F_x}e^{iq_{e1}^{\uparrow}(x+\mathcal{D}/2)}+a_{he}^{\prime\downarrow\downarrow}\varphi_{2}^{F_x}e^{iq_{e1}^{\downarrow}(x+\mathcal{D}/2)}+b_{he}^{\prime\downarrow\uparrow}\varphi_{1}^{F_x}e^{-iq_{e1}^{\uparrow}x}+b_{he}^{\prime\downarrow\downarrow}\varphi_{2}^{F_x}e^{-iq_{e1}^{\downarrow}x}\\+c_{hh}^{\prime\downarrow\uparrow}\varphi_{3}^{F_x}e^{iq_{h1}^{\uparrow}x}+c_{hh}^{\prime\downarrow\downarrow}\varphi_{4}^{F_x}e^{iq_{h1}^{\downarrow}x} +d_{hh}^{\prime\downarrow\uparrow}\varphi_{3}^{F_x}e^{-iq_{h1}^{\uparrow}(x+\mathcal{D}/2)}+d_{hh}^{\prime\downarrow\downarrow}\varphi_{4}^{F_x}e^{-iq_{h1}^{\downarrow}(x+\mathcal{D}/2)}\,, & -\mathcal{D}/2<x<0\\
e_{he}^{\prime\downarrow\uparrow}\varphi_{1}^{F_y}e^{-iq_{e2}^{\uparrow}(x-\mathcal{D}/4)}+e_{he}^{\prime\downarrow\downarrow}\varphi_{2}^{F_y}e^{-iq_{e2}^{\downarrow}(x-\mathcal{D}/4)}+f_{he}^{\prime\downarrow\uparrow}\varphi_{1}^{F_y}e^{iq_{e2}^{\uparrow}x}+f_{he}^{\prime\downarrow\downarrow}\varphi_{2}^{F_y}e^{iq_{e2}^{\downarrow}x}\\+g_{hh}^{\prime\downarrow\uparrow}\varphi_{3}^{F_y}e^{-iq_{h2}^{\uparrow}x}+g_{hh}^{\prime\downarrow\downarrow}\varphi_{4}^{F_y}e^{-iq_{h2}^{\downarrow}x}
+h_{hh}^{\prime\downarrow\uparrow}\varphi_{3}^{F_y}e^{iq_{h2}^{\uparrow}(x-\mathcal{D}/4)}+h_{hh}^{\prime\downarrow\downarrow}\varphi_{4}^{F_y}e^{iq_{h2}^{\downarrow}(x-\mathcal{D}/4)}\,, & 0<x<\mathcal{D}/4\\
m_{he}^{\prime\downarrow\uparrow}\varphi_{1}^{F_z}e^{-iq_{e3}^{\uparrow}(x-\mathcal{D}/2)}+m_{he}^{\prime\downarrow\downarrow}\varphi_{2}^{F_z}e^{-iq_{e3}^{\downarrow}(x-\mathcal{D}/2)}+n_{he}^{\prime\downarrow\uparrow}\varphi_{1}^{F_z}e^{iq_{e3}^{\uparrow}(x-\mathcal{D}/4)}+n_{he}^{\prime\downarrow\downarrow}\varphi_{2}^{F_z}e^{iq_{e3}^{\downarrow}(x-\mathcal{D}/4)}\\+o_{hh}^{\prime\downarrow\uparrow}\varphi_{3}^{F_z}e^{-iq_{h3}^{\uparrow}(x-\mathcal{D}/4)}+o_{hh}^{\prime\downarrow\downarrow}\varphi_{4}^{F_z}e^{-iq_{h3}^{\downarrow}(x-\mathcal{D}/4)}+p_{hh}^{\prime\downarrow\uparrow}\varphi_{3}^{F_z}e^{iq_{h3}^{\uparrow}(x-\mathcal{D}/2)}+p_{hh}^{\prime\downarrow\downarrow}\varphi_{4}^{F_z}e^{iq_{h3}^{\downarrow}(x-\mathcal{D}/2)}\,, & \mathcal{D}/4<x<\mathcal{D}/2\\
t_{he}^{\prime\downarrow\uparrow}\varphi_{1}^{S_{R}}e^{iq_{e}^{S}(x-\mathcal{D}/2)}+t_{he}^{\prime\downarrow\downarrow}\varphi_{2}^{S_{R}}e^{iq_{e}^{S}(x-\mathcal{D}/2)}+t_{hh}^{\prime\downarrow\uparrow}\varphi_{3}^{S_{R}}e^{-iq_{h}^{S}(x-\mathcal{D}/2)}+t_{hh}^{\prime\downarrow\downarrow}\varphi_{4}^{S_{R}}e^{-iq_{h}^{S}(x-\mathcal{D}/2)}\,, & x>\mathcal{D}/2
\end{cases}\\
\phi_{5}(x)&=\begin{cases}
\bar{t}_{ee}^{\prime\uparrow\uparrow}\varphi_{1}^{S_L}e^{-iq_{e}^{S}(x+\mathcal{D}/2)}+\bar{t}_{ee}^{\prime\uparrow\downarrow}\varphi_{2}^{S_L}e^{-iq_{e}^{S}(x+\mathcal{D}/2)}+\bar{t}_{eh}^{\prime\uparrow\uparrow}\varphi_{3}^{S_L}e^{iq_{h}^{S}(x+\mathcal{D}/2)}+\bar{t}_{eh}^{\prime\uparrow\downarrow}\varphi_{4}^{S_L}e^{iq_{h}^{S}(x+\mathcal{D}/2)}\,, & x<-\mathcal{D}/2\\
\bar{a}_{ee}^{\prime\uparrow\uparrow}\varphi_{1}^{F_x}e^{iq_{e1}^{\uparrow}(x+\mathcal{D}/2)}+\bar{a}_{ee}^{\prime\uparrow\downarrow}\varphi_{2}^{F_x}e^{iq_{e1}^{\downarrow}(x+\mathcal{D}/2)}+\bar{b}_{ee}^{\prime\uparrow\uparrow}\varphi_{1}^{F_x}e^{-iq_{e1}^{\uparrow}x}+\bar{b}_{ee}^{\prime\uparrow\downarrow}\varphi_{2}^{F_x}e^{-iq_{e1}^{\downarrow}x}\\+\bar{c}_{eh}^{\prime\uparrow\uparrow}\varphi_{3}^{F_x}e^{iq_{h1}^{\uparrow}x}+\bar{c}_{eh}^{\prime\uparrow\downarrow}\varphi_{4}^{F_x}e^{iq_{h1}^{\downarrow}x}
+\bar{d}_{eh}^{\prime\uparrow\uparrow}\varphi_{3}^{F_x}e^{-iq_{h1}^{\uparrow}(x+\mathcal{D}/2)}+\bar{d}_{eh}^{\prime\uparrow\downarrow}\varphi_{4}^{F_x}e^{-iq_{h1}^{\downarrow}(x+\mathcal{D}/2)}\,, & -\mathcal{D}/2<x<0\\
\bar{e}_{ee}^{\prime\uparrow\uparrow}\varphi_{1}^{F_y}e^{-iq_{e2}^{\uparrow}(x-\mathcal{D}/4)}+\bar{e}_{ee}^{\prime\uparrow\downarrow}\varphi_{2}^{F_y}e^{-iq_{e2}^{\downarrow}(x-\mathcal{D}/4)}+\bar{f}_{ee}^{\prime\uparrow\uparrow}\varphi_{1}^{F_y}e^{iq_{e2}^{\uparrow}x}+\bar{f}_{ee}^{\prime\uparrow\downarrow}\varphi_{2}^{F_y}e^{iq_{e2}^{\downarrow}x}\\+\bar{g}_{eh}^{\prime\uparrow\uparrow}\varphi_{3}^{F_y}e^{-iq_{h2}^{\uparrow}x}+\bar{g}_{eh}^{\prime\uparrow\downarrow}\varphi_{4}^{F_y}e^{-iq_{h2}^{\downarrow}x}+\bar{h}_{eh}^{\prime\uparrow\uparrow}\varphi_{3}^{F_y}e^{iq_{h2}^{\uparrow}(x-\mathcal{D}/4)}+\bar{h}_{eh}^{\prime\uparrow\downarrow}\varphi_{4}^{F_y}e^{iq_{h2}^{\downarrow}(x-\mathcal{D}/4)}\,, & 0<x<\mathcal{D}/4\\
\bar{m}_{ee}^{\prime\uparrow\uparrow}\varphi_{1}^{F_z}e^{-iq_{e3}^{\uparrow}(x-\mathcal{D}/2)}+\bar{m}_{ee}^{\prime\uparrow\downarrow}\varphi_{2}^{F_z}e^{-iq_{e3}^{\downarrow}(x-\mathcal{D}/2)}+\bar{n}_{ee}^{\prime\uparrow\uparrow}\varphi_{1}^{F_z}e^{iq_{e3}^{\uparrow}(x-\mathcal{D}/4)}+\bar{n}_{ee}^{\prime\uparrow\downarrow}\varphi_{2}^{F_z}e^{iq_{e3}^{\downarrow}(x-\mathcal{D}/4)}\\+\bar{o}_{eh}^{\prime\uparrow\uparrow}\varphi_{3}^{F_z}e^{-iq_{h3}^{\uparrow}(x-\mathcal{D}/4)}+\bar{o}_{eh}^{\prime\uparrow\downarrow}\varphi_{4}^{F_z}e^{-iq_{h3}^{\downarrow}(x-\mathcal{D}/4)}+\bar{p}_{eh}^{\prime\uparrow\uparrow}\varphi_{3}^{F_z}e^{iq_{h3}^{\uparrow}(x-\mathcal{D}/2)}+\bar{p}_{eh}^{\prime\uparrow\downarrow}\varphi_{4}^{F_z}e^{iq_{h3}^{\downarrow}(x-\mathcal{D}/2)}\,, & \mathcal{D}/4<x<\mathcal{D}/2\\
\varphi_{1}^{S_{R}}e^{-iq_{e}^{S}(x-\mathcal{D}/2)}+\bar{r}_{eh}^{\prime\uparrow\uparrow}\varphi_{3}^{S_{R}}e^{-iq_{h}^{S}(x-\mathcal{D}/2)}+\bar{r}_{eh}^{\prime\uparrow\downarrow}\varphi_{4}^{S_{R}}e^{-iq_{h}^{S}(x-\mathcal{D}/2)}+\bar{r}_{ee}^{\prime\uparrow\uparrow}\varphi_{1}^{S_{R}}e^{iq_{e}^{S}(x-\mathcal{D}/2)}\\+\bar{r}_{ee}^{\prime\uparrow\downarrow}\varphi_{2}^{S_{R}}e^{iq_{e}^{S}(x-\mathcal{D}/2)}\,, & x>\mathcal{D}/2
\end{cases}\\
\phi_{6}(x)&=\begin{cases}
\bar{t}_{ee}^{\prime\downarrow\uparrow}\varphi_{1}^{S_L}e^{-iq_{e}^{S}(x+\mathcal{D}/2)}+\bar{t}_{ee}^{\prime\downarrow\downarrow}\varphi_{2}^{S_L}e^{-iq_{e}^{S}(x+\mathcal{D}/2)}+\bar{t}_{eh}^{\prime\downarrow\uparrow}\varphi_{3}^{S_L}e^{iq_{h}^{S}(x+\mathcal{D}/2)}+\bar{t}_{eh}^{\prime\downarrow\downarrow}\varphi_{4}^{S_L}e^{iq_{h}^{S}(x+\mathcal{D}/2)}\,, & x<-\mathcal{D}/2\\
\bar{a}_{ee}^{\prime\downarrow\uparrow}\varphi_{1}^{F_x}e^{iq_{e1}^{\uparrow}(x+\mathcal{D}/2)}+\bar{a}_{ee}^{\prime\downarrow\downarrow}\varphi_{2}^{F_x}e^{iq_{e1}^{\downarrow}(x+\mathcal{D}/2)}+\bar{b}_{ee}^{\prime\downarrow\uparrow}\varphi_{1}^{F_x}e^{-iq_{e1}^{\uparrow}x}+\bar{b}_{ee}^{\prime\downarrow\downarrow}\varphi_{2}^{F_x}e^{-iq_{e1}^{\downarrow}x}\\+\bar{c}_{eh}^{\prime\downarrow\uparrow}\varphi_{3}^{F_x}e^{iq_{h1}^{\uparrow}x}+\bar{c}_{eh}^{\prime\downarrow\downarrow}\varphi_{4}^{F_x}e^{iq_{h1}^{\downarrow}x}
+\bar{d}_{eh}^{\prime\downarrow\uparrow}\varphi_{3}^{F_x}e^{-iq_{h1}^{\uparrow}(x+\mathcal{D}/2)}+\bar{d}_{eh}^{\prime\downarrow\downarrow}\varphi_{4}^{F_x}e^{-iq_{h1}^{\downarrow}(x+\mathcal{D}/2)}\,, & -\mathcal{D}/2<x<0\\
\bar{e}_{ee}^{\prime\downarrow\uparrow}\varphi_{1}^{F_y}e^{-iq_{e2}^{\uparrow}(x-\mathcal{D}/4)}+\bar{e}_{ee}^{\prime\downarrow\downarrow}\varphi_{2}^{F_y}e^{-iq_{e2}^{\downarrow}(x-\mathcal{D}/4)}+\bar{f}_{ee}^{\prime\downarrow\uparrow}\varphi_{1}^{F_y}e^{iq_{e2}^{\uparrow}x}+\bar{f}_{ee}^{\prime\downarrow\downarrow}\varphi_{2}^{F_y}e^{iq_{e2}^{\downarrow}x}\\+\bar{g}_{eh}^{\prime\downarrow\uparrow}\varphi_{3}^{F_y}e^{-iq_{h2}^{\uparrow}x}+\bar{g}_{eh}^{\prime\downarrow\downarrow}\varphi_{4}^{F_y}e^{-iq_{h2}^{\downarrow}x}+\bar{h}_{eh}^{\prime\downarrow\uparrow}\varphi_{3}^{F_y}e^{iq_{h2}^{\uparrow}(x-\mathcal{D}/4)}+\bar{h}_{eh}^{\prime\downarrow\downarrow}\varphi_{4}^{F_y}e^{iq_{h2}^{\downarrow}(x-\mathcal{D}/4)}\,, & 0<x<\mathcal{D}/4\\
\bar{m}_{ee}^{\prime\downarrow\uparrow}\varphi_{1}^{F_z}e^{-iq_{e3}^{\uparrow}(x-\mathcal{D}/2)}+\bar{m}_{ee}^{\prime\downarrow\downarrow}\varphi_{2}^{F_z}e^{-iq_{e3}^{\downarrow}(x-\mathcal{D}/2)}+\bar{n}_{ee}^{\prime\downarrow\uparrow}\varphi_{1}^{F_z}e^{iq_{e3}^{\uparrow}(x-\mathcal{D}/4)}+\bar{n}_{ee}^{\prime\downarrow\downarrow}\varphi_{2}^{F_z}e^{iq_{e3}^{\downarrow}(x-\mathcal{D}/4)}\\+\bar{o}_{eh}^{\prime\downarrow\uparrow}\varphi_{3}^{F_z}e^{-iq_{h3}^{\uparrow}(x-\mathcal{D}/4)}+\bar{o}_{eh}^{\prime\downarrow\downarrow}\varphi_{4}^{F_z}e^{-iq_{h3}^{\downarrow}(x-\mathcal{D}/4)}+\bar{p}_{eh}^{\prime\downarrow\uparrow}\varphi_{3}^{F_z}e^{iq_{h3}^{\uparrow}(x-\mathcal{D}/2)}+\bar{p}_{eh}^{\prime\downarrow\downarrow}\varphi_{4}^{F_z}e^{iq_{h3}^{\downarrow}(x-\mathcal{D}/2)}\,, & \mathcal{D}/4<x<\mathcal{D}/2\\
\varphi_{2}^{S_{R}}e^{-iq_{e}^{S}(x-\mathcal{D}/2)}+\bar{r}_{eh}^{\prime\downarrow\uparrow}\varphi_{3}^{S_{R}}e^{-iq_{h}^{S}(x-\mathcal{D}/2)}+\bar{r}_{eh}^{\prime\downarrow\downarrow}\varphi_{4}^{S_{R}}e^{-iq_{h}^{S}(x-\mathcal{D}/2)}+\bar{r}_{ee}^{\prime\downarrow\uparrow}\varphi_{1}^{S_{R}}e^{iq_{e}^{S}(x-\mathcal{D}/2)}\\+\bar{r}_{ee}^{\prime\downarrow\downarrow}\varphi_{2}^{S_{R}}e^{iq_{e}^{S}(x-\mathcal{D}/2)}\,, & x>\mathcal{D}/2
\end{cases}\\
\phi_{7}(x)&=\begin{cases}
\bar{t}_{he}^{\prime\uparrow\uparrow}\varphi_{1}^{S_L}e^{-iq_{e}^{S}(x+\mathcal{D}/2)}+\bar{t}_{he}^{\prime\uparrow\downarrow}\varphi_{2}^{S_L}e^{-iq_{e}^{S}(x+\mathcal{D}/2)}+\bar{t}_{hh}^{\prime\uparrow\uparrow}\varphi_{3}^{S_L}e^{iq_{h}^{S}(x+\mathcal{D}/2)}+\bar{t}_{hh}^{\prime\uparrow\downarrow}\varphi_{4}^{S_L}e^{iq_{h}^{S}(x+\mathcal{D}/2)}\,, & x<-\mathcal{D}/2\\
\bar{a}_{he}^{\prime\uparrow\uparrow}\varphi_{1}^{F_x}e^{iq_{e1}^{\uparrow}(x+\mathcal{D}/2)}+\bar{a}_{he}^{\prime\uparrow\downarrow}\varphi_{2}^{F_x}e^{iq_{e1}^{\downarrow}(x+\mathcal{D}/2)}+\bar{b}_{he}^{\prime\uparrow\uparrow}\varphi_{1}^{F_x}e^{-iq_{e1}^{\uparrow}x}+\bar{b}_{he}^{\prime\uparrow\downarrow}\varphi_{2}^{F_x}e^{-iq_{e1}^{\downarrow}x}\\+\bar{c}_{hh}^{\prime\uparrow\uparrow}\varphi_{3}^{F_x}e^{iq_{h1}^{\uparrow}x}+\bar{c}_{hh}^{\prime\uparrow\downarrow}\varphi_{4}^{F_x}e^{iq_{h1}^{\downarrow}x}
+\bar{d}_{hh}^{\prime\uparrow\uparrow}\varphi_{3}^{F_x}e^{-iq_{h1}^{\uparrow}(x+\mathcal{D}/2)}+\bar{d}_{hh}^{\prime\uparrow\downarrow}\varphi_{4}^{F_x}e^{-iq_{h1}^{\downarrow}(x+\mathcal{D}/2)}\,, & -\mathcal{D}/2<x<0\\
\bar{e}_{he}^{\prime\uparrow\uparrow}\varphi_{1}^{F_y}e^{-iq_{e2}^{\uparrow}(x-\mathcal{D}/4)}+\bar{e}_{he}^{\prime\uparrow\downarrow}\varphi_{2}^{F_y}e^{-iq_{e2}^{\downarrow}(x-\mathcal{D}/4)}+\bar{f}_{he}^{\prime\uparrow\uparrow}\varphi_{1}^{F_y}e^{iq_{e2}^{\uparrow}x}+\bar{f}_{he}^{\prime\uparrow\downarrow}\varphi_{2}^{F_y}e^{iq_{e2}^{\downarrow}x}\\+\bar{g}_{hh}^{\prime\uparrow\uparrow}\varphi_{3}^{F_y}e^{-iq_{h2}^{\uparrow}x}+\bar{g}_{hh}^{\prime\uparrow\downarrow}\varphi_{4}^{F_y}e^{-iq_{h2}^{\downarrow}x}+\bar{h}_{hh}^{\prime\uparrow\uparrow}\varphi_{3}^{F_y}e^{iq_{h2}^{\uparrow}(x-\mathcal{D}/4)}+\bar{h}_{hh}^{\prime\uparrow\downarrow}\varphi_{4}^{F_y}e^{iq_{h2}^{\downarrow}(x-\mathcal{D}/4)}\,, & 0<x<\mathcal{D}/4\\
\bar{m}_{he}^{\prime\uparrow\uparrow}\varphi_{1}^{F_z}e^{-iq_{e3}^{\uparrow}(x-\mathcal{D}/2)}+\bar{m}_{he}^{\prime\uparrow\downarrow}\varphi_{2}^{F_z}e^{-iq_{e3}^{\downarrow}(x-\mathcal{D}/2)}+\bar{n}_{he}^{\prime\uparrow\uparrow}\varphi_{1}^{F_z}e^{iq_{e3}^{\uparrow}(x-\mathcal{D}/4)}+\bar{n}_{he}^{\prime\uparrow\downarrow}\varphi_{2}^{F_z}e^{iq_{e3}^{\downarrow}(x-\mathcal{D}/4)}\\+\bar{o}_{hh}^{\prime\uparrow\uparrow}\varphi_{3}^{F_z}e^{-iq_{h3}^{\uparrow}(x-\mathcal{D}/4)}+\bar{o}_{hh}^{\prime\uparrow\downarrow}\varphi_{4}^{F_z}e^{-iq_{h3}^{\downarrow}(x-\mathcal{D}/4)}+\bar{p}_{hh}^{\prime\uparrow\uparrow}\varphi_{3}^{F_z}e^{iq_{h3}^{\uparrow}(x-\mathcal{D}/2)}+\bar{p}_{hh}^{\prime\uparrow\downarrow}\varphi_{4}^{F_z}e^{iq_{h3}^{\downarrow}(x-\mathcal{D}/2)}\,, & \mathcal{D}/4<x<\mathcal{D}/2\\
\varphi_{3}^{S_{R}}e^{iq_{h}^{S}(x-\mathcal{D}/2)}+\bar{r}_{hh}^{\prime\uparrow\uparrow}\varphi_{3}^{S_{R}}e^{-iq_{h}^{S}(x-\mathcal{D}/2)}+\bar{r}_{hh}^{\prime\uparrow\downarrow}\varphi_{4}^{S_{R}}e^{-iq_{h}^{S}(x-\mathcal{D}/2)}+\bar{r}_{he}^{\prime\uparrow\uparrow}\varphi_{1}^{S_{R}}e^{iq_{e}^{S}(x-\mathcal{D}/2)}\\+\bar{r}_{he}^{\prime\uparrow\downarrow}\varphi_{2}^{S_{R}}e^{iq_{e}^{S}(x-\mathcal{D}/2)}\,, & x>\mathcal{D}/2
\end{cases}\\
\end{split}\nonumber
\end{equation}
\begin{equation}
\begin{split}
\phi_{8}(x)&=\begin{cases}
\bar{t}_{he}^{\prime\downarrow\uparrow}\varphi_{1}^{S_L}e^{-iq_{e}^{S}(x+\mathcal{D}/2)}+\bar{t}_{he}^{\prime\downarrow\downarrow}\varphi_{2}^{S_L}e^{-iq_{e}^{S}(x+\mathcal{D}/2)}+\bar{t}_{hh}^{\prime\downarrow\uparrow}\varphi_{3}^{S_L}e^{iq_{h}^{S}(x+\mathcal{D}/2)}+\bar{t}_{hh}^{\prime\downarrow\downarrow}\varphi_{4}^{S_L}e^{iq_{h}^{S}(x+\mathcal{D}/2)}\,, & x<-\mathcal{D}/2\\
\bar{a}_{he}^{\prime\downarrow\uparrow}\varphi_{1}^{F_x}e^{iq_{e1}^{\uparrow}(x+\mathcal{D}/2)}+\bar{a}_{he}^{\prime\downarrow\downarrow}\varphi_{2}^{F_x}e^{iq_{e1}^{\downarrow}(x+\mathcal{D}/2)}+\bar{b}_{he}^{\prime\downarrow\uparrow}\varphi_{1}^{F_x}e^{-iq_{e1}^{\uparrow}x}+\bar{b}_{he}^{\prime\downarrow\downarrow}\varphi_{2}^{F_x}e^{-iq_{e1}^{\downarrow}x}\\+\bar{c}_{hh}^{\prime\downarrow\uparrow}\varphi_{3}^{F_x}e^{iq_{h1}^{\uparrow}x}+\bar{c}_{hh}^{\prime\downarrow\downarrow}\varphi_{4}^{F_x}e^{iq_{h1}^{\downarrow}x}
+\bar{d}_{hh}^{\prime\downarrow\uparrow}\varphi_{3}^{F_x}e^{-iq_{h1}^{\uparrow}(x+\mathcal{D}/2)}+\bar{d}_{hh}^{\prime\downarrow\downarrow}\varphi_{4}^{F_x}e^{-iq_{h1}^{\downarrow}(x+\mathcal{D}/2)}\,, & -\mathcal{D}/2<x<0\\
\bar{e}_{he}^{\prime\downarrow\uparrow}\varphi_{1}^{F_y}e^{-iq_{e2}^{\uparrow}(x-\mathcal{D}/4)}+\bar{e}_{he}^{\prime\downarrow\downarrow}\varphi_{2}^{F_y}e^{-iq_{e2}^{\downarrow}(x-\mathcal{D}/4)}+\bar{f}_{he}^{\prime\downarrow\uparrow}\varphi_{1}^{F_y}e^{iq_{e2}^{\uparrow}x}+\bar{f}_{he}^{\prime\downarrow\downarrow}\varphi_{2}^{F_y}e^{iq_{e2}^{\downarrow}x}\\+\bar{g}_{hh}^{\prime\downarrow\uparrow}\varphi_{3}^{F_y}e^{-iq_{h2}^{\uparrow}x}+\bar{g}_{hh}^{\prime\downarrow\downarrow}\varphi_{4}^{F_y}e^{-iq_{h2}^{\downarrow}x}
+\bar{h}_{hh}^{\prime\downarrow\uparrow}\varphi_{3}^{F_y}e^{iq_{h2}^{\uparrow}(x-\mathcal{D}/4)}+\bar{h}_{hh}^{\prime\downarrow\downarrow}\varphi_{4}^{F_y}e^{iq_{h2}^{\downarrow}(x-\mathcal{D}/4)}\,, & 0<x<\mathcal{D}/4\\
\bar{m}_{he}^{\prime\downarrow\uparrow}\varphi_{1}^{F_z}e^{-iq_{e3}^{\uparrow}(x-\mathcal{D}/2)}+\bar{m}_{he}^{\prime\downarrow\downarrow}\varphi_{2}^{F_z}e^{-iq_{e3}^{\downarrow}(x-\mathcal{D}/2)}+\bar{n}_{he}^{\prime\downarrow\uparrow}\varphi_{1}^{F_z}e^{iq_{e3}^{\uparrow}(x-\mathcal{D}/4)}+\bar{n}_{he}^{\prime\downarrow\downarrow}\varphi_{2}^{F_z}e^{iq_{e3}^{\downarrow}(x-\mathcal{D}/4)}\\+\bar{o}_{hh}^{\prime\downarrow\uparrow}\varphi_{3}^{F_z}e^{-iq_{h3}^{\uparrow}(x-\mathcal{D}/4)}+\bar{o}_{hh}^{\prime\downarrow\downarrow}\varphi_{4}^{F_z}e^{-iq_{h3}^{\downarrow}(x-\mathcal{D}/4)}+\bar{p}_{hh}^{\prime\downarrow\uparrow}\varphi_{3}^{F_z}e^{iq_{h3}^{\uparrow}(x-\mathcal{D}/2)}+\bar{p}_{hh}^{\prime\downarrow\downarrow}\varphi_{4}^{F_z}e^{iq_{h3}^{\downarrow}(x-\mathcal{D}/2)}\,, & \mathcal{D}/4<x<\mathcal{D}/2\\
\varphi_{4}^{S_{R}}e^{iq_{h}^{S}(x-\mathcal{D}/2)}+\bar{r}_{hh}^{\prime\downarrow\uparrow}\varphi_{3}^{S_{R}}e^{-iq_{h}^{S}(x-\mathcal{D}/2)}+\bar{r}_{hh}^{\prime\downarrow\downarrow}\varphi_{4}^{S_{R}}e^{-iq_{h}^{S}(x-\mathcal{D}/2)}+\bar{r}_{he}^{\prime\downarrow\uparrow}\varphi_{1}^{S_{R}}e^{iq_{e}^{S}(x-\mathcal{D}/2)}\\+\bar{r}_{he}^{\prime\downarrow\downarrow}\varphi_{2}^{S_{R}}e^{iq_{e}^{S}(x-\mathcal{D}/2)}\,, & x>\mathcal{D}/2
\end{cases}
\end{split}
\label{wavv}
\end{equation}
where the expressions for $\varphi_{i}^{S_L}$, $\varphi_{i}^{F_x}$, $\varphi_{i}^{F_y}$, $\varphi_{i}^{S_R}$, $u_0$, $v_0$, $q_{e,h}^{S}$ are provided below Eq.~\eqref{wav} and
{$\varphi_{1}^{F_z}=\begin{pmatrix}
   1\\
   0\\
   0\\
   0                      \end{pmatrix}$, $\varphi_{2}^{F_z}=\begin{pmatrix}
   0\\
   1\\
   0\\
   0                      \end{pmatrix}
$, $\varphi_{3}^{F_z}=\begin{pmatrix}
   0\\
   0\\
   1\\
   0                    \end{pmatrix}
$, $\varphi_{4}^{F_z}=\begin{pmatrix}
   0\\
   0\\
   0\\
   1                     \end{pmatrix}$.}
Further in Eq.~\eqref{wavv}, the terms $r_{pq}^{\prime\sigma\sigma'}$ and $\bar{r}_{pq}^{\prime\sigma\sigma'}$ denote the reflection amplitudes in left and right superconductors, respectively, while the terms $\bar{t}_{pq}^{\prime\sigma\sigma'}$ and $t_{pq}^{\prime\sigma\sigma'}$ represent the corresponding transmission amplitudes, with $\sigma,\sigma'\in\{\uparrow,\downarrow\}$ and $p,q\in\{e,h\}$. $q_{e1,h1}^{\sigma}=\sqrt{\frac{2m^{*}}{\hbar^2}\big(E_F\pm\omega+\rho_\sigma m_1\big)}$ are the wavevectors in $F_x$ layer, $q_{e2,h2}^{\sigma}=\sqrt{\frac{2m^{*}}{\hbar^2}\big(E_F\pm\omega+\rho_\sigma m_2\big)}$ represent the wavevectors in $F_y$ layer and finally $q_{e3,h3}^{\sigma}=\sqrt{\frac{2m^{*}}{\hbar^2}\big(E_F\pm\omega+\rho_\sigma m_3\big)}$ denote the wavevectors in $F_z$ layer. After diagonalizing the Hamiltonian $\Big(H_{BdG}^{\mbox{S-$F_x$-$F_y$-$F_z$-S}}\Big)^{*}(-k)$ instead of $H_{BdG}^{\mbox{S-$F_x$-$F_y$-$F_z$-S}}(k)$ (see Eq.~\eqref{hmfff}), we will obtain conjugate processes $\tilde{\phi}_{l}$ necessary to form $\Gamma^{r}(x,\chi,\omega)$ in next section. We note that $\tilde{\varphi_{i}}^{F_z}=\varphi_{i}^{F_z}$, {and the expressions for $\tilde{\varphi_{i}}^{S_L}$, $\tilde{\varphi_{i}}^{F_x}$, $\tilde{\varphi_{i}}^{F_y}$, $\tilde{\varphi_{i}}^{S_R}$ are provided below Eq.~\eqref{wav}.}

The boundary conditions at the S-$F_x$ interface ($x=-\mathcal{D}/2$) are:
\begin{equation}
\label{eq8}
\phi_{l}|_{x<-\mathcal{D}/2}=\phi_{l}|_{-\mathcal{D}/2<x<0}\,\,\,\,\,\mbox{and}\,\,\,\,\,\frac{d\phi_{l}|_{-\mathcal{D}/2<x<0}}{dx}-\frac{d\phi_{l}|_{x<-\mathcal{D}/2}}{dx}=\frac{2m^{*}\mathcal{B}_1}{\hbar^2}\phi_{l}(x=-\mathcal{D}/2).
\end{equation}
Similarly, the boundary conditions at the $F_x$-$F_y$ interface ($x=0$) are:
\begin{equation}
\phi_{l}|_{-\mathcal{D}/2<x<0}=\phi_{l}|_{0<x<\mathcal{D}/4}\,\,\,\,\,\mbox{and}\,\,\,\,\,\frac{d\phi_{l}|_{0<x<\mathcal{D}/4}}{dx}-\frac{d\phi_{l}|_{-\mathcal{D}/2<x<0}}{dx}=\frac{2m^{*}\mathcal{B}_3}{\hbar^2}\phi_{l}(x=0).
\end{equation}
The boundary conditions at the $F_y$-$F_z$ interface ($x=\mathcal{D}/4$) are:
\begin{equation}
\phi_{l}|_{0<x<\mathcal{D}/4}=\phi_{l}|_{\mathcal{D}/4<x<\mathcal{D}/2}\,\,\,\,\,\mbox{and}\,\,\,\,\,\frac{d\phi_{l}|_{\mathcal{D}/4<x<\mathcal{D}/2}}{dx}-\frac{d\phi_{l}|_{0<x<\mathcal{D}/4}}{dx}=\frac{2m^{*}\mathcal{B}_4}{\hbar^2}\phi_{l}(x=\mathcal{D}/4).
\end{equation}
Finally, the boundary conditions at the $F_z$-S interface ($x=\mathcal{D}/2$) are:
\begin{equation}
\label{eq11}
\phi_{l}|_{\mathcal{D}/4<x<\mathcal{D}/2}=\phi_{l}|_{x>\mathcal{D}/2}\,\,\,\,\,\mbox{and}\,\,\,\,\,\frac{d\phi_{l}|_{x>\mathcal{D}/2}}{dx}-\frac{d\phi_{l}|_{\mathcal{D}/4<x<\mathcal{D}/2}}{dx}=\frac{2m^{*}\mathcal{B}_2}{\hbar^2}\phi_{l}(x=\mathcal{D}/2).
\end{equation}
Solving the boundary conditions (Eqs.~\eqref{eq8}-\eqref{eq11}), a total of $32$ equations are obtained for each kind of incident process described in Eq.~\eqref{wavv}. The scattering amplitudes, such as $r_{pq}^{\prime\sigma\sigma'}$, $\bar{r}_{pq}^{\prime\sigma\sigma'}$, $\bar{t}_{pq}^{\prime\sigma\sigma'}$, $t_{pq}^{\prime\sigma\sigma'}$, are derived by solving these $32$ equations. $s_{1}=r_{eh}^{\prime\uparrow\downarrow}$, $s_{2}= r_{eh}^{\prime\downarrow\uparrow}$, $s_{3} = r_{he}^{\prime\uparrow\downarrow}$, and $s_{4} = r_{he}^{\prime\downarrow\uparrow}$, obtained by putting the wavefunctions
$\phi_{1}$, $\phi_{2}$, $\phi_{3}$, and $\phi_{4}$ into Eqs.~\eqref{eq8}-\eqref{eq11} are used to compute the Josephson current via Eq.~\eqref{eq3}.

{\subsection{Long junction limit}
We diagonalize the BdG Hamiltonians (Eqs.~\eqref{hmfxfy} and \eqref{hmfxfyfz}) to derive the wavefunctions corresponding to different scattering processes in the distinct regions of the bilayer S-$F_x$-$F_y$-S and trilayer S-$F_x$-$F_y$-$F_z$-S JJs, as depicted in
Figs.~12(a) and 12(b), respectively. The wavefunctions for bilayer S-$F_x$-$F_y$-S JJ are as follows:
\begin{equation}
\begin{split}
\phi_{1}(x)&=\begin{cases}
\varphi_{1}^{S_L}e^{iq_{e}^{S}x}+\tilde{r}_{eh}^{\uparrow\uparrow}\varphi_{3}^{S_L}e^{iq_{h}^{S}x}+\tilde{r}_{eh}^{\uparrow\downarrow}\varphi_{4}^{S_L}e^{iq_{h}^{S}x}+\tilde{r}_{ee}^{\uparrow\uparrow}\varphi_{1}^{S_L}e^{-iq_{e}^{S}x}+\tilde{r}_{ee}^{\uparrow\downarrow}\varphi_{2}^{S_L}e^{-iq_{e}^{S}x}\,, & x<0\\
\tilde{a}_{ee}^{\uparrow\uparrow}\varphi_{1}^{F_x}e^{iq_{e1}^{\uparrow}x}+\tilde{a}_{ee}^{\uparrow\downarrow}\varphi_{2}^{F_x}e^{iq_{e1}^{\downarrow}x}+\tilde{b}_{ee}^{\uparrow\uparrow}\varphi_{1}^{F_x}e^{-iq_{e1}^{\uparrow}(x-\mathcal{D}_{1})}+\tilde{b}_{ee}^{\uparrow\downarrow}\varphi_{2}^{F_x}e^{-iq_{e1}^{\downarrow}(x-\mathcal{D}_{1})}\\+\tilde{c}_{eh}^{\uparrow\uparrow}\varphi_{3}^{F_x}e^{iq_{h1}^{\uparrow}(x-\mathcal{D}_{1})}+\tilde{c}_{eh}^{\uparrow\downarrow}\varphi_{4}^{F_x}e^{iq_{h1}^{\downarrow}(x-\mathcal{D}_{1})}
+\tilde{d}_{eh}^{\uparrow\uparrow}\varphi_{3}^{F_x}e^{-iq_{h1}^{\uparrow}x}+\tilde{d}_{eh}^{\uparrow\downarrow}\varphi_{4}^{F_x}e^{-iq_{h1}^{\downarrow}x}\,, & 0<x<\mathcal{D}_{1}\\
\tilde{e}_{ee}^{\uparrow\uparrow}\varphi_{1}^{F_y}e^{iq_{e2}^{\uparrow}(x-\mathcal{D}_{1})}+\tilde{e}_{ee}^{\uparrow\downarrow}\varphi_{2}^{F_y}e^{iq_{e2}^{\downarrow}(x-\mathcal{D}_{1})}+\tilde{f}_{ee}^{\uparrow\uparrow}\varphi_{1}^{F_y}e^{-iq_{e2}^{\uparrow}(x-\mathcal{D}^{\prime})}+\tilde{f}_{ee}^{\uparrow\downarrow}\varphi_{2}^{F_y}e^{-iq_{e2}^{\downarrow}(x-\mathcal{D}^{\prime})}\\+\tilde{g}_{eh}^{\uparrow\uparrow}\varphi_{3}^{F_y}e^{iq_{h2}^{\uparrow}(x-\mathcal{D}^{\prime})}+\tilde{g}_{eh}^{\uparrow\downarrow}\varphi_{4}^{F_y}e^{iq_{h2}^{\downarrow}(x-\mathcal{D}^{\prime})}+\tilde{h}_{eh}^{\uparrow\uparrow}\varphi_{3}^{F_y}e^{-iq_{h2}^{\uparrow}(x-\mathcal{D}_{1})}+\tilde{h}_{eh}^{\uparrow\downarrow}\varphi_{4}^{F_y}e^{-iq_{h2}^{\downarrow}(x-\mathcal{D}_{1})}\,, & \mathcal{D}_{1}<x<\mathcal{D}^{\prime}\\
\tilde{t}_{ee}^{\uparrow\uparrow}\varphi_{1}^{S_{R}}e^{iq_{e}^{S}(x-\mathcal{D}^{\prime})}+\tilde{t}_{ee}^{\uparrow\downarrow}\varphi_{2}^{S_{R}}e^{iq_{e}^{S}(x-\mathcal{D}^{\prime})}+\tilde{t}_{eh}^{\uparrow\uparrow}\varphi_{3}^{S_{R}}e^{-iq_{h}^{S}(x-\mathcal{D}^{\prime})}+\tilde{t}_{eh}^{\uparrow\downarrow}\varphi_{4}^{S_{R}}e^{-iq_{h}^{S}(x-\mathcal{D}^{\prime})}\,, & x>\mathcal{D}^{\prime}
\end{cases}\\
\phi_{2}(x)&=\begin{cases}
\varphi_{2}^{S_L}e^{iq_{e}^{S}x}+\tilde{r}_{eh}^{\downarrow\uparrow}\varphi_{3}^{S_L}e^{iq_{h}^{S}x}+\tilde{r}_{eh}^{\downarrow\downarrow}\varphi_{4}^{S_L}e^{iq_{h}^{S}x}+\tilde{r}_{ee}^{\downarrow\uparrow}\varphi_{1}^{S_L}e^{-iq_{e}^{S}x}+\tilde{r}_{ee}^{\downarrow\downarrow}\varphi_{2}^{S_L}e^{-iq_{e}^{S}x}\,, & x<0\\
\tilde{a}_{ee}^{\downarrow\uparrow}\varphi_{1}^{F_x}e^{iq_{e1}^{\uparrow}x}+\tilde{a}_{ee}^{\downarrow\downarrow}\varphi_{2}^{F_x}e^{iq_{e1}^{\downarrow}x}+\tilde{b}_{ee}^{\downarrow\uparrow}\varphi_{1}^{F_x}e^{-iq_{e1}^{\uparrow}(x-\mathcal{D}_{1})}+\tilde{b}_{ee}^{\downarrow\downarrow}\varphi_{2}^{F_x}e^{-iq_{e1}^{\downarrow}(x-\mathcal{D}_{1})}\\+\tilde{c}_{eh}^{\downarrow\uparrow}\varphi_{3}^{F_x}e^{iq_{h1}^{\uparrow}(x-\mathcal{D}_{1})}+\tilde{c}_{eh}^{\downarrow\downarrow}\varphi_{4}^{F_x}e^{iq_{h1}^{\downarrow}(x-\mathcal{D}_{1})} +\tilde{d}_{eh}^{\downarrow\uparrow}\varphi_{3}^{F_x}e^{-iq_{h1}^{\uparrow}x}+\tilde{d}_{eh}^{\downarrow\downarrow}\varphi_{4}^{F_x}e^{-iq_{h1}^{\downarrow}x}\,, &  0<x<\mathcal{D}_{1}\\
\tilde{e}_{ee}^{\downarrow\uparrow}\varphi_{1}^{F_y}e^{iq_{e2}^{\uparrow}(x-\mathcal{D}_{1})}+\tilde{e}_{ee}^{\downarrow\downarrow}\varphi_{2}^{F_y}e^{iq_{e2}^{\downarrow}(x-\mathcal{D}_{1})}+\tilde{f}_{ee}^{\downarrow\uparrow}\varphi_{1}^{F_y}e^{-iq_{e2}^{\uparrow}(x-\mathcal{D}^{\prime})}+\tilde{f}_{ee}^{\downarrow\downarrow}\varphi_{2}^{F_y}e^{-iq_{e2}^{\downarrow}(x-\mathcal{D}^{\prime})}\\+\tilde{g}_{eh}^{\downarrow\uparrow}\varphi_{3}^{F_y}e^{iq_{h2}^{\uparrow}(x-\mathcal{D}^{\prime})}+\tilde{g}_{eh}^{\downarrow\downarrow}\varphi_{4}^{F_y}e^{iq_{h2}^{\downarrow}(x-\mathcal{D}^{\prime})}+\tilde{h}_{eh}^{\downarrow\uparrow}\varphi_{3}^{F_y}e^{-iq_{h2}^{\uparrow}(x-\mathcal{D}_{1})}+\tilde{h}_{eh}^{\downarrow\downarrow}\varphi_{4}^{F_y}e^{-iq_{h2}^{\downarrow}(x-\mathcal{D}_{1})}\,, & \mathcal{D}_{1}<x<\mathcal{D}^{\prime}\\
\tilde{t}_{ee}^{\downarrow\uparrow}\varphi_{1}^{S_{R}}e^{iq_{e}^{S}(x-\mathcal{D}^{\prime})}+\tilde{t}_{ee}^{\downarrow\downarrow}\varphi_{2}^{S_{R}}e^{iq_{e}^{S}(x-\mathcal{D}^{\prime})}+\tilde{t}_{eh}^{\downarrow\uparrow}\varphi_{3}^{S_{R}}e^{-iq_{h}^{S}(x-\mathcal{D}^{\prime})}+\tilde{t}_{eh}^{\downarrow\downarrow}\varphi_{4}^{S_{R}}e^{-iq_{h}^{S}(x-\mathcal{D}^{\prime})}\,, & x>\mathcal{D}^{\prime}
\end{cases}\\
\phi_{3}(x)&=\begin{cases}
\varphi_{3}^{S_L}e^{-iq_{h}^{S}x}+\tilde{r}_{hh}^{\uparrow\uparrow}\varphi_{3}^{S_L}e^{iq_{h}^{S}x}+\tilde{r}_{hh}^{\uparrow\downarrow}\varphi_{4}^{S_L}e^{iq_{h}^{S}x}+\tilde{r}_{he}^{\uparrow\uparrow}\varphi_{1}^{S_L}e^{-iq_{e}^{S}x}+\tilde{r}_{he}^{\uparrow\downarrow}\varphi_{2}^{S_L}e^{-iq_{e}^{S}x}\,, & x<0\\
\tilde{a}_{he}^{\uparrow\uparrow}\varphi_{1}^{F_x}e^{iq_{e1}^{\uparrow}x}+\tilde{a}_{he}^{\uparrow\downarrow}\varphi_{2}^{F_x}e^{iq_{e1}^{\downarrow}x}+\tilde{b}_{he}^{\uparrow\uparrow}\varphi_{1}^{F_x}e^{-iq_{e1}^{\uparrow}(x-\mathcal{D}_{1})}+\tilde{b}_{he}^{\uparrow\downarrow}\varphi_{2}^{F_x}e^{-iq_{e1}^{\downarrow}(x-\mathcal{D}_{1})}\\+\tilde{c}_{hh}^{\uparrow\uparrow}\varphi_{3}^{F_x}e^{iq_{h1}^{\uparrow}(x-\mathcal{D}_{1})}+\tilde{c}_{hh}^{\uparrow\downarrow}\varphi_{4}^{F_x}e^{iq_{h1}^{\downarrow}(x-\mathcal{D}_{1})} +\tilde{d}_{hh}^{\uparrow\uparrow}\varphi_{3}^{F_x}e^{-iq_{h1}^{\uparrow}x}+\tilde{d}_{hh}^{\uparrow\downarrow}\varphi_{4}^{F_x}e^{-iq_{h1}^{\downarrow}x}\,, & 0<x<\mathcal{D}_{1}\\
\tilde{e}_{he}^{\uparrow\uparrow}\varphi_{1}^{F_y}e^{iq_{e2}^{\uparrow}(x-\mathcal{D}_{1})}+\tilde{e}_{he}^{\uparrow\downarrow}\varphi_{2}^{F_y}e^{iq_{e2}^{\downarrow}(x-\mathcal{D}_{1})}+\tilde{f}_{he}^{\uparrow\uparrow}\varphi_{1}^{F_y}e^{-iq_{e2}^{\uparrow}(x-\mathcal{D}^{\prime})}+\tilde{f}_{he}^{\uparrow\downarrow}\varphi_{2}^{F_y}e^{-iq_{e2}^{\downarrow}(x-\mathcal{D}^{\prime})}\\+\tilde{g}_{hh}^{\uparrow\uparrow}\varphi_{3}^{F_y}e^{iq_{h2}^{\uparrow}(x-\mathcal{D}^{\prime})}+\tilde{g}_{hh}^{\uparrow\downarrow}\varphi_{4}^{F_y}e^{iq_{h2}^{\downarrow}(x-\mathcal{D}^{\prime})}+\tilde{h}_{hh}^{\uparrow\uparrow}\varphi_{3}^{F_y}e^{-iq_{h2}^{\uparrow}(x-\mathcal{D}_{1})}+\tilde{h}_{hh}^{\uparrow\downarrow}\varphi_{4}^{F_y}e^{-iq_{h2}^{\downarrow}(x-\mathcal{D}_{1})}\,, & \mathcal{D}_{1}<x<\mathcal{D}^{\prime}\\
\tilde{t}_{he}^{\uparrow\uparrow}\varphi_{1}^{S_{R}}e^{iq_{e}^{S}(x-\mathcal{D}^{\prime})}+\tilde{t}_{he}^{\uparrow\downarrow}\varphi_{2}^{S_{R}}e^{iq_{e}^{S}(x-\mathcal{D}^{\prime})}+\tilde{t}_{hh}^{\uparrow\uparrow}\varphi_{3}^{S_{R}}e^{-iq_{h}^{S}(x-\mathcal{D}^{\prime})}+\tilde{t}_{hh}^{\uparrow\downarrow}\varphi_{4}^{S_{R}}e^{-iq_{h}^{S}(x-\mathcal{D}^{\prime})}\,, & x>\mathcal{D}^{\prime}
\end{cases}\\
\phi_{4}(x)&=\begin{cases}
\varphi_{4}^{S_L}e^{-iq_{h}^{S}x}+\tilde{r}_{hh}^{\downarrow\uparrow}\varphi_{3}^{S_L}e^{iq_{h}^{S}x}+\tilde{r}_{hh}^{\downarrow\downarrow}\varphi_{4}^{S_L}e^{iq_{h}^{S}x}+\tilde{r}_{he}^{\prime\downarrow\uparrow}\varphi_{1}^{S_L}e^{-iq_{e}^{S}x}+\tilde{r}_{he}^{\downarrow\downarrow}\varphi_{2}^{S_L}e^{-iq_{e}^{S}x}\,, & x<0\\
\tilde{a}_{he}^{\downarrow\uparrow}\varphi_{1}^{F_x}e^{iq_{e1}^{\uparrow}x}+\tilde{a}_{he}^{\downarrow\downarrow}\varphi_{2}^{F_x}e^{iq_{e1}^{\downarrow}x}+\tilde{b}_{he}^{\downarrow\uparrow}\varphi_{1}^{F_x}e^{-iq_{e1}^{\uparrow}(x-\mathcal{D}_{1})}+\tilde{b}_{he}^{\prime\downarrow\downarrow}\varphi_{2}^{F_x}e^{-iq_{e1}^{\downarrow}(x-\mathcal{D}_{1})}\\+\tilde{c}_{hh}^{\downarrow\uparrow}\varphi_{3}^{F_x}e^{iq_{h1}^{\uparrow}(x-\mathcal{D}_{1})}+\tilde{c}_{hh}^{\downarrow\downarrow}\varphi_{4}^{F_x}e^{iq_{h1}^{\downarrow}(x-\mathcal{D}_{1})} +\tilde{d}_{hh}^{\downarrow\uparrow}\varphi_{3}^{F_x}e^{-iq_{h1}^{\uparrow}x}+\tilde{d}_{hh}^{\downarrow\downarrow}\varphi_{4}^{F_x}e^{-iq_{h1}^{\downarrow}x}\,, & 0<x<\mathcal{D}_{1}\\
\tilde{e}_{he}^{\downarrow\uparrow}\varphi_{1}^{F_y}e^{iq_{e2}^{\uparrow}(x-\mathcal{D}_{1})}+\tilde{e}_{he}^{\downarrow\downarrow}\varphi_{2}^{F_y}e^{iq_{e2}^{\downarrow}(x-\mathcal{D}_{1})}+\tilde{f}_{he}^{\downarrow\uparrow}\varphi_{1}^{F_y}e^{-iq_{e2}^{\uparrow}(x-\mathcal{D}^{\prime})}+\tilde{f}_{he}^{\downarrow\downarrow}\varphi_{2}^{F_y}e^{-iq_{e2}^{\downarrow}(x-\mathcal{D}^{\prime})}\\+\tilde{g}_{hh}^{\downarrow\uparrow}\varphi_{3}^{F_y}e^{iq_{h2}^{\uparrow}(x-\mathcal{D}^{\prime})}+\tilde{g}_{hh}^{\downarrow\downarrow}\varphi_{4}^{F_y}e^{iq_{h2}^{\downarrow}(x-\mathcal{D}^{\prime})}
+\tilde{h}_{hh}^{\downarrow\uparrow}\varphi_{3}^{F_y}e^{-iq_{h2}^{\uparrow}(x-\mathcal{D}_{1})}+\tilde{h}_{hh}^{\downarrow\downarrow}\varphi_{4}^{F_y}e^{-iq_{h2}^{\downarrow}(x-\mathcal{D}_{1})}\,, & \mathcal{D}_{1}<x<\mathcal{D}^{\prime}\\
\tilde{t}_{he}^{\downarrow\uparrow}\varphi_{1}^{S_{R}}e^{iq_{e}^{S}(x-\mathcal{D}^{\prime})}+\tilde{t}_{he}^{\downarrow\downarrow}\varphi_{2}^{S_{R}}e^{iq_{e}^{S}(x-\mathcal{D}^{\prime})}+\tilde{t}_{hh}^{\downarrow\uparrow}\varphi_{3}^{S_{R}}e^{-iq_{h}^{S}(x-\mathcal{D}^{\prime})}+\tilde{t}_{hh}^{\downarrow\downarrow}\varphi_{4}^{S_{R}}e^{-iq_{h}^{S}(x-\mathcal{D}^{\prime})}\,, & x>\mathcal{D}^{\prime}
\end{cases}\\
\phi_{5}(x)&=\begin{cases}
\tilde{t}_{ee}^{\prime\uparrow\uparrow}\varphi_{1}^{S_L}e^{-iq_{e}^{S}x}+\tilde{t}_{ee}^{\prime\uparrow\downarrow}\varphi_{2}^{S_L}e^{-iq_{e}^{S}x}+\tilde{t}_{eh}^{\prime\uparrow\uparrow}\varphi_{3}^{S_L}e^{iq_{h}^{S}x}+\tilde{t}_{eh}^{\prime\uparrow\downarrow}\varphi_{4}^{S_L}e^{iq_{h}^{S}x}\,, & x<0\\
\tilde{a}_{ee}^{\prime\uparrow\uparrow}\varphi_{1}^{F_x}e^{iq_{e1}^{\uparrow}x}+\tilde{a}_{ee}^{\prime\uparrow\downarrow}\varphi_{2}^{F_x}e^{iq_{e1}^{\downarrow}x}+\tilde{b}_{ee}^{\prime\uparrow\uparrow}\varphi_{1}^{F_x}e^{-iq_{e1}^{\uparrow}(x-\mathcal{D}_{1})}+\tilde{b}_{ee}^{\prime\uparrow\downarrow}\varphi_{2}^{F_x}e^{-iq_{e1}^{\downarrow}(x-\mathcal{D}_{1})}\\+\tilde{c}_{eh}^{\prime\uparrow\uparrow}\varphi_{3}^{F_x}e^{iq_{h1}^{\uparrow}(x-\mathcal{D}_{1})}+\tilde{c}_{eh}^{\prime\uparrow\downarrow}\varphi_{4}^{F_x}e^{iq_{h1}^{\downarrow}(x-\mathcal{D}_{1})}
+\tilde{d}_{eh}^{\prime\uparrow\uparrow}\varphi_{3}^{F_x}e^{-iq_{h1}^{\uparrow}x}+\tilde{d}_{eh}^{\prime\uparrow\downarrow}\varphi_{4}^{F_x}e^{-iq_{h1}^{\downarrow}x}\,, & 0<x<\mathcal{D}_{1}\\
\tilde{e}_{ee}^{\prime\uparrow\uparrow}\varphi_{1}^{F_y}e^{iq_{e2}^{\uparrow}(x-\mathcal{D}_{1})}+\tilde{e}_{ee}^{\prime\uparrow\downarrow}\varphi_{2}^{F_y}e^{iq_{e2}^{\downarrow}(x-\mathcal{D}_{1})}+\tilde{f}_{ee}^{\prime\uparrow\uparrow}\varphi_{1}^{F_y}e^{-iq_{e2}^{\uparrow}(x-\mathcal{D}^{\prime})}+\tilde{f}_{ee}^{\prime\uparrow\downarrow}\varphi_{2}^{F_y}e^{-iq_{e2}^{\downarrow}(x-\mathcal{D}^{\prime})}\\+\tilde{g}_{eh}^{\prime\uparrow\uparrow}\varphi_{3}^{F_y}e^{iq_{h2}^{\uparrow}(x-\mathcal{D}^{\prime})}+\tilde{g}_{eh}^{\prime\uparrow\downarrow}\varphi_{4}^{F_y}e^{iq_{h2}^{\downarrow}(x-\mathcal{D}^{\prime})}+\tilde{h}_{eh}^{\prime\uparrow\uparrow}\varphi_{3}^{F_y}e^{-iq_{h2}^{\uparrow}(x-\mathcal{D}_{1})}+\tilde{h}_{eh}^{\prime\uparrow\downarrow}\varphi_{4}^{F_y}e^{-iq_{h2}^{\downarrow}(x-\mathcal{D}_{1})}\,, & \mathcal{D}_{1}<x<\mathcal{D}^{\prime}\\
\varphi_{1}^{S_{R}}e^{-iq_{e}^{S}(x-\mathcal{D}^{\prime})}+\tilde{r}_{eh}^{\prime\uparrow\uparrow}\varphi_{3}^{S_{R}}e^{-iq_{h}^{S}(x-\mathcal{D}^{\prime})}+\tilde{r}_{eh}^{\prime\uparrow\downarrow}\varphi_{4}^{S_{R}}e^{-iq_{h}^{S}(x-\mathcal{D}^{\prime})}+\tilde{r}_{ee}^{\prime\uparrow\uparrow}\varphi_{1}^{S_{R}}e^{iq_{e}^{S}(x-\mathcal{D}^{\prime})}\\+\tilde{r}_{ee}^{\prime\uparrow\downarrow}\varphi_{2}^{S_{R}}e^{iq_{e}^{S}(x-\mathcal{D}^{\prime})}\,, & x>\mathcal{D}^{\prime}
\end{cases}\\
\phi_{6}(x)&=\begin{cases}
\tilde{t}_{ee}^{\prime\downarrow\uparrow}\varphi_{1}^{S_L}e^{-iq_{e}^{S}x}+\tilde{t}_{ee}^{\prime\downarrow\downarrow}\varphi_{2}^{S_L}e^{-iq_{e}^{S}x}+\tilde{t}_{eh}^{\prime\downarrow\uparrow}\varphi_{3}^{S_L}e^{iq_{h}^{S}x}+\tilde{t}_{eh}^{\prime\downarrow\downarrow}\varphi_{4}^{S_L}e^{iq_{h}^{S}x}\,, & x<0\\
\tilde{a}_{ee}^{\prime\downarrow\uparrow}\varphi_{1}^{F_x}e^{iq_{e1}^{\uparrow}x}+\tilde{a}_{ee}^{\prime\downarrow\downarrow}\varphi_{2}^{F_x}e^{iq_{e1}^{\downarrow}x}+\tilde{b}_{ee}^{\prime\downarrow\uparrow}\varphi_{1}^{F_x}e^{-iq_{e1}^{\uparrow}(x-\mathcal{D}_{1})}+\tilde{b}_{ee}^{\prime\downarrow\downarrow}\varphi_{2}^{F_x}e^{-iq_{e1}^{\downarrow}(x-\mathcal{D}_{1})}\\+\tilde{c}_{eh}^{\prime\downarrow\uparrow}\varphi_{3}^{F_x}e^{iq_{h1}^{\uparrow}(x-\mathcal{D}_{1})}+\tilde{c}_{eh}^{\prime\downarrow\downarrow}\varphi_{4}^{F_x}e^{iq_{h1}^{\downarrow}(x-\mathcal{D}_{1})}
+\tilde{d}_{eh}^{\prime\downarrow\uparrow}\varphi_{3}^{F_x}e^{-iq_{h1}^{\uparrow}x}+\tilde{d}_{eh}^{\prime\downarrow\downarrow}\varphi_{4}^{F_x}e^{-iq_{h1}^{\downarrow}x}\,, & 0<x<\mathcal{D}_{1}\\
\tilde{e}_{ee}^{\prime\downarrow\uparrow}\varphi_{1}^{F_y}e^{iq_{e2}^{\uparrow}(x-\mathcal{D}_{1})}+\tilde{e}_{ee}^{\prime\downarrow\downarrow}\varphi_{2}^{F_y}e^{iq_{e2}^{\downarrow}(x-\mathcal{D}_{1})}+\tilde{f}_{ee}^{\prime\downarrow\uparrow}\varphi_{1}^{F_y}e^{-iq_{e2}^{\uparrow}(x-\mathcal{D}^{\prime})}+\tilde{f}_{ee}^{\prime\downarrow\downarrow}\varphi_{2}^{F_y}e^{-iq_{e2}^{\downarrow}(x-\mathcal{D}^{\prime})}\\+\tilde{g}_{eh}^{\prime\downarrow\uparrow}\varphi_{3}^{F_y}e^{iq_{h2}^{\uparrow}(x-\mathcal{D}^{\prime})}+\tilde{g}_{eh}^{\prime\downarrow\downarrow}\varphi_{4}^{F_y}e^{iq_{h2}^{\downarrow}(x-\mathcal{D}^{\prime})}+\tilde{h}_{eh}^{\prime\downarrow\uparrow}\varphi_{3}^{F_y}e^{-iq_{h2}^{\uparrow}(x-\mathcal{D}_{1})}+\tilde{h}_{eh}^{\prime\downarrow\downarrow}\varphi_{4}^{F_y}e^{-iq_{h2}^{\downarrow}(x-\mathcal{D}_{1})}\,, & \mathcal{D}_{1}<x<\mathcal{D}^{\prime}\\
\varphi_{2}^{S_{R}}e^{-iq_{e}^{S}(x-\mathcal{D}^{\prime})}+\tilde{r}_{eh}^{\prime\downarrow\uparrow}\varphi_{3}^{S_{R}}e^{-iq_{h}^{S}(x-\mathcal{D}^{\prime})}+\tilde{r}_{eh}^{\prime\downarrow\downarrow}\varphi_{4}^{S_{R}}e^{-iq_{h}^{S}(x-\mathcal{D}^{\prime})}+\tilde{r}_{ee}^{\prime\downarrow\uparrow}\varphi_{1}^{S_{R}}e^{iq_{e}^{S}(x-\mathcal{D}^{\prime})}\\+\tilde{r}_{ee}^{\prime\downarrow\downarrow}\varphi_{2}^{S_{R}}e^{iq_{e}^{S}(x-\mathcal{D}^{\prime})}\,, & x>\mathcal{D}^{\prime}
\end{cases}\\
\end{split}\nonumber
\end{equation}
\newpage
\begin{equation}
\begin{split}
\phi_{7}(x)&=\begin{cases}
\tilde{t}_{he}^{\prime\uparrow\uparrow}\varphi_{1}^{S_L}e^{-iq_{e}^{S}x}+\tilde{t}_{he}^{\prime\uparrow\downarrow}\varphi_{2}^{S_L}e^{-iq_{e}^{S}x}+\tilde{t}_{hh}^{\prime\uparrow\uparrow}\varphi_{3}^{S_L}e^{iq_{h}^{S}x}+\tilde{t}_{hh}^{\prime\uparrow\downarrow}\varphi_{4}^{S_L}e^{iq_{h}^{S}x}\,, & x<0\\
\tilde{a}_{he}^{\prime\uparrow\uparrow}\varphi_{1}^{F_x}e^{iq_{e1}^{\uparrow}x}+\tilde{a}_{he}^{\prime\uparrow\downarrow}\varphi_{2}^{F_x}e^{iq_{e1}^{\downarrow}x}+\tilde{b}_{he}^{\prime\uparrow\uparrow}\varphi_{1}^{F_x}e^{-iq_{e1}^{\uparrow}(x-\mathcal{D}_{1})}+\tilde{b}_{he}^{\prime\uparrow\downarrow}\varphi_{2}^{F_x}e^{-iq_{e1}^{\downarrow}(x-\mathcal{D}_{1})}\\+\tilde{c}_{hh}^{\prime\uparrow\uparrow}\varphi_{3}^{F_x}e^{iq_{h1}^{\uparrow}(x-\mathcal{D}_{1})}+\tilde{c}_{hh}^{\prime\uparrow\downarrow}\varphi_{4}^{F_x}e^{iq_{h1}^{\downarrow}(x-\mathcal{D}_{1})}
+\tilde{d}_{hh}^{\prime\uparrow\uparrow}\varphi_{3}^{F_x}e^{-iq_{h1}^{\uparrow}x}+\tilde{d}_{hh}^{\prime\uparrow\downarrow}\varphi_{4}^{F_x}e^{-iq_{h1}^{\downarrow}x}\,, & 0<x<\mathcal{D}_{1}\\
\tilde{e}_{he}^{\prime\uparrow\uparrow}\varphi_{1}^{F_y}e^{iq_{e2}^{\uparrow}(x-\mathcal{D}_{1})}+\tilde{e}_{he}^{\prime\uparrow\downarrow}\varphi_{2}^{F_y}e^{iq_{e2}^{\downarrow}(x-\mathcal{D}_{1})}+\tilde{f}_{he}^{\prime\uparrow\uparrow}\varphi_{1}^{F_y}e^{-iq_{e2}^{\uparrow}(x-\mathcal{D}^{\prime})}+\tilde{f}_{he}^{\prime\uparrow\downarrow}\varphi_{2}^{F_y}e^{-iq_{e2}^{\downarrow}(x-\mathcal{D}^{\prime})}\\+\tilde{g}_{hh}^{\prime\uparrow\uparrow}\varphi_{3}^{F_y}e^{iq_{h2}^{\uparrow}(x-\mathcal{D}^{\prime})}+\tilde{g}_{hh}^{\prime\uparrow\downarrow}\varphi_{4}^{F_y}e^{iq_{h2}^{\downarrow}(x-\mathcal{D}^{\prime})}+\tilde{h}_{hh}^{\prime\uparrow\uparrow}\varphi_{3}^{F_y}e^{-iq_{h2}^{\uparrow}(x-\mathcal{D}_{1})}+\tilde{h}_{hh}^{\prime\uparrow\downarrow}\varphi_{4}^{F_y}e^{-iq_{h2}^{\downarrow}(x-\mathcal{D}_{1})}\,, & \mathcal{D}_{1}<x<\mathcal{D}^{\prime}\\
\varphi_{3}^{S_{R}}e^{iq_{h}^{S}(x-\mathcal{D}^{\prime})}+\tilde{r}_{hh}^{\prime\uparrow\uparrow}\varphi_{3}^{S_{R}}e^{-iq_{h}^{S}(x-\mathcal{D}^{\prime})}+\tilde{r}_{hh}^{\prime\uparrow\downarrow}\varphi_{4}^{S_{R}}e^{-iq_{h}^{S}(x-\mathcal{D}^{\prime})}+\tilde{r}_{he}^{\prime\uparrow\uparrow}\varphi_{1}^{S_{R}}e^{iq_{e}^{S}(x-\mathcal{D}^{\prime})}\\+\tilde{r}_{he}^{\prime\uparrow\downarrow}\varphi_{2}^{S_{R}}e^{iq_{e}^{S}(x-\mathcal{D}^{\prime})}\,, & x>\mathcal{D}^{\prime}
\end{cases}\\
\phi_{8}(x)&=\begin{cases}
\tilde{t}_{he}^{\prime\downarrow\uparrow}\varphi_{1}^{S_L}e^{-iq_{e}^{S}x}+\tilde{t}_{he}^{\prime\downarrow\downarrow}\varphi_{2}^{S_L}e^{-iq_{e}^{S}x}+\tilde{t}_{hh}^{\prime\downarrow\uparrow}\varphi_{3}^{S_L}e^{iq_{h}^{S}x}+\tilde{t}_{hh}^{\prime\downarrow\downarrow}\varphi_{4}^{S_L}e^{iq_{h}^{S}x}\,, & x<0\\
\tilde{a}_{he}^{\prime\downarrow\uparrow}\varphi_{1}^{F_x}e^{iq_{e1}^{\uparrow}x}+\tilde{a}_{he}^{\prime\downarrow\downarrow}\varphi_{2}^{F_x}e^{iq_{e1}^{\downarrow}x}+\tilde{b}_{he}^{\prime\downarrow\uparrow}\varphi_{1}^{F_x}e^{-iq_{e1}^{\uparrow}(x-\mathcal{D}_{1})}+\tilde{b}_{he}^{\prime\downarrow\downarrow}\varphi_{2}^{F_x}e^{-iq_{e1}^{\downarrow}(x-\mathcal{D}_{1})}\\+\tilde{c}_{hh}^{\prime\downarrow\uparrow}\varphi_{3}^{F_x}e^{iq_{h1}^{\uparrow}(x-\mathcal{D}_{1})}+\tilde{c}_{hh}^{\prime\downarrow\downarrow}\varphi_{4}^{F_x}e^{iq_{h1}^{\downarrow}(x-\mathcal{D}_{1})}
+\tilde{d}_{hh}^{\prime\downarrow\uparrow}\varphi_{3}^{F_x}e^{-iq_{h1}^{\uparrow}x}+\tilde{d}_{hh}^{\prime\downarrow\downarrow}\varphi_{4}^{F_x}e^{-iq_{h1}^{\downarrow}x}\,, & 0<x<\mathcal{D}_{1}\\
\tilde{e}_{he}^{\prime\downarrow\uparrow}\varphi_{1}^{F_y}e^{iq_{e2}^{\uparrow}(x-\mathcal{D}_{1})}+\tilde{e}_{he}^{\prime\downarrow\downarrow}\varphi_{2}^{F_y}e^{iq_{e2}^{\downarrow}(x-\mathcal{D}_{1})}+\tilde{f}_{he}^{\prime\downarrow\uparrow}\varphi_{1}^{F_y}e^{-iq_{e2}^{\uparrow}(x-\mathcal{D}^{\prime})}+\tilde{f}_{he}^{\prime\downarrow\downarrow}\varphi_{2}^{F_y}e^{-iq_{e2}^{\downarrow}(x-\mathcal{D}^{\prime})}\\+\tilde{g}_{hh}^{\prime\downarrow\uparrow}\varphi_{3}^{F_y}e^{iq_{h2}^{\uparrow}(x-\mathcal{D}^{\prime})}+\tilde{g}_{hh}^{\prime\downarrow\downarrow}\varphi_{4}^{F_y}e^{iq_{h2}^{\downarrow}(x-\mathcal{D}^{\prime})}
+\tilde{h}_{hh}^{\prime\downarrow\uparrow}\varphi_{3}^{F_y}e^{-iq_{h2}^{\uparrow}(x-\mathcal{D}_{1})}+\tilde{h}_{hh}^{\prime\downarrow\downarrow}\varphi_{4}^{F_y}e^{-iq_{h2}^{\downarrow}(x-\mathcal{D}_{1})}\,, & \mathcal{D}_{1}<x<\mathcal{D}^{\prime}\\
\varphi_{4}^{S_{R}}e^{iq_{h}^{S}(x-\mathcal{D}^{\prime})}+\tilde{r}_{hh}^{\prime\downarrow\uparrow}\varphi_{3}^{S_{R}}e^{-iq_{h}^{S}(x-\mathcal{D}^{\prime})}+\tilde{r}_{hh}^{\prime\downarrow\downarrow}\varphi_{4}^{S_{R}}e^{-iq_{h}^{S}(x-\mathcal{D}^{\prime})}+\tilde{r}_{he}^{\prime\downarrow\uparrow}\varphi_{1}^{S_{R}}e^{iq_{e}^{S}(x-\mathcal{D}^{\prime})}\\+\tilde{r}_{he}^{\prime\downarrow\downarrow}\varphi_{2}^{S_{R}}e^{iq_{e}^{S}(x-\mathcal{D}^{\prime})}\,, & x>\mathcal{D}^{\prime}
\end{cases}
\end{split}
\label{wav1}
\end{equation}
where the expressions for $\varphi_{i}^{S_L}$, $\varphi_{i}^{F_x}$, $\varphi_{i}^{F_y}$, $\varphi_{i}^{S_R}$, $u_0$, $v_0$, $q_{e,h}^{S}$, $q_{e1,h1}^{\sigma}$, $q_{e2,h2}^{\sigma}$ are provided below Eq.~\eqref{wav}. Further in Eq.~\eqref{wav1}, the terms $\tilde{r}_{pq}^{\sigma\sigma'}$ and $\tilde{r}_{pq}^{\prime\sigma\sigma'}$ denote the reflection amplitudes in left and right superconductors, respectively, while the terms $\tilde{t}_{pq}^{\sigma\sigma'}$ and $\tilde{t}_{pq}^{\prime\sigma\sigma'}$ represent the corresponding transmission amplitudes, with $\sigma,\sigma'\in\{\uparrow,\downarrow\}$ and $p,q\in\{e,h\}$. After diagonalizing the Hamiltonian $\Big(\mathcal{H}_{BdG}^{\mbox{S-$F_x$-$F_y$-S}}\Big)^{*}(-k)$ instead of $\mathcal{H}_{BdG}^{\mbox{S-$F_x$-$F_y$-S}}(k)$ (see Eq.~\eqref{hmfxfy}), we will obtain conjugate processes $\tilde{\phi}_{l}$ $(l=1,2,...,8)$ necessary to form $\Gamma^{r}(x,\chi,\omega)$ in next section. The expressions for $\tilde{\varphi}_{i}^{S_L}$, $\tilde{\varphi}_{i}^{S_R}$, $\tilde{\varphi}_{i}^{F_x}$, $\tilde{\varphi}_{i}^{F_y}$ are provided below Eq.~\eqref{wav}.}

{The boundary conditions at the S-$F_x$ interface ($x=0$) are:
\begin{equation}
\label{eqq36}
\phi_{l}|_{x<0}=\phi_{l}|_{0<x<\mathcal{D}_{1}}\,\,\,\,\,\mbox{and}\,\,\,\,\,\frac{d\phi_{l}|_{0<x<\mathcal{D}_{1}}}{dx}-\frac{d\phi_{l}|_{x<0}}{dx}=\frac{2m^{*}\mathcal{B}_{1}}{\hbar^2}\phi_{l}(x=0).
\end{equation}
Similarly, the boundary conditions at the $F_x$-$F_y$ interface ($x=\mathcal{D}_{1}$) are:
\begin{equation}
\label{eqq37}
\phi_{l}|_{0<x<\mathcal{D}_{1}}=\phi_{l}|_{\mathcal{D}_{1}<x<\mathcal{D}^{\prime}}\,\,\,\,\,\mbox{and}\,\,\,\,\
\frac{d\phi_{l}|_{\mathcal{D}_{1}<x<\mathcal{D}^{\prime}}}{dx}-\frac{d\phi_{l}|_{0<x<\mathcal{D}_{1}}}{dx}=\frac{2m^{*}\mathcal{B}}{\hbar^2}\phi_{l}(x=\mathcal{D}_{1}).
\end{equation}
Finally, the boundary conditions at the $F_y$-S interface ($x=\mathcal{D}^{\prime}$) are:
\begin{equation}
\label{eqq38}
\phi_{l}|_{\mathcal{D}_{1}<x<\mathcal{D}^{\prime}}=\phi_{l}|_{x>\mathcal{D}^{\prime}}\,\,\,\,\,\mbox{and}\,\,\,\,\,\frac{d\phi_{l}|_{x>\mathcal{D}^{\prime}}}{dx}-\frac{d\phi_{l}|_{\mathcal{D}_{1}<x<\mathcal{D}^{\prime}}}{dx}=\frac{2m^{*}\mathcal{B}_{2}}{\hbar^2}\phi_{l}(x=\mathcal{D}^{\prime}).
\end{equation}
Solving the boundary conditions (Eqs.~\eqref{eqq36}-\eqref{eqq38}), a total of $24$ equations are determined for each kind of incident process described in Eq.~\eqref{wav1}. The scattering amplitudes, such as $\tilde{r}_{pq}^{\sigma\sigma'}$, $\tilde{r}_{pq}^{\prime\sigma\sigma'}$, $\tilde{t}_{pq}^{\prime\sigma\sigma'}$, $\tilde{t}_{pq}^{\sigma\sigma'}$, are obtained by solving these $24$ equations. $s_{1}=\tilde{r}_{eh}^{\uparrow\downarrow}$, $s_{2}=\tilde{r}_{eh}^{\downarrow\uparrow}$, $s_{3}=\tilde{r}_{he}^{\uparrow\downarrow}$, and $s_{4} = \tilde{r}_{he}^{\downarrow\uparrow}$, determined by putting the wavefunctions
$\phi_{1}$, $\phi_{2}$, $\phi_{3}$, and $\phi_{4}$ into Eqs.~\eqref{eqq36}-\eqref{eqq38} are used to compute the Josephson current via Eq.~\eqref{eq3}.}

{The wavefunctions for trilayer S-$F_x$-$F_y$-$F_z$-S JJ are as follows:
\begin{equation}
\begin{split}
\phi_{1}(x)&=\begin{cases}
\varphi_{1}^{S_L}e^{iq_{e}^{S}x}+{r}_{eh}^{\prime\prime\uparrow\uparrow}\varphi_{3}^{S_L}e^{iq_{h}^{S}x}+{r}_{eh}^{\prime\prime\uparrow\downarrow}\varphi_{4}^{S_L}e^{iq_{h}^{S}x}+{r}_{ee}^{\prime\prime\uparrow\uparrow}\varphi_{1}^{S_L}e^{-iq_{e}^{S}x}+{r}_{ee}^{\prime\prime\uparrow\downarrow}\varphi_{2}^{S_L}e^{-iq_{e}^{S}x}\,, & x<0\\
{a}_{ee}^{\prime\prime\uparrow\uparrow}\varphi_{1}^{F_x}e^{iq_{e1}^{\uparrow}x}+{a}_{ee}^{\prime\prime\uparrow\downarrow}\varphi_{2}^{F_x}e^{iq_{e1}^{\downarrow}x}+{b}_{ee}^{\prime\prime\uparrow\uparrow}\varphi_{1}^{F_x}e^{-iq_{e1}^{\uparrow}(x-\mathcal{D}_{1})}+{b}_{ee}^{\prime\prime\uparrow\downarrow}\varphi_{2}^{F_x}e^{-iq_{e1}^{\downarrow}(x-\mathcal{D}_{1})}\\+{c}_{eh}^{\prime\prime\uparrow\uparrow}\varphi_{3}^{F_x}e^{iq_{h1}^{\uparrow}(x-\mathcal{D}_{1})}+{c}_{eh}^{\prime\prime\uparrow\downarrow}\varphi_{4}^{F_x}e^{iq_{h1}^{\downarrow}(x-\mathcal{D}_{1})}
+{d}_{eh}^{\prime\prime\uparrow\uparrow}\varphi_{3}^{F_x}e^{-iq_{h1}^{\uparrow}x}+{d}_{eh}^{\prime\prime\uparrow\downarrow}\varphi_{4}^{F_x}e^{-iq_{h1}^{\downarrow}x}\,, & 0<x<\mathcal{D}_{1}\\
{e}_{ee}^{\prime\prime\uparrow\uparrow}\varphi_{1}^{F_y}e^{iq_{e2}^{\uparrow}(x-\mathcal{D}_{1})}+{e}_{ee}^{\prime\prime\uparrow\downarrow}\varphi_{2}^{F_y}e^{iq_{e2}^{\downarrow}(x-\mathcal{D}_{1})}+{f}_{ee}^{\prime\prime\uparrow\uparrow}\varphi_{1}^{F_y}e^{-iq_{e2}^{\uparrow}(x-\mathcal{D}^{\prime})}+{f}_{ee}^{\prime\prime\uparrow\downarrow}\varphi_{2}^{F_y}e^{-iq_{e2}^{\downarrow}(x-\mathcal{D}^{\prime})}\\+{g}_{eh}^{\prime\prime\uparrow\uparrow}\varphi_{3}^{F_y}e^{iq_{h2}^{\uparrow}(x-\mathcal{D}^{\prime})}+{g}_{eh}^{\prime\prime\uparrow\downarrow}\varphi_{4}^{F_y}e^{iq_{h2}^{\downarrow}(x-\mathcal{D}^{\prime})}+{h}_{eh}^{\prime\prime\uparrow\uparrow}\varphi_{3}^{F_y}e^{-iq_{h2}^{\uparrow}(x-\mathcal{D}_{1})}+{h}_{eh}^{\prime\prime\uparrow\downarrow}\varphi_{4}^{F_y}e^{-iq_{h2}^{\downarrow}(x-\mathcal{D}_{1})}\,, & \mathcal{D}_{1}<x<\mathcal{D}^{\prime}\\
m_{ee}^{\prime\prime\uparrow\uparrow}\varphi_{1}^{F_z}e^{iq_{e3}^{\uparrow}(x-\mathcal{D}^{\prime})}+m_{ee}^{\prime\prime\uparrow\downarrow}\varphi_{2}^{F_z}e^{iq_{e3}^{\downarrow}(x-\mathcal{D}^{\prime})}+n_{ee}^{\prime\prime\uparrow\uparrow}\varphi_{1}^{F_z}e^{-iq_{e3}^{\uparrow}(x-\mathcal{D}^{\prime\prime})}+n_{ee}^{\prime\prime\uparrow\downarrow}\varphi_{2}^{F_z}e^{-iq_{e3}^{\downarrow}(x-\mathcal{D}^{\prime\prime})}\\+o_{eh}^{\prime\prime\uparrow\uparrow}\varphi_{3}^{F_z}e^{iq_{h3}^{\uparrow}(x-\mathcal{D}^{\prime\prime})}+o_{eh}^{\prime\prime\uparrow\downarrow}\varphi_{4}^{F_z}e^{iq_{h3}^{\downarrow}(x-\mathcal{D}^{\prime\prime})}
+p_{eh}^{\prime\prime\uparrow\uparrow}\varphi_{3}^{F_z}e^{-iq_{h3}^{\uparrow}(x-\mathcal{D}^{\prime})}+p_{eh}^{\prime\prime\uparrow\downarrow}\varphi_{4}^{F_z}e^{-iq_{h3}^{\downarrow}(x-\mathcal{D}^{\prime})}\,, & \mathcal{D}^{\prime}<x<\mathcal{D}^{\prime\prime}\\
{t}_{ee}^{\prime\prime\uparrow\uparrow}\varphi_{1}^{S_{R}}e^{iq_{e}^{S}(x-\mathcal{D}^{\prime\prime})}+{t}_{ee}^{\prime\prime\uparrow\downarrow}\varphi_{2}^{S_{R}}e^{iq_{e}^{S}(x-\mathcal{D}^{\prime\prime})}+{t}_{eh}^{\prime\prime\uparrow\uparrow}\varphi_{3}^{S_{R}}e^{-iq_{h}^{S}(x-\mathcal{D}^{\prime\prime})}+{t}_{eh}^{\prime\prime\uparrow\downarrow}\varphi_{4}^{S_{R}}e^{-iq_{h}^{S}(x-\mathcal{D}^{\prime\prime})}\,, & x>\mathcal{D}^{\prime\prime}
\end{cases}\\
\end{split}\nonumber
\end{equation}
\newpage
\begin{equation}
\begin{split}
\phi_{2}(x)&=\begin{cases}
\varphi_{2}^{S_L}e^{iq_{e}^{S}x}+{r}_{eh}^{\prime\prime\downarrow\uparrow}\varphi_{3}^{S_L}e^{iq_{h}^{S}x}+{r}_{eh}^{\prime\prime\downarrow\downarrow}\varphi_{4}^{S_L}e^{iq_{h}^{S}x}+{r}_{ee}^{\prime\prime\downarrow\uparrow}\varphi_{1}^{S_L}e^{-iq_{e}^{S}x}+{r}_{ee}^{\prime\prime\downarrow\downarrow}\varphi_{2}^{S_L}e^{-iq_{e}^{S}x}\,, & x<0\\
{a}_{ee}^{\prime\prime\downarrow\uparrow}\varphi_{1}^{F_x}e^{iq_{e1}^{\uparrow}x}+{a}_{ee}^{\prime\prime\downarrow\downarrow}\varphi_{2}^{F_x}e^{iq_{e1}^{\downarrow}x}+{b}_{ee}^{\prime\prime\downarrow\uparrow}\varphi_{1}^{F_x}e^{-iq_{e1}^{\uparrow}(x-\mathcal{D}_{1})}+{b}_{ee}^{\prime\prime\downarrow\downarrow}\varphi_{2}^{F_x}e^{-iq_{e1}^{\downarrow}(x-\mathcal{D}_{1})}\\+{c}_{eh}^{\prime\prime\downarrow\uparrow}\varphi_{3}^{F_x}e^{iq_{h1}^{\uparrow}(x-\mathcal{D}_{1})}+{c}_{eh}^{\prime\prime\downarrow\downarrow}\varphi_{4}^{F_x}e^{iq_{h1}^{\downarrow}(x-\mathcal{D}_{1})} +{d}_{eh}^{\prime\prime\downarrow\uparrow}\varphi_{3}^{F_x}e^{-iq_{h1}^{\uparrow}x}+{d}_{eh}^{\prime\prime\downarrow\downarrow}\varphi_{4}^{F_x}e^{-iq_{h1}^{\downarrow}x}\,, &  0<x<\mathcal{D}_{1}\\
{e}_{ee}^{\prime\prime\downarrow\uparrow}\varphi_{1}^{F_y}e^{iq_{e2}^{\uparrow}(x-\mathcal{D}_{1})}+{e}_{ee}^{\prime\prime\downarrow\downarrow}\varphi_{2}^{F_y}e^{iq_{e2}^{\downarrow}(x-\mathcal{D}_{1})}+{f}_{ee}^{\prime\prime\downarrow\uparrow}\varphi_{1}^{F_y}e^{-iq_{e2}^{\uparrow}(x-\mathcal{D}^{\prime})}+{f}_{ee}^{\prime\prime\downarrow\downarrow}\varphi_{2}^{F_y}e^{-iq_{e2}^{\downarrow}(x-\mathcal{D}^{\prime})}\\+{g}_{eh}^{\prime\prime\downarrow\uparrow}\varphi_{3}^{F_y}e^{iq_{h2}^{\uparrow}(x-\mathcal{D}^{\prime})}+{g}_{eh}^{\prime\prime\downarrow\downarrow}\varphi_{4}^{F_y}e^{iq_{h2}^{\downarrow}(x-\mathcal{D}^{\prime})}+{h}_{eh}^{\prime\prime\downarrow\uparrow}\varphi_{3}^{F_y}e^{-iq_{h2}^{\uparrow}(x-\mathcal{D}_{1})}+{h}_{eh}^{\prime\prime\downarrow\downarrow}\varphi_{4}^{F_y}e^{-iq_{h2}^{\downarrow}(x-\mathcal{D}_{1})}\,, & \mathcal{D}_{1}<x<\mathcal{D}^{\prime}\\
m_{ee}^{\prime\prime\downarrow\uparrow}\varphi_{1}^{F_z}e^{iq_{e3}^{\uparrow}(x-\mathcal{D}^{\prime})}+m_{ee}^{\prime\prime\downarrow\downarrow}\varphi_{2}^{F_z}e^{iq_{e3}^{\downarrow}(x-\mathcal{D}^{\prime})}+n_{ee}^{\prime\prime\downarrow\uparrow}\varphi_{1}^{F_z}e^{-iq_{e3}^{\uparrow}(x-\mathcal{D}^{\prime\prime})}+n_{ee}^{\prime\prime\downarrow\downarrow}\varphi_{2}^{F_z}e^{-iq_{e3}^{\downarrow}(x-\mathcal{D}^{\prime\prime})}\\+o_{eh}^{\prime\prime\downarrow\uparrow}\varphi_{3}^{F_z}e^{iq_{h3}^{\uparrow}(x-\mathcal{D}^{\prime\prime})}+o_{eh}^{\prime\prime\downarrow\downarrow}\varphi_{4}^{F_z}e^{iq_{h3}^{\downarrow}(x-\mathcal{D}^{\prime\prime})}
+p_{eh}^{\prime\prime\downarrow\uparrow}\varphi_{3}^{F_z}e^{-iq_{h3}^{\uparrow}(x-\mathcal{D}^{\prime})}+p_{eh}^{\prime\prime\downarrow\downarrow}\varphi_{4}^{F_z}e^{-iq_{h3}^{\downarrow}(x-\mathcal{D}^{\prime})}\,, & \mathcal{D}^{\prime}<x<\mathcal{D}^{\prime\prime}\\
{t}_{ee}^{\prime\prime\downarrow\uparrow}\varphi_{1}^{S_{R}}e^{iq_{e}^{S}(x-\mathcal{D}^{\prime\prime})}+{t}_{ee}^{\prime\prime\downarrow\downarrow}\varphi_{2}^{S_{R}}e^{iq_{e}^{S}(x-\mathcal{D}^{\prime\prime})}+{t}_{eh}^{\prime\prime\downarrow\uparrow}\varphi_{3}^{S_{R}}e^{-iq_{h}^{S}(x-\mathcal{D}^{\prime\prime})}+{t}_{eh}^{\prime\prime\downarrow\downarrow}\varphi_{4}^{S_{R}}e^{-iq_{h}^{S}(x-\mathcal{D}^{\prime\prime})}\,, & x>\mathcal{D}^{\prime\prime}
\end{cases}\\
\phi_{3}(x)&=\begin{cases}
\varphi_{3}^{S_L}e^{-iq_{h}^{S}x}+{r}_{hh}^{\prime\prime\uparrow\uparrow}\varphi_{3}^{S_L}e^{iq_{h}^{S}x}+{r}_{hh}^{\prime\prime\uparrow\downarrow}\varphi_{4}^{S_L}e^{iq_{h}^{S}x}+{r}_{he}^{\prime\prime\uparrow\uparrow}\varphi_{1}^{S_L}e^{-iq_{e}^{S}x}+{r}_{he}^{\prime\prime\uparrow\downarrow}\varphi_{2}^{S_L}e^{-iq_{e}^{S}x}\,, & x<0\\
{a}_{he}^{\prime\prime\uparrow\uparrow}\varphi_{1}^{F_x}e^{iq_{e1}^{\uparrow}x}+{a}_{he}^{\prime\prime\uparrow\downarrow}\varphi_{2}^{F_x}e^{iq_{e1}^{\downarrow}x}+{b}_{he}^{\prime\prime\uparrow\uparrow}\varphi_{1}^{F_x}e^{-iq_{e1}^{\uparrow}(x-\mathcal{D}_{1})}+{b}_{he}^{\prime\prime\uparrow\downarrow}\varphi_{2}^{F_x}e^{-iq_{e1}^{\downarrow}(x-\mathcal{D}_{1})}\\+{c}_{hh}^{\prime\prime\uparrow\uparrow}\varphi_{3}^{F_x}e^{iq_{h1}^{\uparrow}(x-\mathcal{D}_{1})}+{c}_{hh}^{\prime\prime\uparrow\downarrow}\varphi_{4}^{F_x}e^{iq_{h1}^{\downarrow}(x-\mathcal{D}_{1})} +{d}_{hh}^{\prime\prime\uparrow\uparrow}\varphi_{3}^{F_x}e^{-iq_{h1}^{\uparrow}x}+{d}_{hh}^{\prime\prime\uparrow\downarrow}\varphi_{4}^{F_x}e^{-iq_{h1}^{\downarrow}x}\,, & 0<x<\mathcal{D}_{1}\\
{e}_{he}^{\prime\prime\uparrow\uparrow}\varphi_{1}^{F_y}e^{iq_{e2}^{\uparrow}(x-\mathcal{D}_{1})}+{e}_{he}^{\prime\prime\uparrow\downarrow}\varphi_{2}^{F_y}e^{iq_{e2}^{\downarrow}(x-\mathcal{D}_{1})}+{f}_{he}^{\prime\prime\uparrow\uparrow}\varphi_{1}^{F_y}e^{-iq_{e2}^{\uparrow}(x-\mathcal{D}^{\prime})}+{f}_{he}^{\prime\prime\uparrow\downarrow}\varphi_{2}^{F_y}e^{-iq_{e2}^{\downarrow}(x-\mathcal{D}^{\prime})}\\+{g}_{hh}^{\prime\prime\uparrow\uparrow}\varphi_{3}^{F_y}e^{iq_{h2}^{\uparrow}(x-\mathcal{D}^{\prime})}+{g}_{hh}^{\prime\prime\uparrow\downarrow}\varphi_{4}^{F_y}e^{iq_{h2}^{\downarrow}(x-\mathcal{D}^{\prime})}+{h}_{hh}^{\prime\prime\uparrow\uparrow}\varphi_{3}^{F_y}e^{-iq_{h2}^{\uparrow}(x-\mathcal{D}_{1})}+{h}_{hh}^{\prime\prime\uparrow\downarrow}\varphi_{4}^{F_y}e^{-iq_{h2}^{\downarrow}(x-\mathcal{D}_{1})}\,, & \mathcal{D}_{1}<x<\mathcal{D}^{\prime}\\
m_{he}^{\prime\prime\uparrow\uparrow}\varphi_{1}^{F_z}e^{iq_{e3}^{\uparrow}(x-\mathcal{D}^{\prime})}+m_{he}^{\prime\prime\uparrow\downarrow}\varphi_{2}^{F_z}e^{iq_{e3}^{\downarrow}(x-\mathcal{D}^{\prime})}+n_{he}^{\prime\prime\uparrow\uparrow}\varphi_{1}^{F_z}e^{-iq_{e3}^{\uparrow}(x-\mathcal{D}^{\prime\prime})}+n_{he}^{\prime\prime\uparrow\downarrow}\varphi_{2}^{F_z}e^{-iq_{e3}^{\downarrow}(x-\mathcal{D}^{\prime\prime})}\\+o_{hh}^{\prime\prime\uparrow\uparrow}\varphi_{3}^{F_z}e^{iq_{h3}^{\uparrow}(x-\mathcal{D}^{\prime\prime})}+o_{hh}^{\prime\prime\uparrow\downarrow}\varphi_{4}^{F_z}e^{iq_{h3}^{\downarrow}(x-\mathcal{D}^{\prime\prime})}
+p_{hh}^{\prime\prime\uparrow\uparrow}\varphi_{3}^{F_z}e^{-iq_{h3}^{\uparrow}(x-\mathcal{D}^{\prime})}+p_{hh}^{\prime\prime\uparrow\downarrow}\varphi_{4}^{F_z}e^{-iq_{h3}^{\downarrow}(x-\mathcal{D}^{\prime})}\,, & \mathcal{D}^{\prime}<x<\mathcal{D}^{\prime\prime}\\
{t}_{he}^{\prime\prime\uparrow\uparrow}\varphi_{1}^{S_{R}}e^{iq_{e}^{S}(x-\mathcal{D}^{\prime\prime})}+{t}_{he}^{\prime\prime\uparrow\downarrow}\varphi_{2}^{S_{R}}e^{iq_{e}^{S}(x-\mathcal{D}^{\prime\prime})}+{t}_{hh}^{\prime\prime\uparrow\uparrow}\varphi_{3}^{S_{R}}e^{-iq_{h}^{S}(x-\mathcal{D}^{\prime\prime})}+{t}_{hh}^{\prime\prime\uparrow\downarrow}\varphi_{4}^{S_{R}}e^{-iq_{h}^{S}(x-\mathcal{D}^{\prime\prime})}\,, & x>\mathcal{D}^{\prime\prime}
\end{cases}\\
\phi_{4}(x)&=\begin{cases}
\varphi_{4}^{S_L}e^{-iq_{h}^{S}x}+{r}_{hh}^{\prime\prime\downarrow\uparrow}\varphi_{3}^{S_L}e^{iq_{h}^{S}x}+{r}_{hh}^{\prime\prime\downarrow\downarrow}\varphi_{4}^{S_L}e^{iq_{h}^{S}x}+{r}_{he}^{\prime\prime\downarrow\uparrow}\varphi_{1}^{S_L}e^{-iq_{e}^{S}x}+{r}_{he}^{\prime\prime\downarrow\downarrow}\varphi_{2}^{S_L}e^{-iq_{e}^{S}x}\,, & x<0\\
{a}_{he}^{\prime\prime\downarrow\uparrow}\varphi_{1}^{F_x}e^{iq_{e1}^{\uparrow}x}+{a}_{he}^{\prime\prime\downarrow\downarrow}\varphi_{2}^{F_x}e^{iq_{e1}^{\downarrow}x}+{b}_{he}^{\prime\prime\downarrow\uparrow}\varphi_{1}^{F_x}e^{-iq_{e1}^{\uparrow}(x-\mathcal{D}_{1})}+{b}_{he}^{\prime\prime\downarrow\downarrow}\varphi_{2}^{F_x}e^{-iq_{e1}^{\downarrow}(x-\mathcal{D}_{1})}\\+{c}_{hh}^{\prime\prime\downarrow\uparrow}\varphi_{3}^{F_x}e^{iq_{h1}^{\uparrow}(x-\mathcal{D}_{1})}+{c}_{hh}^{\prime\prime\downarrow\downarrow}\varphi_{4}^{F_x}e^{iq_{h1}^{\downarrow}(x-\mathcal{D}_{1})} +{d}_{hh}^{\prime\prime\downarrow\uparrow}\varphi_{3}^{F_x}e^{-iq_{h1}^{\uparrow}x}+{d}_{hh}^{\prime\prime\downarrow\downarrow}\varphi_{4}^{F_x}e^{-iq_{h1}^{\downarrow}x}\,, & 0<x<\mathcal{D}_{1}\\
{e}_{he}^{\prime\prime\downarrow\uparrow}\varphi_{1}^{F_y}e^{iq_{e2}^{\uparrow}(x-\mathcal{D}_{1})}+{e}_{he}^{\prime\prime\downarrow\downarrow}\varphi_{2}^{F_y}e^{iq_{e2}^{\downarrow}(x-\mathcal{D}_{1})}+{f}_{he}^{\prime\prime\downarrow\uparrow}\varphi_{1}^{F_y}e^{-iq_{e2}^{\uparrow}(x-\mathcal{D}^{\prime})}+{f}_{he}^{\prime\prime\downarrow\downarrow}\varphi_{2}^{F_y}e^{-iq_{e2}^{\downarrow}(x-\mathcal{D}^{\prime})}\\+{g}_{hh}^{\prime\prime\downarrow\uparrow}\varphi_{3}^{F_y}e^{iq_{h2}^{\uparrow}(x-\mathcal{D}^{\prime})}+{g}_{hh}^{\prime\prime\downarrow\downarrow}\varphi_{4}^{F_y}e^{iq_{h2}^{\downarrow}(x-\mathcal{D}^{\prime})}
+\tilde{h}_{\prime\prime hh}^{\downarrow\uparrow}\varphi_{3}^{F_y}e^{-iq_{h2}^{\uparrow}(x-\mathcal{D}_{1})}+{h}_{hh}^{\prime\prime\downarrow\downarrow}\varphi_{4}^{F_y}e^{-iq_{h2}^{\downarrow}(x-\mathcal{D}_{1})}\,, & \mathcal{D}_{1}<x<\mathcal{D}^{\prime}\\
m_{he}^{\prime\prime\downarrow\uparrow}\varphi_{1}^{F_z}e^{iq_{e3}^{\uparrow}(x-\mathcal{D}^{\prime})}+m_{he}^{\prime\prime\downarrow\downarrow}\varphi_{2}^{F_z}e^{iq_{e3}^{\downarrow}(x-\mathcal{D}^{\prime})}+n_{he}^{\prime\prime\downarrow\uparrow}\varphi_{1}^{F_z}e^{-iq_{e3}^{\uparrow}(x-\mathcal{D}^{\prime\prime})}+n_{he}^{\prime\prime\downarrow\downarrow}\varphi_{2}^{F_z}e^{-iq_{e3}^{\downarrow}(x-\mathcal{D}^{\prime\prime})}\\+o_{hh}^{\prime\prime\downarrow\uparrow}\varphi_{3}^{F_z}e^{iq_{h3}^{\uparrow}(x-\mathcal{D}^{\prime\prime})}+o_{hh}^{\prime\prime\downarrow\downarrow}\varphi_{4}^{F_z}e^{iq_{h3}^{\downarrow}(x-\mathcal{D}^{\prime\prime})}
+p_{hh}^{\prime\prime\downarrow\uparrow}\varphi_{3}^{F_z}e^{-iq_{h3}^{\uparrow}(x-\mathcal{D}^{\prime})}+p_{hh}^{\prime\prime\downarrow\downarrow}\varphi_{4}^{F_z}e^{-iq_{h3}^{\downarrow}(x-\mathcal{D}^{\prime})}\,, & \mathcal{D}^{\prime}<x<\mathcal{D}^{\prime\prime}\\
{t}_{he}^{\prime\prime\downarrow\uparrow}\varphi_{1}^{S_{R}}e^{iq_{e}^{S}(x-\mathcal{D}^{\prime\prime})}+{t}_{he}^{\prime\prime\downarrow\downarrow}\varphi_{2}^{S_{R}}e^{iq_{e}^{S}(x-\mathcal{D}^{\prime\prime})}+{t}_{hh}^{\prime\prime\downarrow\uparrow}\varphi_{3}^{S_{R}}e^{-iq_{h}^{S}(x-\mathcal{D}^{\prime\prime})}+{t}_{hh}^{\prime\prime\downarrow\downarrow}\varphi_{4}^{S_{R}}e^{-iq_{h}^{S}(x-\mathcal{D}^{\prime\prime})}\,, & x>\mathcal{D}^{\prime\prime}
\end{cases}\\
\phi_{5}(x)&=\begin{cases}
\tilde{t}_{ee}^{\prime\prime\uparrow\uparrow}\varphi_{1}^{S_L}e^{-iq_{e}^{S}x}+\tilde{t}_{ee}^{\prime\prime\uparrow\downarrow}\varphi_{2}^{S_L}e^{-iq_{e}^{S}x}+\tilde{t}_{eh}^{\prime\prime\uparrow\uparrow}\varphi_{3}^{S_L}e^{iq_{h}^{S}x}+\tilde{t}_{eh}^{\prime\prime\uparrow\downarrow}\varphi_{4}^{S_L}e^{iq_{h}^{S}x}\,, & x<0\\
\tilde{a}_{ee}^{\prime\prime\uparrow\uparrow}\varphi_{1}^{F_x}e^{iq_{e1}^{\uparrow}x}+\tilde{a}_{ee}^{\prime\prime\uparrow\downarrow}\varphi_{2}^{F_x}e^{iq_{e1}^{\downarrow}x}+\tilde{b}_{ee}^{\prime\prime\uparrow\uparrow}\varphi_{1}^{F_x}e^{-iq_{e1}^{\uparrow}(x-\mathcal{D}_{1})}+\tilde{b}_{ee}^{\prime\prime\uparrow\downarrow}\varphi_{2}^{F_x}e^{-iq_{e1}^{\downarrow}(x-\mathcal{D}_{1})}\\+\tilde{c}_{eh}^{\prime\prime\uparrow\uparrow}\varphi_{3}^{F_x}e^{iq_{h1}^{\uparrow}(x-\mathcal{D}_{1})}+\tilde{c}_{eh}^{\prime\prime\uparrow\downarrow}\varphi_{4}^{F_x}e^{iq_{h1}^{\downarrow}(x-\mathcal{D}_{1})}
+\tilde{d}_{eh}^{\prime\prime\uparrow\uparrow}\varphi_{3}^{F_x}e^{-iq_{h1}^{\uparrow}x}+\tilde{d}_{eh}^{\prime\prime\uparrow\downarrow}\varphi_{4}^{F_x}e^{-iq_{h1}^{\downarrow}x}\,, & 0<x<\mathcal{D}_{1}\\
\tilde{e}_{ee}^{\prime\prime\uparrow\uparrow}\varphi_{1}^{F_y}e^{iq_{e2}^{\uparrow}(x-\mathcal{D}_{1})}+\tilde{e}_{ee}^{\prime\prime\uparrow\downarrow}\varphi_{2}^{F_y}e^{iq_{e2}^{\downarrow}(x-\mathcal{D}_{1})}+\tilde{f}_{ee}^{\prime\prime\uparrow\uparrow}\varphi_{1}^{F_y}e^{-iq_{e2}^{\uparrow}(x-\mathcal{D}^{\prime})}+\tilde{f}_{ee}^{\prime\prime\uparrow\downarrow}\varphi_{2}^{F_y}e^{-iq_{e2}^{\downarrow}(x-\mathcal{D}^{\prime})}\\+\tilde{g}_{eh}^{\prime\prime\uparrow\uparrow}\varphi_{3}^{F_y}e^{iq_{h2}^{\uparrow}(x-\mathcal{D}^{\prime})}+\tilde{g}_{eh}^{\prime\prime\uparrow\downarrow}\varphi_{4}^{F_y}e^{iq_{h2}^{\downarrow}(x-\mathcal{D}^{\prime})}+\tilde{h}_{eh}^{\prime\prime\uparrow\uparrow}\varphi_{3}^{F_y}e^{-iq_{h2}^{\uparrow}(x-\mathcal{D}_{1})}+\tilde{h}_{eh}^{\prime\prime\uparrow\downarrow}\varphi_{4}^{F_y}e^{-iq_{h2}^{\downarrow}(x-\mathcal{D}_{1})}\,, & \mathcal{D}_{1}<x<\mathcal{D}^{\prime}\\
\tilde{m}_{ee}^{\prime\prime\uparrow\uparrow}\varphi_{1}^{F_z}e^{iq_{e3}^{\uparrow}(x-\mathcal{D}^{\prime})}+\tilde{m}_{ee}^{\prime\prime\uparrow\downarrow}\varphi_{2}^{F_z}e^{iq_{e3}^{\downarrow}(x-\mathcal{D}^{\prime})}+\tilde{n}_{ee}^{\prime\prime\uparrow\uparrow}\varphi_{1}^{F_z}e^{-iq_{e3}^{\uparrow}(x-\mathcal{D}^{\prime\prime})}+\tilde{n}_{ee}^{\prime\prime\uparrow\downarrow}\varphi_{2}^{F_z}e^{-iq_{e3}^{\downarrow}(x-\mathcal{D}^{\prime\prime})}\\+\tilde{o}_{eh}^{\prime\prime\uparrow\uparrow}\varphi_{3}^{F_z}e^{iq_{h3}^{\uparrow}(x-\mathcal{D}^{\prime\prime})}+\tilde{o}_{eh}^{\prime\prime\uparrow\downarrow}\varphi_{4}^{F_z}e^{iq_{h3}^{\downarrow}(x-\mathcal{D}^{\prime\prime})}
+\tilde{p}_{eh}^{\prime\prime\uparrow\uparrow}\varphi_{3}^{F_z}e^{-iq_{h3}^{\uparrow}(x-\mathcal{D}^{\prime})}+\tilde{p}_{eh}^{\prime\prime\uparrow\downarrow}\varphi_{4}^{F_z}e^{-iq_{h3}^{\downarrow}(x-\mathcal{D}^{\prime})}\,, & \mathcal{D}^{\prime}<x<\mathcal{D}^{\prime\prime}\\
\varphi_{1}^{S_{R}}e^{-iq_{e}^{S}(x-\mathcal{D}^{\prime\prime})}+\tilde{r}_{eh}^{\prime\prime\uparrow\uparrow}\varphi_{3}^{S_{R}}e^{-iq_{h}^{S}(x-\mathcal{D}^{\prime\prime})}+\tilde{r}_{eh}^{\prime\prime\uparrow\downarrow}\varphi_{4}^{S_{R}}e^{-iq_{h}^{S}(x-\mathcal{D}^{\prime\prime})}+\tilde{r}_{ee}^{\prime\prime\uparrow\uparrow}\varphi_{1}^{S_{R}}e^{iq_{e}^{S}(x-\mathcal{D}^{\prime\prime})}\\+\tilde{r}_{ee}^{\prime\prime\uparrow\downarrow}\varphi_{2}^{S_{R}}e^{iq_{e}^{S}(x-\mathcal{D}^{\prime\prime})}\,, & x>\mathcal{D}^{\prime\prime}
\end{cases}\\
\phi_{6}(x)&=\begin{cases}
\tilde{t}_{ee}^{\prime\prime\downarrow\uparrow}\varphi_{1}^{S_L}e^{-iq_{e}^{S}x}+\tilde{t}_{ee}^{\prime\prime\downarrow\downarrow}\varphi_{2}^{S_L}e^{-iq_{e}^{S}x}+\tilde{t}_{eh}^{\prime\prime\downarrow\uparrow}\varphi_{3}^{S_L}e^{iq_{h}^{S}x}+\tilde{t}_{eh}^{\prime\prime\downarrow\downarrow}\varphi_{4}^{S_L}e^{iq_{h}^{S}x}\,, & x<0\\
\tilde{a}_{ee}^{\prime\prime\downarrow\uparrow}\varphi_{1}^{F_x}e^{iq_{e1}^{\uparrow}x}+\tilde{a}_{ee}^{\prime\prime\downarrow\downarrow}\varphi_{2}^{F_x}e^{iq_{e1}^{\downarrow}x}+\tilde{b}_{ee}^{\prime\prime\downarrow\uparrow}\varphi_{1}^{F_x}e^{-iq_{e1}^{\uparrow}(x-\mathcal{D}_{1})}+\tilde{b}_{ee}^{\prime\prime\downarrow\downarrow}\varphi_{2}^{F_x}e^{-iq_{e1}^{\downarrow}(x-\mathcal{D}_{1})}\\+\tilde{c}_{eh}^{\prime\prime\downarrow\uparrow}\varphi_{3}^{F_x}e^{iq_{h1}^{\uparrow}(x-\mathcal{D}_{1})}+\tilde{c}_{eh}^{\prime\prime\downarrow\downarrow}\varphi_{4}^{F_x}e^{iq_{h1}^{\downarrow}(x-\mathcal{D}_{1})}
+\tilde{d}_{eh}^{\prime\prime\downarrow\uparrow}\varphi_{3}^{F_x}e^{-iq_{h1}^{\uparrow}x}+\tilde{d}_{eh}^{\prime\prime\downarrow\downarrow}\varphi_{4}^{F_x}e^{-iq_{h1}^{\downarrow}x}\,, & 0<x<\mathcal{D}_{1}\\
\tilde{e}_{ee}^{\prime\prime\downarrow\uparrow}\varphi_{1}^{F_y}e^{iq_{e2}^{\uparrow}(x-\mathcal{D}_{1})}+\tilde{e}_{ee}^{\prime\prime\downarrow\downarrow}\varphi_{2}^{F_y}e^{iq_{e2}^{\downarrow}(x-\mathcal{D}_{1})}+\tilde{f}_{ee}^{\prime\prime\downarrow\uparrow}\varphi_{1}^{F_y}e^{-iq_{e2}^{\uparrow}(x-\mathcal{D}^{\prime})}+\tilde{f}_{ee}^{\prime\prime\downarrow\downarrow}\varphi_{2}^{F_y}e^{-iq_{e2}^{\downarrow}(x-\mathcal{D}^{\prime})}\\+\tilde{g}_{eh}^{\prime\prime\downarrow\uparrow}\varphi_{3}^{F_y}e^{iq_{h2}^{\uparrow}(x-\mathcal{D}^{\prime})}+\tilde{g}_{eh}^{\prime\prime\downarrow\downarrow}\varphi_{4}^{F_y}e^{iq_{h2}^{\downarrow}(x-\mathcal{D}^{\prime})}+\tilde{h}_{eh}^{\prime\prime\downarrow\uparrow}\varphi_{3}^{F_y}e^{-iq_{h2}^{\uparrow}(x-\mathcal{D}_{1})}+\tilde{h}_{eh}^{\prime\prime\downarrow\downarrow}\varphi_{4}^{F_y}e^{-iq_{h2}^{\downarrow}(x-\mathcal{D}_{1})}\,, & \mathcal{D}_{1}<x<\mathcal{D}^{\prime}\\
\tilde{m}_{ee}^{\prime\prime\downarrow\uparrow}\varphi_{1}^{F_z}e^{iq_{e3}^{\uparrow}(x-\mathcal{D}^{\prime})}+\tilde{m}_{ee}^{\prime\prime\downarrow\downarrow}\varphi_{2}^{F_z}e^{iq_{e3}^{\downarrow}(x-\mathcal{D}^{\prime})}+\tilde{n}_{ee}^{\prime\prime\downarrow\uparrow}\varphi_{1}^{F_z}e^{-iq_{e3}^{\uparrow}(x-\mathcal{D}^{\prime\prime})}+\tilde{n}_{ee}^{\prime\prime\downarrow\downarrow}\varphi_{2}^{F_z}e^{-iq_{e3}^{\downarrow}(x-\mathcal{D}^{\prime\prime})}\\+\tilde{o}_{eh}^{\prime\prime\downarrow\uparrow}\varphi_{3}^{F_z}e^{iq_{h3}^{\uparrow}(x-\mathcal{D}^{\prime\prime})}+\tilde{o}_{eh}^{\prime\prime\downarrow\downarrow}\varphi_{4}^{F_z}e^{iq_{h3}^{\downarrow}(x-\mathcal{D}^{\prime\prime})}
+\tilde{p}_{eh}^{\prime\prime\downarrow\uparrow}\varphi_{3}^{F_z}e^{-iq_{h3}^{\uparrow}(x-\mathcal{D}^{\prime})}+\tilde{p}_{eh}^{\prime\prime\downarrow\downarrow}\varphi_{4}^{F_z}e^{-iq_{h3}^{\downarrow}(x-\mathcal{D}^{\prime})}\,, & \mathcal{D}^{\prime}<x<\mathcal{D}^{\prime\prime}\\
\varphi_{2}^{S_{R}}e^{-iq_{e}^{S}(x-\mathcal{D}^{\prime\prime})}+\tilde{r}_{eh}^{\prime\prime\downarrow\uparrow}\varphi_{3}^{S_{R}}e^{-iq_{h}^{S}(x-\mathcal{D}^{\prime\prime})}+\tilde{r}_{eh}^{\prime\prime\downarrow\downarrow}\varphi_{4}^{S_{R}}e^{-iq_{h}^{S}(x-\mathcal{D}^{\prime\prime})}+\tilde{r}_{ee}^{\prime\prime\downarrow\uparrow}\varphi_{1}^{S_{R}}e^{iq_{e}^{S}(x-\mathcal{D}^{\prime\prime})}\\+\tilde{r}_{ee}^{\prime\prime\downarrow\downarrow}\varphi_{2}^{S_{R}}e^{iq_{e}^{S}(x-\mathcal{D}^{\prime\prime})}\,, & x>\mathcal{D}^{\prime\prime}
\end{cases}\\
\end{split}\nonumber
\end{equation}
\begin{equation}
\begin{split}
\phi_{7}(x)&=\begin{cases}
\tilde{t}_{he}^{\prime\prime\uparrow\uparrow}\varphi_{1}^{S_L}e^{-iq_{e}^{S}x}+\tilde{t}_{he}^{\prime\prime\uparrow\downarrow}\varphi_{2}^{S_L}e^{-iq_{e}^{S}x}+\tilde{t}_{hh}^{\prime\prime\uparrow\uparrow}\varphi_{3}^{S_L}e^{iq_{h}^{S}x}+\tilde{t}_{hh}^{\prime\prime\uparrow\downarrow}\varphi_{4}^{S_L}e^{iq_{h}^{S}x}\,, & x<0\\
\tilde{a}_{he}^{\prime\prime\uparrow\uparrow}\varphi_{1}^{F_x}e^{iq_{e1}^{\uparrow}x}+\tilde{a}_{he}^{\prime\prime\uparrow\downarrow}\varphi_{2}^{F_x}e^{iq_{e1}^{\downarrow}x}+\tilde{b}_{he}^{\prime\prime\uparrow\uparrow}\varphi_{1}^{F_x}e^{-iq_{e1}^{\uparrow}(x-\mathcal{D}_{1})}+\tilde{b}_{he}^{\prime\prime\uparrow\downarrow}\varphi_{2}^{F_x}e^{-iq_{e1}^{\downarrow}(x-\mathcal{D}_{1})}\\+\tilde{c}_{hh}^{\prime\prime\uparrow\uparrow}\varphi_{3}^{F_x}e^{iq_{h1}^{\uparrow}(x-\mathcal{D}_{1})}+\tilde{c}_{hh}^{\prime\prime\uparrow\downarrow}\varphi_{4}^{F_x}e^{iq_{h1}^{\downarrow}(x-\mathcal{D}_{1})}
+\tilde{d}_{hh}^{\prime\prime\uparrow\uparrow}\varphi_{3}^{F_x}e^{-iq_{h1}^{\uparrow}x}+\tilde{d}_{hh}^{\prime\prime\uparrow\downarrow}\varphi_{4}^{F_x}e^{-iq_{h1}^{\downarrow}x}\,, & 0<x<\mathcal{D}_{1}\\
\tilde{e}_{he}^{\prime\prime\uparrow\uparrow}\varphi_{1}^{F_y}e^{iq_{e2}^{\uparrow}(x-\mathcal{D}_{1})}+\tilde{e}_{he}^{\prime\prime\uparrow\downarrow}\varphi_{2}^{F_y}e^{iq_{e2}^{\downarrow}(x-\mathcal{D}_{1})}+\tilde{f}_{he}^{\prime\prime\uparrow\uparrow}\varphi_{1}^{F_y}e^{-iq_{e2}^{\uparrow}(x-\mathcal{D}^{\prime})}+\tilde{f}_{he}^{\prime\prime\uparrow\downarrow}\varphi_{2}^{F_y}e^{-iq_{e2}^{\downarrow}(x-\mathcal{D}^{\prime})}\\+\tilde{g}_{hh}^{\prime\prime\uparrow\uparrow}\varphi_{3}^{F_y}e^{iq_{h2}^{\uparrow}(x-\mathcal{D}^{\prime})}+\tilde{g}_{hh}^{\prime\prime\uparrow\downarrow}\varphi_{4}^{F_y}e^{iq_{h2}^{\downarrow}(x-\mathcal{D}^{\prime})}+\tilde{h}_{hh}^{\prime\prime\uparrow\uparrow}\varphi_{3}^{F_y}e^{-iq_{h2}^{\uparrow}(x-\mathcal{D}_{1})}+\tilde{h}_{hh}^{\prime\prime\uparrow\downarrow}\varphi_{4}^{F_y}e^{-iq_{h2}^{\downarrow}(x-\mathcal{D}_{1})}\,, & \mathcal{D}_{1}<x<\mathcal{D}^{\prime}\\
\tilde{m}_{he}^{\prime\prime\uparrow\uparrow}\varphi_{1}^{F_z}e^{iq_{e3}^{\uparrow}(x-\mathcal{D}^{\prime})}+\tilde{m}_{he}^{\prime\prime\uparrow\downarrow}\varphi_{2}^{F_z}e^{iq_{e3}^{\downarrow}(x-\mathcal{D}^{\prime})}+\tilde{n}_{he}^{\prime\prime\uparrow\uparrow}\varphi_{1}^{F_z}e^{-iq_{e3}^{\uparrow}(x-\mathcal{D}^{\prime\prime})}+\tilde{n}_{he}^{\prime\prime\uparrow\downarrow}\varphi_{2}^{F_z}e^{-iq_{e3}^{\downarrow}(x-\mathcal{D}^{\prime\prime})}\\+\tilde{o}_{hh}^{\prime\prime\uparrow\uparrow}\varphi_{3}^{F_z}e^{iq_{h3}^{\uparrow}(x-\mathcal{D}^{\prime\prime})}+\tilde{o}_{hh}^{\prime\prime\uparrow\downarrow}\varphi_{4}^{F_z}e^{iq_{h3}^{\downarrow}(x-\mathcal{D}^{\prime\prime})}
+\tilde{p}_{hh}^{\prime\prime\uparrow\uparrow}\varphi_{3}^{F_z}e^{-iq_{h3}^{\uparrow}(x-\mathcal{D}^{\prime})}+\tilde{p}_{hh}^{\prime\prime\uparrow\downarrow}\varphi_{4}^{F_z}e^{-iq_{h3}^{\downarrow}(x-\mathcal{D}^{\prime})}\,, & \mathcal{D}^{\prime}<x<\mathcal{D}^{\prime\prime}\\
\varphi_{3}^{S_{R}}e^{iq_{h}^{S}(x-\mathcal{D}^{\prime\prime})}+\tilde{r}_{hh}^{\prime\prime\uparrow\uparrow}\varphi_{3}^{S_{R}}e^{-iq_{h}^{S}(x-\mathcal{D}^{\prime\prime})}+\tilde{r}_{hh}^{\prime\prime\uparrow\downarrow}\varphi_{4}^{S_{R}}e^{-iq_{h}^{S}(x-\mathcal{D}^{\prime\prime})}+\tilde{r}_{he}^{\prime\prime\uparrow\uparrow}\varphi_{1}^{S_{R}}e^{iq_{e}^{S}(x-\mathcal{D}^{\prime\prime})}\\+\tilde{r}_{he}^{\prime\prime\uparrow\downarrow}\varphi_{2}^{S_{R}}e^{iq_{e}^{S}(x-\mathcal{D}^{\prime\prime})}\,, & x>\mathcal{D}^{\prime\prime}
\end{cases}\\
\phi_{8}(x)&=\begin{cases}
\tilde{t}_{he}^{\prime\prime\downarrow\uparrow}\varphi_{1}^{S_L}e^{-iq_{e}^{S}x}+\tilde{t}_{he}^{\prime\prime\downarrow\downarrow}\varphi_{2}^{S_L}e^{-iq_{e}^{S}x}+\tilde{t}_{hh}^{\prime\prime\downarrow\uparrow}\varphi_{3}^{S_L}e^{iq_{h}^{S}x}+\tilde{t}_{hh}^{\prime\prime\downarrow\downarrow}\varphi_{4}^{S_L}e^{iq_{h}^{S}x}\,, & x<0\\
\tilde{a}_{he}^{\prime\prime\downarrow\uparrow}\varphi_{1}^{F_x}e^{iq_{e1}^{\uparrow}x}+\tilde{a}_{he}^{\prime\prime\downarrow\downarrow}\varphi_{2}^{F_x}e^{iq_{e1}^{\downarrow}x}+\tilde{b}_{he}^{\prime\prime\downarrow\uparrow}\varphi_{1}^{F_x}e^{-iq_{e1}^{\uparrow}(x-\mathcal{D}_{1})}+\tilde{b}_{he}^{\prime\prime\downarrow\downarrow}\varphi_{2}^{F_x}e^{-iq_{e1}^{\downarrow}(x-\mathcal{D}_{1})}\\+\tilde{c}_{hh}^{\prime\prime\downarrow\uparrow}\varphi_{3}^{F_x}e^{iq_{h1}^{\uparrow}(x-\mathcal{D}_{1})}+\tilde{c}_{hh}^{\prime\prime\downarrow\downarrow}\varphi_{4}^{F_x}e^{iq_{h1}^{\downarrow}(x-\mathcal{D}_{1})}
+\tilde{d}_{hh}^{\prime\prime\downarrow\uparrow}\varphi_{3}^{F_x}e^{-iq_{h1}^{\uparrow}x}+\tilde{d}_{hh}^{\prime\prime\downarrow\downarrow}\varphi_{4}^{F_x}e^{-iq_{h1}^{\downarrow}x}\,, & 0<x<\mathcal{D}_{1}\\
\tilde{e}_{he}^{\prime\prime\downarrow\uparrow}\varphi_{1}^{F_y}e^{iq_{e2}^{\uparrow}(x-\mathcal{D}_{1})}+\tilde{e}_{he}^{\prime\prime\downarrow\downarrow}\varphi_{2}^{F_y}e^{iq_{e2}^{\downarrow}(x-\mathcal{D}_{1})}+\tilde{f}_{he}^{\prime\prime\downarrow\uparrow}\varphi_{1}^{F_y}e^{-iq_{e2}^{\uparrow}(x-\mathcal{D}^{\prime})}+\tilde{f}_{he}^{\prime\prime\downarrow\downarrow}\varphi_{2}^{F_y}e^{-iq_{e2}^{\downarrow}(x-\mathcal{D}^{\prime})}\\+\tilde{g}_{hh}^{\prime\prime\downarrow\uparrow}\varphi_{3}^{F_y}e^{iq_{h2}^{\uparrow}(x-\mathcal{D}^{\prime})}+\tilde{g}_{hh}^{\prime\prime\downarrow\downarrow}\varphi_{4}^{F_y}e^{iq_{h2}^{\downarrow}(x-\mathcal{D}^{\prime})}
+\tilde{h}_{hh}^{\prime\prime\downarrow\uparrow}\varphi_{3}^{F_y}e^{-iq_{h2}^{\uparrow}(x-\mathcal{D}_{1})}+\tilde{h}_{hh}^{\prime\prime\downarrow\downarrow}\varphi_{4}^{F_y}e^{-iq_{h2}^{\downarrow}(x-\mathcal{D}_{1})}\,, & \mathcal{D}_{1}<x<\mathcal{D}^{\prime}\\
\tilde{m}_{he}^{\prime\prime\downarrow\uparrow}\varphi_{1}^{F_z}e^{iq_{e3}^{\uparrow}(x-\mathcal{D}^{\prime})}+\tilde{m}_{he}^{\prime\prime\downarrow\downarrow}\varphi_{2}^{F_z}e^{iq_{e3}^{\downarrow}(x-\mathcal{D}^{\prime})}+\tilde{n}_{he}^{\prime\prime\downarrow\uparrow}\varphi_{1}^{F_z}e^{-iq_{e3}^{\uparrow}(x-\mathcal{D}^{\prime\prime})}+\tilde{n}_{he}^{\prime\prime\downarrow\downarrow}\varphi_{2}^{F_z}e^{-iq_{e3}^{\downarrow}(x-\mathcal{D}^{\prime\prime})}\\+\tilde{o}_{hh}^{\prime\prime\downarrow\uparrow}\varphi_{3}^{F_z}e^{iq_{h3}^{\uparrow}(x-\mathcal{D}^{\prime\prime})}+\tilde{o}_{hh}^{\prime\prime\downarrow\downarrow}\varphi_{4}^{F_z}e^{iq_{h3}^{\downarrow}(x-\mathcal{D}^{\prime\prime})}
+\tilde{p}_{hh}^{\prime\prime\downarrow\uparrow}\varphi_{3}^{F_z}e^{-iq_{h3}^{\uparrow}(x-\mathcal{D}^{\prime})}+\tilde{p}_{hh}^{\prime\prime\downarrow\downarrow}\varphi_{4}^{F_z}e^{-iq_{h3}^{\downarrow}(x-\mathcal{D}^{\prime})}\,, & \mathcal{D}^{\prime}<x<\mathcal{D}^{\prime\prime}\\
\varphi_{4}^{S_{R}}e^{iq_{h}^{S}(x-\mathcal{D}^{\prime\prime})}+\tilde{r}_{hh}^{\prime\prime\downarrow\uparrow}\varphi_{3}^{S_{R}}e^{-iq_{h}^{S}(x-\mathcal{D}^{\prime\prime})}+\tilde{r}_{hh}^{\prime\prime\downarrow\downarrow}\varphi_{4}^{S_{R}}e^{-iq_{h}^{S}(x-\mathcal{D}^{\prime\prime})}+\tilde{r}_{he}^{\prime\prime\downarrow\uparrow}\varphi_{1}^{S_{R}}e^{iq_{e}^{S}(x-\mathcal{D}^{\prime\prime})}\\+\tilde{r}_{he}^{\prime\prime\downarrow\downarrow}\varphi_{2}^{S_{R}}e^{iq_{e}^{S}(x-\mathcal{D}^{\prime\prime})}\,, & x>\mathcal{D}^{\prime\prime}
\end{cases}
\end{split}
\label{wav2}
\end{equation}
where the expressions for $\varphi_{i}^{S_L}$, $\varphi_{i}^{F_x}$, $\varphi_{i}^{F_y}$, $\varphi_{i}^{S_R}$, $u_0$, $v_0$, $q_{e,h}^{S}$ are provided below Eq.~\eqref{wav}, while the expressions for $\varphi_{i}^{F_z}$, $q_{e1,h1}^{\sigma}$, $q_{e2,h2}^{\sigma}$, $q_{e3,h4}^{\sigma}$ are given below Eq.~\eqref{wavv}. Further in Eq.~\eqref{wav2}, the terms ${r}_{pq}^{\prime\prime\sigma\sigma'}$ and $\tilde{r}_{pq}^{\prime\prime\sigma\sigma'}$ denote the reflection amplitudes in left and right superconductors, respectively, while the terms ${t}_{pq}^{\prime\prime\sigma\sigma'}$ and $\tilde{t}_{pq}^{\prime\prime\sigma\sigma'}$ represent the corresponding transmission amplitudes, with $\sigma,\sigma'\in\{\uparrow,\downarrow\}$ and $p,q\in\{e,h\}$. After diagonalizing the Hamiltonian $\Big(\mathcal{H}_{BdG}^{\mbox{S-$F_x$-$F_y$-$F_z$-S}}\Big)^{*}(-k)$ instead of $\mathcal{H}_{BdG}^{\mbox{S-$F_x$-$F_y$-$F_z$-S}}(k)$ (see Eq.~\eqref{hmfxfyfz}), we will obtain conjugate processes $\tilde{\phi}_{l}$ $(l=1,2,...,8)$ necessary to form $\Gamma^{r}(x,\chi,\omega)$ in next section. The expressions for $\tilde{\varphi}_{i}^{S_L}$, $\tilde{\varphi}_{i}^{S_R}$, $\tilde{\varphi}_{i}^{F_x}$, $\tilde{\varphi}_{i}^{F_y}$, $\tilde{\varphi}_{i}^{N}$ are provided below Eqs.~\eqref{wav} and \eqref{wavv}.}

{The boundary conditions at the S-$F_x$ interface ($x=0$) are:
\begin{equation}
\label{eqq40}
\phi_{l}|_{x<0}=\phi_{l}|_{0<x<\mathcal{D}_{1}}\,\,\,\,\,\mbox{and}\,\,\,\,\,\frac{d\phi_{l}|_{0<x<\mathcal{D}_{1}}}{dx}-\frac{d\phi_{l}|_{x<0}}{dx}=\frac{2m^{*}\mathcal{B}_{1}}{\hbar^2}\phi_{l}(x=0).
\end{equation}
Similarly, the boundary conditions at the $F_x$-$F_y$ interface ($x=\mathcal{D}_{1}$) are:
\begin{equation}
\label{eqq41}
\phi_{l}|_{0<x<\mathcal{D}_{1}}=\phi_{l}|_{\mathcal{D}_{1}<x<\mathcal{D}^{\prime}}\,\,\,\,\,\mbox{and}\,\,\,\,\
\frac{d\phi_{l}|_{\mathcal{D}_{1}<x<\mathcal{D}^{\prime}}}{dx}-\frac{d\phi_{l}|_{0<x<\mathcal{D}_{1}}}{dx}=\frac{2m^{*}\mathcal{B}_{3}}{\hbar^2}\phi_{l}(x=\mathcal{D}_{1}).
\end{equation}
The boundary conditions at the $F_y$-$F_z$ interface ($x=\mathcal{D}^{\prime}$) are:
\begin{equation}
\label{eqq42}
\phi_{l}|_{\mathcal{D}_{1}<x<\mathcal{D}^{\prime}}=\phi_{l}|_{\mathcal{D}^{\prime}<x<\mathcal{D}^{\prime\prime}}\,\,\,\,\,\mbox{and}\,\,\,\,\
\frac{d\phi_{l}|_{\mathcal{D}^{\prime}<x<\mathcal{D}^{\prime\prime}}}{dx}-\frac{d\phi_{l}|_{\mathcal{D}_{1}<x<\mathcal{D}^{\prime}}}{dx}=\frac{2m^{*}\mathcal{B}_{4}}{\hbar^2}\phi_{l}(x=\mathcal{D}^{\prime}).
\end{equation}
Finally, the boundary conditions at the $F_z$-S interface ($x=\mathcal{D}^{\prime\prime}$) are:
\begin{equation}
\label{eqq43}
\phi_{l}|_{\mathcal{D}^{\prime}<x<\mathcal{D}^{\prime\prime}}=\phi_{l}|_{x>\mathcal{D}^{\prime\prime}}\,\,\,\,\,\mbox{and}\,\,\,\,\,\frac{d\phi_{l}|_{x>\mathcal{D}^{\prime\prime}}}{dx}-\frac{d\phi_{l}|_{\mathcal{D}^{\prime}<x<\mathcal{D}^{\prime\prime}}}{dx}=\frac{2m^{*}\mathcal{B}_{2}}{\hbar^2}\phi_{l}(x=\mathcal{D}^{\prime\prime}).
\end{equation}
By applying the above boundary conditions at $x=0$, $x=\mathcal{D}_{1}$, $x=\mathcal{D}^{\prime}$ and $x=\mathcal{D}^{\prime\prime}$, a total of $32$ equations are derived for each type of incident process described in Eq.~\eqref{wav2}. The scattering amplitudes, such as ${r}_{pq}^{\prime\prime\sigma\sigma'}$, $\tilde{r}_{pq}^{\prime\prime\sigma\sigma'}$, $\tilde{t}_{pq}^{\prime\prime\sigma\sigma'}$, ${t}_{pq}^{\prime\prime\sigma\sigma'}$, are derived by solving these $24$ equations. $s_{1}={r}_{eh}^{\prime\prime\uparrow\downarrow}$, $s_{2}={r}_{eh}^{\prime\prime\downarrow\uparrow}$, $s_{3}={r}_{he}^{\prime
\prime\uparrow\downarrow}$, and $s_{4}={r}_{he}^{\prime\prime\downarrow\uparrow}$, obtained by putting the wavefunctions
$\phi_{1}$, $\phi_{2}$, $\phi_{3}$, and $\phi_{4}$ into Eqs.~\eqref{eqq40}-\eqref{eqq43} are used to calculate the Josephson current via Eq.~\eqref{eq3}.}
\setcounter{subsection}{0}
\section*{Appendix B: Green's functions}
In this Appendix, we provide the detailed derivation of the Green's functions. Using the methodology outlined in Refs.~\cite{cayy,amb}, the retarded Green's function is expressed as:
\begin{equation}
\label{RGF}
\begin{split}
\Gamma^{r}(x,\chi,\omega)=
\begin{cases}
\phi_{1}(x)[\alpha_{11}\tilde{\phi}_{5}^{T}(\chi)+\alpha_{12}\tilde{\phi}_{6}^{T}(\chi)+\alpha_{13}\tilde{\phi}_{7}^{T}(\chi)+\alpha_{14}\tilde{\phi}_{8}^{T}(\chi)]\\
+
\phi_{2}(x)[\alpha_{21}\tilde{\phi}_{5}^{T}(\chi)+\alpha_{22}\tilde{\phi}_{6}^{T}(\chi)+\alpha_{23}\tilde{\phi}_{7}^{T}(\chi)+\alpha_{24}\tilde{\phi}_{8}^{T}(\chi)]\\
+
\phi_{3}(x)[\alpha_{31}\tilde{\phi}_{5}^{T}(\chi)+\alpha_{32}\tilde{\phi}_{6}^{T}(\chi)+\alpha_{33}\tilde{\phi}_{7}^{T}(\chi)+\alpha_{34}\tilde{\phi}_{8}^{T}(\chi)]\\
+
\phi_{4}(x)[\alpha_{41}\tilde{\phi}_{5}^{T}(\chi)+\alpha_{42}\tilde{\phi}_{6}^{T}(\chi)+\alpha_{43}\tilde{\phi}_{7}^{T}(\chi)+\alpha_{44}\tilde{\phi}_{8}^{T}(\chi)]
\,,\quad x>\chi&\\
\phi_{5}(x)[\beta_{11}\tilde{\phi}_{1}^{T}(\chi)+\beta_{12}\tilde{\phi}_{2}^{T}(\chi)+\beta_{13}\tilde{\phi}_{3}^{T}(\chi)+\beta_{14}\tilde{\phi}_{4}^{T}(\chi)]\\
+\phi_{6}(x)[\beta_{21}\tilde{\phi}_{1}^{T}(\chi)+\beta_{22}\tilde{\phi}_{2}^{T}(\chi)+\beta_{23}\tilde{\phi}_{3}^{T}(\chi)+\beta_{24}\tilde{\phi}_{4}^{T}(\chi)]\\
+\phi_{7}(x)[\beta_{31}\tilde{\phi}_{1}^{T}(\chi)+\beta_{32}\tilde{\phi}_{2}^{T}(\chi)+\beta_{33}\tilde{\phi}_{3}^{T}(\chi)+\beta_{34}\tilde{\phi}_{4}^{T}(\chi)]\\
+\phi_{8}(x)[\beta_{41}\tilde{\phi}_{1}^{T}(\chi)+\beta_{42}\tilde{\phi}_{2}^{T}(\chi)+\beta_{43}\tilde{\phi}_{3}^{T}(\chi)+\beta_{44}\tilde{\phi}_{4}^{T}(\chi)]\,, \quad x<\chi&
\end{cases}
\end{split}
\end{equation}
In Eq.~\eqref{RGF}, the coefficients $\alpha_{ab}$ and $\beta_{cd}$ (where $a,b\in{1,2,3,4}$ and $c,d\in{1,2,3,4}$) are derived from the equations of motion for $\Gamma^{r}(x,\chi,\omega)$,
\begin{equation}
[\omega-H_{BdG}(x)]\Gamma^{r}(x,\chi,\omega)=\delta(x-\chi).
\label{rgf1}
\end{equation}
By integrating Eq.~\eqref{rgf1} with respect to $x$ over a small region surrounding $x=\chi$, we obtain
\begin{equation}
\label{conditionGRSO}
[\Gamma^{r}(x>\chi)]_{x=\chi}=[\Gamma^{r}(x<\chi)]_{x=\chi}\,\,\, \mbox{and}\,\,\,
\Bigg[\frac{d}{dx}\Gamma^{r}(x>\chi)\Bigg]_{x=\chi}-\Bigg[\frac{d}{dx}\Gamma^{r}(x<\chi)\Bigg]_{x=\chi}=\gamma\tau_{z}\sigma_{0},
\end{equation}
where $\gamma=2m^{*}/\hbar^2$ and $\tau_{i}$, $\sigma_{i}$ represent Pauli matrices in particle-hole and spin spaces respectively. To determine the coefficients $\alpha_{ab}$ and $\beta_{cd}$, we begin by substituting the wavefunctions from Eq.~\eqref{wav} into Eq.~\eqref{RGF} for the {S-$F_x$-$F_y$-S} JJ and from Eq.~\eqref{wavv} into Eq.~\eqref{RGF} for the S-$F_x$-$F_y$-$F_z$-S JJ. Next, the expressions of $\Gamma^{r}$ for $x>\chi$ and $x<\chi$, obtained from Eq.~\eqref{RGF}, are substituted into Eq.~\eqref{conditionGRSO}, yielding a system of $32$ equations. Solving this system allows us to express the coefficients $\alpha_{ab}$ and $\beta_{cd}$ in terms of different scattering amplitudes, including $r_{pq}^{\sigma\sigma'}$, $\bar{r}_{pq}^{\sigma\sigma'}$, $\bar{t}_{pq}^{\sigma\sigma'}$, $t_{pq}^{\sigma\sigma'}$ for the {S-$F_x$-$F_y$-S} JJ and $r_{pq}^{\prime\sigma\sigma'}$, $\bar{r}_{pq}^{\prime\sigma\sigma'}$, $\bar{t}_{pq}^{\prime\sigma\sigma'}$, $t_{pq}^{\prime\sigma\sigma'}$ for the S-$F_x$-$F_y$-$F_z$-S JJ.
$\Gamma^{r}$ can be represented as
\begin{equation}
\label{GF}
\Gamma^{r}(x,\chi,\omega)=
\begin{pmatrix}
\Gamma^{r}_{ee}&\Gamma^{r}_{eh}\\
\Gamma^{r}_{he}&\Gamma^{r}_{hh}
\end{pmatrix},
\end{equation}
where $\Gamma^{r}_{ee}$, $\Gamma^{r}_{eh}$, $\Gamma^{r}_{he}$, $\Gamma^{r}_{hh}$ are matrices. Each element of $\Gamma^{r}(x,\chi,\omega)$ under the influence of spin-flip scattering is given as
\begin{equation}\label{gr}
 \Gamma^{r}_{pq}(x,\chi,\omega)=\begin{pmatrix}                      [\Gamma^{r}_{pq}]_{\uparrow\uparrow} &[\Gamma^{r}_{pq}]_{\uparrow\downarrow}\\
                      [\Gamma^{r}_{pq}]_{\downarrow\uparrow}&[\Gamma^{r}_{pq}]_{\downarrow\downarrow}                     \end{pmatrix},\,\,\mbox{with}\,\, p,q\in\{e,h\}.
\end{equation}

We substitute the values of the coefficients $\alpha_{ab}$ and $\beta_{cd}$ into Eq.~\eqref{RGF} and use Eqs.~\eqref{GF} and \eqref{gr} to compute each element of the retarded Green's function $\Gamma^{r}$, where the scattering amplitudes $r_{pq}^{\sigma\sigma'}$ are obtained using Eqs.~\eqref{eqq8}-\eqref{eqq10} for the {S-$F_x$-$F_y$-S} JJ and the scattering amplitudes $r_{pq}^{\prime\sigma\sigma'}$ are determined using Eqs.~\eqref{eq8}-\eqref{eq11} for the S-$F_x$-$F_y$-$F_z$-S JJ.

\subsection{Bilayer {S-$F_x$-$F_y$-S} Josephson junction}
In case of {S-$F_x$-$F_y$-S} JJ, in the left superconducting region, for $\Gamma^{r}_{eh}$ we find
\begin{equation}
\label{supampp}
\begin{split}
[\Gamma^{r}_{eh}]_{\uparrow\uparrow}&=\frac{\gamma}{2i(u_0^2-v_0^2)}\Bigg[\frac{r_{ee}^{\downarrow\uparrow}e^{-iq_{e}^{S}(x+\chi+\mathcal{D})}u_0v_0+r_{eh}^{\downarrow\downarrow}e^{i(q_{h}^{S}(x+\frac{\mathcal{D}}{2})-q_{e}^{S}(\chi+\frac{\mathcal{D}}{2}))}v_0^2}{q_{e}^{S}}\\&+\frac{r_{he}^{\uparrow\uparrow}e^{i(q_{h}^{S}(\chi+\frac{\mathcal{D}}{2})-q_{e}^{S}(x+\frac{\mathcal{D}}{2}))}u_0^2+r_{hh}^{\uparrow\downarrow}e^{iq_{h}^{S}(x+\chi+\mathcal{D})}u_0v_0}{q_{h}^{S}}\Bigg],
\end{split}\nonumber
\end{equation}
\begin{equation}
\begin{split}
[\Gamma^{r}_{eh}]_{\downarrow\downarrow}&=-\frac{\gamma}{2i(u_0^2-v_0^2)}\Bigg[\frac{r_{ee}^{\uparrow\downarrow}e^{-iq_{e}^{S}(x+\chi+\mathcal{D})}u_0v_0+r_{eh}^{\uparrow\uparrow}e^{i(q_{h}^{S}(x+\frac{\mathcal{D}}{2})-q_{e}^{S}(\chi+\frac{\mathcal{D}}{2}))}v_0^2}{q_{e}^{S}}\\&+\frac{r_{he}^{\downarrow\downarrow}e^{i(q_{h}^{S}(\chi+\frac{\mathcal{D}}{2})-q_{e}^{S}(x+\frac{\mathcal{D}}{2}))}u_0^2+r_{hh}^{\downarrow\uparrow}e^{iq_{h}^{S}(x+\chi+\mathcal{D})}u_0v_0}{q_{h}^{S}}\Bigg],\\
[\Gamma^{r}_{eh}]_{\uparrow\downarrow}&=\frac{\gamma}{2i(u_0^2-v_0^2)}\Bigg[\frac{e^{iq_{e}^{S}|x-\chi|}u_0v_0+r_{ee}^{\uparrow\uparrow}e^{-iq_{e}^{S}(x+\chi+\mathcal{D})}u_0v_0+r_{eh}^{\uparrow\downarrow}e^{i(q_{h}^{S}(x+\frac{\mathcal{D}}{2})-q_{e}^{S}(\chi+\frac{\mathcal{D}}{2}))}v_0^2}{q_{e}^{S}}\\&+\frac{e^{-iq_{h}^{S}|x-\chi|}u_0v_0+r_{he}^{\downarrow\uparrow}e^{i(q_{h}^{S}(\chi+\frac{\mathcal{D}}{2})-q_{e}^{S}(x+\frac{\mathcal{D}}{2}))}u_0^2+r_{hh}^{\downarrow\downarrow}e^{iq_{h}^{S}(x+\chi+\mathcal{D})}u_0v_0}{q_{h}^{S}}\Bigg],\\
[\Gamma^{r}_{eh}]_{\downarrow\uparrow}&=-\frac{\gamma}{2i(u_0^2-v_0^2)}\Bigg[\frac{e^{iq_{e}^{S}|x-\chi|}u_0v_0+r_{ee}^{\downarrow\downarrow}e^{-iq_{e}^{S}(x+\chi+\mathcal{D})}u_0v_0+r_{eh}^{\downarrow\uparrow}e^{i(q_{h}^{S}(x+\frac{\mathcal{D}}{2})-q_{e}^{S}(\chi+\frac{\mathcal{D}}{2}))}v_0^2}{q_{e}^{S}}\\&+\frac{e^{-iq_{h}^{S}|x-\chi|}u_0v_0+r_{he}^{\uparrow\downarrow}e^{i(q_{h}^{S}(\chi+\frac{\mathcal{D}}{2})-q_{e}^{S}(x+\frac{\mathcal{D}}{2}))}u_0^2+r_{hh}^{\uparrow\uparrow}e^{iq_{h}^{S}(x+\chi+\mathcal{D})}u_0v_0}{q_{h}^{S}}\Bigg].
\end{split}
\end{equation}

\subsection{Trilayer S-$F_x$-$F_y$-$F_z$-S Josephson junction}
In case of S-$F_x$-$F_y$-$F_z$-S JJ, in the left superconducting region, for $\Gamma^{r}_{eh}$ we find
\begin{equation}
\label{suppamp}
\begin{split}
[\Gamma^{r}_{eh}]_{\uparrow\uparrow}&=\frac{\gamma}{2i(u_0^2-v_0^2)}\Bigg[\frac{r_{ee}^{\prime\downarrow\uparrow}e^{-iq_{e}^{S}(x+\chi+\mathcal{D})}u_0v_0+r_{eh}^{\prime\downarrow\downarrow}e^{i(q_{h}^{S}(x+\frac{\mathcal{D}}{2})-q_{e}^{S}(\chi+\frac{\mathcal{D}}{2}))}v_0^2}{q_{e}^{S}}\\&+\frac{r_{he}^{\prime\uparrow\uparrow}e^{i(q_{h}^{S}(\chi+\frac{\mathcal{D}}{2})-q_{e}^{S}(x+\frac{\mathcal{D}}{2}))}u_0^2+r_{hh}^{\prime\uparrow\downarrow}e^{iq_{h}^{S}(x+\chi+\mathcal{D})}u_0v_0}{q_{h}^{S}}\Bigg],\\
[\Gamma^{r}_{eh}]_{\downarrow\downarrow}&=-\frac{\gamma}{2i(u_0^2-v_0^2)}\Bigg[\frac{r_{ee}^{\prime\uparrow\downarrow}e^{-iq_{e}^{S}(x+\chi+\mathcal{D})}u_0v_0+r_{eh}^{\prime\uparrow\uparrow}e^{i(q_{h}^{S}(x+\frac{\mathcal{D}}{2})-q_{e}^{S}(\chi+\frac{\mathcal{D}}{2}))}v_0^2}{q_{e}^{S}}\\&+\frac{r_{he}^{\prime\downarrow\downarrow}e^{i(q_{h}^{S}(\chi+\frac{\mathcal{D}}{2})-q_{e}^{S}(x+\frac{\mathcal{D}}{2}))}u_0^2+r_{hh}^{\prime\downarrow\uparrow}e^{iq_{h}^{S}(x+\chi+\mathcal{D})}u_0v_0}{q_{h}^{S}}\Bigg],\\
[\Gamma^{r}_{eh}]_{\uparrow\downarrow}&=\frac{\gamma}{2i(u_0^2-v_0^2)}\Bigg[\frac{e^{iq_{e}^{S}|x-\chi|}u_0v_0+r_{ee}^{\prime\uparrow\uparrow}e^{-iq_{e}^{S}(x+\chi+\mathcal{D})}u_0v_0+r_{eh}^{\prime\uparrow\downarrow}e^{i(q_{h}^{S}(x+\frac{\mathcal{D}}{2})-q_{e}^{S}(\chi+\frac{\mathcal{D}}{2}))}v_0^2}{q_{e}^{S}}\\&+\frac{e^{-iq_{h}^{S}|x-\chi|}u_0v_0+r_{he}^{\prime\downarrow\uparrow}e^{i(q_{h}^{S}(\chi+\frac{\mathcal{D}}{2})-q_{e}^{S}(x+\frac{\mathcal{D}}{2}))}u_0^2+r_{hh}^{\prime\downarrow\downarrow}e^{iq_{h}^{S}(x+\chi+\mathcal{D})}u_0v_0}{q_{h}^{S}}\Bigg],\\
[\Gamma^{r}_{eh}]_{\downarrow\uparrow}&=-\frac{\gamma}{2i(u_0^2-v_0^2)}\Bigg[\frac{e^{iq_{e}^{S}|x-\chi|}u_0v_0+r_{ee}^{\prime\downarrow\downarrow}e^{-iq_{e}^{S}(x+\chi+\mathcal{D})}u_0v_0+r_{eh}^{\prime\downarrow\uparrow}e^{i(q_{h}^{S}(x+\frac{\mathcal{D}}{2})-q_{e}^{S}(\chi+\frac{\mathcal{D}}{2}))}v_0^2}{q_{e}^{S}}\\&+\frac{e^{-iq_{h}^{S}|x-\chi|}u_0v_0+r_{he}^{\prime\uparrow\downarrow}e^{i(q_{h}^{S}(\chi+\frac{\mathcal{D}}{2})-q_{e}^{S}(x+\frac{\mathcal{D}}{2}))}u_0^2+r_{hh}^{\prime\uparrow\uparrow}e^{iq_{h}^{S}(x+\chi+\mathcal{D})}u_0v_0}{q_{h}^{S}}\Bigg].
\end{split}
\end{equation}
\setcounter{subsection}{0}
\section*{APPENDIX C: Analytical expressions for superconducting pairing amplitudes}
In this Appendix, we provide the analytical expressions for even- and odd-$\omega$ SS, EST, and MST superconducting pairing amplitudes in our two setups: (A) Bilayer S-$F_x$-$F_y$-S JJ, and (B) Trilayer S-$F_x$-$F_y$-$F_z$-S JJ.

\subsection{Bilayer {S-$F_x$-$F_y$-S} Josephson junction}
For SS superconducting pairing, in the left superconducting region, we find
\begin{eqnarray}
\label{ESE}
f_{0}^{E}(x,\chi,\omega)&=&\frac{\gamma u_0 v_0}{2i(u_0^2-v_0^2)}\Bigg[\frac{e^{iq_{e}^{S}|x-\chi|}}{q_{e}^{S}}+\frac{e^{-iq_{h}^{S}|x-\chi|}}{q_{h}^{S}}\Bigg]+
\frac{\gamma u_0 v_0}{4i(u_0^2-v_0^2)}\Bigg[\frac{(r_{ee}^{\uparrow\uparrow}+r_{ee}^{\downarrow\downarrow})e^{-iq_{e}^{S}(x+\chi+\mathcal{D})}}{q_{e}^{S}}+\frac{(r_{hh}^{\uparrow\uparrow}+r_{hh}^{\downarrow\downarrow})e^{iq_{h}^{S}(x+\chi+\mathcal{D})}}{q_{h}^{S}}\Bigg]\nonumber\\&&
+\frac{\gamma }{8i(u_0^2-v_0^2)}\Big(e^{i(q_h^S(x+\frac{\mathcal{D}}{2})-q_e^S(\chi+\frac{\mathcal{D}}{2}))}+e^{i(q_h^S(\chi+\frac{\mathcal{D}}{2})-q_e^S(x+\frac{\mathcal{D}}{2}))}\Big)\Bigg[\frac{(r_{eh}^{\uparrow\downarrow}+r_{eh}^{\downarrow\uparrow})v_0^2}{q_{e}^{S}}+\frac{(r_{he}^{\uparrow\downarrow}+r_{he}^{\downarrow\uparrow})u_0^2}{q_{h}^{S}}\Bigg],\\
f_{0}^{O}(x,\chi,\omega)&=&\frac{\gamma}{8i(u_0^2-v_0^2)}\Big(e^{i(q_h^S(x+\frac{\mathcal{D}}{2})-q_e^S(\chi+\frac{\mathcal{D}}{2}))}-e^{i(q_h^S(\chi+\frac{\mathcal{D}}{2})-q_e^S(x+\frac{\mathcal{D}}{2}))}\Big)\Bigg[\frac{(r_{eh}^{\uparrow\downarrow}+r_{eh}^{\downarrow\uparrow})v_0^2}{q_{e}^{S}}-\frac{(r_{he}^{\downarrow\uparrow}+r_{he}^{\uparrow\downarrow})u_0^2}{q_{h}^{S}}\Bigg],
\label{OSO}
\end{eqnarray}
where $r_{ee}^{\uparrow\uparrow}$, $r_{ee}^{\downarrow\downarrow}$, $r_{hh}^{\uparrow\uparrow}$, $r_{hh}^{\downarrow\downarrow}$ denote the normal reflection amplitudes without flip in left superconductor, while $r_{eh}^{\uparrow\downarrow}$, $r_{eh}^{\downarrow\uparrow}$, $r_{he}^{\uparrow\downarrow}$, $r_{he}^{\downarrow\uparrow}$ represent the corresponding Andreev reflection amplitudes without flip in left superconductor. {At $x=\chi$, odd-$\omega$ SS pairing vanishes, while even-$\omega$ SS pairing is finite and given as,
\begin{eqnarray}
\label{ESE-L}
f_{0}^{E,L}(x,\omega)&=&\frac{\gamma u_0 v_0}{2i(u_0^2-v_0^2)}\Bigg[\frac{1}{q_{e}^{S}}+\frac{1}{q_{h}^{S}}\Bigg]+
\frac{\gamma u_0 v_0}{4i(u_0^2-v_0^2)}\Bigg[\frac{(r_{ee}^{\uparrow\uparrow}+r_{ee}^{\downarrow\downarrow})e^{-iq_{e}^{S}(2x+\mathcal{D})}}{q_{e}^{S}}+\frac{(r_{hh}^{\uparrow\uparrow}+r_{hh}^{\downarrow\downarrow})e^{iq_{h}^{S}(2x+\mathcal{D})}}{q_{h}^{S}}\Bigg]\nonumber\\&&
+\frac{\gamma e^{i(q_{h}^{S}-q_{e}^{S})(x+\frac{\mathcal{D}}{2})}}{4i(u_0^2-v_0^2)}\Bigg[\frac{(r_{eh}^{\uparrow\downarrow}+r_{eh}^{\downarrow\uparrow})v_0^2}{q_{e}^{S}}+\frac{(r_{he}^{\uparrow\downarrow}+r_{he}^{\downarrow\uparrow})u_0^2}{q_{h}^{S}}\Bigg].
\end{eqnarray}}
For EST superconducting pairing, we find
\begin{eqnarray}
\label{ETO-equal-upup}
f_{\uparrow\uparrow}^{E}(x,\chi,\omega)&=&\frac{\gamma}{4i(u_0^2-v_0^2)}\Big(e^{i(q_h^S(\chi+\frac{\mathcal{D}}{2})-q_e^S(x+\frac{\mathcal{D}}{2}))}-e^{i(q_h^S(x+\frac{\mathcal{D}}{2})-q_e^S(\chi+\frac{\mathcal{D}}{2}))}\Big)\Bigg[\frac{r_{he}^{\uparrow\uparrow}u_0^2}{q_h^S}-\frac{r_{eh}^{\downarrow\downarrow}v_0^2}{q_e^S}\Bigg],\\
f_{\uparrow\uparrow}^{O}(x,\chi,\omega)&=&\frac{\gamma u_0v_0}{2i(u_0^2-v_0^2)}\Bigg[\frac{r_{hh}^{\uparrow\downarrow}e^{iq_h^{S}(x+\chi+\mathcal{D})}}{q_h^{S}}+\frac{r_{ee}^{\downarrow\uparrow}e^{-iq_e^{S}(x+\chi+\mathcal{D})}}{q_e^{S}}\Bigg]+\frac{\gamma}{4i(u_0^2-v_0^2)}\Bigg[\frac{r_{he}^{\uparrow\uparrow}u_0^2}{q_h^{S}}+\frac{r_{eh}^{\downarrow\downarrow} v_0^2}{q_e^S}\Bigg]\Big(e^{i(q_h^S(\chi+\frac{\mathcal{D}}{2})-q_e^S(x+\frac{\mathcal{D}}{2}))}\nonumber\\&&+e^{i(q_h^S(x+\frac{\mathcal{D}}{2})-q_e^S(\chi+\frac{\mathcal{D}}{2}))}\Big),\\
f_{\downarrow\downarrow}^{E}(x,\chi,\omega)&=&\frac{\gamma}{4i(u_0^2-v_0^2)}\Big(e^{i(q_h^S(x+\frac{\mathcal{D}}{2})-q_e^S(\chi+\frac{\mathcal{D}}{2}))}-e^{i(q_h^S(\chi+\frac{\mathcal{D}}{2})-q_e^S(x+\frac{\mathcal{D}}{2}))}\Big)\Bigg[\frac{r_{he}^{\downarrow\downarrow}u_0^2}{q_h^S}-\frac{r_{eh}^{\uparrow\uparrow}v_0^2}{q_e^S}\Bigg],\\
f_{\downarrow\downarrow}^{O}(x,\chi,\omega)&=&-\frac{\gamma u_0v_0}{2i(u_0^2-v_0^2)}\Bigg[\frac{r_{hh}^{\downarrow\uparrow}e^{iq_h^{S}(x+\chi+\mathcal{D})}}{q_h^{S}}+\frac{r_{ee}^{\uparrow\downarrow}e^{-iq_e^{S}(x+\chi+\mathcal{D})}}{q_e^{S}}\Bigg]-\frac{\gamma}{4i(u_0^2-v_0^2)}\Bigg[\frac{r_{he}^{\downarrow\downarrow}u_0^2}{q_h^{S}}+\frac{r_{eh}^{\uparrow\uparrow} v_0^2}{q_e^S}\Bigg]\Big(e^{i(q_h^S(\chi+\frac{\mathcal{D}}{2})-q_e^S(x+\frac{\mathcal{D}}{2}))}\nonumber\\&&+e^{i(q_h^S(x+\frac{\mathcal{D}}{2})-q_e^S(\chi+\frac{\mathcal{D}}{2}))}\Big),
\label{OTE-equal-downdown}
\end{eqnarray}
where $r_{ee}^{\uparrow\downarrow}$, $r_{ee}^{\downarrow\uparrow}$, $r_{hh}^{\uparrow\downarrow}$, $r_{hh}^{\downarrow\uparrow}$ represent the spin-flip normal reflection amplitudes in left superconductor, while $r_{eh}^{\uparrow\uparrow}$, $r_{eh}^{\downarrow\downarrow}$, $r_{he}^{\uparrow\uparrow}$, and $r_{he}^{\downarrow\downarrow}$ denote the corresponding spin-flip Andreev reflection amplitudes in left superconductor. {At $x=\chi$, even-$\omega$ EST pairing vanishes, while odd-$\omega$ EST pairing is finite and given as,
\begin{eqnarray}
\label{OTE-equal-upup-L}
f_{\uparrow\uparrow}^{O,L}(x,\omega)&=&\frac{\gamma u_0v_0}{2i(u_0^2-v_0^2)}\Bigg[\frac{r_{hh}^{\uparrow\downarrow}e^{iq_h^{S}(2x+\mathcal{D})}}{q_h^{S}}+\frac{r_{ee}^{\downarrow\uparrow}e^{-iq_e^{S}(2x+\mathcal{D})}}{q_e^{S}}\Bigg]+\frac{\gamma e^{i(q_{h}^{S}-q_{e}^{S})(x+\frac{\mathcal{D}}{2})}}{2i(u_0^2-v_0^2)}\Bigg[\frac{r_{he}^{\uparrow\uparrow}u_0^2}{q_h^{S}}+\frac{r_{eh}^{\downarrow\downarrow} v_0^2}{q_e^S}\Bigg],\\
f_{\downarrow\downarrow}^{O,L}(x,\omega)&=&-\frac{\gamma u_0v_0}{2i(u_0^2-v_0^2)}\Bigg[\frac{r_{hh}^{\downarrow\uparrow}e^{iq_h^{S}(2x+\mathcal{D})}}{q_h^{S}}+\frac{r_{ee}^{\uparrow\downarrow}e^{-iq_e^{S}(2x+\mathcal{D})}}{q_e^{S}}\Bigg]-\frac{\gamma e^{i(q_{h}^{S}-q_{e}^{S})(x+\frac{\mathcal{D}}{2})}}{2i(u_0^2-v_0^2)}\Bigg[\frac{r_{he}^{\downarrow\downarrow}u_0^2}{q_h^{S}}+\frac{r_{eh}^{\uparrow\uparrow} v_0^2}{q_e^S}\Bigg].
\label{OTE-equal-downdown-L}
\end{eqnarray}}
For MST superconducting pairing, we find
\begin{eqnarray}
f_3^{E}(x,\chi,\omega)&=&\frac{\gamma}{8i(u_0^2-v_0^2)}\Big(e^{i(q_h^S(x+\frac{\mathcal{D}}{2})-q_e^S(\chi+\frac{\mathcal{D}}{2}))}-e^{i(q_h^S(\chi+\frac{\mathcal{D}}{2})-q_e^S(x+\frac{\mathcal{D}}{2}))}\Big)\Bigg[\frac{(r_{eh}^{\uparrow\downarrow}-r_{eh}^{\downarrow\uparrow})v_0^2}{q_{e}^{S}}+\frac{(r_{he}^{\uparrow\downarrow}-r_{he}^{\downarrow\uparrow})u_0^2}{q_h^{S}}\Bigg],\\
f_3^{O}(x,\chi,\omega)&=&\frac{\gamma u_0v_0}{4i(u_0^2-v_0^2)}\Bigg[\frac{(r_{ee}^{\uparrow\uparrow}-r_{ee}^{\downarrow\downarrow})e^{-iq_e^S(x+\chi+\mathcal{D})}}{q_e^S}-\frac{(r_{hh}^{\uparrow\uparrow}-r_{hh}^{\downarrow\downarrow})e^{iq_h^S(x+\chi+\mathcal{D})}}{q_h^S}\Bigg]+\frac{\gamma}{8i(u_0^2-v_0^2)}\Bigg[\frac{(r_{eh}^{\uparrow\downarrow}-r_{eh}^{\downarrow\uparrow})v_0^2}{q_{e}^{S}}\nonumber\\&&+\frac{(r_{he}^{\downarrow\uparrow}-r_{he}^{\uparrow\downarrow})u_0^2}{q_{h}^S}\Bigg]\Big(e^{i(q_h^S(x+\frac{\mathcal{D}}{2})-q_e^S(\chi+\frac{\mathcal{D}}{2}))}+e^{i(q_h^S(\chi+\frac{\mathcal{D}}{2})-q_e^S(x+\frac{\mathcal{D}}{2}))}\Big).
\end{eqnarray}
{At $x=\chi$, even-$\omega$ MST pairing vanishes, while odd-$\omega$ MST pairing is finite and given as,
\begin{eqnarray}
f_3^{O,L}(x,\omega)&=&\frac{\gamma u_0v_0}{4i(u_0^2-v_0^2)}\Bigg[\frac{(r_{ee}^{\uparrow\uparrow}-r_{ee}^{\downarrow\downarrow})e^{-iq_e^S(2x+\mathcal{D})}}{q_e^S}-\frac{(r_{hh}^{\uparrow\uparrow}-r_{hh}^{\downarrow\downarrow})e^{iq_h^S(2x+\mathcal{D})}}{q_h^S}\Bigg]+\frac{\gamma e^{i(q_{h}^{S}-q_{e}^{S})(x+\frac{\mathcal{D}}{2})}}{4i(u_0^2-v_0^2)}\Bigg[\frac{(r_{eh}^{\uparrow\downarrow}-r_{eh}^{\downarrow\uparrow})v_0^2}{q_{e}^{S}}\nonumber\\&&+\frac{(r_{he}^{\downarrow\uparrow}-r_{he}^{\uparrow\downarrow})u_0^2}{q_{h}^S}\Bigg].
\end{eqnarray}}
\subsection{Trilayer S-$F_x$-$F_y$-$F_z$-S Josephson junction}
For even- and odd-$\omega$ SS superconducting pairing, in the left superconducting region, we obtain
\begin{eqnarray}
\label{ESE-J}
f_{0}^{E}(x,\chi,\omega)&=&\frac{\gamma u_0 v_0}{2i(u_0^2-v_0^2)}\Bigg[\frac{e^{iq_{e}^{S}|x-\chi|}}{q_{e}^{S}}+\frac{e^{-iq_{h}^{S}|x-\chi|}}{q_{h}^{S}}\Bigg]+
\frac{\gamma u_0 v_0}{4i(u_0^2-v_0^2)}\Bigg[\frac{(r_{ee}^{\prime\uparrow\uparrow}+r_{ee}^{\prime\downarrow\downarrow})e^{-iq_{e}^{S}(x+\chi+\mathcal{D})}}{q_{e}^{S}}+\frac{(r_{hh}^{\prime\uparrow\uparrow}+r_{hh}^{\prime\downarrow\downarrow})e^{iq_{h}^{S}(x+\chi+\mathcal{D})}}{q_{h}^{S}}\Bigg]\nonumber\\&&+\frac{\gamma }{8i(u_0^2-v_0^2)}\Big(e^{i(q_h^S(x+\frac{\mathcal{D}}{2})-q_e^S(\chi+\frac{\mathcal{D}}{2}))}+e^{i(q_h^S(\chi+\frac{\mathcal{D}}{2})-q_e^S(x+\frac{\mathcal{D}}{2}))}\Big)\Bigg[\frac{(r_{eh}^{\prime\uparrow\downarrow}+r_{eh}^{\prime\downarrow\uparrow})v_0^2}{q_{e}^{S}}+\frac{(r_{he}^{\prime\uparrow\downarrow}+r_{he}^{\prime\downarrow\uparrow})u_0^2}{q_{h}^{S}}\Bigg],\\
f_{0}^{O}(x,\chi,\omega)&=&\frac{\gamma}{8i(u_0^2-v_0^2)}\Big(e^{i(q_h^S(x+\frac{\mathcal{D}}{2})-q_e^S(\chi+\frac{\mathcal{D}}{2}))}-e^{i(q_h^S(\chi+\frac{\mathcal{D}}{2})-q_e^S(x+\frac{\mathcal{D}}{2}))}\Big)\Bigg[\frac{(r_{eh}^{\prime\uparrow\downarrow}+r_{eh}^{\prime\downarrow\uparrow})v_0^2}{q_{e}^{S}}-\frac{(r_{he}^{\prime\downarrow\uparrow}+r_{he}^{\prime\uparrow\downarrow})u_0^2}{q_{h}^{S}}\Bigg],
\label{OSO-J}
\end{eqnarray}
where $r_{ee}^{\prime\uparrow\uparrow}$, $r_{ee}^{\prime\downarrow\downarrow}$, $r_{hh}^{\prime\uparrow\uparrow}$, $r_{hh}^{\prime\downarrow\downarrow}$ denote the normal reflection amplitudes without flip in left superconductor, while $r_{eh}^{\prime\uparrow\downarrow}$, $r_{eh}^{\prime\downarrow\uparrow}$, $r_{he}^{\prime\uparrow\downarrow}$, $r_{he}^{\prime\downarrow\uparrow}$ represent the corresponding Andreev reflection amplitudes without flip in left superconductor. {At $x=\chi$, the odd-$\omega$ SS pairing vanishes, whereas the even-$\omega$ SS pairing remains finite and is given by
\begin{eqnarray}
\label{ESE-J-L}
f_{0}^{E,L}(x,\omega)&=&\frac{\gamma u_0 v_0}{2i(u_0^2-v_0^2)}\Bigg[\frac{1}{q_{e}^{S}}+\frac{1}{q_{h}^{S}}\Bigg]+
\frac{\gamma u_0 v_0}{4i(u_0^2-v_0^2)}\Bigg[\frac{(r_{ee}^{\prime\uparrow\uparrow}+r_{ee}^{\prime\downarrow\downarrow})e^{-iq_{e}^{S}(2x+\mathcal{D})}}{q_{e}^{S}}+\frac{(r_{hh}^{\prime\uparrow\uparrow}+r_{hh}^{\prime\downarrow\downarrow})e^{iq_{h}^{S}(2x+\mathcal{D})}}{q_{h}^{S}}\Bigg]\nonumber\\&&
+\frac{\gamma e^{i(q_{h}^{S}-q_{e}^{S})(x+\frac{\mathcal{D}}{2})}}{4i(u_0^2-v_0^2)}\Bigg[\frac{(r_{eh}^{\prime\uparrow\downarrow}+r_{eh}^{\prime\downarrow\uparrow})v_0^2}{q_{e}^{S}}+\frac{(r_{he}^{\prime\uparrow\downarrow}+r_{he}^{\prime\downarrow\uparrow})u_0^2}{q_{h}^{S}}\Bigg].
\end{eqnarray}}
For even- and odd-$\omega$ EST superconducting pairing, we obtain
\begin{eqnarray}
\label{ETO-equal-upup-J}
f_{\uparrow\uparrow}^{E}(x,\chi,\omega)&=&\frac{\gamma}{4i(u_0^2-v_0^2)}\Big(e^{i(q_h^S(\chi+\frac{\mathcal{D}}{2})-q_e^S(x+\frac{\mathcal{D}}{2}))}-e^{i(q_h^S(x+\frac{\mathcal{D}}{2})-q_e^S(\chi+\frac{\mathcal{D}}{2}))}\Big)\Bigg[\frac{r_{he}^{\prime\uparrow\uparrow}u_0^2}{q_h^S}-\frac{r_{eh}^{\prime\downarrow\downarrow}v_0^2}{q_e^S}\Bigg],\\
f_{\uparrow\uparrow}^{O}(x,\chi,\omega)&=&\frac{\gamma u_0v_0}{2i(u_0^2-v_0^2)}\Bigg[\frac{r_{hh}^{\prime\uparrow\downarrow}e^{iq_h^{S}(x+\chi+\mathcal{D})}}{q_h^{S}}+\frac{r_{ee}^{\prime\downarrow\uparrow}e^{-iq_e^{S}(x+\chi+\mathcal{D})}}{q_e^{S}}\Bigg]+\frac{\gamma}{4i(u_0^2-v_0^2)}\Bigg[\frac{r_{he}^{\prime\uparrow\uparrow}u_0^2}{q_h^{S}}+\frac{r_{eh}^{\prime\downarrow\downarrow} v_0^2}{q_e^S}\Bigg]\Big(e^{i(q_h^S(\chi+\frac{\mathcal{D}}{2})-q_e^S(x+\frac{\mathcal{D}}{2}))}\nonumber\\&&+e^{i(q_h^S(x+\frac{\mathcal{D}}{2})-q_e^S(\chi+\frac{\mathcal{D}}{2}))}\Big),\\
f_{\downarrow\downarrow}^{E}(x,\chi,\omega)&=&\frac{\gamma}{4i(u_0^2-v_0^2)}\Big(e^{i(q_h^S(x+\frac{\mathcal{D}}{2})-q_e^S(\chi+\frac{\mathcal{D}}{2}))}-e^{i(q_h^S(\chi+\frac{\mathcal{D}}{2})-q_e^S(x+\frac{\mathcal{D}}{2}))}\Big)\Bigg[\frac{r_{he}^{\prime\downarrow\downarrow}u_0^2}{q_h^S}-\frac{r_{eh}^{\prime\uparrow\uparrow}v_0^2}{q_e^S}\Bigg],
\end{eqnarray}
\begin{eqnarray}
f_{\downarrow\downarrow}^{O}(x,\chi,\omega)&=&-\frac{\gamma u_0v_0}{2i(u_0^2-v_0^2)}\Bigg[\frac{r_{hh}^{\prime\downarrow\uparrow}e^{iq_h^{S}(x+\chi+\mathcal{D})}}{q_h^{S}}+\frac{r_{ee}^{\prime\uparrow\downarrow}e^{-iq_e^{S}(x+\chi+\mathcal{D})}}{q_e^{S}}\Bigg]-\frac{\gamma}{4i(u_0^2-v_0^2)}\Bigg[\frac{r_{he}^{\prime\downarrow\downarrow}u_0^2}{q_h^{S}}+\frac{r_{eh}^{\prime\uparrow\uparrow} v_0^2}{q_e^S}\Bigg]\Big(e^{i(q_h^S(\chi+\frac{\mathcal{D}}{2})-q_e^S(x+\frac{\mathcal{D}}{2}))}\nonumber\\&&+e^{i(q_h^S(x+\frac{\mathcal{D}}{2})-q_e^S(\chi+\frac{\mathcal{D}}{2}))}\Big),
\label{OTE-equal-downdown-J}
\end{eqnarray}
where $r_{ee}^{\prime\uparrow\downarrow}$, $r_{ee}^{\prime\downarrow\uparrow}$, $r_{hh}^{\prime\uparrow\downarrow}$, $r_{hh}^{\prime\downarrow\uparrow}$ denote the spin-flip normal reflection amplitudes in left superconductor, while $r_{eh}^{\prime\uparrow\uparrow}$, $r_{eh}^{\prime\downarrow\downarrow}$, $r_{he}^{\prime\uparrow\uparrow}$, and $r_{he}^{\prime\downarrow\downarrow}$ represent the corresponding spin-flip Andreev reflection amplitudes in left superconductor. {At $x=\chi$, the even-$\omega$ EST pairing vanishes, whereas the odd-$\omega$ EST pairing remains finite and is given by
\begin{eqnarray}
\label{OTE-equal-upup-J-L}
f_{\uparrow\uparrow}^{O,L}(x,\omega)&=&\frac{\gamma u_0v_0}{2i(u_0^2-v_0^2)}\Bigg[\frac{r_{hh}^{\prime\uparrow\downarrow}e^{iq_h^{S}(2x+\mathcal{D})}}{q_h^{S}}+\frac{r_{ee}^{\prime\downarrow\uparrow}e^{-iq_e^{S}(2x+\mathcal{D})}}{q_e^{S}}\Bigg]+\frac{\gamma e^{i(q_{h}^{S}-q_{e}^{S})(x+\frac{\mathcal{D}}{2})}}{2i(u_0^2-v_0^2)}\Bigg[\frac{r_{he}^{\prime\uparrow\uparrow}u_0^2}{q_h^{S}}+\frac{r_{eh}^{\prime\downarrow\downarrow} v_0^2}{q_e^S}\Bigg],\\
f_{\downarrow\downarrow}^{O,L}(x,\omega)&=&-\frac{\gamma u_0v_0}{2i(u_0^2-v_0^2)}\Bigg[\frac{r_{hh}^{\prime\downarrow\uparrow}e^{iq_h^{S}(2x+\mathcal{D})}}{q_h^{S}}+\frac{r_{ee}^{\prime\uparrow\downarrow}e^{-iq_e^{S}(2x+\mathcal{D})}}{q_e^{S}}\Bigg]-\frac{\gamma e^{i(q_{h}^{S}-q_{e}^{S})(x+\frac{\mathcal{D}}{2})}}{2i(u_0^2-v_0^2)}\Bigg[\frac{r_{he}^{\prime\downarrow\downarrow}u_0^2}{q_h^{S}}+\frac{r_{eh}^{\prime\uparrow\uparrow} v_0^2}{q_e^S}\Bigg].
\label{OTE-equal-downdown-J-L}
\end{eqnarray}}
For even- and odd-$\omega$ MST superconducting pairing, we obtain
\begin{eqnarray}
f_3^{E}(x,\chi,\omega)&=&\frac{\gamma}{8i(u_0^2-v_0^2)}\Big(e^{i(q_h^S(x+\frac{\mathcal{D}}{2})-q_e^S(\chi+\frac{\mathcal{D}}{2}))}-e^{i(q_h^S(\chi+\frac{\mathcal{D}}{2})-q_e^S(x+\frac{\mathcal{D}}{2}))}\Big)\Bigg[\frac{(r_{eh}^{\prime\uparrow\downarrow}-r_{eh}^{\prime\downarrow\uparrow})v_0^2}{q_{e}^{S}}+\frac{(r_{he}^{\prime\uparrow\downarrow}-r_{he}^{\prime\downarrow\uparrow})u_0^2}{q_h^{S}}\Bigg],\\
f_3^{O}(x,\chi,\omega)&=&\frac{\gamma u_0v_0}{4i(u_0^2-v_0^2)}\Bigg[\frac{(r_{ee}^{\prime\uparrow\uparrow}-r_{ee}^{\prime\downarrow\downarrow})e^{-iq_e^S(x+\chi+\mathcal{D})}}{q_e^S}-\frac{(r_{hh}^{\prime\uparrow\uparrow}-r_{hh}^{\prime\downarrow\downarrow})e^{iq_h^S(x+\chi+\mathcal{D})}}{q_h^S}\Bigg]+\frac{\gamma}{8i(u_0^2-v_0^2)}\Bigg[\frac{(r_{eh}^{\prime\uparrow\downarrow}-r_{eh}^{\prime\downarrow\uparrow})v_0^2}{q_{e}^{S}}\nonumber\\&&+\frac{(r_{he}^{\prime\downarrow\uparrow}-r_{he}^{\prime\uparrow\downarrow})u_0^2}{q_{h}^S}\Bigg]\Big(e^{i(q_h^S(x+\frac{\mathcal{D}}{2})-q_e^S(\chi+\frac{\mathcal{D}}{2}))}+e^{i(q_h^S(\chi+\frac{\mathcal{D}}{2})-q_e^S(x+\frac{\mathcal{D}}{2}))}\Big).
\end{eqnarray}
{At $x=\chi$, the even-$\omega$ MST pairing vanishes, whereas the odd-$\omega$ MST pairing remains finite and is given by
\begin{eqnarray}
f_3^{O,L}(x,\omega)&=&\frac{\gamma u_0v_0}{4i(u_0^2-v_0^2)}\Bigg[\frac{(r_{ee}^{\prime\uparrow\uparrow}-r_{ee}^{\prime\downarrow\downarrow})e^{-iq_e^S(2x+\mathcal{D})}}{q_e^S}-\frac{(r_{hh}^{\prime\uparrow\uparrow}-r_{hh}^{\prime\downarrow\downarrow})e^{iq_h^S(2x+\mathcal{D})}}{q_h^S}\Bigg]+\frac{\gamma e^{i(q_{h}^{S}-q_{e}^{S})(x+\frac{\mathcal{D}}{2})}}{4i(u_0^2-v_0^2)}\Bigg[\frac{(r_{eh}^{\prime\uparrow\downarrow}-r_{eh}^{\prime\downarrow\uparrow})v_0^2}{q_{e}^{S}}\nonumber\\&&+\frac{(r_{he}^{\prime\downarrow\uparrow}-r_{he}^{\prime\uparrow\downarrow})u_0^2}{q_{h}^S}\Bigg].
\end{eqnarray}}
\setcounter{subsection}{0}
{\section*{Appendix D: Spatial Dependence of Non-local odd-frequency spin-triplet pairing}}
In this Appendix, we examine the impact of anomalous Josephson current on the spatial dependence of non-local odd-$\omega$ ST pairing in the left superconducting region for both tunneling and transparent interfaces between ferromagnetic layers.

\subsection{{Tunneling ferromagnetic interfaces}}
We first consider tunneling ferromagnetic interfaces for the setups shown in Figs.~1(a) and 1(b).

\subsubsection{Bilayer S-$F_x$-$F_y$-S Josephson junction}
As discussed in {Sec.~IV.A.1}, the anomalous Josephson current vanishes in the {S-$F_x$-$F_y$-S} JJ due to the preservation of inversion symmetry. In Fig.~24, we plot the absolute values of the non-local ($x\neq\chi$) even- and odd-$\omega$ SS, EST, and MST pairing amplitudes versus position $x$ in the left superconducting region for the {S-$F_x$-$F_y$-S} JJ, for both {transparent} and disordered {S-$F_x$ and $F_y$-S interfaces.} We see that even- and odd-$\omega$ SS, EST and MST pairings are finite, and exhibit an oscillatory decay in the left superconducting region. This is seen for junctions both {transparent} and {disordered ferromagnet-superconductor interfaces.} It is to be noted that disorder enhances the pairing magnitudes.
\begin{figure*}[ht!]
\centering{\includegraphics[width=0.95\textwidth]{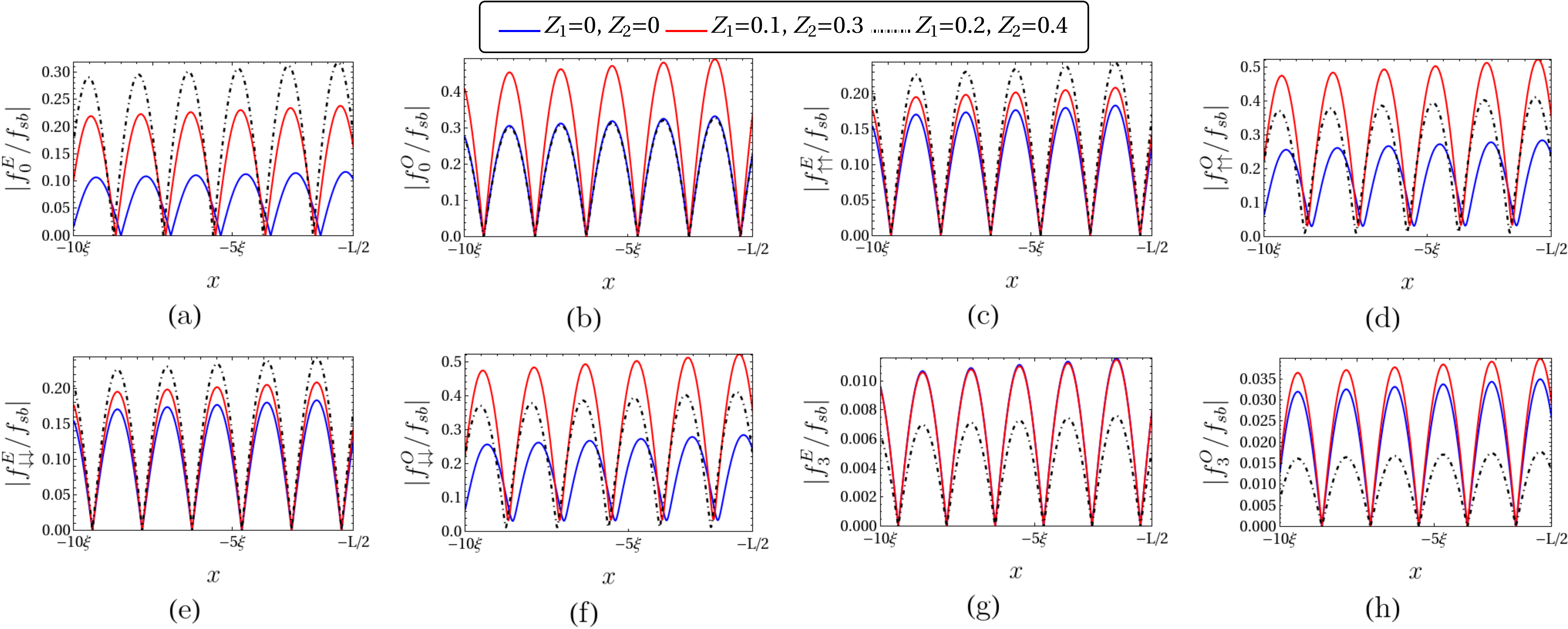}}
\vskip -0.1 in \caption{\small \sl Absolute values of the non-local ($x\neq\chi$) even- and odd-$\omega$ SS (a,b), EST (c,d,e,f), and MST (g,h) pairing amplitudes in
the left superconducting region vs. position $x$ for {S-$F_x$-$F_y$-S} JJ, considering both {transparent} and {disordered S-$F_x$ and $F_y$-S interfaces in the short junction limit.} Parameters: $\varphi=0$, $k_{F}\mathcal{D}=1.5\pi$, $k_{F}\xi=2$, $\chi=0$, $Z=2.5$, $m_1=m_2=0.5E_F$, $E_F=100\Delta_0$, $T/T_c=0.002$, $k_{B}T/\Delta_0=0.001$.}
\end{figure*}

\subsubsection{Trilayer S-F$_x$-F$_y$-F$_z$-S Josephson junction}
To examine the effect of anomalous Josephson current on the spatial dependence of {non-local} odd-$\omega$ ST pairing, we consider S-$F_x$-$F_y$-$F_z$-S JJ, where the anomalous Josephson current exists. In Fig.~25, we plot the absolute values of the non-local ($x\neq\chi$) even- and odd-$\omega$ SS, EST, and MST pairing amplitudes versus position $x$ in the left superconducting region for S-$F_x$-$F_y$-$F_z$-S JJ, considering both {transparent} and
disordered {S-$F_x$ and $F_z$-S interfaces.} We see that even- and odd-$\omega$ SS and ST pairings exhibit an oscillatory decay in the left superconducting region, which is also noticed in Fig.~24 for {S-$F_x$-$F_y$-S} JJ.
This behavior persists in junctions regardless of {disordered ferromagnet-superconductor interfaces,} especially for EST pairing, while MST pairing experiences significant suppression under strong disorder. Thus, {non-local} odd-$\omega$ ST pairing exhibits similar behavior regardless of the presence or absence of anomalous Josephson current {for tunneling ferromagnetic interfaces}, indicating that the two are mutually exclusive effects.
\begin{figure*}[ht!]
\centering{\includegraphics[width=0.95\textwidth]{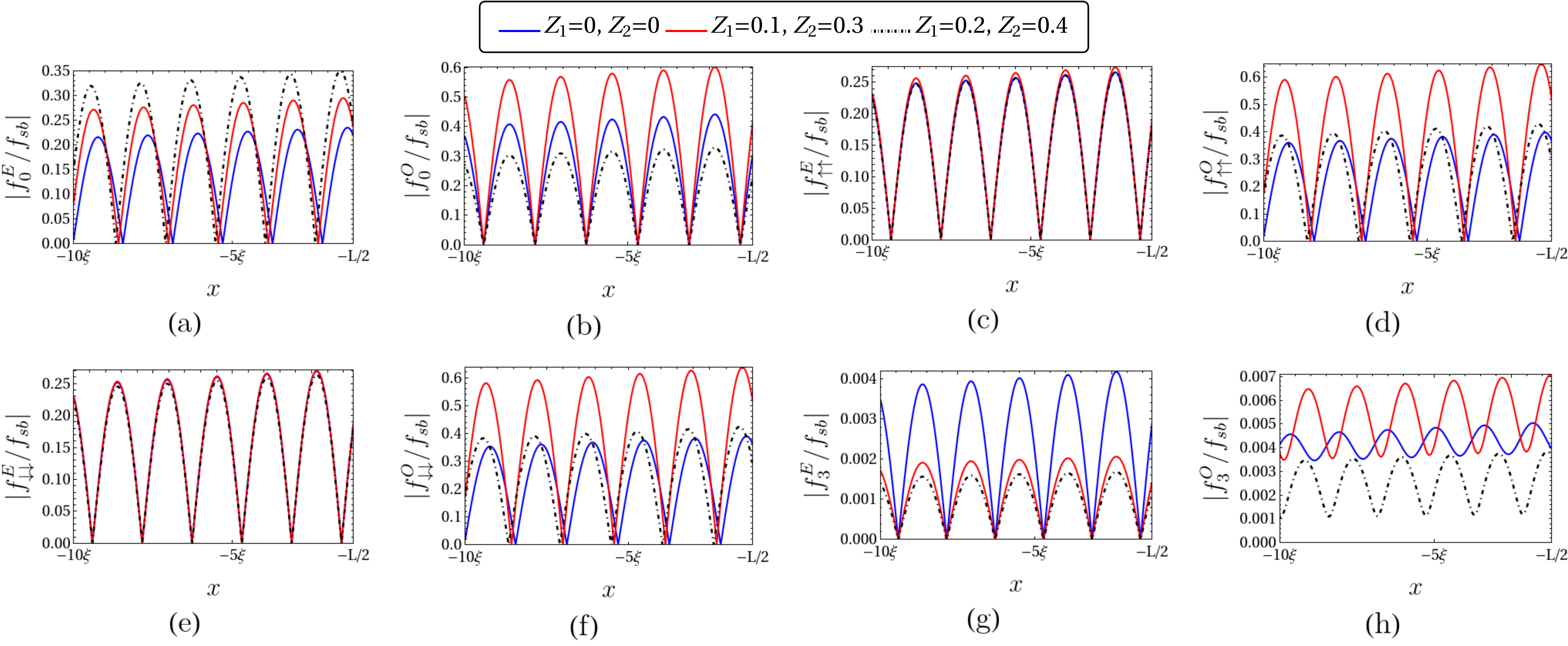}}
\vskip -0.1 in \caption{\small \sl Absolute values of the non-local ($x\neq\chi$) even- and odd-$\omega$ SS (a,b), EST (c,d,e,f), and MST (g,h) pairing amplitudes in the left superconducting region vs. position $x$ for S-$F_x$-$F_y$-$F_z$-S JJ, considering both {transparent} and disordered {S-$F_x$ and $F_z$-S interfaces in the short junction limit.} Parameters: $\varphi=0$, $k_{F}\mathcal{D}=1.5\pi$, $k_{F}\xi=2$, $\chi=0$, $Z_3=Z_4=1.58$, $m_1=m_2=0.5E_F$, $E_F=100\Delta_0$, $T/T_c=0.002$, $k_{B}T/\Delta_0=0.001$.}
\end{figure*}

\subsection{{Transparent ferromagnetic interfaces}}
Here, we consider transparent ferromagnetic interfaces for the two setups shown in Figs.~1(a) and 1(b).

\subsubsection{Bilayer S-$F_x$-$F_y$-S Josephson junction}
In Fig.~26, we plot the absolute values of the non-local ($x\neq\chi$) even- and odd-$\omega$ SS, EST, and MST pairing amplitudes versus position $x$ in the left superconducting
region for {S-$F_x$-$F_y$-S} JJ, considering both {transparent} and disordered {S-$F_x$ and $F_y$-S interfaces.} Even- and odd-$\omega$ SS, EST and MST pairings exhibit an oscillatory decay in the left superconducting region.
\begin{figure*}[ht!]
\centering{\includegraphics[width=0.95\textwidth]{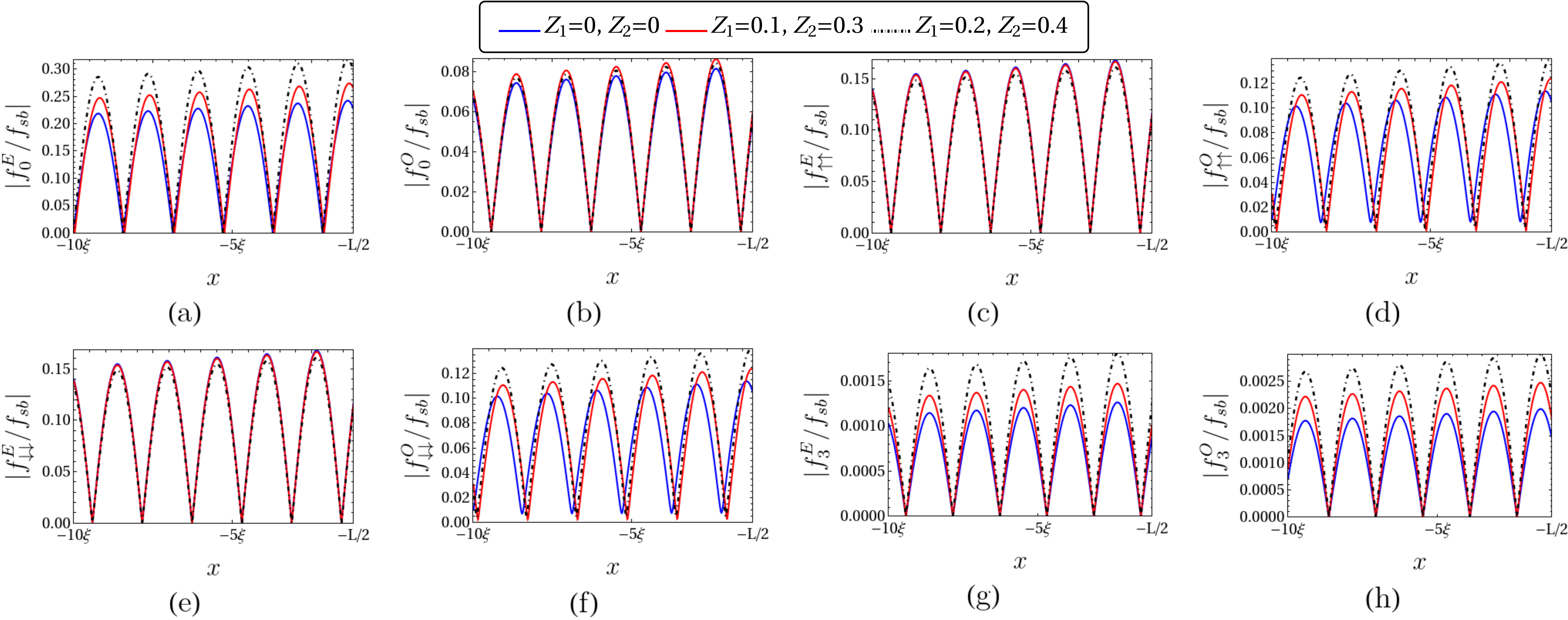}}
\vskip -0.1 in \caption{\small \sl Absolute values of the non-local ($x\neq\chi$) even- and odd-$\omega$ SS (a,b), EST (c,d,e,f), and MST (g,h) pairing amplitudes in the left superconducting region vs. position $x$ for {S-$F_x$-$F_y$-S} JJ, considering both {transparent} and disordered {S-$F_x$ and $F_y$-S interfaces in the short junction limit.} Parameters: $\varphi=0$, $k_{F}\mathcal{D}=1.5\pi$, $k_{F}\xi=2$, $\chi=0$, $Z=0$, $m_1=m_2=0.66E_F$, $E_F=100\Delta_0$, $T/T_c=0.002$, $k_{B}T/\Delta_0=0.001$.}
\end{figure*}

\subsubsection{Trilayer S-$F_x$-$F_y$-$F_z$-S Josephson junction}
To check the impact of anomalous Josephson current on the spatial dependence of {non-local} odd-$\omega$ ST pairing, Fig.~27 shows the non-local ($x\neq\chi$) even- and odd-$\omega$ SS, EST, and MST pairing magnitudes, plotted versus position $x$ in the left superconducting region for the S-$F_x$-$F_y$-$F_z$-S JJ, considering both {transparent} and  disordered {S-$F_x$ and $F_z$-S interfaces.} We notice that even- and odd-$\omega$ SS, EST and MST pairings exhibit an oscillatory decay, consistent with the results shown in Fig.~26 for the {S-$F_x$-$F_y$-S} JJ, with a small increase for odd-$\omega$ SS pairing and even- and odd-$\omega$ EST pairings cases. This behavior persists in junctions regardless
of {disordered ferromagnet-superconductor interfaces. Non-local} odd-$\omega$ ST pairing shows similar behavior irrespective of the presence or absence of anomalous Josephson current {for transparent ferromagnetic interfaces}, indicating that the two phenomena are again mutually exclusive.
\begin{figure*}[ht!]
\centering{\includegraphics[width=0.95\textwidth]{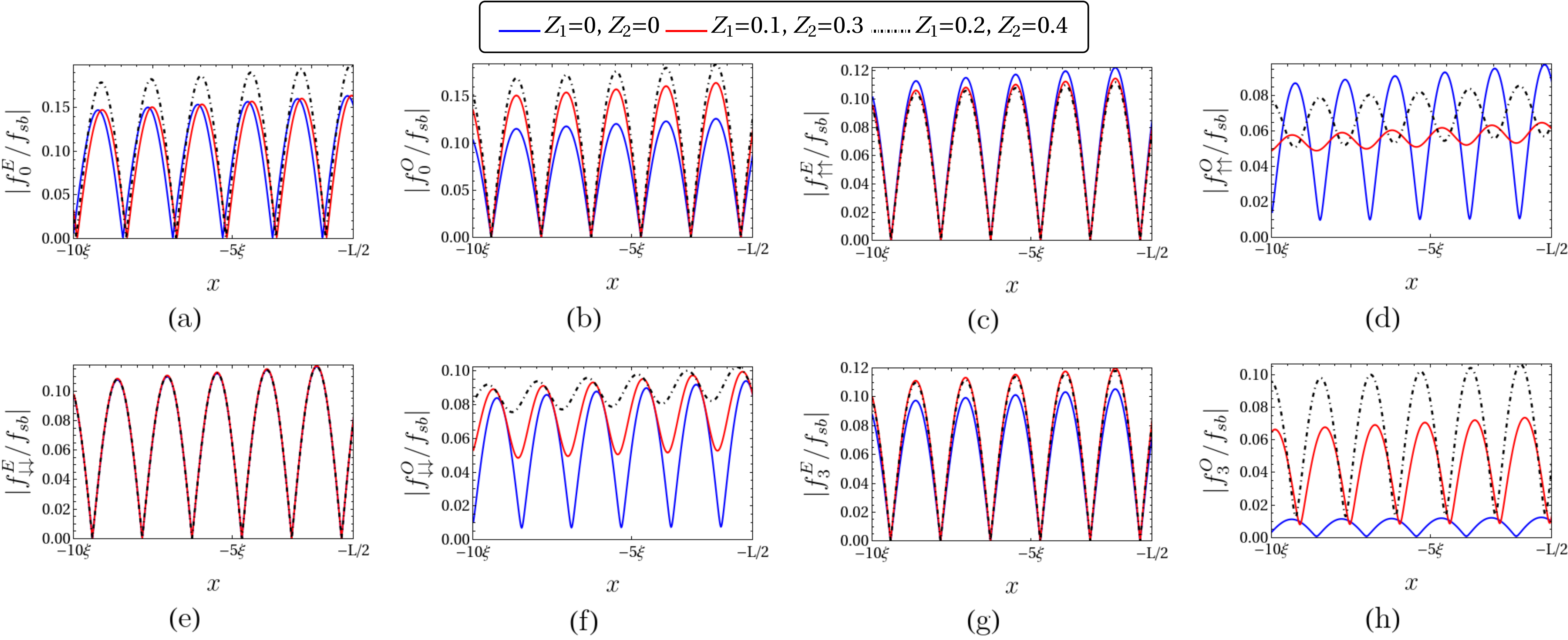}}
\vskip -0.1 in \caption{\small \sl Absolute values of the non-local ($x\neq\chi$) even- and odd-$\omega$ SS (a,b), EST (c,d,e,f), and MST (g,h) pairing amplitudes in the left superconducting region vs. position $x$ for S-$F_x$-$F_y$-$F_z$-S JJ, considering both {transparent} and disordered {S-$F_x$ and $F_z$-S interfaces in the short junction limit.} Parameters: $\varphi=0$, $k_{F}\mathcal{D}=1.5\pi$, $k_{F}\xi=2$, $\chi=0$, $Z_3=Z_4=0$, $m_1=m_2=m_3=0.66E_F$, $E_F=100\Delta_0$, $T/T_c=0.002$, $k_{B}T/\Delta_0=0.001$.}
\end{figure*}
\setcounter{subsection}{0}
{\section*{Appendix E: comparing anomalous Josephson current with pairing amplitudes}
In this Appendix, we compare anomalous Josephson current, local and non-local pairing amplitudes between bilayer S-$F_x$-$F_y$-S, and trilayer S-$F_x$-$F_y$-$F_z$-S JJs in both short and long junction limits.
\subsection{Short junction limit}
In Table II, we compare the anomalous Josephson current, as well as even- and odd-$\omega$ SS, EST, and MST pairing amplitudes, across bilayer S-$F_x$-$F_y$-S and trilayer S-$F_x$-$F_y$-$F_z$-S JJs for both tunneling and transparent interfaces between the ferromagnetic
layers in the short junction limit. We consider both local and non-local pairing amplitudes. At magnetization $m=0$, the anomalous Josephson current vanishes in all setups. For the trilayer S-$F_x$-$F_y$-$F_z$-S JJ, the anomalous Josephson current is finite and exhibits peaks at nonzero magnetization for tunneling ferromagnetic interfaces, as shown in Fig.~4(a), whereas for the transparent ferromagnetic interfaces it shows both zeros and peaks at $m\neq0$, see Fig.~9(a). In contrast to anomalous Josephson current, the local even-$\omega$ SS pairing exhibits a peak at $m=0$ for both tunneling and transparent ferromagnetic interfaces, as shown in Figs.~2(a) and 7(a). For tunneling ferromagnetic interfaces at $m\neq0$, we observe zeros irrespective of the presence of a finite anomalous Josephson current, see Figs.~2(a) and 5(a). The local odd-$\omega$ EST and MST pairings vanish at $m=0$, similar to anomalous Josephson current. Local odd-$\omega$ EST and MST pairings exhibit peaks at $m\neq0$ in all setups, see Figs.~2, 5, 7 and 10. Our results indicate that the anomalous Josephson current has no or at best a marginal effect on local odd-$\omega$ ST pairing for both tunneling and transparent ferromagnetic interfaces.}

{The non-local even-$\omega$ SS pairing remains finite, showing a peak at $m=0$ for both tunneling and transparent ferromagnetic interfaces, as shown in Figs.~3(a) and 8(a), wherein anomalous Josephson current vanishes; while for tunneling ferromagnetic interfaces at $m\neq0$, one sees zeros regardless of finite anomalous Josephson current, see Figs.~3(a) and 6(a). Non-local odd-$\omega$ SS pairing, on the other hand, shows a dip at $m=0$ and exhibits peaks at $m\neq0$ for both tunneling and transparent ferromagnetic interfaces irrespective of whether anomalous Josephson current flows or not, see Figs.~3(b), 6(b), 8(b) and 11(b).
\begin{table}[ht!]
\caption{{Comparing anomalous Josephson current, even- and odd-$\omega$ SS, EST, and MST pairing amplitudes between bilayer S-$F_x$-$F_y$-S, and trilayer S-$F_x$-$F_y$-$F_z$-S JJs in the \textbf{short junction limit.} $F_x$-$F_y$ and $F_y$-$F_z$ interfaces can be either transparent or tunneling; while S-$F_x$ and $F_y$-S or $F_z$-S interfaces can be either transparent or disordered. Bold text indicates differences between anomalous Josephson current and SS/ST pairing, while regular font indicates similarities.}}
{
\begin{tabular}
{|p{1.8cm}|p{1.97cm}|p{6.6cm}|p{6.9cm}|}
\hline
& Ferromagnetic interfaces & \centering{Bilayer S-$F_x$-$F_y$-S JJ} & \hspace{1.9cm}Trilayer S-$F_x$-$F_y$-$F_z$-S JJ\\
\hline
\multirow{2}{2.2cm}{Anomalous Josephson current} & Tunnel & \centering{Absent} & Absent at $m=0$. Exists with peaks at $m\neq0$. No zeros at $m\neq0$, Fig.~4(a)\\\cline{2-4}
& Transparent & \centering{Absent} & Absent at $m=0$. Exists with peaks \& zeros at $m\neq0$, Fig.~9(a)\\\hline
\multicolumn{4}{|c|}{LOCAL PAIRING AMPLITUDES}\\\hline
\multirow{2}{2.2cm} {Even-$\omega$ SS} & Tunnel & \textbf{shows peak at $\bf{m=0}$ and zeros at $\bf{m\neq0}$, Fig.~2(a)} & \bf{shows peak at $\bf{m=0}$ and zeros at $\bf{m\neq0}$, Fig.~5(a)}\\\cline{2-4}
& Transparent & \textbf{shows peak at $\bf{m=0}$ and no peak or zeros at $\bf{m\neq0}$, Fig.~7(a)} & \textbf{shows peak at $\bf{m=0}$ and no peaks or zeros at $\bf{m\neq0}$, Fig.~10(a)}\\\hline
\multirow{2}{2.2cm}{Odd-$\omega$ EST} & Tunnel & \textbf{Absent at $\bf{m=0}$, shows peaks at $\bf{m\neq 0}$, Figs.~2(b,c)} & shows peaks at $m\neq0$, Figs.~5(b,c)\\\cline{2-4}
& Transparent & \textbf{shows peaks at $\bf{m\neq0}$, Figs.~7(b,c)} & \textbf{ shows peaks at $\bf{m\neq0}$, Figs.~10(b,c)}\\\hline
\multirow{2}{2.2cm}{Odd-$\omega$ MST} & Tunnel & \textbf{shows peaks at $\bf{m\neq0}$, Fig.~2(d)} & shows peaks at $m\neq0$, Fig.~5(d)\\\cline{2-4}
& Transparent & \textbf{shows both  peaks \& zeros at $\bf{m\neq0}$, Fig.~7(d)} & \textbf{ shows peaks at $\bf{m\neq}0$, Fig.~10(d)}\\\hline
\multicolumn{4}{|c|}{NON-LOCAL PAIRING AMPLITUDES}\\\hline
\multirow{2}{2.2cm}{Even-$\omega$ SS} & Tunnel & \textbf{shows peak at $\bf{m=0}$ and zeros at $\bf{m\neq0}$, Fig.~3(a)} & \textbf{shows peak at $\bf{m=0}$ and zeros at $\bf{m\neq0}$, Fig.~6(a)}\\\cline{2-4}
& Transparent & \textbf{shows peak at $\bf{m=0}$ and no peaks or zeros at $\bf{m\neq0}$, Fig.~8(a)} & \textbf{shows peak at $\bf{m=0}$ and no peaks or zeros at $\bf{m\neq0}$, Fig.~11(a)}\\\hline
\multirow{2}{2.2cm}{Odd-$\omega$ SS} & Tunnel & \textbf{shows dip at $\bf{m=0}$ and peaks at $\bf{m\neq0}$, Fig.~3(b)} & \textbf{shows dip at $\bf{m=0}$ and peaks at $\bf{m\neq0}$, Fig.~6(b)}\\\cline{2-4}
& Transparent & \textbf{shows dip at $\bf{m=0}$ and peaks at $\bf{m\neq0}$, Fig.~8(b)} & \textbf{shows dip at $\bf{m=0}$ and peaks at $\bf{m\neq0}$, Fig.~11(b)} \\\hline
\multirow{2}{2.2cm}{Even-$\omega$ EST} & Tunnel & \textbf{shows both peaks \& dips at $\bf{m\neq0}$, Figs.~3(c,e)} & shows peaks at $m\neq0$, Figs.~6(c,e)\\\cline{2-4}
& Transparent & \textbf{Absent at $\bf{m=0}$. shows peaks at $\bf{m\neq0}$, Figs.~8(c,e)} & \textbf{shows peaks at $\bf{m\neq0}$, Figs.~11(c,e)}\\\hline
\multirow{2}{2.2cm}{Odd-$\omega$ EST} & Tunnel & \textbf{shows peaks at $\bf{m\neq}0$, Figs.~3(d,f)} & shows peaks at $m\neq0$, Figs.~6(d,f)\\\cline{2-4}
& Transparent & \textbf{shows  peaks at $\bf{m\neq0}$, Figs.~8(d,f)} & \textbf{ shows peaks at $\bf{m\neq0}$, Figs.~11(d,f)}\\\hline
\multirow{2}{2.2cm}{Even-$\omega$ MST} & Tunnel & \textbf{shows peaks at $\bf{m\neq0}$, Fig.~3(g)} & shows peaks at $m\neq0$, Fig.~6(g)\\\cline{2-4}
& Transparent & \textbf{shows peaks \& zeros at $\bf{m\neq0}$, Fig.~8(g)} & \textbf{ peaks at $\bf{m\neq0}$, Fig.~11(g)}\\\hline
\multirow{2}{2.2cm}{Odd-$\omega$ MST} & Tunnel & \textbf{shows peaks at $\bf{m\neq0}$, Fig.~3(h)} & shows peaks at $m\neq0$, Fig.~6(h) \\\cline{2-4}
& Transparent & \textbf{shows peaks \& zeros at $\bf{m\neq0}$, Fig.~8(h)} & \textbf{shows peaks at $\bf{m\neq0}$, Fig.~11(h)} \\\hline
\end{tabular}}
\end{table}
Non-local even- and odd-$\omega$ EST and MST pairings vanish at $m=0$, similar to the anomalous Josephson current. Non-local even-$\omega$ EST pairing exhibits both peaks and dips at $m\neq0$ in the absence of anomalous Josephson current for tunneling ferromagnetic interfaces, as shown in Figs.~3(c,e), while for transparent ferromagnetic interfaces it shows only peaks, see Figs.~8(c,e). For finite anomalous Josephson current, non-local even-$\omega$ EST pairing exhibits peaks at $m\neq0$, as shown in Figs.~6(c,e) and 11(c,e). Non-local odd-$\omega$ EST and non-local even- and odd-$\omega$ MST pairings show peaks for $m\neq0$ in all setups, see Figs.~3, 6, 8 and 11. Our results indicate that anomalous Josephson current has no influence on non-local odd-$\omega$ ST pairing for either tunneling or transparent ferromagnetic interfaces.
\subsection{Long junction limit}
In Table III, we compare the anomalous Josephson current, as well as even- and odd-$\omega$ SS, EST, and MST pairing
\begin{table*}[ht!]
\caption{{Comparing anomalous Josephson current, even- and odd-$\omega$ SS, EST, and MST pairing amplitudes between bilayer S-$F_x$-$F_y$-S, and trilayer S-$F_x$-$F_y$-$F_z$-S JJs in \textbf{long junction limit.} $F_x$-$F_y$ and $F_y$-$F_z$ interfaces can be either transparent or tunneling; while S-$F_x$ and $F_y$-S or $F_z$-S interfaces can be either transparent or disordered. Text in bold indicates differences between anomalous Josephson current and SS/ST pairing, while text in regular font indicates similarity.}}
{
\begin{tabular}
{|p{1.8cm}|p{1.97cm}|p{6.6cm}|p{6.9cm}|}
\hline
& Ferromagnetic interfaces & \centering{Bilayer S-$F_x$-$F_y$-S JJ} & \hspace{1.9cm}Trilayer S-$F_x$-$F_y$-$F_z$-S JJ\\
\hline
\multirow{2}{2.2cm}{Anomalous Josephson current} & Tunnel & \centering{Absent} & shows peaks at $\mathcal{D}_{1}\approx5\pi, 15\pi$ and zeros at $\mathcal{D}_{1}\approx10\pi, 20\pi$, Fig.~15
\\\cline{2-4}
& Transparent & \centering{Absent} & Exhibits rapid oscillations, Fig.~21\\\hline
\multicolumn{4}{|c|}{LOCAL PAIRING AMPLITUDES}\\\hline
\multirow{2}{2.2cm} {Even-$\omega$ SS} & Tunnel & \textbf{shows dip at $\bf{\mathcal{D}_{1}\approx10\pi}$, Fig.~13(a)} & \textbf{shows peak at $\bf{\mathcal{D}_{1}\approx10\pi}$, zeros at $\bf{\mathcal{D}_{1}\approx5\pi, 15\pi}$, for zero disorder, and oscillations for finite disorder, Fig.~16(a)}\\\cline{2-4}
& Transparent & \textbf{shows dip at $\bf{\mathcal{D}_{1}\approx10\pi}$, Fig.~19(a)} & \textbf{shows peak at $\bf{\mathcal{D}_{1}\approx10\pi}$ and zeros at $\bf{\mathcal{D}_{1}\approx5\pi, 15\pi}$, Fig.~22(a)}\\\hline
\multirow{2}{2.2cm}{Odd-$\omega$ EST} & Tunnel & \textbf{shows peaks at $\bf{\mathcal{D}_{1}\approx5\pi, 15\pi}$ and zeros at $\bf{\mathcal{D}_{1}\approx10\pi, 20\pi}$, Figs.~13(b,c)} &
shows peaks at ${\mathcal{D}_{1}\approx5\pi, 15\pi}$ and zeros at ${\mathcal{D}_{1}\approx10\pi, 20\pi}$, Figs.~16(b,c)\\\cline{2-4}
& Transparent & \textbf{shows peaks at $\bf{\mathcal{D}_{1}\approx5\pi, 15\pi}$ and zeros at $\bf{\mathcal{D}_{1}\approx10\pi, 20\pi}$, Figs.~19(b,c)} & \textbf{shows peaks at $\bf{\mathcal{D}_{1}\approx5\pi, 15\pi}$ and zeros at $\bf{\mathcal{D}_{1}\approx10\pi, 20\pi}$, Figs.~22(b,c)}\\\hline
\multirow{2}{2.2cm}{Odd-$\omega$ MST} & Tunnel & \textbf{Finite and does not show any peaks, Fig.~13(d)} & \textbf{shows peak at $\bf{\mathcal{D}_{1}\approx10\pi}$ and zeros at $\bf{\mathcal{D}_{1}\approx5\pi, 15\pi}$, Fig.~16(d)}\\\cline{2-4}
& Transparent & \textbf{Finite and does not show any peaks, Fig.~19(d)} & \textbf{shows peak at $\bf{\mathcal{D}_{1}\approx10\pi}$ and zeros at $\bf{\mathcal{D}_{1}\approx5\pi, 15\pi}$, Fig.~22(d)}\\\hline
\multicolumn{4}{|c|}{NON-LOCAL PAIRING AMPLITUDES}\\\hline
\multirow{2}{2.2cm} {Even-$\omega$ SS} & Tunnel & \textbf{shows peak at $\bf{\mathcal{D}_{1}\approx10\pi}$, zeros at $\bf{\mathcal{D}_{1}\approx4\pi,16\pi}$, for zero disorder, and oscillations for finite disorder, Fig.~14(a)} & \textbf{Exhibits rapid oscillations, Fig.~18(a)}\\\cline{2-4}
& Transparent & \textbf{shows peak at $\bf{\mathcal{D}_{1}\approx10\pi}$, Fig.~20(a)} & Exhibits rapid oscillations, Fig.~23(a)\\\hline
\multirow{2}{2.2cm} {Odd-$\omega$ SS} & Tunnel & \textbf{shows peak at $\bf{\mathcal{D}_{1}\approx10\pi}$ and zeros at $\bf{\mathcal{D}_{1}\approx3\pi,17\pi}$, Fig.~14(b)} & \textbf{shows peak at $\bf{\mathcal{D}_{1}\approx10\pi}$ and zeros at $\bf{\mathcal{D}_{1}\approx5\pi,15\pi}$, Fig.~18(b)}\\\cline{2-4}
& Transparent & \textbf{shows peak at $\bf{\mathcal{D}_{1}\approx10\pi}$ and zeros at $\bf{\mathcal{D}_{1}\approx3\pi,17\pi}$, Fig.~20(b)} & \textbf{shows peak at $\bf{\mathcal{D}_{1}\approx10\pi}$ and zeros at $\bf{\mathcal{D}_{1}\approx5\pi,15\pi}$, Fig.~23(b)}\\\hline
\multirow{2}{2.2cm}{Even-$\omega$ EST} & Tunnel & \textbf{shows peaks at $\bf{\mathcal{D}_{1}\approx5\pi,15\pi}$ and zeros at $\bf{\mathcal{D}_{1}\approx10\pi,20\pi}$, Figs.~14(c,e)} &
shows peaks at $\mathcal{D}_{1}\approx5\pi,15\pi$ and zeros at $\mathcal{D}_{1}\approx10\pi,20\pi$, Figs.~18(c,e)\\\cline{2-4}
& Transparent & \textbf{shows peaks at $\bf{\mathcal{D}_{1}\approx5\pi,15\pi}$ and zeros at $\bf{\mathcal{D}_{1}\approx10\pi,20\pi}$, Figs.~20(c,e)} & \textbf{shows peaks at $\bf{\mathcal{D}_{1}\approx5\pi,15\pi}$ and zeros at $\bf{\mathcal{D}_{1}\approx10\pi,20\pi}$, Figs.~23(c,e)}\\\hline
\multirow{2}{2.2cm}{Odd-$\omega$ EST} & Tunnel & \textbf{shows peaks at $\bf{\mathcal{D}_{1}\approx5\pi,15\pi}$ and zeros at $\bf{\mathcal{D}_{1}\approx10\pi}$, Figs.~14(d,f)} &
shows peaks at $\mathcal{D}_{1}\approx5\pi,15\pi$ and zero at $\mathcal{D}_{1}\approx10\pi$, Figs.~18(d,f)\\\cline{2-4}
& Transparent & \textbf{shows peaks at $\bf{\mathcal{D}_{1}\approx5\pi,15\pi}$, Figs.~20(d,f)} & \textbf{shows peaks at $\bf{\mathcal{D}_{1}\approx5\pi,15\pi}$, Figs.~23(d,f)}\\\hline
\multirow{2}{2.2cm}{Even-$\omega$ MST} & Tunnel & \textbf{Finite and does not show any peaks, Fig.~14(g)} & \textbf{Exhibits rapid oscillations, Fig.~18(g)}\\\cline{2-4}
& Transparent & \textbf{shows peak at $\bf{\mathcal{D}_{1}\approx10\pi}$, Fig.~20(g)} & \textbf{shows peak at $\bf{\mathcal{D}_{1}\approx10\pi}$ and zeros at $\bf{\mathcal{D}_{1}\approx5\pi,15\pi}$, Fig.~23(g)}\\\hline
\multirow{2}{2.2cm}{Odd-$\omega$ MST} & Tunnel & \textbf{Finite and does not show any peaks, Fig.~14(h)} & \textbf{Exhibits rapid oscillations, Fig.~18(h)}\\\cline{2-4}
& Transparent & \textbf{Exhibits rapid oscillations, Fig.~20(h)} & Exhibits rapid oscillations, Fig.~23(h)\\\hline
\end{tabular}}
\end{table*}
amplitudes between bilayer S-$F_x$-$F_y$-S, and trilayer S-$F_x$-$F_y$-$F_z$-S JJs for both tunneling and transparent ferromagnetic interfaces in the long junction limit. In a bilayer S-$F_x$-$F_y$-S JJ, the anomalous Josephson current vanishes. In contrast, for a trilayer S-$F_x$-$F_y$-$F_z$-S JJ, the anomalous Josephson current is finite and exhibits peaks at $\mathcal{D}_{1}\approx5\pi,15\pi$ for tunneling ferromagnetic interfaces, see Fig.~15, while for transparent ferromagnetic interfaces it exhibits a rapid oscillatory behavior, as shown in Fig.~21. Anomalous Josephson current vanishes at $\mathcal{D}_{1}\approx10\pi,20\pi$ for tunneling ferromagnetic interfaces. The local even-$\omega$ SS pairing is finite and displays a dip at $\mathcal{D}_{1}\approx10\pi$ for a bilayer S-$F_x$-$F_y$-S JJ for both tunneling and transparent ferromagnetic interface, see Figs.~13(a) and 19(a). However, for a trilayer S-$F_x$-$F_y$-$F_z$-S JJ, the local even-$\omega$ SS pairing vanishes at $\mathcal{D}_{1}\approx5\pi,15\pi$, while it develops a peak at $\mathcal{D}_{1}\approx10\pi$, in absence of disorder, as shown in Figs.~16(a) and 22(a). Local even-$\omega$ SS pairing exhibits oscillatory behavior due to disordered ferromagnet-superconductor interfaces. Local odd-$\omega$ EST pairing vanishes at $\mathcal{D}_{1}\approx10\pi,20\pi$, while it exhibits peaks at $\mathcal{D}_{1}\approx5\pi,15\pi$, regardless of the presence of anomalous Josephson current for tunneling or transparent ferromagnetic interfaces, see Figs.~13(b,c), 16(b,c), 19(b,c) and 22(b,c). Local odd-$\omega$ MST pairing is finite and does not show any peak in absence of anomalous Josephson current, as shown in Figs.~13(d) and 19(d). In presence of anomalous Josephson current, local odd-$\omega$ MST pairing vanishes at $\mathcal{D}_{1}\approx5\pi,15\pi$, while it shows a peak at $\mathcal{D}_{1}\approx10\pi$, see Figs.~16(d) and 22(d). Thus, we can conclude that anomalous Josephson current has no or at best a marginal influence on local odd-$\omega$ EST pairing.}

{The non-local even-$\omega$ SS pairing vanishes at $\mathcal{D}_{1}\approx4\pi, 16\pi$, and it exhibits a peak at $\mathcal{D}_{1}\approx10\pi$ for a bilayer S-$F_x$-$F_y$-S JJ for tunneling ferromagnetic interfaces, and for transparent ferromagnet-superconductor interfaces, while for disordered ferromagnet-superconductor interfaces, it shows oscillatory behavior, see Fig.~14(a). For transparent ferromagnetic interfaces, the non-local even-$\omega$ SS pairing shows a peak at $\mathcal{D}_{1}\approx10\pi$, as shown in Fig.~20(a). However, for trilayer S-$F_x$-$F_y$-$F_z$-S JJ, non-local even-$\omega$ SS pairing exhibits rapid oscillations for both tunneling and transparent ferromagnetic interfaces, see Figs.~18(a) and 23(a). Non-local odd-$\omega$ SS pairing vanishes at $\mathcal{D}_{1}\approx3\pi,17\pi$, while it exhibits a peak at $\mathcal{D}_{1}\approx10\pi$ for a bilayer S-$F_x$-$F_y$-S JJ for both tunneling and transparent ferromagnetic interface, as shown in Figs.~14(b) and 20(b). For a trilayer S-$F_x$-$F_y$-$F_z$-S JJ, non-local odd-$\omega$ SS pairing vanishes at $\mathcal{D}_{1}\approx5\pi,15\pi$, while it exhibits a peak at $\mathcal{D}_{1}\approx10\pi$ for both tunneling and transparent ferromagnetic interfaces, see Figs.~18(b) and 23(b). Non-local even-$\omega$ EST pairing vanishes at $\mathcal{D}_{1}\approx10\pi,20\pi$, while it exhibits peaks at $\mathcal{D}_{1}\approx5\pi,15\pi$, regardless of the presence of anomalous Josephson current for both tunneling and transparent ferromagnetic interfaces, as shown in Figs.~14(c,e), 18(c,e), 20(c,e) and 23(c,e). Non-local odd-$\omega$ EST pairing vanishes at $\mathcal{D}_{1}\approx10\pi$, while it exhibits peaks at $\mathcal{D}_{1}\approx5\pi,15\pi$ irrespective of the presence of anomalous Josephson current for tunneling interfaces between the ferromagnetic layers, see Figs.~14(d,f) and 18(d,f). However, for transparent ferromagnetic interfaces, non-local odd-$\omega$ EST pairing only exhibits peaks at $\mathcal{D}_{1}\approx5\pi,15\pi$, as shown in Figs.~20(d,f) and 23(d,f). Non-local even-$\omega$ MST pairing does not exhibit peaks for a bilayer S-$F_x$-$F_y$-S JJ for tunneling ferromagnetic interface, see Fig.~14(g), however for transparent ferromagnetic interface, it shows a peak at $\mathcal{D}_{1}\approx10\pi$, as shown in Fig.~20(g). For a trilayer S-$F_x$-$F_y$-$F_z$-S JJ, non-local even-$\omega$ MST pairing exhibits rapid oscillations for tunneling ferromagnetic interfaces, see Fig.~18(g), while for transparent ferromagnetic interfaces, it exhibits a peak at $\mathcal{D}_{1}\approx10\pi$ and vanishes at $\mathcal{D}_{1}\approx5\pi,15\pi$, as shown in Fig.~23(g). Non-local odd-$\omega$ MST pairing does not show any peak for tunneling ferromagnetic interface in the absence of anomalous Josephson current, see Fig.~14(h), while in the presence of anomalous Josephson current it exhibits rapid oscillations, as shown in Fig.~18(h). For transparent ferromagnetic interfaces, non-local odd-$\omega$ MST pairing exhibits rapid oscillations irrespective of the presence of anomalous Josephson current, see Figs.~20(h) and 23(h). These findings indicate that the anomalous Josephson current plays no role in non-local odd-$\omega$ EST pairing. Therefore, the main conclusion of our work is again vindicated in the long junction limit too.}

\end{widetext}

\end{document}